\documentclass[AMA,STIX1COL]{WileyNJD-v2}

\articletype{PREPRINT}%

\usepackage{amsmath,bm}
\usepackage{caption}
\usepackage{subcaption}
\usepackage{hyperref}
\usepackage{url}
\usepackage{soul}
\usepackage{accents}
\usepackage{pdfpages}
\usepackage{ifthen,changepage}

\newcommand{\velbf}[0]{\mathbf{v}}
\newcommand{\vel}[0]{\mathrm{v}}
\newcommand{\disbf}[0]{\mathbf{d}}
\newcommand{\dis}[0]{\mathrm{d}}
\newcommand{\posbf}[0]{\mathbf{x}}
\newcommand{\pos}[0]{\mathrm{x}}
\newcommand{\pre}[0]{\mathrm{p}}
\newcommand{\prebf}[0]{\mathbf{p}}
\newcommand{\npa}[1]{\text{\hl{${#1}$}}}
\newcommand*{\dt}[1]{%
  \accentset{\mbox{\large\bfseries .}}{#1}}
\newcommand*{\ddt}[1]{%
  \accentset{\mbox{\large\bfseries ..}}{#1}}  
\newcommand{\acc}[0]{\dt{\mathrm{v}}} 
\newcommand{\accbf}[0]{\dt{\mathbf{v}}} 
\newcommand{\IntVol}[1]{\int_{\Omega_e} {#1} \,d\Omega }
\newcommand{\Nb}[0]{\mathbf{N}}
\newcommand{\IntSur}[1]{\int_{\Gamma_t} {#1} \,d\Gamma }
\newcommand{\GAa}[0]{\text{GA-I}^{q_{n+\alpha}}_{\color{blue}{p_{n+\alpha}}}}
\newcommand{\GAb}[0]{\text{GA-I}^{q_{n+1}}_{\color{blue}{p_{n+\alpha}}}}
\newcommand{\GAc}[0]{\text{GA-II}_{\color{blue}{p_{n+\alpha}}}}
\newcommand{\GAd}[0]{\text{GA-I}^{q_{n+\alpha}}_{\color{red}{p_{n+1}}}}
\newcommand{\GAe}[0]{\text{GA-I}^{q_{n+1}}_{\color{red}{p_{n+1}}}}
\newcommand{\GAf}[0]{\text{GA-II}_{\color{red}{p_{n+1}}}}

\newcommand{\review}[1]{{\color{black}{#1}}}
\received{26 April 2016}
\revised{6 June 2016}
\accepted{6 June 2016}

\begin{document}

\title{Generalized-$\alpha$ scheme in the PFEM for velocity-pressure and displacement-pressure formulations of the incompressible Navier-Stokes equations}

\author[1]{Eduardo FERNÁNDEZ$^{1,*}$}

\author[1]{Simon FÉVRIER$^1$}

\author[1]{Martin LACROIX$^1$}

\author[1]{Romain BOMAN$^1$}

\author[1]{Jean-Philippe PONTHOT$^1$}

\authormark{FERNÁDEZ \textsc{et al}}

\address[1]{\orgdiv{Aerospace and Mechanical Engineering Department}, \orgname{University of Liège}, \orgaddress{\country{Belgium}}}

\corres{Quartier Polytech 1, B52/3, Allee de la Decouverte 13A, 4000 Liege, Belgium. *\email{efsanchez@uliege.be}}

\abstract[Summary]{ 
Despite the increasing use of the Particle Finite Element Method (PFEM) in fluid flow simulation and the outstanding success of the Generalized-$\alpha$ time integration method, very little discussion has been devoted to their combined performance. This work aims to contribute in this regard by addressing three main aspects. Firstly, it includes a detailed implementation analysis of the Generalized-$\alpha$ method in PFEM. The work recognizes and compares different implementation approaches from the literature, which differ mainly in the terms that are $\alpha$-interpolated (state variables or forces of momentum equation) and the type of treatment for the pressure in the time integration scheme. Secondly, the work compares the performance of the Generalized-$\alpha$ method against the Backward Euler and Newmark schemes for the solution of the incompressible Navier-Stokes equations. Thirdly, the study is enriched by considering not only the classical velocity-pressure formulation but also the displacement-pressure formulation that is gaining interest in the fluid-structure interaction field. The work is carried out using various 2D and 3D benchmark problems such as the fluid sloshing, the solitary wave propagation, the flow around a cylinder, and the collapse of a cylindrical water column.
}


\maketitle

\section{Introduction}\label{sec:Introduction}

% =========== Introduction to PFEM ==============
The Particle Finite Element Method (PFEM) has drawn attention of the simulation community due to the possibility to formulate fluid flow equations in a Lagrangian framework, allowing the use of classical Lagrangian FEM and easing tracking of fluid boundaries, even in case of large and unpredictable boundary motions due to efficient remeshing algorithms \citep{onate2004particle}. The method has been extended to various materials and multi-physics problems with moving domains, such as plasticity \citep{carbonell2020modelling}, fluid-structure interaction \citep{cerquaglia2019fully,meduri2022lagrangian} and phase change \citep{bobach2021phase} among others \citep{sengani2020review,cremonesi2020state}. 

% ======= PFEM & Time Integration schemes =======
In the PFEM, Lagrangian-based governing equations are solved using spatial FEM-Galerkin and temporal discretizations. The first one subdivides the computational domain in a number of finite elements to approximate the solution of partial differential equations governing the body motion. This procedure, applied to the momentum equation of a Newtonian fluid, leads to the following expression:
\begin{equation} \label{EQ:SemiDiscrete_Momentum}
		\bm{f}^{\mathrm{dyn}}(\accbf,\posbf,t) + \bm{f}^{\mathrm{int}}(\velbf,\prebf,\posbf,t) = \bm{f}^{\mathrm{ext}}(\posbf,t)
\end{equation}
\vspace{1mm}

\noindent where $\bm{f}^{\mathrm{dyn}}$, $\bm{f}^{\mathrm{int}}$ and $\bm{f}^{\mathrm{ext}}$ represent, respectively, dynamic (inertial or d'Alembert forces), internal and external forces acting on the nodes defining the finite elements. These forces depend on the nodal acceleration ($\acc$), velocity ($\vel$), position ($\pos$), pressure ($\pre$), and time ($t$). The momentum equation is complemented with the continuity equation to form the set of Navier-Stokes equations. In addition, the transient problem requires a temporal discretization to approximate time variation of state variables. For example, starting from a known configuration of the body at time $t_n$, the updated condition of the fluid at $t_{n+1}$ is governed by the discretized momentum equation:
\begin{equation} \label{EQ:Discrete_Momentum}
		\bm{f}^{\mathrm{dyn}}(\accbf_{n+1},\posbf_{n+1}) + \bm{f}^{\mathrm{int}}(\velbf_{n+1},\prebf_{n+1},\posbf_{n+1}) = \bm{f}^{\mathrm{ext}}(\posbf_{n+1})
\end{equation}

\noindent where the subscript $n$+$1$ indicates that variables are defined at time $t_{n+1}$. To relate state variables with their time derivatives ($\acc_{n+1}$, $\vel_{n+1}$ and $\pos_{n+1}$), a time integration scheme must be used. For example, common schemes in the PFEM literature are the implicit Backward Euler \citep{cerquaglia2019fully, bobach2021phase, idelsohn2004particle, idelsohn2008unified, idelsohn2010challenge, ryzhakov2012improving, zhu2014modeling}, Trapezoidal \citep{onate2014lagrangian,franci2015effect}, Newmark \citep{franci2020lagrangian,franci2021lagrangian} and Newmark-Bossak \citep{ryzhakov2010monolithic,ryzhakov2014two,ryzhakov2017modified,ryzhakov2017fast}. Although explicit schemes can also be found \citep{cremonesi2017explicit}. However, one of the best performing time integration schemes reported in the computational dynamics literature, and which has not been assessed in the PFEM context yet, is the Generalized-$\alpha$ method. This time integration scheme was proposed by Chung and Hulbert \citep{chung1993time} for solving dynamic equations in solid mechanics. Later, Jansen et.al.\citep{jansen2000generalized} extended the idea to fluid mechanics. Since then, several works have demonstrated outstanding performance of the Generalized-$\alpha$ method for solving the incompressible Navier-Stokes equations, mostly using Eulerian finite element discretization \cite{dettmer2003analysis}. The essential feature of the Generalized-$\alpha$ (GA) method is that it writes state equations at times $t_{n+\alpha_f}$ and $t_{n+\alpha_m}$, instead of doing so at time $t_{n+1}$. For this method, the momentum equation is as follows:
\begin{equation}\label{EQ:MomentumEquation_Forces}
\bm{f}^\mathrm{dyn}_{n+\alpha_m} + \bm{f}^\mathrm{int}_{n+\alpha_f} - \bm{f}^\mathrm{ext}_{n+\alpha_f} = 0 
\end{equation}
\vspace{1mm}

For solving Eq.~\eqref{EQ:MomentumEquation_Forces}, two approaches can be identified in the literature, which are denoted as GA-I and GA-II in this work. The first one (GA-I) is the most reported one for fluid dynamics and follows the original contribution of Chung and Hulbert \citep{chung1993time} and Jansen et.al.\citep{jansen2000generalized}. GA-I computes nodal acceleration, velocity, position and pressure at $t_{n+\alpha_m}$ or $t_{n+\alpha_f}$ assuming a linear combination between those at $t_n$ and $t_{n+1}$, as follows: 
\begin{subequations} \label{EQ:StateVariables_AlphaMethod_all}
\begin{align} \label{EQ:StateVariables_AlphaMethod_acceleration}
	\acc_{n+\alpha_m} &= (1-\alpha_m) \: \acc_{n} + \alpha_m \: \acc_{n+1} \\[1ex]  
	\mathrm{x}_{n+\alpha_m} &= (1-\alpha_m) \: {\mathrm{x}}_{n} + \alpha_m \: {\mathrm{x}}_{n+1} \\[1ex] 
	\label{EQ:StateVariables_AlphaMethod_vel}
	\mathrm{v}_{n+\alpha_f} &= (1-\alpha_f) \: {\mathrm{v}}_{n} + \alpha_f \: {\mathrm{v}}_{n+1}
\\[1ex] \label{EQ:StateVariables_AlphaMethod_dis}
	\mathrm{x}_{n+\alpha_f} &= (1-\alpha_f) \: {\mathrm{x}}_{n} + \alpha_f \: {\mathrm{x}}_{n+1}
\\[1ex] \label{EQ:StateVariables_AlphaMethod_pressure}
	\mathrm{p}_{n+\alpha_f} &= (1-\alpha_f) \: {\mathrm{p}}_{n} + \alpha_f \: {\mathrm{p}}_{n+1}	
\end{align}
\end{subequations}

\noindent where $\alpha_m$ and $\alpha_f$ are user-defined parameters. In this way, the momentum Eq.~\eqref{EQ:MomentumEquation_Forces} is written as:
\begin{equation}\label{EQ:MomentumEquation_Forces_GAI}
\bm{f}^{\mathrm{dyn}}(\accbf_{n+\alpha_m},\posbf_{n+\alpha_m}) + \bm{f}^{\mathrm{int}}(\velbf_{n+\alpha_f},\prebf_{n+\alpha_f},\posbf_{n+\alpha_f}) = \bm{f}^{\mathrm{ext}}(\posbf_{n+\alpha_f})
\end{equation}
\vspace{1mm}

Unavoidably, the GA-I approach (Eq.~\ref{EQ:MomentumEquation_Forces_GAI}) results in a system of equations with constitutive matrices that must be computed at times $t_{n+\alpha_m}$ and $t_{n+\alpha_f}$. The second implementation approach (GA-II) follows the rationale of Hilber-Hughes-Taylor (HHT-$\alpha$)\citep{hilber1977improved} and Wood-Bossak-Zienkiewicz (WBZ-$\alpha$)\citep{wood1980alpha}, who paved the way to the Generalized-$\alpha$ method. Their idea is to formulate dynamic, internal and external forces of the momentum equation at $t_{n+\alpha_m}$ and $t_{n+\alpha_f}$ using an interpolation of forces between $t_n$ and $t_{n+1}$, as follows:
\begin{subequations} \label{EQ:GA_II_parametrization}
\begin{align}
\label{EQ:GA_II_parametrization_a}
\bm{f}^\mathrm{dyn}_{n+\alpha_m} &= (1-\alpha_m)\:\bm{f}^\mathrm{dyn}_{n} + \alpha_m \: \bm{f}^\mathrm{dyn}_{n+1}
\\[1ex] 
\bm{f}^\mathrm{int}_{n+\alpha_f} &= (1-\alpha_f)\:\bm{f}^\mathrm{int}_{n} + \alpha_f \: \bm{f}^\mathrm{int}_{n+1}
\\[1ex]
\bm{f}^\mathrm{ext}_{n+\alpha_f} &= (1-\alpha_f)\:\bm{f}^\mathrm{ext}_{n} + \alpha_f \: \bm{f}^\mathrm{ext}_{n+1} 
\label{EQ:GA_II_parametrization_c}
\end{align}
\end{subequations}
\vspace{1mm}

\begin{figure}[t] \captionsetup[sub]{font=normalsize}
	\centering 
	\includegraphics[width=0.85\linewidth]{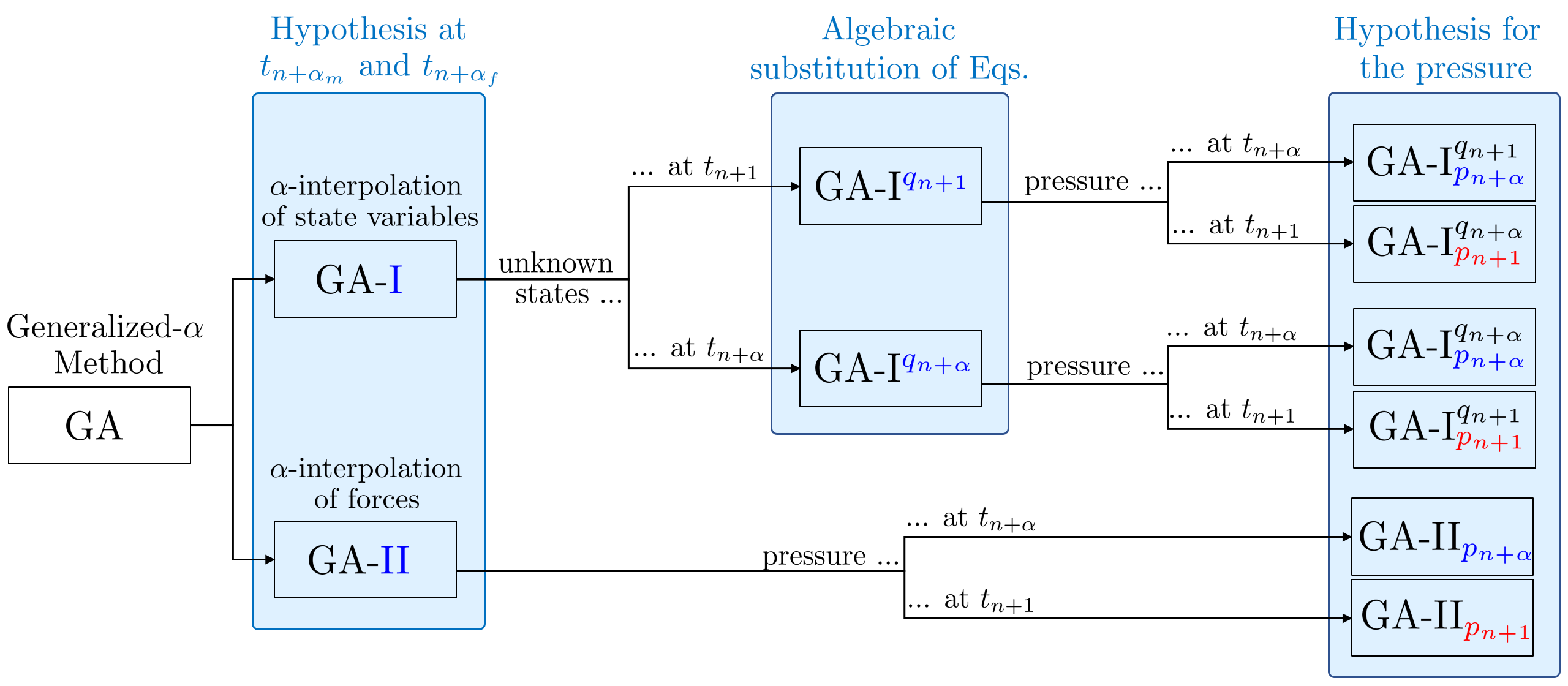}
	\caption{Classification of the different implementation approaches of the Generalized-$\alpha$ method for solving the incompressible Navier-Stokes equations, which results in six different implementations of the algorithm.}
	\label{Fig:GA_Approaches}
\end{figure}

When replacing Eq.~\eqref{EQ:GA_II_parametrization} into the momentum balance Eq.~\eqref{EQ:MomentumEquation_Forces}, GA-II yields a system of equations with state variables and constitutive matrices at $t_{n}$ and $t_{n+1}$. This offers advantages over the implementation approach GA-I. For instance, it avoids computationally expensive algorithms or constitutive matrices at $t_{n+\alpha_f}$ and $t_{n+\alpha_m}$, such as in problems involving contact and/or plasticity \citep{graillet1999numerical,ponthot1999efficient}. Possibly for this reason, the GA-II strategy can be easily found in solid mechanics, but hardly in Newtonian-based fluid mechanics. From now on, various implementation approaches will be introduced, which are diagrammed in Fig.~\ref{Fig:GA_Approaches} for the sake of clarity. 

In addition, it can be identified that Eqs.~\eqref{EQ:StateVariables_AlphaMethod_acceleration}-\eqref{EQ:StateVariables_AlphaMethod_pressure} are used in two ways in the literature. The first one consists of substituting Eq.~\eqref{EQ:StateVariables_AlphaMethod_all} in the momentum equation to obtain a system whose unknowns are the velocity and pressure at $t_{n+1}$ (see, for example, Vald\'{e}s V\'{a}zquez \citep{valdes2007nonlinear}). This implementation approach is denoted GA-I$^{q_{n+1}}$ in this work. The second approach consists of solving state equations directly at $t_{n+\alpha_f}$ to obtain the velocity and pressure at $t_{n+\alpha_f}$. Then, using Eqs.~\eqref{EQ:StateVariables_AlphaMethod_acceleration}-\eqref{EQ:StateVariables_AlphaMethod_pressure} to retrieve kinematic variables and pressure at $t_{n+1}$ (see for example Jansen et.al.\citep{jansen2000generalized}). This implementation approach is denoted GA-I$^{q_{n+\alpha}}$ in this work, as illustrated in the second classification level of Fig.~\ref{Fig:GA_Approaches}.

Another implementation detail that deserves attention is the way in which pressure is treated in the time integration scheme. For example, two different approaches can be identified in the literature for the incompressible Navier-Stokes equations. The first one follows the principle of Jansen et.al.\citep{jansen2000generalized}, in which the pressure is set at $t_{n+\alpha_f}$ in the momentum equation using Eq.~\eqref{EQ:StateVariables_AlphaMethod_pressure}. The second approach uses the pressure at $t_{n+1}$ (or sets $\alpha_f=1$ in Eq.~\ref{EQ:StateVariables_AlphaMethod_pressure} while keeping $\alpha_f \neq 1$ in Eqs.~\ref{EQ:StateVariables_AlphaMethod_vel}-\ref{EQ:StateVariables_AlphaMethod_dis}). These approaches are denoted as GA-I$_{\color{blue}p_{n+\alpha}}$ and GA-I$_{\color{red}{p_{n+1}}}$, respectively. Combining these with the other above-mentioned implementation approaches, 6 different schemes of the alpha-generalized method are eventually derived from the literature for a velocity-pressure formulation:  $\GAa$, $\GAb$,  $\GAd$, $\GAe$, $\GAc$, $\GAf$, which are illustrated in the third classification level of Fig.~\ref{Fig:GA_Approaches}. For instance, $\GAd$ is used in \citep{bazilevs2007variational}, $\GAa$ in \citep{jansen2000generalized,gravemeier2011algebraic,lovric2018new}, and $\GAe$ in \citep{dettmer2003analysis,valdes2007nonlinear,saksono2007adaptive}. The other schemes, to the best of our knowledge, could not be recognized in the fluid mechanics literature but they were derived accordingly to become part of this study.

Despite the popularity of the Generalized-$\alpha$ method in solid and fluid dynamics, the different approaches mentioned above have not been explored in such detail as to recognize in advance which one suits better the PFEM. One of the few works addressing such a variety of approaches is the recent work of Liu et.al.\citep{liu2021note}. The authors use stable B-spline (NURBS) elements for Eulerian spatial discretization of Navier-Stokes equations to analyse the effect of using the pressure at $t_{n+1}$ instead of $t_{n+\alpha}$ (GA-I{$_{\color{red}p_{n+1}}$} versus GA-I{$_{\color{blue}p_{n+\alpha}}$). The authors show that GA-I$_{\color{red}p_{n+1}}$ reduces the performance of the Generalized-$\alpha$ method since only first-order accuracy is achieved, at least for the pressure. However, since the present work uses a spatial Lagrangian PFEM discretization stabilized with the Pressure-Stabilizing Petrov-Galerkin (PSPG) formulation, it is not possible to guarantee a priori that using the pressure at $p_{n+\alpha_f}$ will lead to better performance than using it at $p_{n+1}$. The argument is that, depending on the implementation approach or on the formulation (velocity-based, or displacement-based), it is possible to obtain a pressure term, $p_n$, acting on the right-hand side of the momentum equation. This could result in spurious oscillations of the external force vector due to numerical instabilities inherent to the spatial discretization, and would be a reason to avoid the $\alpha$-interpolation of the pressure.

% === PFEM & displacement-pressure formulation ==
One of the advantages provided by the PFEM is the possibility to identify evolving fluid boundaries easily, which facilitates numerical schemes for fluid-structure interaction, among others. In this context, advantages have been observed in formulating the Navier-Stokes equations with respect to nodal displacements and pressures instead of the classical velocity-pressure approach, since it facilitates unified formulations for fluids and solids. Some of the few authors introducing a displacement-pressure formulation in PFEM for the incompressible Navier-Stokes equations are Ryzhakov et.al.\citep{ryzhakov2017fast}, who use the Newmark-Bossak scheme when producing results but the Backward Euler for description of their methodology. In the Lagrangian-based FEM for fluid dynamics, the works of Radovitzky and Ortiz \citep{radovitzky1998lagrangian} and Avancini and Sanches\citep{avancini2020total} present displacement-pressure \review{and position-pressure formulations, respectively}. Notably, both works use the Newmark's method as time integration scheme. Avancini and Sanches\citep{avancini2020total} present a total Lagrangian formulation without remeshing while Radovitzky and Ortiz\citep{radovitzky1998lagrangian} use an advancing front method as remeshing technique, which constraints fluid topology. Thus, both works experience difficulties in capturing wave breaking and splashing phenomena, which is an essential feature of PFEM. To the best of our knowledge, no authors in Lagrangian-based fluid mechanics, let alone in PFEM, include the Generalized-$\alpha$ method in a displacement-based formulation. 

% ============== contribution ===================
This work aims to contribute to the PFEM performance for fluid flow simulations by introducing the Generalized-$\alpha$ time integration scheme in the formulation. In view of the foregoing discussion, this work is focused on five aspects. Firstly, to incorporate the Generalized-$\alpha$ time integration scheme into PFEM for solving the incompressible Navier-Stokes equations in an updated Lagrangian framework. Secondly, to include all the implementation approaches in order to identify differences or equivalences. Thirdly, to verify whether or not omitting the intermediate pressure ($p_{n+\alpha}$) in the time integration scheme leads to a worse performance of the Generalized-$\alpha$ method. Fourthly, to compare the performance of the Generalized-$\alpha$ method with the popular choices in the PFEM literature, such as the Backward Euler and Newmark schemes. Fifthly and lastly, to carry out the analysis for both velocity-based and displacement-based formulations, given the interest of the community in the latter for fluid-structure interaction simulation. The study is carried out using \review{incompressible Newtonian fluids and} well-reported benchmarks as the flow around a cylinder, the fluid sloshing, the solitary wave propagation, and the collapse of a 3D cylindrical water column. 

Consistent with the CFD literature, results indicate that the Generalized-$\alpha$ method in PFEM outperforms the Backward Euler and Newmark schemes as it does not suffer from excessive numerical damping for large time steps and exhibits less spurious oscillations than the classical Trapezoidal rule. In turn, it is observed that implementation approaches assuming an interpolation of state variables (GA-I) or of equilibrium forces (GA-II) lead to similar results in the PFEM, \review{at least for incompressible Newtonian fluids}. Regarding the treatment of the pressure, a reduction in performance of the Generalized-$\alpha$ scheme was observed when the pressure is imposed at $t_{n+1}$ instead of $t_{n+\alpha_f}$, which is in line with the observations of Liu et.al\citep{liu2021note}. Although two cases must be distinguished. The first one leads to a large degradation of the time integration scheme and results from integrating the pressure gradient in the current configuration (at $t_{n+1}$). In contrast, the second case integrates the pressure gradient in the intermediate configuration (at $t_{n+\alpha_f}$), which leads to a minor degradation of the Generalized-$\alpha$ performance. Finally, findings are likewise valid for the displacement-based formulation.  

% ================ outline ======================
The reminder of this manuscript is organized as follows. Section \ref{sec:2} introduces the discretized Navier-Stokes equations for an incompressible Newtonian fluid. Section \ref{sec:3} presents the studied time integration schemes including all implementation approaches of the Generalized-$\alpha$ method and describes the PFEM implementation of this work. The benchmark problems, results and discussions are provided in section \ref{sec:4}. Finally, section \ref{sec:5} gathers the conclusions of this work.

\section{Space discretization}\label{sec:2}

In a simplified fashion, the Particle Finite Element Method (PFEM) is composed of two phases. Given a spatio-temporal FEM discretization of the fluid at time t$_n$, the first step is to solve the Lagrangian system of equations using FEM to obtain the nodal velocity and pressure at t$_{n+1}$. Then, nodal acceleration and position are updated consistently with the time integration scheme. The second stage in PFEM consists of remeshing the fluid domain in its updated configuration to improve mesh quality when necessary. In the following, the FEM part is presented for an incompressible Newtonian fluid. For a global overview of PFEM on the simulation of other materials, the reader is referred to\citep{cremonesi2020state}.

\subsection{Governing equations}

Incompressible Navier-Stokes equations can be obtained from the momentum and mass conservation equations for a continuum. For a given time $t$ and current configuration $\Omega(t)$, these equations read as follows:
\begin{subequations} \label{EQ:NS_Equations_VP}
\begin{align} 
\rho \: \acc - \mu \Delta \vel	+ \nabla \pre & = \mathrm{f} \;\;\;,\;\; \text{in \;} \Omega(t)  \label{EQ:NS_Equations_VP_a}
\\[1ex] 
\nabla \cdot \vel & = 0  \;\;\;,\;\; \text{in \;} \Omega(t) 
\end{align}
\end{subequations}

\review
{
\noindent where $\rho$ is the density, $\vel = [\vel_x, \vel_y, \vel_z]$ is the velocity vector, $\Delta$ is the Laplacian operator, and $\mathrm{f}$ is the body force vector. The point over the velocity ($\dt{\text{\small{$\square$}}}$) represents the Lagrangian time derivative, thus $\acc$ is the acceleration vector. The pressure and dynamic viscosity that define the Cauchy stress tensor of a Newtonian fluid are denoted, respectively, as $\pre$ and $\mu$ in Eq.~\eqref{EQ:NS_Equations_VP_a}.

The system of Eqs.~\eqref{EQ:NS_Equations_VP} must be completed with Dirichlet and Neumann boundary conditions to obtain a velocity-pressure formulation of the Navier-Stokes equations. This formulation arises due to the constitutive law that defines the Cauchy stress tensor as a function of pressure and velocity. Note that such tensor can also be defined in terms of displacement and pressure, knowing that:
\begin{equation} \label{EQ:displacement_definition} 
\vel = \dt{\dis}  \;\;\;\; \text{and} \;\;\;\; \acc = \ddt{\dis}
\end{equation}
\\
\noindent where $\dis = [\dis_x , \dis_y , \dis_z]$ is the displacement vector defined as the difference between the current ($\pos$) and reference ($\pos_0$) position ($\dis = \pos - \pos_0$). Using Eq.~\eqref{EQ:displacement_definition} in \eqref{EQ:NS_Equations_VP}, the following displacement-based formulation is obtained:
\begin{subequations} \label{EQ:NS_Equations_DP}
\begin{align} 
\rho \: \ddt{\dis} - \mu \Delta \dt{\dis} 	+ \nabla \pre & = \mathrm{f} \;\;\;,\;\; \text{in \;} \Omega(t) 
\\[1ex] 
\nabla \cdot \dt{\dis} & = 0  \;\;\;,\;\; \text{in \;} \Omega(t)
\end{align}
\end{subequations}
\vspace{1mm}
}

Since equations are written in a Lagrangian formalism, no convective effects have to be taken into account. Thus, the same FEM-Galerkin procedure can be applied to Eq.~\eqref{EQ:NS_Equations_VP} or Eq.~\eqref{EQ:NS_Equations_DP} to obtain the semi-discrete Navier-Stokes equations

\subsection{Finite element formulation}

To facilitate the remeshing process required in PFEM, equal order linear interpolation is used for both the pressure and kinematic variables. For this, triangular (2D) and tetrahedral (3D) elements are used in the spatial discretization of the governing equations. Inside finite elements, state variables are interpolated as:
\begin{equation} \label{EQ:StatesParametrization}
\dis = \Nb_v \; \bar{\dis}
\;\;\; , \;\;\;
\vel = \Nb_v \; \bar{\vel}
\;\;\; , \;\;\;
\pre = \Nb_p \; \bar{\pre}
\end{equation}
\noindent where $\Nb_v$ is a matrix and $\Nb_p$ is a vector containing linear shape functions for kinematic variables and pressure, respectively. Bar symbol in Eq.~\eqref{EQ:StatesParametrization} denotes variables at the nodes of finite elements. For a triangular (2D) element with nodes labeled as 1, 2 and 3, the aforementioned entities are defined as: 

\begin{subequations}
\begin{gather}
\begin{align}
\Nb_v = 
\left[
\begin{matrix}
N_1 & N_2 & N_3 & 0 & 0 & 0
\\[1ex]
0 & 0 & 0 & N_1 & N_2 & N_3 
\end{matrix}
\right]
\;\;\;\;\;,\;\;\;\;\;
\Nb_p = 
\left[
\begin{matrix}
N_1 & N_2 & N_3 
\end{matrix}
\right]
\end{align}
\\[2ex]
\bar{\dis}
=
\left[
\begin{matrix}
\dis_{x,1} &\: \dis_{x,2} &\: \dis_{x,3} &\: \dis_{y,1} &\: \dis_{y,2} \:& \dis_{y,3}
\end{matrix}
\right]^\intercal
\;\;\;,\;\;\;
\bar{\vel}
=
\left[
\begin{matrix}
\vel_{x,1} &\: \vel_{x,2} &\: \vel_{x,3} &\: \vel_{y,1} &\: \vel_{y,2} \:& \vel_{y,3}
\end{matrix}
\right]^\intercal
\;\;\;,\;\;\;
\bar{\pre}
=
\left[
\begin{matrix}
\pre_{1} &\: \pre_{2} &\: \pre_{3}
\end{matrix}
\right]^\intercal
\end{gather}
\end{subequations}
\vspace{1mm}

The space-discretized momentum and continuity equations are written as follows:
\begin{subequations} \label{EQ:MomentumEquation_Space_VP_final}
\begin{align}
\mathbf{M} \: \dt{\bar{\velbf}} + \mathbf{K} \: \bar{\velbf} - \mathbf{D}^\intercal 
\: \bar{\prebf} & = \bar{\mathbf{f}}  \label{EQ:MomentumEquation_Space_VP_final_a}
\\[2ex]
\mathbf{C} \: \dt{\bar{\velbf}} + \mathbf{D} \: \bar{\velbf} + \mathbf{L}
\: \bar{\prebf} & = \bar{\mathbf{h}}  \label{EQ:MomentumEquation_Space_VP_final_b}
\end{align}
\end{subequations}  

\vspace{2mm}\noindent and for a displacement-based formulation, the system becomes:
\begin{subequations} \label{EQ:MomentumEquation_Space_DP_final}
\begin{align}
\mathbf{M} \: \ddt{\bar{\disbf}} + \mathbf{K} \: \dt{\bar{\disbf}} - \mathbf{D}^\intercal 
\: \bar{\prebf} & = \bar{\mathbf{f}} \label{EQ:MomentumEquation_Space_DP_final_a}
\\[2ex]
\mathbf{C} \: \ddt{\bar{\disbf}} + \mathbf{D} \: \dt{\bar{\disbf}} + \mathbf{L}
\: \bar{\prebf} & = \bar{\mathbf{h}} \label{EQ:MomentumEquation_Space_DP_final_b}
\end{align}
\end{subequations}

In the momentum equation (Eqs.~\ref{EQ:MomentumEquation_Space_VP_final_a} and \ref{EQ:MomentumEquation_Space_DP_final_a}), $\mathbf{M}$, $\mathbf{K}$ and $\mathbf{D}$ are, respectively, the mass matrix, the matrix containing the viscous terms, and the gradient matrix, all computed on the current configuration $\Omega(t)$.  $\bar{\mathbf{f}}$ is a vector containing the body forces and surface tractions, denoted as $\bar{b}$ and $\bar{\mathrm{t}}$, respectively. At elemental level, these entities are defined as:

\begin{equation}\label{EQ:matrixTerms}
	\mathbf{M}_e = \IntVol{ \rho \, \Nb^\intercal_v \, \Nb_v} 
	\;\;\; , \;\;\;
	\mathbf{K}_e = \IntVol{ \mathbf{B}^\intercal \, \mathbf{m}_v \, \mathbf{B}} 
	\;\;\; , \;\;\;
	\mathbf{D}_e = 
\IntVol{ \Nb_p^\intercal \, \mathbf{m}_p \, \mathbf{B}  }
	\;\;\; , \;\;\;
	\bar{\mathbf{f}}_e= \IntVol{\rho \, \Nb_v^\intercal \, \bar{\bm{b}}  } + \IntSur{\Nb_v^\intercal \bar{\mathbf{t}}}
\end{equation}
\vspace{1mm}

\noindent where $\mathbf{B}$ is a matrix containing the gradient of shape functions for the velocity, $\mathbf{m}_v$ is a matrix with material viscous properties, and $\mathbf{m}_p$ is a vector that sorts matrix B for the pressure gradient. In 2D, these entities are defined as:

\begin{equation}\label{EQ:matrixTerms_2}
\mathbf{B} = 
	\left[	
	\begin{matrix}
	\displaystyle\frac{\partial N_1}{\partial  x} \quad & \displaystyle\frac{\partial N_2}{\partial x} \quad & \displaystyle\frac{\partial N_3}{\partial x} \quad & 0 & 0 & 0\\[2ex]
	0 & 0 & 0 & \displaystyle\frac{\partial N_1}{\partial y} & \displaystyle\frac{\partial N_2}{\partial y} & \displaystyle\frac{\partial N_3}{\partial y} \\[2ex]
	\displaystyle\frac{\partial N_1}{\partial y} & \displaystyle\frac{\partial N_2}{\partial y} & \displaystyle\frac{\partial N_3}{\partial y} & \displaystyle\frac{\partial N_1}{\partial x} \quad & \displaystyle\frac{\partial N_2}{\partial x} \quad & \displaystyle\frac{\partial N_3}{\partial x} \\ 
	\end{matrix}
	\right]
	\;\;\; , \;\;\;\;\;\;
	\mathbf{m}_v =
	\mu \, 
	\left[	
	\begin{matrix}
	\;\displaystyle\frac{4}{3} \quad & \quad -\displaystyle\frac{2}{3} \quad & \quad 0 \\[2ex]
	-\displaystyle\frac{2}{3} \quad & \quad \;\displaystyle\frac{4}{3} \quad & \quad 0\\[2ex]
	\;0 \quad & \quad \;0 \quad & \quad 1\\ 
	\end{matrix}
	\right] 
	\;\;\; , \;\;\;\;\;\;
	\mathbf{m}_p = [1 \quad 1 \quad 0]
\end{equation}
\vspace{1mm}

\review{
The continuity equations (Eqs.~\ref{EQ:MomentumEquation_Space_VP_final_b} and \ref{EQ:MomentumEquation_Space_DP_final_b}) include the Pressure-Stabilizing Petrov-Galerkin (PSPG) stabilization \citep{hughes1986new}. Its purpose is to circumvent the saddle point problem, where pressure acts as a Lagrange multiplier of the incompressibility constraint \citep{donea2003finite}, and to satisfy the LBB condition (Ladyzhenskaya-Babu$\check{\text{s}}$ka-Brezzi) since equal order linear elements are used to discretize velocity (or displacement) and pressure. At elemental level, the stabilization terms, $\mathbf{C}$, $\mathbf{L}$ and $\bar{\mathbf{h}}$ are defined as:
}
\begin{equation}
	\mathbf{C}_e = \IntVol{ \tau_\mathrm{PSPG} \, \nabla \Nb_p^\intercal \, \Nb_v } 
	\;\;\; , \;\;\;
	\mathbf{L}_e = \IntVol{ \tau_\mathrm{PSPG} \, \frac{1}{\rho} \, \nabla \Nb_p^\intercal \, \nabla \Nb_p} 
	\;\;\; , \;\;\;
	\bar{\mathbf{h}}_e = \IntVol{ \tau_\mathrm{PSPG} \, \rho \, \nabla \Nb_p^\intercal \, \bar{\bm{b}}  } 
\end{equation}

\noindent with $\tau_\mathrm{PSPG}$ computed as\citep{tezduyar1991stabilized}:

\begin{equation}\label{EQ:tauPSPG}
	\tau_\mathrm{PSPG} = 
	\displaystyle\frac{1}{
	\left(
	\frac{4}{\Delta t^2} + \frac{4 \vel^\intercal \vel}{\tilde{h}^2} 
	+ 
	\left( \frac{4\mu}{\rho\:\tilde{h}^2}\right)^2
	\right)^\frac{1}{2}
	}
\end{equation}

\noindent where $\Delta t$ is the time step and $\tilde{h}$ is a characteristic size defined as the circumcircle diameter of the element.

For further implementation details concerning the terms in Eqs.~\eqref{EQ:matrixTerms} -  \eqref{EQ:tauPSPG}, the reader is referred to \citep{fevrier2020travail}. To associate the semi-discretized equations with the introductory discussion of this manuscript, it is pertinent to define the forces involved in the momentum equation. The dynamic ($\bm{f}^\mathrm{dyn}$), internal ($\bm{f}^\mathrm{int}$) and external ($\bm{f}^\mathrm{ext}$) forces are defined as: 
\begin{subequations} \label{EQ:Forces_definition}
\begin{align}
\bm{f}^\mathrm{dyn}(t) & = \mathbf{M} \: \dt{\bar{\velbf}} = \mathbf{M} \: \ddt{\bar{\disbf}}
\\[1ex]
\bm{f}^\mathrm{int}(t) & = \mathbf{K} \: \bar{\velbf} - \mathbf{D}^\intercal\: \bar{\prebf} =   \mathbf{K} \: \dt{\bar{\disbf}} - \mathbf{D}^\intercal\: \bar{\prebf}
\\[1ex]
\bm{f}^\mathrm{ext}(t) & = \bar{\mathbf{f}} 
\end{align}
\end{subequations}

Equations \eqref{EQ:MomentumEquation_Space_VP_final} and \eqref{EQ:MomentumEquation_Space_DP_final} are the final formulas of the momentum and continuity equations that are solved in this work. However, to complete the system, equations for the velocity and displacement time derivatives are needed, which are given by the time integration methods presented below.

\section{Time discretization and integration} \label{sec:3}

To complete the system composed of momentum and continuity equations (Eqs.~\ref{EQ:MomentumEquation_Space_VP_final} and \ref{EQ:MomentumEquation_Space_DP_final}), time $t$ is discretized and time derivatives are integrated over the time step [$t_n$ , $t_{n+1}$] as follows: 
\begin{equation} \label{EQ:TimeIntegration}
    \bar{\vel}_{n+1} = \bar{\vel}_{n} + \displaystyle \int_{t_n}^{t_{n+1}} \dt{\bar{\vel}}(t) \:dt 
    \;\;\;\;, \;\;\;\;\;\;
    \bar{\dis}_{n+1} = \bar{\mathrm{x}}_{n+1} - \bar{\mathrm{x}}_n =  \displaystyle \int_{t_n}^{t_{n+1}} \dt{\bar{\dis}}(t) \:d t 
\end{equation}

\noindent where $\bar{\mathrm{x}}_{n+1}$ and $\bar{\mathrm{x}}_n$ are the nodal positions at $t_{n+1}$ and $t_n$, respectively. However, as $\dt{\bar{\vel}}(t)$ and $\dt{\bar{\dis}}(t)$ are unknown functions, they must be approximated leading to different time integration schemes. The approximate acceleration and velocity functions are denoted by $\dt{v}$ and $v$, respectively. These hold for the time step and are integrated as:
\begin{equation} \label{EQ:TimeIntegration_approx}
    \bar{\vel}_{n+1} = \bar{\vel}_{n} + \displaystyle \int_{0}^{\Delta t} \dt{v}(\tau) \:d\tau 
\;\;\;\;, \;\;\;\;\;\;
    \bar{\dis}_{n+1} = \displaystyle \int_{0}^{\Delta t} v(\tau) \:d \tau 
\end{equation}

\noindent where $\Delta t = t_{n+1} - t_n$ is the time step and $\tau$ is time during a time step. In the following, two set of approximation formulas are presented, the Backward Euler and Newmark. For now on, equations deal with nodal variables only, so the bar symbol will be skipped for the sake of simplicity.

\subsection{Backward Euler}

The Backward Euler time integration scheme defines the unknown acceleration function $\dt{v}(\tau)$ as constant and equal to the acceleration of the current time step $\acc_{n+1}$, leading to :
\begin{equation} \label{EQ:TimeIntegration_acc_BE}
  \begin{matrix}
   \mathrm{v}_{n+1} &= \mathrm{v}_{n} + \displaystyle \int_{0}^{\Delta t} \acc_{n+1} \:d\tau = \mathrm{v}_{n} +  \Delta t \: \acc_{n+1}
  \end{matrix}   
\end{equation}

Similarly, Backward Euler can be applied for integrating the velocity and obtain the displacement as $\mathrm{d}_{n+1} = \Delta t \: \vel_{n+1}$. Nevertheless, for consistency with Newmark's method that is presented later, displacement is obtained from integrating the acceleration twice, as follows:
\begin{align} \label{EQ:TimeIntegration_dis_BE}
\vel(\tau) &= \vel_n + \displaystyle \int_{0}^{\tau} \acc_{n+1} dt = \vel_n + \tau \: \acc_{n+1} 
\\[2ex]
\dis_{n+1} &= \displaystyle \int_{0}^{\Delta t} \vel(\tau) \:d \tau = \displaystyle \int_{0}^{\Delta t} \left[ \mathrm{v}_{n} + \tau \: \acc_{n+1} \right] \:d \tau = \mathrm{v}_{n} \: \Delta t +   \frac{\Delta t^2}{2} \: \acc_{n+1} 
\end{align}
\vspace{1mm}

Kinematic relationships provided by the time integration scheme must be used in the discretized momentum and continuity equations, either in the velocity-pressure system (Eq.~\ref{EQ:MomentumEquation_Space_VP_final}) or displacement-pressure system (Eq.~\ref{EQ:MomentumEquation_Space_DP_final}). In this work, both cases are considered. However, the algebraic substitution \review{is not detailed to avoid overextending the manuscript}. The resulting momentum equations are summarized in the first row of tables provided in appendices \ref{TAB:Velocity_Pressure} and \ref{TAB:Displacement_Pressure}.

\subsection{Newmark's Method}\label{sec:Newmark}

This method\citep{newmark1959method} can be thought of as one that approximates the unknown acceleration function $\dt{v}(\tau)$ by a constant acceleration equal to a linear combination between the current and previous acceleration, as follows:
\begin{equation} \label{EQ:Newmark_Velocity_new}
	\dt{v}(\tau) = (1-\gamma) \: \acc_{n} +  \gamma \: \acc_{n+1} 
\end{equation}
\noindent where $\gamma$ is an interpolation parameter. Using Eq.~\eqref{EQ:Newmark_Velocity_new} in \eqref{EQ:TimeIntegration_approx}, the well-known Newmark equation for nodal velocities is obtained:
\begin{equation} \label{EQ:Newmark_Velocity}
		\mathrm{v}_{n+1} = \mathrm{v}_{n} + \displaystyle \int_{0}^{\Delta t} \left[(1-\gamma) \: \acc_{n} +  \gamma \: \acc_{n+1}\right] \:d\tau 
		= \mathrm{v}_{n} + (1-\gamma)\: \Delta t \:\acc_{n} +  \gamma \: \Delta t \: \acc_{n+1} 
\end{equation}
\vspace{1mm}

To compute nodal displacements, the Newmark's method similarly approximates the acceleration function $\dt{v}(\tau)$ by an interpolation between the current and previous acceleration, but now controlled by a parameter $2\beta$, as follows:
\begin{equation} \label{EQ:Newmark_Acceleration_new}
	\dt{v}(\tau) = (1-2\beta) \: \acc_{n} +  2\beta \: \acc_{n+1}
\end{equation}

This function (Eq.~\ref{EQ:Newmark_Acceleration_new}) is integrated twice to obtain the well-known Newmark equation for nodal displacements:
\begin{align} 
\vel(\tau) &= \vel_n + \displaystyle \int_{0}^{\tau} \left[ (1-2\beta) \: \acc_{n} + 2 \beta  \: \acc_{n+1} \right] \: dt 
= \mathrm{v}_{n} + (1-2\beta) \: \tau \:\acc_{n} +  2\beta \: \tau \: \acc_{n+1} 
\\[2ex]
\mathrm{d}_{n+1} &=  \displaystyle \int_{0}^{\Delta t} \left[ \mathrm{v}_{n} + (1-2\beta) \: \tau \:\acc_{n} +  2\beta \: \tau \: \acc_{n+1} \right] \:d \tau 
= \Delta t \: \vel_{n} +  \frac{1-2\beta}{2} \Delta t^2 \:\acc_{n} +  \beta \Delta t^2 \: \acc_{n+1} \label{EQ:Newmark_displacement}
\end{align}
\vspace{1mm}

The expressions given by Newmark's method allow to complete the system composed by the momentum and continuity equation, either in the velocity-based (Eq.~\ref{EQ:MomentumEquation_Space_VP_final}) or in the displacement-based (Eq.~\ref{EQ:MomentumEquation_Space_DP_final}) formulation. To avoid overextending the manuscript, the procedure is shown below only for the displacement-based formulation. The reader is referred to Appendix \ref{TAB:Velocity_Pressure} (second row of table) to obtain the momentum equation for the velocity-based formulation.

Using Eq.~\eqref{EQ:Newmark_Velocity} to isolate the acceleration:
\begin{equation} \label{EQ:Newmark_acc}
		\acc_{n+1} = \frac{\vel_{n+1} - \vel_n}{\gamma \Delta t} - \frac{1-\gamma}{\gamma} \acc_n 
\end{equation}

\noindent and doing so also in Eq.~\eqref{EQ:Newmark_displacement}:
\begin{equation} \label{EQ:Newmark_vel}
		\acc_{n+1} = \frac{\dis_{n+1}}{\beta \Delta t^2} - \frac{\vel_n}{\beta \Delta t} - \frac{1 - 2\beta}{2\beta} \acc_n 
\end{equation}
\vspace{1mm}

\noindent velocity in terms of displacement can be obtained using Eqs.~\eqref{EQ:Newmark_acc} and \eqref{EQ:Newmark_vel}:
\begin{equation} \label{EQ:Newmark_dis}
		\vel_{n+1} = \frac{\gamma}{\beta \Delta t} \dis_{n+1} + \frac{\beta - \gamma}{\beta} \vel_n + \frac{2\beta - \gamma}{2\beta} \Delta t \acc_n  
\end{equation}
\vspace{1mm}

Having expressed nodal acceleration and velocity in terms of displacement, these are replaced in Eq.~\eqref{EQ:MomentumEquation_Space_DP_final} to give:
\begin{subequations} \label{EQ:MomentumEquation_NM_DP}
\begin{align}
\mathbf{M}_{n+1}  
\left(
\frac{\disbf_{n+1}}{\beta \Delta t^2} - \frac{\velbf_n}{\beta \Delta t} - \frac{1 - 2\beta}{2\beta} \accbf_n
\right)
+ \mathbf{K}_{n+1} 
\left(
\frac{\gamma}{\beta \Delta t} \disbf_{n+1} + \frac{\beta - \gamma}{\beta} \velbf_n + \frac{2\beta - \gamma}{2\beta} \Delta t \accbf_n
\right)
- \mathbf{D}^\intercal_{n+1} 
\: {\prebf}_{n+1} & = {\mathbf{f}}_{n+1} \;\;,
\\[2ex]
\mathbf{C}_{n+1} 
\left(
\frac{\disbf_{n+1}}{\beta \Delta t^2} - \frac{\velbf_n}{\beta \Delta t} - \frac{1 - 2\beta}{2\beta} \accbf_n
\right)
+ \mathbf{D}_{n+1} 
\left(
\frac{\gamma}{\beta \Delta t} \disbf_{n+1} + \frac{\beta - \gamma}{\beta} \velbf_n + \frac{2\beta - \gamma}{2\beta} \Delta t \accbf_n
\right)
+ \mathbf{L}_{n+1} 
\: {\prebf}_{n+1} & = {\mathbf{h}}_{n+1} \;\;,
\end{align}
\end{subequations} 
\vspace{1mm}

Arranging terms in a matrix form leads to:
\begin{equation} \label{EQ:Equtions_from_NM}
\left[
\begin{matrix}
	\mathbf{M}_{n+1} \: \displaystyle\frac{1}{\beta \Delta t^2} + \mathbf{K}_{n+1} \: \displaystyle\frac{\gamma}{\beta \Delta t}  & \hspace{5mm} - \mathbf{D}^\intercal_{n+1} 
	\\[3ex]
	\mathbf{C}_{n+1} \: \displaystyle\frac{1}{\beta \Delta t^2} + \mathbf{D}_{n+1} \: \displaystyle\frac{\gamma}{\beta \Delta t} & \hspace{5mm}  \mathbf{L}_{n+1}
\end{matrix}
\right]
\;
\left[
\begin{matrix}
	\mathbf{d}_{n+1}
	\\[4ex]
	\mathbf{p}_{n+1}
\end{matrix}
\right]
=
\left[
\begin{matrix}
	\mathbf{f}_{n+1} + \mathbf{M}_{n+1} 
	\left( 
	\displaystyle\frac{\mathbf{v}_{n}}{\beta\Delta t} +
	\frac{1-2\beta}{2\beta} \accbf_n
	\right)
	+
	\mathbf{K}_{n+1}
	\left(
	\displaystyle\frac{\gamma-\beta}{\gamma} \velbf_n
	+
	\displaystyle\frac{\gamma-2\beta}{2\beta} \Delta t \accbf_n
	\right)
	\\[3ex]
	\mathbf{h}_{n+1} + \mathbf{C}_{n+1} 
	\left( 
	\displaystyle\frac{\mathbf{v}_{n}}{\beta\Delta t} +
	\frac{1-2\beta}{2\beta} \accbf_n
	\right)
	+
	\mathbf{D}_{n+1}
	\left(
	\displaystyle\frac{\gamma-\beta}{\gamma} \velbf_n
	+
	\displaystyle\frac{\gamma-2\beta}{2\beta} \Delta t \accbf_n
	\right)
\end{matrix}
\right]
\end{equation}
\vspace{1mm}

Eq.~\eqref{EQ:Equtions_from_NM} is the final expresion of the discretized Navier-Stokes equations solved in this work using the Newmark's Method in a displacement-pressure formulation. The momentum equation is also given in Appendix \ref{TAB:Displacement_Pressure} (second row of table) to ease comparison with equations obtained with the other time integration schemes.

The $\gamma$ and $\beta$ parameters of Newmark's method allow to control accuracy and numerical damping, two competing criteria that require the user to find a compromise. Notably, $\gamma = 1$ and $\beta = 0.5$ lead to the highly-damped Backward Euler method defined previously, and $\gamma = 0.5$ and $\beta = 0.25$ results in the classical undamped (but second order accurate) Trapezoidal rule. When comparing tables from Appendices \ref{TAB:Velocity_Pressure} and \ref{TAB:Displacement_Pressure}, it can be observed that the velocity-pressure formulation does not contain the $\beta$ parameter, since this is assigned to the displacement term. For this reason, velocity-pressure formulations in Eulerian-based frameworks do not invoke such a parameter, as can be appreciated in the manuscript of Jansen et.al.\citep{jansen2000generalized}. Thus, to interpolate the Backward Euler and Trapezoidal rule through the Newmark's method, we resort to the linear stability analysis developed for solid mechanics \citep{belytschko2014nonlinear,geradin2014mechanical}. This establishes that, for a linear system, unconditional stability is ensured if:
\begin{equation} \label{EQ:Newmark_Parametrization}
		\gamma \geq \frac{1}{2} \;\;\;\; \text{and} \;\;\;\; \beta \geq \frac{1}{4} \left( \gamma + \frac{1}{2}\right)^2 
\end{equation}
\vspace{1mm}

For the numerical examples of this work, $\gamma$ is chosen and $\beta$ is obtained as the smallest value satisfying Eq.~\eqref{EQ:Newmark_Parametrization}. This applies to both the Newmark and the Generalized-$\alpha$ methods.

As discussed in section 1, Backward Euler dominates the PFEM literature presumably for its ease of implementation, and few works can be found using the Newmark's method\cite{franci2016unified}. However, one of the most successful time integration schemes in solid and fluid mechanics, which has not been assessed in the PFEM context to the best of our knowledge, is the Generalized-$\alpha$ presented below.

\subsection{Generalized-$\alpha$ method}

Unlike the Backward Euler and Newmark formulas that complete the system composed of the momentum and continuity equations, the Generalized-$\alpha$ method seeks to increase accuracy of the time integration scheme without compromising numerical damping for high frequencies. This method is based on the Newmark's method and was proposed by Chung and Hulbert\citep{chung1993time} in the context of solid mechanics and then extended to fluid mechanics by Jansen et.al\citep{jansen2000generalized}. The Generalized-$\alpha$ method consists of posing the system of equations for a time between $t_n$ and $t_{n+1}$. The time of the dynamic, internal and external forces are defined differently, as this provides control over accuracy and numerical dissipation of high frequency modes. Commonly, dynamic forces are writen at $t_{n+\alpha_m}$, while internal and external forces at $t_{n+\alpha_f}$. These intermediate times are expressed in terms of weighting parameters $\alpha_m$ and $\alpha_f$, as follows:
\begin{equation}  \label{EQ:timeIntermediate}
\begin{matrix}
t_{n+\alpha_m} = (1-\alpha_m) \: t_n + \alpha_m \: t_{n+1} = t_n + \alpha_m \: \Delta t \\[2ex]
t_{n+\alpha_f} = (1-\alpha_f) \: t_n + \alpha_f \: t_{n+1} = t_n + \alpha_f \: \Delta t
\end{matrix}
\end{equation}

As discussed in the introduction of this manuscript, there are two implementation approaches of Eq.~\eqref{EQ:MomentumEquation_Forces} in the literature, which are denoted as GA-I and GA-II. The first assumes that state variables scale linearly between $t_n$ and $t_{n+1}$, which leads to the Eqs.~\eqref{EQ:StateVariables_AlphaMethod_acceleration}-\eqref{EQ:StateVariables_AlphaMethod_pressure}. The second assumes a linear variation for the equilibrium forces, leading to Eqs.~\eqref{EQ:GA_II_parametrization_a}-\eqref{EQ:GA_II_parametrization_c}. It can be easily demonstrated that under specific conditions, both approaches are equivalent. For example, if mass conservation is assumed: $\mathbf{M}_{n} = \mathbf{M}_{n+\alpha_m} = \mathbf{M}_{n+1}$, then,

\begin{subequations}\label{EQ:EqualDynamicForces}  
\begin{align}
\bm{f}^\mathrm{dyn}_{n+\alpha_m} & = \mathbf{M}_{n+\alpha_m} \accbf_{n+\alpha} 
\\[1ex]
& = \mathbf{M}_{n+\alpha_m} ((1-\alpha_m)\accbf_n + \alpha_m \accbf_{n+1}) \hspace{2.3cm} \text{(GA-I assumption)}
\\[1ex]
& = (1-\alpha_m)\:\mathbf{M}_{n+\alpha_m} \accbf_n + \alpha_m\:\mathbf{M}_{n+\alpha_m} \accbf_{n+1}
\\[1ex]
& = (1-\alpha_m)\:\mathbf{M}_{n} \velbf_n + \alpha_m\:\mathbf{M}_{n+1} \velbf_{n+1} \hspace{2.1cm} \text{(because of mass conservation)}
\\[1ex]
& = (1-\alpha_m)\:\bm{f}^\mathrm{dyn}_{n} + \alpha_m\:\bm{f}^\mathrm{dyn}_{n+1} \hspace{3cm} \text{(GA-II assumption)}
\end{align}
\end{subequations} 
\vspace{1mm}

This shows that scaling the acceleration (GA-I) or the dynamic force (GA-II) at $t_{n+\alpha_m}$ is equivalent under mass conservation.  Regarding the internal force, similar development can be performed to show that there is a similarity between GA-I and GA-II. Taking the viscous forces for analysis: 
\begin{subequations}\label{EQ:GAI_GAII_K}
\begin{align} 
\overbrace{
\mathbf{K}_{n+\alpha_f} \:((1-\alpha_f) \: \velbf_n + \alpha_f \velbf_{n+1})
}^\text{GA-I} 
& & \approx &  &
\overbrace{
 (1-\alpha_f) \: \mathbf{K}_{n} \:  \velbf_{n} +   \alpha_f \: \mathbf{K}_{n+1} \:  \velbf_{n+1}
}^\text{GA-II}
\\[2ex]
 \npa{(1-\alpha_f)} \: \mathbf{K}_{n+\alpha_f} \:  \npa{\velbf_n} +  \npa{\alpha_f} \mathbf{K}_{n+\alpha_f} \:  \npa{\velbf_{n+1}} 
& & \approx &  &
 \npa{(1-\alpha_f)} \: \mathbf{K}_{n} \:  \npa{\velbf_{n}} +   \npa{\alpha_f} \: \mathbf{K}_{n+1} \:  \npa{\velbf_{n+1}}\label{EQ:GAI_GAII_K_b}
\end{align}
\end{subequations}
\vspace{1mm}

\noindent where highlighted terms are those appearing on both sides of Eq.~\eqref{EQ:GAI_GAII_K_b}. Note that GA-I computes matrix $\mathbf{K}$ in $\Omega(t_{n+\alpha_f})$, while in GA-II does it in $\Omega(t_{n})$ and $\Omega(t_{n+1})$. These tend to the same matrix in case of small deformation where $\Omega(t_{n}) \approx \Omega(t_{n+\alpha_f}) \approx \Omega(t_{n+1})$, or when time step tends to zero, since if $t_{n+1} \to t_n \Rightarrow t_{n+\alpha_f} \to t_n$. Thus, the Generalized-$\alpha$ time integration method should produce similar results under GA-I or GA-II schemes in problems featuring small deformations or using small time steps. \review{Also, for linear problems, scaling state variables (GA-I) or forces in the momentum equation (GA-II) is equivalent. As Newtonian fluids assume a linear relationship between shear stress and shear rate, GA-I and GA-II are expected to provide similar results in this work.} Equations using GA-I for solving the incompressible Navier-Stokes equations (Eqs.~\ref{EQ:MomentumEquation_Space_VP_final} and \ref{EQ:MomentumEquation_Space_DP_final}) are presented below.

\subsubsection{GA-I}

To implement GA-I, it suffices to replace into Eq.~\eqref{EQ:MomentumEquation_Forces} the equations developed for the incompressible Newtonian fluid considered in this work. In a velocity-pressure formulation, Eq.~\eqref{EQ:MomentumEquation_Forces} reads as follows:
\begin{equation} \label{EQ:MomentumEquation_AlphaMethod}
\mathbf{M}_{n+\alpha_m}  \accbf_{n+\alpha_m} + \mathbf{K}_{n+\alpha_f} \velbf_{n+\alpha_f} - \mathbf{D}^\intercal_{n+\alpha_f} \: \mathbf{p}_{n+\alpha_f} = \mathbf{f}_{n+\alpha_f}
\end{equation}

State variables $\acc_{n+\alpha_m}$, $\vel_{n+\alpha_f}$ and $\pre_{n+\alpha_f}$ are defined by the Eqs.~\eqref{EQ:StateVariables_AlphaMethod_acceleration}, \eqref{EQ:StateVariables_AlphaMethod_vel} and \eqref{EQ:StateVariables_AlphaMethod_pressure}, which depend on the state variables at time $t_n$ and $t_{n+1}$. The Generalized-$\alpha$ method uses the Newmark formulas (Eqs.~\ref{EQ:Newmark_Velocity} and \ref{EQ:Newmark_displacement}) to relate state variables at $t_{n+1}$ and their time derivatives. As mentioned in the introduction, this set of equations (Eqs.~\eqref{EQ:StateVariables_AlphaMethod_acceleration}-\eqref{EQ:StateVariables_AlphaMethod_pressure}, \eqref{EQ:Newmark_Velocity} and \eqref{EQ:Newmark_displacement}) are used in two ways in the literature, which were denoted by GA-I$^{q_{n+\alpha}}$ and GA-I$^{q_{n+1}}$. The system of equations obtained with these two approaches are presented below. 

\paragraph{\underline{GA-I$^{q_{n+\alpha}}$}}

This approach writes the unknowns of the momentum equation at times $t_{n+\alpha_f}$. In a velocity-pressure formulation, Eq.~\eqref{EQ:MomentumEquation_AlphaMethod} is expressed in terms of $\mathrm{v}_{n+\alpha_f}$ and $p_{n+\alpha_f}$. To do so, Eqs.~\eqref{EQ:Newmark_Velocity}, \eqref{EQ:StateVariables_AlphaMethod_acceleration} and \eqref{EQ:StateVariables_AlphaMethod_vel} can be used to get the relationship between $\acc_{n+\alpha_m}$ and $\vel_{n+\alpha_f}$ while eliminating $\acc_{n+1}$ and $\vel_{n+1}$, which is:
\begin{equation} \label{EQ:AccAlphaM}
\acc_{n + \alpha_m} = \frac{\alpha_m}{\gamma \: \alpha_f \: \Delta t } \left( \mathrm{v}_{n + \alpha_f} -  \mathrm{v}_n \right) + \left( 1 - \frac{\alpha_m}{\gamma} \right) \acc_n
\end{equation}

Replacing \eqref{EQ:AccAlphaM} into \eqref{EQ:MomentumEquation_AlphaMethod} and arranging terms, the following momentum equation is obtained:

\begin{equation} \label{EQ:MomentumEquation_AlphaMethod_I_a}
\left(\mathbf{M}_{n+\alpha_m} \: \frac{\alpha_m}{\gamma \: \alpha_f \: \Delta t } + \mathbf{K}_{n+\alpha_f} \right) \mathbf{v}_{n+\alpha_f} - \mathbf{D}^\intercal_{n+\alpha_f} \mathbf{p}_{n+\alpha_f} = \mathbf{f}_{n+\alpha_f} + \mathbf{M}_{n+\alpha_m}  \left( \frac{\alpha_m}{\gamma \: \alpha_f \: \Delta t } \: \mathrm{v}_n  - \frac{\gamma - \alpha_m}{\gamma} \: \dot{\mathrm{v}}_n \right) 
\end{equation}
\vspace{1mm}

Applying PSGP stabilization, the system of equations to be solved becomes:
\begin{equation} \label{EQ:MomentumEquation_Space_Time_Alpha_1_a}
\begin{matrix}
\left[
\begin{matrix}
	\mathbf{M}_{{n+\alpha_m}} \: \displaystyle\frac{{\alpha_m}}{{\alpha_f}\gamma\Delta t} + \mathbf{K}_{{n+\alpha_f}}  & \hspace{4mm} - \mathbf{D}^\intercal_{{n+\alpha_f}}	\\[4ex]
	\mathbf{C}_{{n+\alpha_m}} \: \displaystyle\frac{{\alpha_m}}{{\alpha_f}\gamma\Delta t} + \mathbf{D}_{{n+\alpha_f}}  & \hspace{4mm}  \mathbf{L}_{{n+\alpha_f}}
\end{matrix}
\right]
\;
\left[
\begin{matrix}
	\mathbf{v}_{{n+\alpha_f}}
	\\[5ex]
	\mathbf{p}_{{n+\alpha_f}}
\end{matrix}
\right] 
= 
\left[
\begin{matrix}
	\mathbf{f}_{{n+\alpha_f}} + \mathbf{M}_{{n+\alpha_m}}
	  \left( \displaystyle\frac{{\alpha_m}}{\gamma \: {\alpha_f} \: \Delta t} \mathbf{v}_\mathrm{n} + \displaystyle\frac{ {\alpha_m} - \gamma }{\gamma} \dot{\mathbf{v}}_\mathrm{n}
	  \right)
	\\[3ex]
	\mathbf{h}_{{n+\alpha_f}} +  \mathbf{C}_{{n+\alpha_m}} 
	  \left( \displaystyle\frac{{\alpha_m}}{\gamma \: {\alpha_f} \: \Delta t} \mathbf{v}_\mathrm{n} + \displaystyle\frac{ {\alpha_m} - \gamma }{\gamma} \dot{\mathbf{v}}_\mathrm{n}
	  \right)
\end{matrix}
\right]
\end{matrix}
\end{equation}
\vspace{1mm}

After solving Eq.~\eqref{EQ:MomentumEquation_Space_Time_Alpha_1_a}, state variables at $t_{n+1}$ are computed with equations given by the GA-I assumption (Eq.~\ref{EQ:StateVariables_AlphaMethod_all}) and Newmark's formulas. Expressions for updating state variables once the system of equations is solved are listed in Appendix \ref{TAB:Update_States} (third row of table). 

For a displacement-based formulation, acceleration $\ddt{\dis}_{n+\alpha_m}$ and velocity $\dt{\dis}_{n+\alpha_f}$ must be expressed in terms of displacements ${\dis}_{n+\alpha_f}$. However, Newmark's formulas relating state variables and their time derivatives are valid for $t_{n+1}$, and not for $t_{n+\alpha}$. As a consequence, algebraic substitution from the available formulas leads to a system whose unknowns are displacement at $t_{n+1}$ and pressure at $t_{n+\alpha_f}$, as shown in Appendix \ref{TAB:Displacement_Pressure} (third row of table). To arrive at such a system of equations, the acceleration $\acc_{n+\alpha_m}$ (equal to $\ddt{\dis}_{n+\alpha_m}$) in terms of displacements $\dis_{n+1}$ is obtained using Eqs.~\eqref{EQ:Newmark_vel} and \eqref{EQ:StateVariables_AlphaMethod_acceleration}, which leads to:
\begin{equation} \label{EQ:displacements_for_alpha}
\acc_{n+\alpha_m} = \left(1 - \frac{\alpha_m}{2\beta} \right) \acc_n 
    - \frac{\alpha_m}{\beta \Delta t} \mathrm{v}_n  + \frac{\alpha_m}{\beta\Delta t^2} \mathrm{d}_{n+1}
\end{equation}
\vspace{1mm}

Then, velocities $\vel_{n+\alpha_f}$ (equal to $\dt{\dis}_{n+\alpha_f}$) in terms of displacements $\dis_{n+1}$ are obtained using Eqs.~\eqref{EQ:Newmark_dis} and \eqref{EQ:StateVariables_AlphaMethod_vel}, which leads to:
\begin{equation} \label{EQ:displacements_for_alpha_b}
\mathrm{v}_{n+\alpha_f} = \left(1 - \frac{\alpha_f \gamma}{\beta} \right) \mathrm{v}_n 
    + \alpha_f \Delta t \left( 1 - \frac{\gamma}{2\beta} \right)  \dot{\mathrm{v}}_n + \frac{\alpha_f \gamma}{\beta \Delta t} \:\mathrm{d}_{n+1}
\end{equation}
\vspace{1mm}

Replacing Eqs.~\eqref{EQ:displacements_for_alpha} and Eqs.~\eqref{EQ:displacements_for_alpha_b} in the momentum equation \eqref{EQ:MomentumEquation_AlphaMethod} and adding the PSPG stabilization, the system of equations becomes:
\begin{equation} \label{EQ:MomentumEquation_Space_Time_Alpha_1_position_a}
\begin{matrix}
\hspace{0mm}
\left[
\begin{matrix}
	\mathbf{M}_{{n+\alpha_m}} \: \frac{{\alpha_m}}{\beta\Delta t^2} + \mathbf{K}_{{n+\alpha_f}}\frac{{\alpha_f}\gamma}{\beta \Delta t}  & \:\:\:\:\:\: - \mathbf{D}^\intercal_{{n+\alpha_f}}
	\\[4ex]
	\mathbf{C}_{{n+\alpha_m}} \: \frac{{\alpha_m}}{\beta\Delta t^2} + \mathbf{D}_{{n+\alpha_f}} \frac{{\alpha_f}\gamma}{\beta \Delta t} & \:\:\:\:\:\: \mathbf{L}_{{n+\alpha_f}}
\end{matrix}
\right]
\;
\left[
\begin{matrix}
	\mathbf{d}_{n+1}
	\\[5ex]
	\mathbf{p}_{{n+\alpha_f}}
\end{matrix}
\right]
=
\hspace{0mm}
\left[
\begin{matrix}
	\mathbf{f}_{{n+\alpha_f}} + \mathbf{M}_{{n+\alpha_m}} \left( \frac{{\alpha_m}}{\beta \Delta t} \mathbf{v}_{n} + \frac{ {\alpha_m} -2\beta}{2\beta}  \dot{\mathbf{v}}_{n} \right) 
	-
	 \mathbf{K}_{{n+\alpha_f}} \left(  \frac{\beta - {\alpha_f}\gamma}{\beta}  \mathbf{v}_{n} + \text{\footnotesize{${\alpha_f}\Delta t$}} \frac{2\beta-\gamma}{2\beta}  \dot{\mathbf{v}}_{n} \right) 
	\\[3ex]
	\mathbf{h}_{{n+\alpha_f}} + \mathbf{C}_{{n+\alpha_m}} \left( \frac{{\alpha_m}}{\beta \Delta t} \mathbf{v}_{n} + \frac{ {\alpha_m} -2\beta}{2\beta}  \dot{\mathbf{v}}_{n} \right) 
	-
	\mathbf{D}_{{n+\alpha_f}} \left(  \frac{\beta - {\alpha_f}\gamma}{\beta}  \mathbf{v}_{n} + \text{\footnotesize{${\alpha_f}\Delta t$}} \frac{2\beta-\gamma}{2\beta}  \dot{\mathbf{v}}_{n} \right) 
\end{matrix}
\right]
\end{matrix}
\end{equation}
\vspace{1mm}

\paragraph{\underline{GA-I$^{q_{n+1}}$}}

Instead of solving the system at $t_{n+\alpha_f}$, the GA-I parameterization can be used to replace terms in the momentum equation to avoid states at $t_{n+\alpha_f}$, as follows: 
\begin{equation} \label{EQ:MomentumEquation_AlphaMethod_2}
\mathbf{M}_{n+\alpha_m}  
\overbrace{((1-\alpha_m) \: \accbf_{n} + \alpha_m \: \accbf_{n+1})}^\text{\normalsize{$\accbf_{n+\alpha_m}$}} 
\:+\: 
\mathbf{K}_{n+\alpha_f} 
\overbrace{((1-\alpha_f) \: \velbf_{n} + \alpha_f \: \velbf_{n+1})}^\text{\normalsize{$\velbf_{n+\alpha_f}$}} 
\:-\:
\mathbf{D}^\intercal_{n+\alpha_f} \: 
\overbrace{((1-\alpha_f) \: \prebf_{n} + \alpha_f \: \prebf_{n+1}	)}^\text{\normalsize{$\prebf_{n+\alpha_f}$}}  
\:= \:
\mathbf{f}_{n+\alpha_f}
\end{equation}
\vspace{1mm}

Arranging states at $t_n$ in the right-hand side:
\begin{equation} \label{EQ:MomentumEquation_AlphaMethod_2}
\mathbf{M}_{n+\alpha_m} \: \alpha_m \: \accbf_{n+1} + \mathbf{K}_{n+\alpha_f} \: \alpha_f \: \velbf_{n+1} - \mathbf{D}^\intercal_{n+\alpha_f} \: \alpha_f \: \prebf_{n+1} = \mathbf{f}_{n+\alpha_f} - \mathbf{M}_{n+\alpha_m}  (1-\alpha_m) \: \accbf_{n} - \mathbf{K}_{n+\alpha_f} (1-\alpha_f) \: \velbf_{n} + \mathbf{D}^\intercal_{n+\alpha_f} \: (1-\alpha_f) \: \prebf_{n}
\end{equation}
\vspace{1mm}

Resorting to Newmark's formulas:
\begin{equation} \label{EQ:MomentumEquation_AlphaMethod_I_replaced}
\begin{matrix}
\hspace{5mm}
\mathbf{M}_{n+\alpha_m} \alpha_m 
\left(
\displaystyle\frac{\velbf_{n+1} - \velbf_n}{\gamma \Delta t} - \displaystyle\frac{1-\gamma}{\gamma}\: \accbf_n
\right)
 + \mathbf{K}_{n+\alpha_f} \alpha_f \mathbf{v}_{n+1} - \mathbf{D}^\intercal_{n+\alpha_f} \alpha_f\: \mathbf{p}_{n+1}
\\[2ex]
&\hspace{-30mm} = \mathbf{f}_{n+\alpha_f} - \mathbf{M}_{n+\alpha_m} (1-\alpha_m) \dot{\mathbf{v}}_{n} - \mathbf{K}_{n+\alpha_f} (1-\alpha_f) {\mathbf{v}}_{n} + \mathbf{D}^\intercal_{n+\alpha_f} (1-\alpha_f) \mathbf{p}_{n}
\end{matrix}
\end{equation}
\vspace{1mm}

Repeating the procedure for the continuity equation and rearranging terms, the following velocity-based system of equations is obtained:
\begin{equation} \label{EQ:MomentumEquation_Space_Time_Alpha_1}
\begin{matrix}
\left[
\begin{matrix}
	\mathbf{M}_{{n+\alpha_m}} \: \displaystyle\frac{{\alpha_m}}{\gamma\Delta t} + \mathbf{K}_{{n+\alpha_f}} {{\alpha_f}}  & \:\:\:\:\:\:-\mathbf{D}^\intercal_{{n+\alpha_f}} {\alpha_f}
	\\[4ex]
	\mathbf{C}_{{n+\alpha_m}} \: \displaystyle\frac{\alpha_m}{\gamma\Delta t} + \mathbf{D}_{{n+\alpha_f}} {\alpha_f} & \:\:\:\:\:\: \mathbf{L}_{{n+\alpha_f}} {\alpha_f}
\end{matrix}
\right]
\;
\left[
\begin{matrix}
	\mathbf{v}_{n+1}
	\\[5ex]
	\mathbf{p}_{n+1}
\end{matrix}
\right] 
= 
\left[
\begin{matrix}
	\mathbf{f}_{{n+\alpha_f}} + \mathbf{M}_{{n+\alpha_m}} 
	  \left( \displaystyle\frac{\alpha_m}{\gamma \Delta t} \mathbf{v}_\mathrm{n} + \displaystyle\frac{{\alpha_m} - \gamma}{\gamma} \accbf_\mathrm{n}
	  \right) 
	  {
	  -
	  (1-\alpha_f) 
      \left(
	  \mathbf{K}_{n+\alpha_f}\mathbf{v}_\mathrm{n} 
	  - \mathbf{D}^\intercal_{n+\alpha_f} \mathbf{p}_\mathrm{n}
	  \right)  
	  }
	\\[3ex]
	\mathbf{h}_{{n+\alpha_f}} +  \mathbf{C}_{{n+\alpha_m}} 
	  \left( \displaystyle\frac{\alpha_m}{\gamma \Delta t} \mathbf{v}_\mathrm{n} + \displaystyle\frac{{\alpha_m} - \gamma}{\gamma} \accbf_\mathrm{n}
	  \right)
	  {
	  - (1-\alpha_f) \left(\mathbf{D}_{n+\alpha_f} \mathbf{v}_\mathrm{n} 
	  + \mathbf{L}_{n+\alpha_f} \mathbf{p}_\mathrm{n} \right)
	  }
\end{matrix}
\right]
\end{matrix}
\end{equation}

The procedure for the displacement-based equation is omitted, although the reader can refer to Appendix \ref{TAB:Displacement_Pressure} (fourth row of table) to obtain the momentum equation in the implementation approach GA-I$^{q_{n+1}}$. Note that in both approaches, GA-I$^{q_{n+\alpha}}$ and GA-I$^{q_{n+1}}$, and in both formulations (velocity-based and displacement-based), matrices are computed at $t_{n+\alpha}$, but GA-I$^{q_{n+1}}$ defines the unknown state variables at $t_{n+1}$. However, the change of variables in GA-I$^{q_{n+1}}$ introduces other terms on the right-hand side, in particular, the pressure term $\mathrm{p}_n$, which is not present in the GA-I$^{q_{n+\alpha}}$, Newmark nor Backward Euler schemes. 

\paragraph{\underline{GA-I$_{p_{n+\alpha}}$ and GA-I$_{p_{n+1}}$}}\vspace{2mm}

As explained in the introduction of this manuscript, another implementation approach commonly seen in the literature is to avoid the pressure at $t_{n+\alpha_f}$ by writing it at $t_{n+1}$. Although placing the pressure at $t_{n+1}$ is not consistent with the principle of the Generalized-$\alpha$ scheme, it is recognized as the most popular choice in the literature\citep{liu2021note}. An argument is that, as the pressure acts as a Lagrange multiplier, it should not be subjected to time integration\citep{saksono2007adaptive}. However, in the way in which time integration schemes were presented in this section, such argument can be questionable. It should be recalled that Backward Euler and Newmark have been introduced as methods to approximate time derivatives of state variables appearing in the momentum equation. However, the incompressible Navier-Stokes equations do not invoke a time derivative of pressure, unless other formulations are used such as the quasi-incompressible Newtonian fluid\citep{onate2014lagrangian}. In other words, per se, there is no need to apply time integration to the pressure. On the other hand, the Generalized-$\alpha$ has been introduced as a method that aims at improving Newmark's accuracy by posing state equations at intermediate times $t_{n+\alpha_m}$ and $t_{n+\alpha_f}$, but it does not present a new approximation formula to the time derivatives of state variables. So, rather than a time integration method, the Generalized-$\alpha$ method can be seen as a numerical scheme that upgrades Newmark's accuracy from first to second order. 

The approach that places pressure at $t_{n+1}$ is likewise included in this work for two reasons. First, for the sake of completeness since it represents a popular choice in the literature. Second, depending on the implementation approach, a pressure term $\pre_n$ is obtained on the right-hand side of the system of equations (compare GA-I$^{q_{n+\alpha}}$ and GA-I$^{q_{n+1}}$). This could result in spurious oscillations of the external force vector due to numerical instabilities inherent to the spatial discretization. If so, taking the pressure out of Generalized-$\alpha$ scheme could prove beneficial in FEM formulations requiring stabilization. 

The implementation approaches using pressure at $t_{n+1}$ and $t_{n+\alpha}$ are denoted by the subscripts {\color{red}$p_{n+1}$} and {\color{blue}$p_{n+\alpha}$}, respectively. In combination with the other two approaches, four implementations of the Generalized-$\alpha$ scheme in a velocity-pressure formulation are obtained: $\GAa$, $\GAb$, $\GAd$ and $\GAe$. For a displacement-pressure formulation, the combination reduces to three approaches: $\GAa$, $\GAb$, and $\GAe$, since $\GAd$ leads to the same system of equations than $\GAe$. Development of equations for the cases imposing pressure at $t_{n+1}$ is omitted. Instead, the reader is provided with the momentum equations in Appendix \ref{TAB:Velocity_Pressure} (fifth and sixth rows of table) for a velocity-based formulation, and in Appendix \ref{TAB:Displacement_Pressure} (fifth row of table) for a displacement-based formulation. Note that in such tables, gradient matrix multiplying the pressure $\pre_{n+1}$ is integrated in the updated configuration $\Omega(t_{n+1})$ to be consistent with the pressure time ($t_{n+1}$) obtained by setting $\alpha_f = 1.0$. That is, discretized momentum equations listed in Appendices \ref{TAB:Velocity_Pressure} and \ref{TAB:Displacement_Pressure} for approaches $\GAd$ and $\GAe$ are obtained from the following momentum equation:
\begin{equation} \label{EQ:MomentumEquation_AlphaMethod_noPre}
\mathbf{M}_{n+\alpha_m}  \accbf_{n+\alpha_m} + \mathbf{K}_{n+\alpha_f} \velbf_{n+\alpha_f} - \mathbf{D}^\intercal_\npa{n+1} \: \mathbf{p}_{n+1} = \mathbf{f}_{n+\alpha_f}
\end{equation}
\vspace{1mm}

An alternative implementation for $\GAd$ and $\GAe$ is to compute the gradient matrix in the intermediate configuration $\Omega(t_{n+\alpha_f})$, as done in Eulerian-based works \citep{valdes2007nonlinear}. This alternative would lead to the following momentum equation:
\begin{equation} \label{EQ:MomentumEquation_AlphaMethod_noPre_2}
\mathbf{M}_{n+\alpha_m}  \accbf_{n+\alpha_m} + \mathbf{K}_{n+\alpha_f} \velbf_{n+\alpha_f} - \mathbf{D}^\intercal_\npa{n+\alpha_f} \: \mathbf{p}_{n+1} = \mathbf{f}_{n+\alpha_f}
\end{equation}

\noindent where the highlighted term is to stand out the difference between Eqs.~\eqref{EQ:MomentumEquation_AlphaMethod_noPre} and \eqref{EQ:MomentumEquation_AlphaMethod_noPre_2}. For the sake of simplicity, the time-discretized equations for Eq.~\eqref{EQ:MomentumEquation_AlphaMethod_noPre_2} are omitted, but they can be easily obtained from Table \ref{TAB:Velocity_Pressure}1 (fifth and sixth rows) and \ref{TAB:Displacement_Pressure}2 (fifth row) by using a gradient matrix computed in $\Omega(t_{n+\alpha_f})$. 

In the following, equations are developed for the GA-II approach  that assumes linear scaling between times $t_n$ and $t_{n+1}$ for the dynamic, internal and external forces.

\subsubsection{GA-II}

The formulas for computing the intermediate forces (Eq.~\ref{EQ:GA_II_parametrization}) are replaced in the momentum equation (Eq.~\ref{EQ:MomentumEquation_Forces}), which leads to:
\begin{equation} 
\alpha_m \: \bm{f}^\mathrm{dyn}_{n+1} + \alpha_f \:\bm{f}^\mathrm{int}_{n+1} = \alpha_f \:\bm{f}^\mathrm{ext}_{n+1} + (1-\alpha_f) \: \bm{f}^\mathrm{ext}_{n} - (1-\alpha_m)\bm{f}^\mathrm{dyn}_{n} - (1-\alpha_f)\bm{f}^\mathrm{int}_{n} 
\end{equation}
\vspace{1mm}

Replacing terms for an incompressible Newtonian fluid, the following discretized momentum equation is obtained:
\begin{equation}  \label{EQ:NewtFluid_Method2}
\begin{matrix} 
\mathbf{M}_{n+1} \alpha_m \dt{\mathbf{v}}_{n+1} + \mathbf{K}_{n+1} \alpha_f \mathbf{v}_{n+1} - \mathbf{D}^\intercal_{n+1} \alpha_f\: \mathbf{p}_{n+1}
= \alpha_f\:\mathbf{f}_{n+1} 
+ 
(1-\alpha_f)\: (\mathbf{f}_{n} - \mathbf{K}_{n} \: {\mathbf{v}}_{n} +  \mathbf{D}^\intercal_{n} \: \mathbf{p}_{n}) - \mathbf{M}_{n} (1-\alpha_m) \dt{\mathbf{v}}_{n}
\end{matrix}
\end{equation}

Then, Newmark formulas are used to get a velocity-pressure formulation. Expressing the nodal acceleration in terms of the nodal velocities:
\begin{equation} \label{EQ:NewtFluid_Method2_2}
\begin{matrix}
\hspace{-10mm}
\mathbf{M}_{n+1} \alpha_m 
  \left(\displaystyle\frac{\mathbf{v}_{n+1}-\mathbf{v}_n}{\gamma \: \Delta t} { -\frac{(1-\gamma)}{\gamma}   \dt{\mathbf{v}}_{n}} 
 \right)
+ \mathbf{K}_{n+1} \alpha_f \mathbf{v}_{n+1} - \mathbf{D}^\intercal_{n+1} \alpha_f\: \mathbf{p}_{n+1}
= \\[3ex]
&\hspace{-10mm} \alpha_f\:\mathbf{f}_{n+1} 
+ (1-\alpha_f)\:(\mathbf{f}_{n} - \mathbf{K}_{n} \: {\mathbf{v}}_{n} + \mathbf{D}^\intercal_{n} \: \mathbf{p}_{n} ) 
- \mathbf{M}_{n} (1-\alpha_m) \dt{\mathbf{v}}_{n} 
\end{matrix}
\end{equation}

\noindent Including the PSPG stabilization and arranging terms:
\begin{equation} \label{EQ:NewtFluid_Method2_3}
\hspace{-1mm}
\begin{matrix}
\left[
\begin{matrix}
	\mathbf{M}_{n+1} \: \displaystyle\frac{\alpha_m}{\gamma\Delta t} + \mathbf{K}_{n+1} {\alpha_f}  & \quad\quad - \mathbf{D}^\intercal_{n+1} \alpha_f
	\\[4ex]
	\mathbf{C}_{n+1} \: \displaystyle\frac{\alpha_m}{\gamma\Delta t} + \mathbf{D}_{n+1} {\alpha_f} &  \quad\quad \mathbf{L}_{n+1} \alpha_f
\end{matrix}
\right]
\;
\left[
\begin{matrix}
	\mathbf{v}_{n+1}
	\\[5ex]
	\mathbf{p}_{n+1}
\end{matrix}
\right] 
= \\[8ex]
&
\hspace{-25mm}
\left[
\begin{matrix}
	\alpha_f\:\mathbf{f}_{n+1} + 
	(1{\scriptstyle - }\alpha_f)\:( \overbrace{ \color{blue} \mathbf{f}_{n} - \mathbf{K}_{n} \mathbf{v}_\mathrm{n} + \mathbf{D}^\intercal_{n}\mathbf{p}_\mathrm{n} }^\text{\normalsize{\color{blue}$\bm{f}^\mathrm{ext}_n - \bm{f}^\mathrm{int}_n$}}) 
	- (1{\scriptstyle - }\alpha_m)
	\overbrace{\color{blue}\mathbf{M}_{n}\dt{\mathbf{v}}_n}^\text{\normalsize{\color{blue}$\bm{f}^\mathrm{dyn}_n$}}   
	  +
	  \mathbf{M}_{n+1}\alpha_m 
	  \left(\frac{1}{\gamma \Delta t} \mathbf{v}_\mathrm{n} + \frac{1 - \gamma}{\gamma} \dt{\mathbf{v}}_\mathrm{n}
	  \right)
	\\[3ex]
	\alpha_f\:\mathbf{h}_{n+1} + 
	(1{\scriptstyle - }\alpha_f)\:({\color{blue}\mathbf{h}_{n} - \mathbf{D}_{n} \mathbf{v}_\mathrm{n} - \mathbf{L}_{n} \mathbf{p}_\mathrm{n}  }) 
	- (1{\scriptstyle - }\alpha_m)\:\:{\color{blue}\mathbf{C}_{n}\dt{\mathbf{v}}_n}
	  \:+
	  \mathbf{C}_{n+1}\alpha_m 
	  \left(\frac{1}{\gamma \Delta t} \mathbf{v}_\mathrm{n} + \frac{1 - \gamma}{\gamma} \dt{\mathbf{v}}_\mathrm{n}
	  \right)
\end{matrix}
\right]
\end{matrix}
\end{equation}
\vspace{2mm}

Notably, Eq.~\eqref{EQ:NewtFluid_Method2_3} does not contain terms at $t_{n+\alpha}$, which is the most distinctive feature with respect to the GA-I equations. The reader can easily verify this by comparing the seventh row of Table \ref{TAB:Velocity_Pressure}1 (appendix \ref{TAB:Velocity_Pressure}) with the previous rows of that table. At first glance, this could be seen as an advantage for computing programming, since there is no need to calculate and store intermediate state variables. However, in the PFEM context, this does not present a major advantage over the GA-I approach. The reason is that the finite element mesh is reconstructed at each time step, so dynamic, internal and external forces from the previous time step must be recalculated in the new mesh (blue terms in Eq.~\ref{EQ:NewtFluid_Method2_3}). This requires to assemble the $\mathbf{K}$, $\mathbf{D}$, $\mathbf{L}$ and $\mathbf{C}$ matrices not only at $t_{n+1}$ but also at $t_n$. If there is no remeshing, then forces at $t_n$ can be stored to avoid additional matrix assembling.

As done for the GA-I approach, equations can be developed for a displacement-pressure formulation, whose momentum equation can be found in the sixth row of Table \ref{TAB:Displacement_Pressure}2 (appendix \ref{TAB:Displacement_Pressure}). In addition, equations resulting from setting the pressure at $t_{n+1}$ instead of $t_{n+\alpha_f}$ are presented in the last rows of Tables \ref{TAB:Velocity_Pressure}1 and \ref{TAB:Displacement_Pressure}2. Those equations assume that gradient of pressure $\pre_{n+1}$ is integrated in the current configuration, as in Eq.~\eqref{EQ:MomentumEquation_AlphaMethod_noPre}. To obtain the expression that integrates the pressure gradient in the intermediate configuration, as in Eq.~\eqref{EQ:MomentumEquation_AlphaMethod_noPre_2}, gradient matrix $\mathbf{D}^\intercal_{n+1}$ must be replaced by $\mathbf{D}^\intercal_{n+\alpha_f}$, which is computed as:
\begin{equation}
	\mathbf{D}^\intercal_{n+\alpha_f} = (1-\alpha_f) \: \mathbf{D}^\intercal_{n} + \alpha_f \: \mathbf{D}^\intercal_{n+1} 
\end{equation}

\subsubsection{Generalized-$\alpha$ : Set of parameters}

Being based on Newmark's method, the Generalized-$\alpha$ method requires 4 parameters to be defined by the user: $\gamma$, $\beta$, $\alpha_m$, and $\alpha_f$. However, these can be expressed as a function of the spectral radius $\rho_\infty$. According to the stability analysis of Jansen et.al.\citep{jansen2000generalized} developed for a linear system, the parameterization that ensures stability and second-order accuracy is as follows: 
\begin{equation} \label{EQ:alpha_parameters}
\alpha_m = \frac{1}{2} \left( \frac{3-\rho_\infty}{1+\rho_\infty} \right) \hspace{2.5mm} , \hspace{2.5mm}
\alpha_f =   \frac{1}{1+\rho_\infty}  \hspace{2.5mm} , \hspace{2.5mm}
\gamma = \frac{1}{2} + \alpha_m - \alpha_f \hspace{2.5mm} , \hspace{2.5mm} 0\leq \rho_\infty \leq 1 
\end{equation}
\vspace{1mm}

Note that formulas in Eq.~\eqref{EQ:alpha_parameters} differ from those of Chung and Hulbert \citep{chung1993time}. On one side because works define $\alpha$ in opposite ways ($\alpha$ weights $t_{n+1}$ in Jansen.et.al.\citep{jansen2000generalized} while $t_n$ in Chung and Hulbert\citep{chung1993time}). Taking this into account, expressions for $\alpha_f$ and $\gamma$ in Eq.~\eqref{EQ:alpha_parameters} turn out to be equivalent in both works. On the other hand, because amplification matrices are different, since Chung and Hulbert\citep{chung1993time} consider displacement, velocity and acceleration, while Jansen et.al.\citep{jansen2000generalized} only accounts for velocity and acceleration. This results in different equations for $\alpha_m$, from which we chose that of Jansen et.al.\citep{jansen2000generalized} since our momentum equation contains only two kinematic variables. \review{ Given the use of a Lagrangian scheme, the $\beta$ variable from Newmark's method must be included. As stated in Section \ref{sec:Newmark}, $\beta$ is parameterized according to Eq.~\eqref{EQ:Newmark_Parametrization}. This would be one distinctive aspect among Lagrangian and Eulerian implementations of the Generalized-$\alpha$ method.}

As highlighted by Jansen et.al.\citep{jansen2000generalized} and Lovri\'{c} et.al.\citep{lovric2018new}, $\rho_\infty = 0.0$ leads to a scheme with strong numerical damping for high frequencies. In terms of spectral stability and for a linear problem, $\rho_\infty = 0.0$ is equivalent to the backward differentiation scheme of second order. To make an analogy with the way time integration formulas have been presented, the second order backward differentiation scheme is obtained by defining the approximate acceleration function as:

\begin{equation}
\dt{v} = \frac{1}{3} \: \frac{\vel_{n} - \vel_{n-1}}{\Delta t} + \frac{2}{3}  \: \acc_{n+1}
\end{equation} 

\noindent which time-integrated results in the following (compare with equation 2.7 of reference \citep{lovric2018new}):
\begin{equation}
\vel_{n+1} = \vel_n + \frac{1}{3} \: (\vel_n - \vel_{n-1}) + \frac{2}{3} \: \Delta t \: \acc_{n+1}
\end{equation}
\vspace{1mm}

On the other hand, $\rho_\infty = 1.0$ preserves all high frequencies and it is equivalent to the midpoint rule. For a linear problem, it is also equivalent to the Trapezoidal rule obtained by setting $\gamma = 0.5$ and $\beta = 0.25$ in Newmark's method.

For the numerical examples of this manuscript, $\rho_\infty$ is imposed and $\alpha_m$, $\alpha_f$ and $\gamma$ are obtained from Eq.~\eqref{EQ:alpha_parameters}, while $\beta$ from Eq.~\eqref{EQ:Newmark_Parametrization}. It is noteworthy that $\rho_\infty = 0.0$ leads to $\alpha_m = 1.5$ and $\alpha_f = 1.0$. In such a case, the GA-I and GA-II approches are identical since external and internal forces are computed at the current time, $\bm{f}^\mathrm{int}_{n+1}$ and $\bm{f}^\mathrm{ext}_{n+1}$, in addition, dynamic forces in these approaches are equivalent when mass conservation holds (as shown in Eq.~\ref{EQ:EqualDynamicForces}). Furthermore, if $\rho_\infty = 0.0$, approaches that impose pressure at $t_{n+1}$ agree with those that impose it at $t_{n+\alpha_f}$. Meaning, all of the aforementioned implementation approaches of the Generalized-$\alpha$ method should produce the same result when $\rho_\infty = 0.0$.

Before reporting numerical comparisons of the different time integration schemes, the following subsection briefly describes our PFEM implementation.

\subsection{PFEM implementation}

The PFEM implementation of this work follows the original idea of Idelsohn et.al.\citep{idelsohn2004particle}, which is based on a remeshing procedure using the Alpha Shape algorithm\citep{edelsbrunner1994three}. Namely, the fluid is initially discretized using a cloud of particles, as shown in Fig.~\ref{Fig:alphaShape_a}. Then a Delaunay triangulation is applied to generate a mesh on the convex hull of the cloud (Fig.~\ref{Fig:alphaShape_b}). Next, the Alpha Shape algorithm is used to remove those elements whose circumcircle radius exceeds a \review{global} characteristic size of the mesh scaled by a $\alpha_\mathrm{shape}$ parameter. In this way, depending on the value of $\alpha_\mathrm{shape}$, large or highly distorted elements are removed from the triangulation, leaving a discretization of the fluid with a clear definition of its boundaries and free surface, as shown in Fig.~\ref{Fig:alphaShape_c}. This mesh is used to discretize the Lagrangian Navier-Stokes equations, whose solution gives the nodal velocity (or displacement) and pressure (Fig.~\ref{Fig:alphaShape_f}). The new fluid position is obtained by the time integration scheme (Fig.~\ref{Fig:alphaShape_e}). Finally, the mesh quality is checked to determine if re-meshing is necessary (Fig.~\ref{Fig:alphaShape_d}) or if the triangulation can be preserved for the next time step. A recent survey on PFEM and more details on the above steps can be found in \citep{cremonesi2020state}.

\begin{figure}[t] \captionsetup[sub]{font=normalsize} \captionsetup[subfigure]{labelformat=empty}
\centering 
	\includegraphics[width=0.90\linewidth]{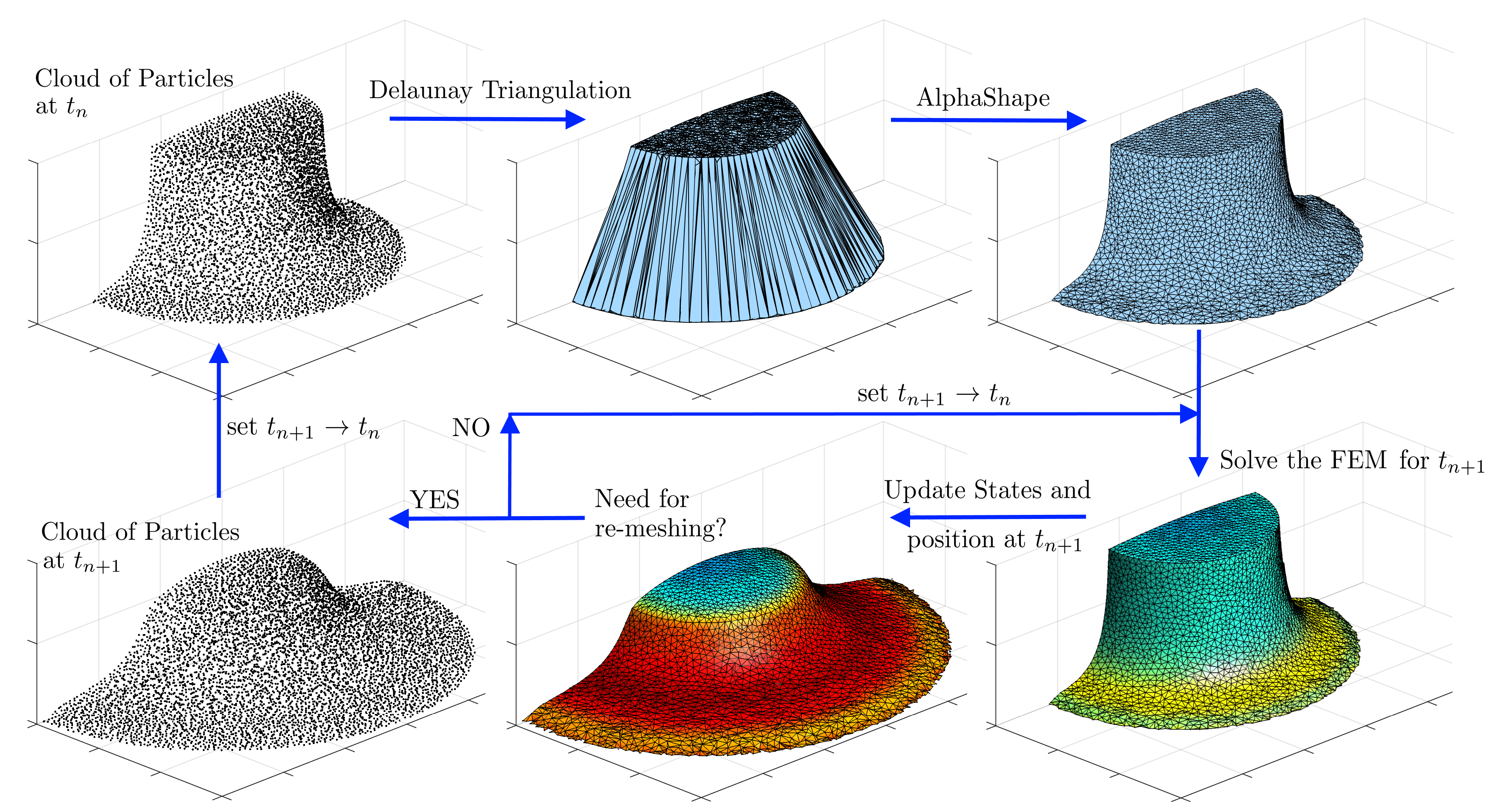}
	\\
	\begin{subfigure}[b]{0.05\textwidth}
		\caption{}
		\label{Fig:alphaShape_a}
	\end{subfigure}
	~
	\begin{subfigure}[b]{0.05\textwidth}	
		\caption{}
		\label{Fig:alphaShape_b}
	\end{subfigure}	
	~
	\begin{subfigure}[b]{0.05\textwidth}
		\caption{}
		\label{Fig:alphaShape_c}
	\end{subfigure}
	~
	\begin{subfigure}[b]{0.05\textwidth}	
		\caption{}
		\label{Fig:alphaShape_d}
	\end{subfigure}	
	~
	\begin{subfigure}[b]{0.05\textwidth}	
		\caption{}
		\label{Fig:alphaShape_e}
	\end{subfigure}	
	~
	\begin{subfigure}[b]{0.05\textwidth}	
		\caption{}
		\label{Fig:alphaShape_f}
	\end{subfigure}	
	\\
	\vspace{-55mm}
	\hspace{-40mm} (\textbf{a}) \hspace{45mm} (\textbf{b}) \hspace{45mm} (\textbf{c})
	\\
	\vspace{39mm}
	\hspace{-40mm} (\textbf{d}) \hspace{47mm} (\textbf{e}) \hspace{47mm} (\textbf{f})
	\\
\caption{Illustration of the PFEM procedure using the collapse of a cylindrical water column problem. (a) Cloud of particles with known state variables at $t_n$. (b) Delaunay triangulation on the entire set of particles. (c) Large and distorted elements are removed from the triangulation using Alpha Shape algorithm\citep{edelsbrunner1994three}. (f) Norm of nodal velocities for illustrating a FEM. (e) Illustration of updating state variables and nodal position. (d) The updated cloud of particles to be used in the next time step. }
\label{Fig:alphaShape}
\end{figure}

The nonlinear system of equations, i.e.~Eq.~\eqref{EQ:Equtions_from_NM}, \eqref{EQ:MomentumEquation_Space_Time_Alpha_1_a}, \eqref{EQ:MomentumEquation_Space_Time_Alpha_1_position_a} or \eqref{EQ:MomentumEquation_Space_Time_Alpha_1}, is assembled in matrix form as $\mathbf{A} \: \mathbf{q} = \mathbf{b}$, where $\mathbf{A}$ is a large matrix storing matrix terms of the momentum and continuity equations computed at time $t_{n+1}$, $\mathbf{q}$ is a vector containing the solution the for state variables, and $\mathbf{b}$ is a vector storing the external forces. Note that all mass matrices are "consistent" and computed at the time specified by the subscript ($t_{n}$, $t_{n+1}$ or $t_{n+\alpha_m}$). The system $\mathbf{A} \: \mathbf{q} = \mathbf{b}$ is solved using the Picard (or fixed-point) algorithm, which sets as predictor the values of state variables of previous time $t_n$. \review{Regarding the stabilization term, $\tau_\mathrm{PSPG}$ is computed using the velocity of the previous iteration of the nonlinear algorithm.} Once the solution of $\mathbf{q}$ is obtained, particle kinematics is updated according to the time integration scheme. Updating equations are summarized in Appendix \ref{TAB:Update_States}.

The success of the PFEM method relies on the use of efficient remeshing algorithms, such as the Delaunay triangulation followed by the Alpha Shape. This process involves a negligible computational cost in 2D and a minor one in 3D. However, the Alpha Shape algorithm brings some issues during remeshing operations. The most notable is the mass variation of the system produced by the discrete addition and removal of finite elements, which can be treated with mesh refinement or tuning of the $\alpha_\mathrm{shape}$ parameter\citep{franci2017effect}. However, examples in the present work are free of such a problem either because they do not require remeshing (such as the fluid sloshing), possess an Eulerian domain (such as the flow around a cylinder), or because they have slip boundary conditions and do not feature fluid splashing (such as the solitary wave propagation and collapse of a water column). The parameter $\alpha_\mathrm{shape}$ is set to 1.25 in all examples of this work.

\review{
PFEM can handle problems with non-uniform mesh size and Eulerian domains, such as the flow around a cylinder problem introduced in the following section. Mesh refinement is achieved by adding and removing particles into the domain according to a size field. For the flow around a cylinder problem (Section \ref{sec:FAC}), the size field is a function of the distance to the cylinder, while for the collapse of a water column problem (Section \ref{sec:CCWC}), the size field is defined in terms of the distance to the free surface. If the size of a finite element is larger than the prescribed size, a particle is added in the center of the element and mesh refinement occurs. The un-refinement process operates in a similar way, but now a particle is removed if the distance to the nearest particle is less than that prescribed in the size field. This addition and removal process imposes upper and lower size limits on the elements depending on the desired size distribution. An illustration of this process for the flow around a cylinder problem is shown in Fig.~\ref{Fig:addition_removal}. In this context, the Alpha Shape must be applied locally, i.e.~the upper bound of the Alpha-Shape algorithm must be defined by a parameter $\alpha_\mathrm{shape}$ scaled by the desired local element size and not by a global reference size as previously mentioned.  
 
To simulate an Eulerian domain (the flow around a cylinder), particles in the inlet and outlet zone are treated in a Lagrangian fashion, i.e., they are allowed to move. However, before proceeding with the remeshing process for the next time step, these particles are repositioned on the inlet and outlet lines. For further details regarding the simulation of Eulerian domains using PFEM, readers are referred to \citep{meduri2019fully,cerquaglia2019development}.
}

\review{
Importantly, the remeshing process in PFEM does not affect the use of the Generalized-$\alpha$ method. Therefore, implementation aspects of the Generalized-$\alpha$ presented for the standard FEM are equally valid in PFEM. The only practical shortcoming is with respect to the GA-II approach, as it requires to recompute the dynamic, internal and external forces of the previous time $t_n$ due to the new nodal connectivity at time $t_{n+1}$. Likewise, drawbacks could arise with material models that require historical variables, however such cases are outside the scope of this work.  
}

\begin{figure}[t] \captionsetup[sub]{font=normalsize} \captionsetup[subfigure]{labelformat=empty}
\centering 
	\includegraphics[width=0.85\linewidth]{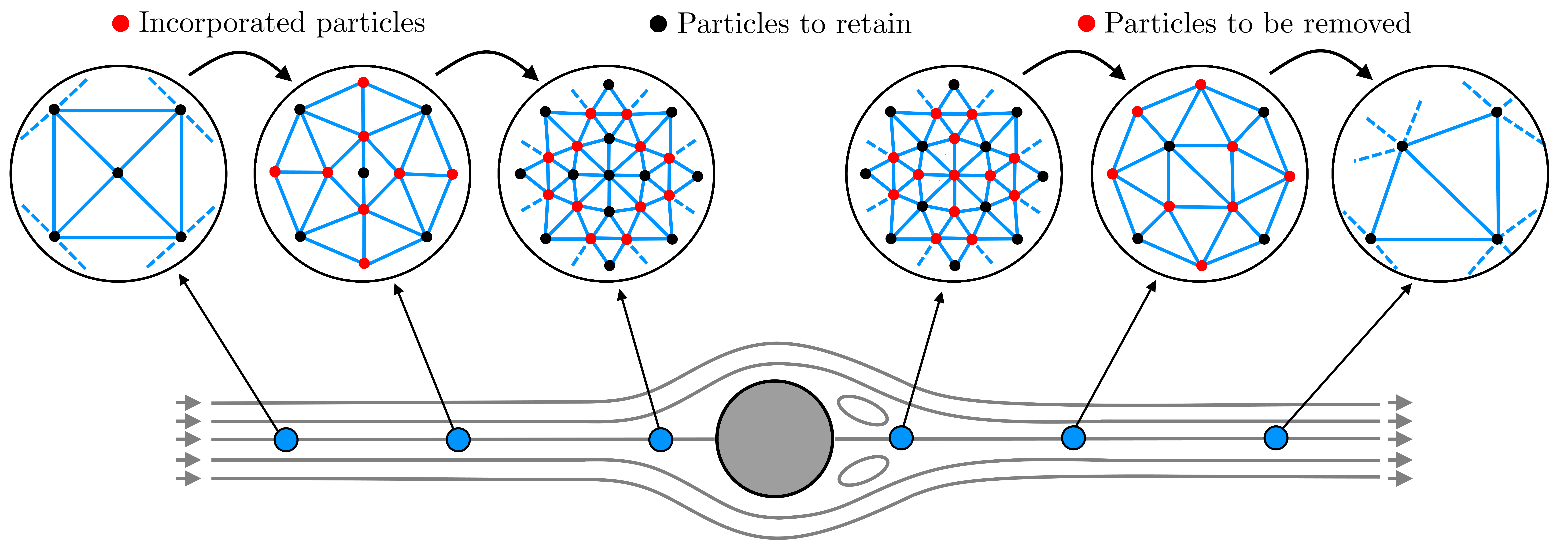}
	\\
\caption{\review{Illustration of particle addition and removal for mesh refinement in PFEM. The flow around a cylinder problem is illustrated. On the left side of the cylinder, particles are added in the center of elements as they approach the cylinder. On the right side, particles are removed if the distance to the nearest particle is less than an imposed threshold.}}
\label{Fig:addition_removal}
\end{figure}

The time integration schemes analyzed in this work have been implemented in an in-house code named PFEM3D \citep{fevrier2020travail}. This is a C++ code that uses the Eigen\citep{eigenweb} library for linear algebra operations, the CGAL\citep{cgal:dy-as3-22a} library for triangulation and Alpha Shape, OpenMP\citep{openmp08} for parallel threading, and Gmsh\citep{geuzaine2009gmsh} for the input/output of finite element meshes.

\section{Numerical Examples}\label{sec:4}

Four distinctive benchmark problems are used in this section for comparing time integration schemes. The first is the sloshing of a fluid characterized by small displacements so no remeshing is needed, \review{which allows to validate the FEM implementation of the time integration schemes.} The second is the solitary wave propagation in a constant depth container, which uses remeshing for minor improvements to the mesh discretizing the wave. The third is the flow around a cylinder, characterized by an Eulerian computational domain that features mesh refinement and significant velocity gradients around the cylinder, so that remeshing is applied at each time step. The fourth is the collapse of a cylindrical water column which, unlike the previous ones, is a 3D problem where time step $\Delta t$ is defined in terms of a maximum CFL (Courant–Friedrichs–Lewy) number.

This section mainly reports simulation data for comparison effects and does not present snapshots of simulations. Instead, the reader is guided to reference \citep{Youtube} for animations of each problem.

\subsection{Small amplitude sloshing}

This free-surface problem consists of simulating the sloshing of a fluid caused by imposing a non-uniform initial elevation of the free surface \citep{ramaswamy1990numerical}. The geometry of the problem is illustrated in Fig.~\ref{Fig:Sloshing_TestCase_a}. The width of the container is $b=1$ m and the initial free surface elevation is defined as:
\begin{equation}
h(x) = H + a \: \mathrm{sin} \left( \frac{\pi(x - x_0)}{b} - \frac{\pi}{2} \right) 
\end{equation}

\noindent where $H$ is the mean height equal to 1.0 m and $a$ is the initial amplitude equal to 0.01 m. A unit gravity acceleration is used, $g$ = 1.0 m/s$^2$, as well as fictitious fluid parameters, $\mu = 0.01$ Pa~s and $\rho$ = 1.0 kg/m$^3$. The fluid domain is discretized with 3073 particles and 5942 elements of 0.02 m average size, as shown in Fig.~\ref{Fig:Sloshing_TestCase_b}. No external pressure is imposed in the free-surface, a free-slip condition is defined on the container walls and non-slip condition in the base. Given the small initial amplitude $a$, elements are subjected to small deformations so that remeshing process is not needed in this problem. 

\begin{figure}[t] \captionsetup[sub]{font=normalsize} \captionsetup[subfigure]{labelformat=empty}
	\centering 
	\begin{subfigure}[b]{0.3\textwidth}
		\includegraphics[width=0.90\linewidth]{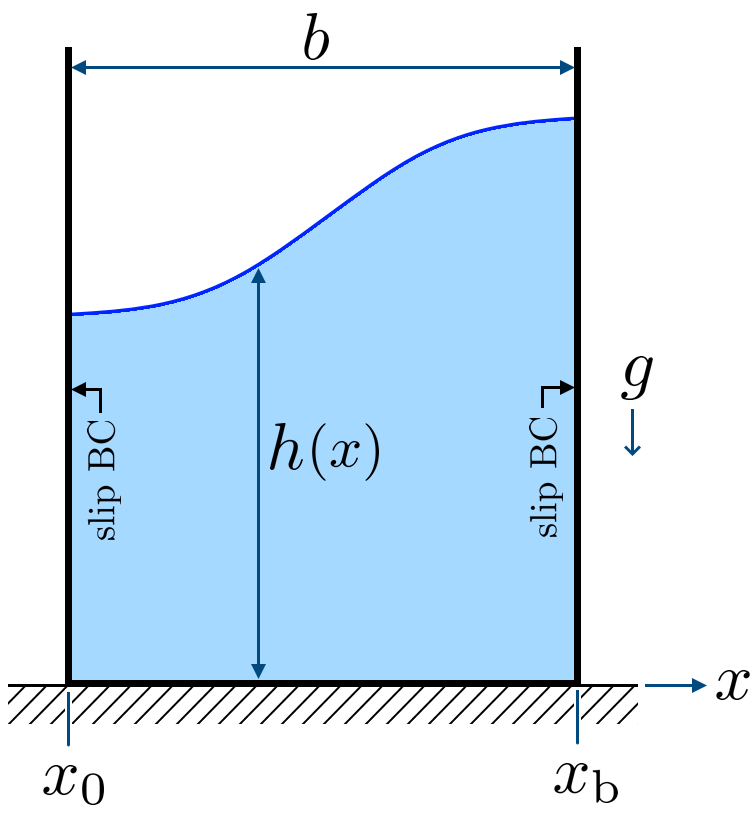}
		\caption{}
		\label{Fig:Sloshing_TestCase_a}
	\end{subfigure}
	~\hspace{5mm}
	\begin{subfigure}[b]{0.30\textwidth}	
		\includegraphics[width=0.85\linewidth]{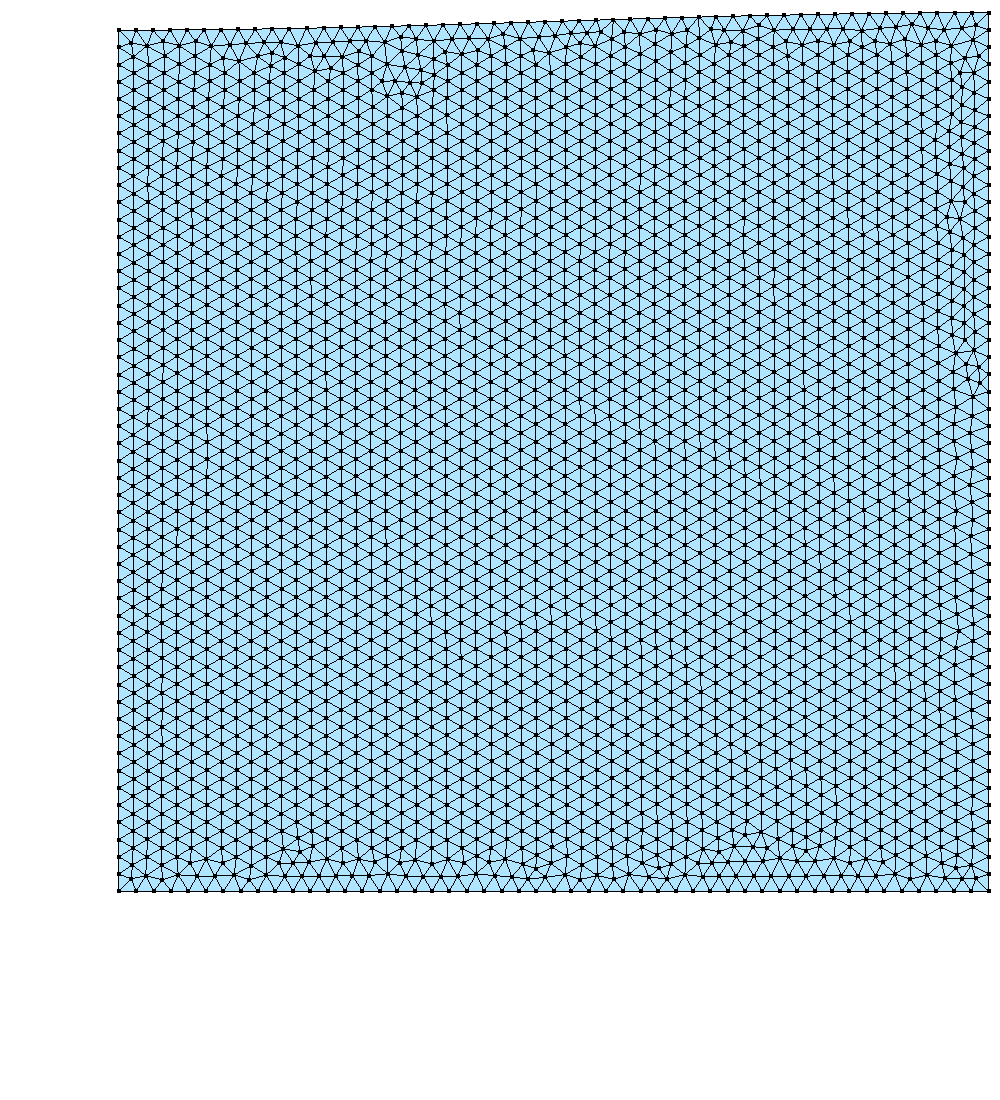}
		\caption{}
		\label{Fig:Sloshing_TestCase_b}
	\end{subfigure}
	\\
	\vspace{-57mm}
	\hspace{-55mm} (\textbf{a}) \hspace{55mm} (\textbf{b})
	\vspace{45mm}
\caption{Small amplitude sloshing. (a) Geometry illustration. (b) Initial finite element discretization}
\label{Fig:Sloshing_TestCase}
\end{figure}

Several sloshing problems are solved with variations in the time step $\Delta t$, which are 0.400, 0.100, 0.025, and 0.00625 s. The criterion for comparison is free surface elevation at coordinate $x_b$ (see Fig.~\ref{Fig:Sloshing_TestCase_a}). Fig.~\ref{Fig:SS_time} plots $h(x_b)$ versus time for different time steps using Backward Euler (Fig.~\ref{Fig:SS_time_a}) and Generalized-$\alpha$ (Fig.~\ref{Fig:SS_time_b}) in a velocity-pressure formulation. Implementation $\GAa$ with a spectral radius $\rho_\infty = 0.0$ is used for Generalized-$\alpha$. Graphs also show a solution obtained with the Trapezoidal rule ($\Delta t = 0.00625$ s). For validating the simulation, extreme points of the oscillatory curve reported by Avancini and Sanches \citep{avancini2020total} are also displayed. 

The strong dependence of Backward Euler on the time step is clear when comparing Figs. \ref{Fig:SS_time_a} and \ref{Fig:SS_time_b}. Even when using a small time step $\Delta t = 0.00625$ it is not possible to match the curve of the Trapezoidal rule. Instead, the Generalized-$\alpha$ method produces curves very close to that of the Trapezoidal rule, even when using a large $\Delta t = 0.100$. To facilitate the comparison between time integration schemes, an error measure is plotted next. The error criterion considers the area comprised between a reference elevation curve $h_\mathrm{ref}(t)$ and that of the simulation $h(x_b,t)$, that is: 
\begin{equation}
A_\mathrm{diff} = \int_0^{20} \left| h_\mathrm{ref}(t) - h(x_b,t) \right|  dt
\end{equation}

\noindent where $A_\mathrm{diff}$ is the error measure. The reference elevation $h_\mathrm{ref}(t)$ represents a solution obtained with a time step $\Delta_t = 0.0015625$ using the Trapezoidal rule.

\begin{figure}[t] \captionsetup[sub]{font=normalsize} \captionsetup[subfigure]{labelformat=empty}
	\centering 
	\begin{subfigure}[b]{0.48\textwidth}
		\includegraphics[trim=3 0 32 17,clip,width=1.00\linewidth]{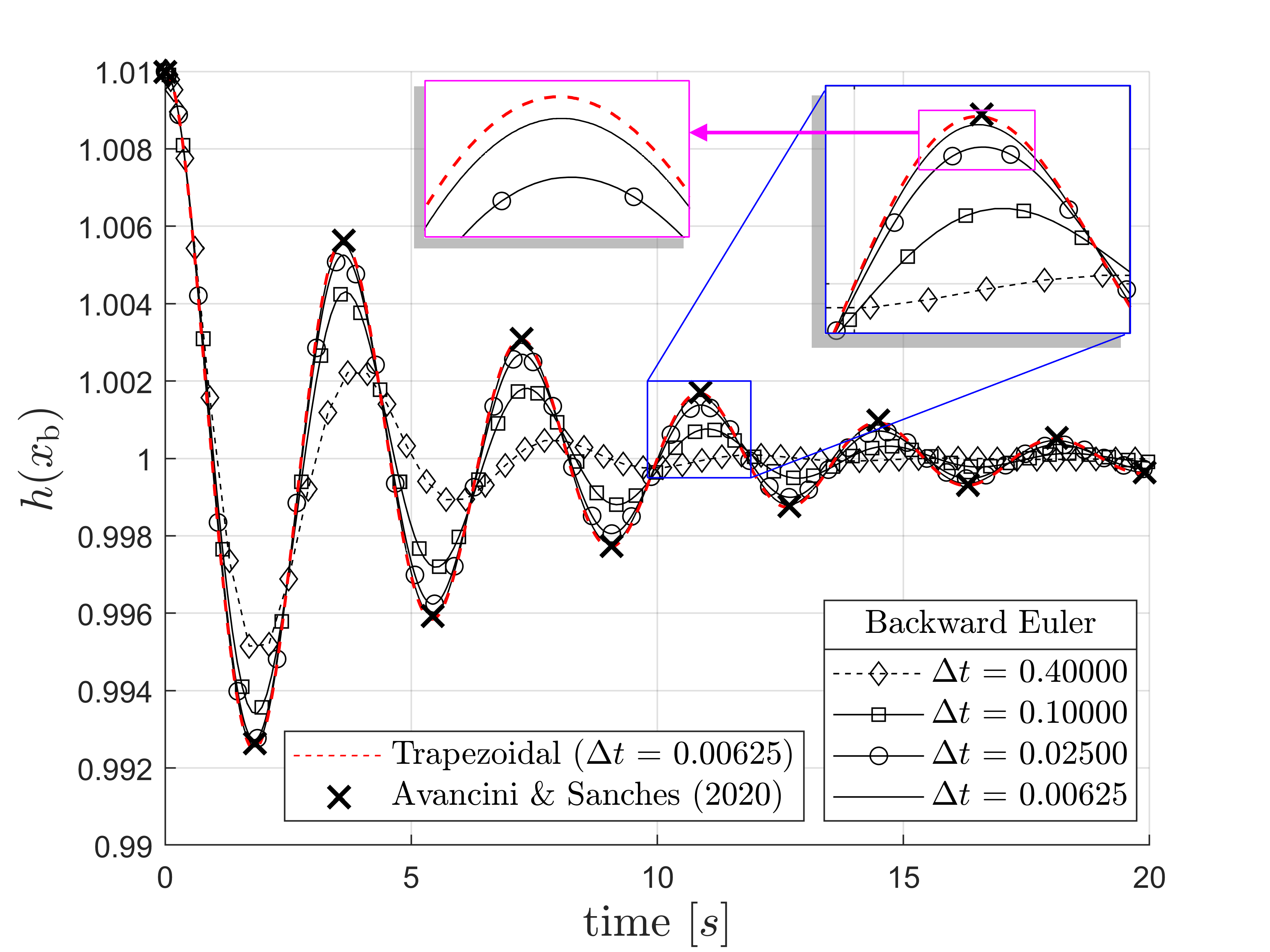}
		\caption{}
		\label{Fig:SS_time_a}
	\end{subfigure}
	~
	\begin{subfigure}[b]{0.48\textwidth}	
		\includegraphics[trim=3 0 32 17,clip,width=1.00\linewidth]{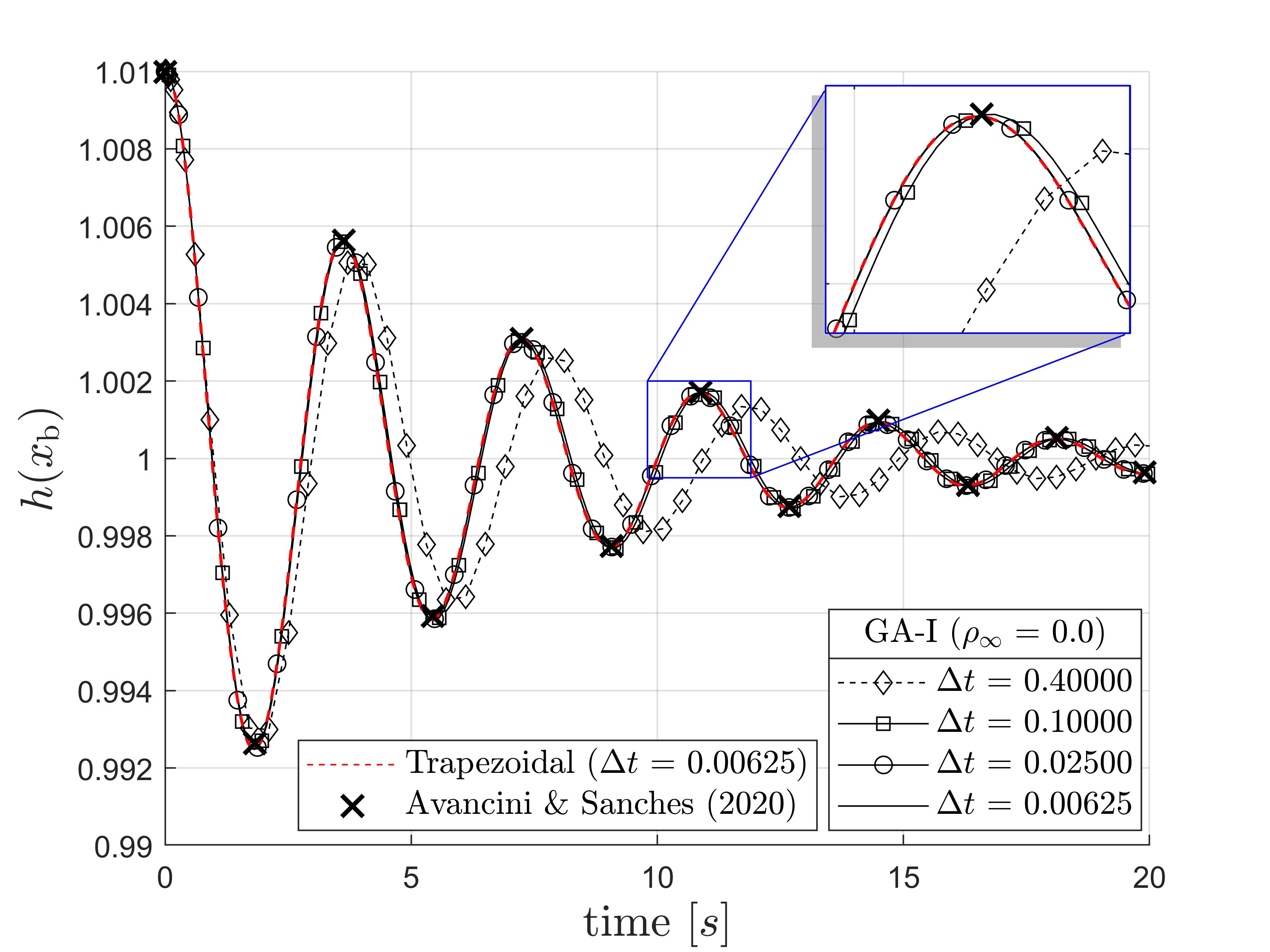}
		\caption{}
		\label{Fig:SS_time_b}
	\end{subfigure}
	\\
	\vspace{-74mm}
	\hspace{-83mm} (\textbf{a}) \hspace{83mm} (\textbf{b})
	\vspace{61mm}
\caption{Results of the fluid sloshing. Elevation of free surface with respect to time. Results obtained using the Trapezoidal rule, (a) Backward Euler and (b) Generalized-$\alpha$ ($\GAa$ and $\rho_\infty = 0.0$).}
\label{Fig:SS_time}
\end{figure}

\begin{figure}[t]\captionsetup[sub]{font=normalsize}\captionsetup[subfigure]{labelformat=empty}
	\centering 
	\begin{subfigure}[b]{0.47\textwidth}
		\includegraphics[trim=5 0 32 17,clip,width=1.00\linewidth]{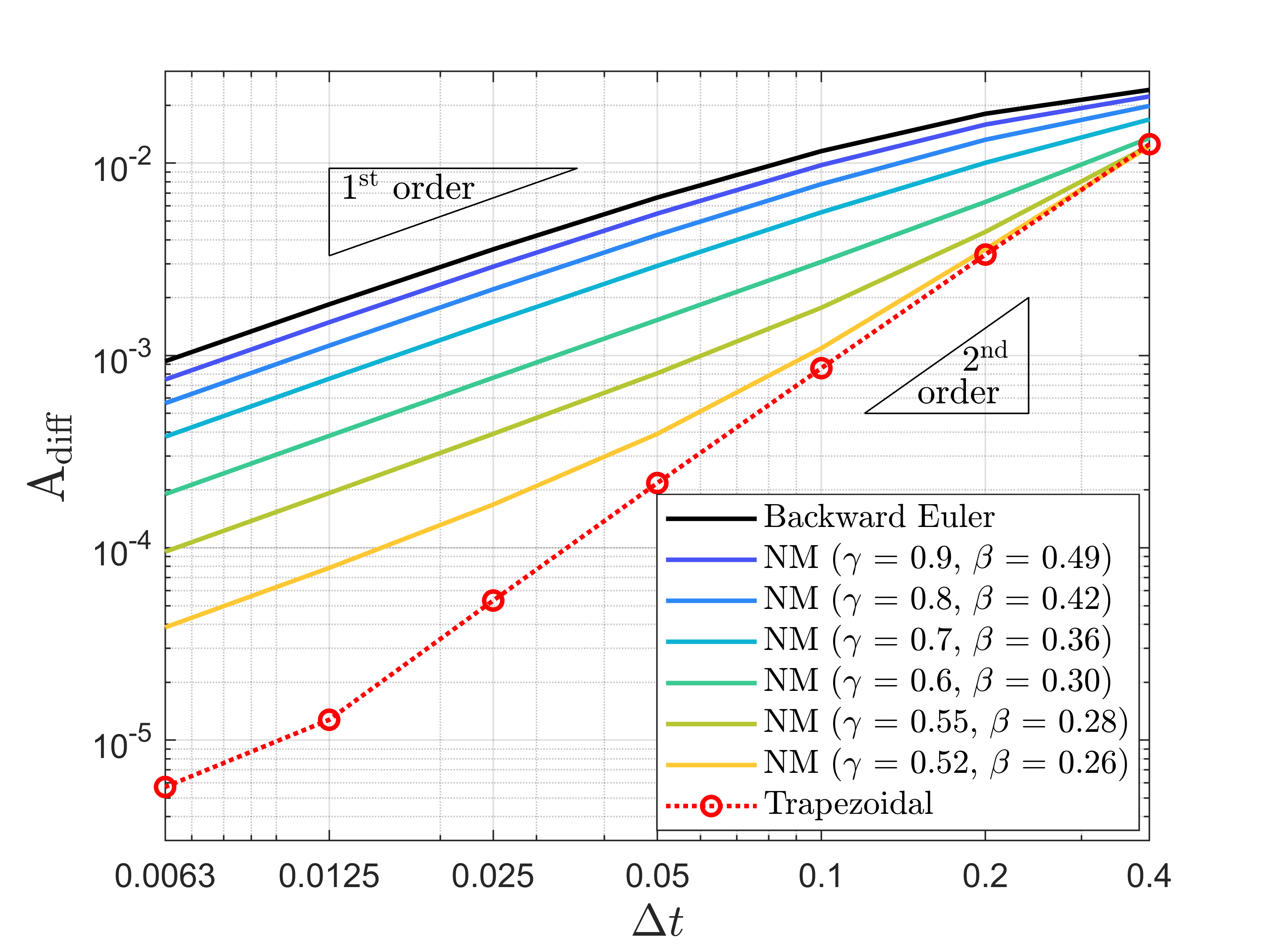}
		\caption{}
		\label{Fig:SS_Error_a}
	\end{subfigure}
	~\hspace{2mm}
	\begin{subfigure}[b]{0.47\textwidth}	
		\includegraphics[trim=5 0 32 17,clip,width=1.00\linewidth]{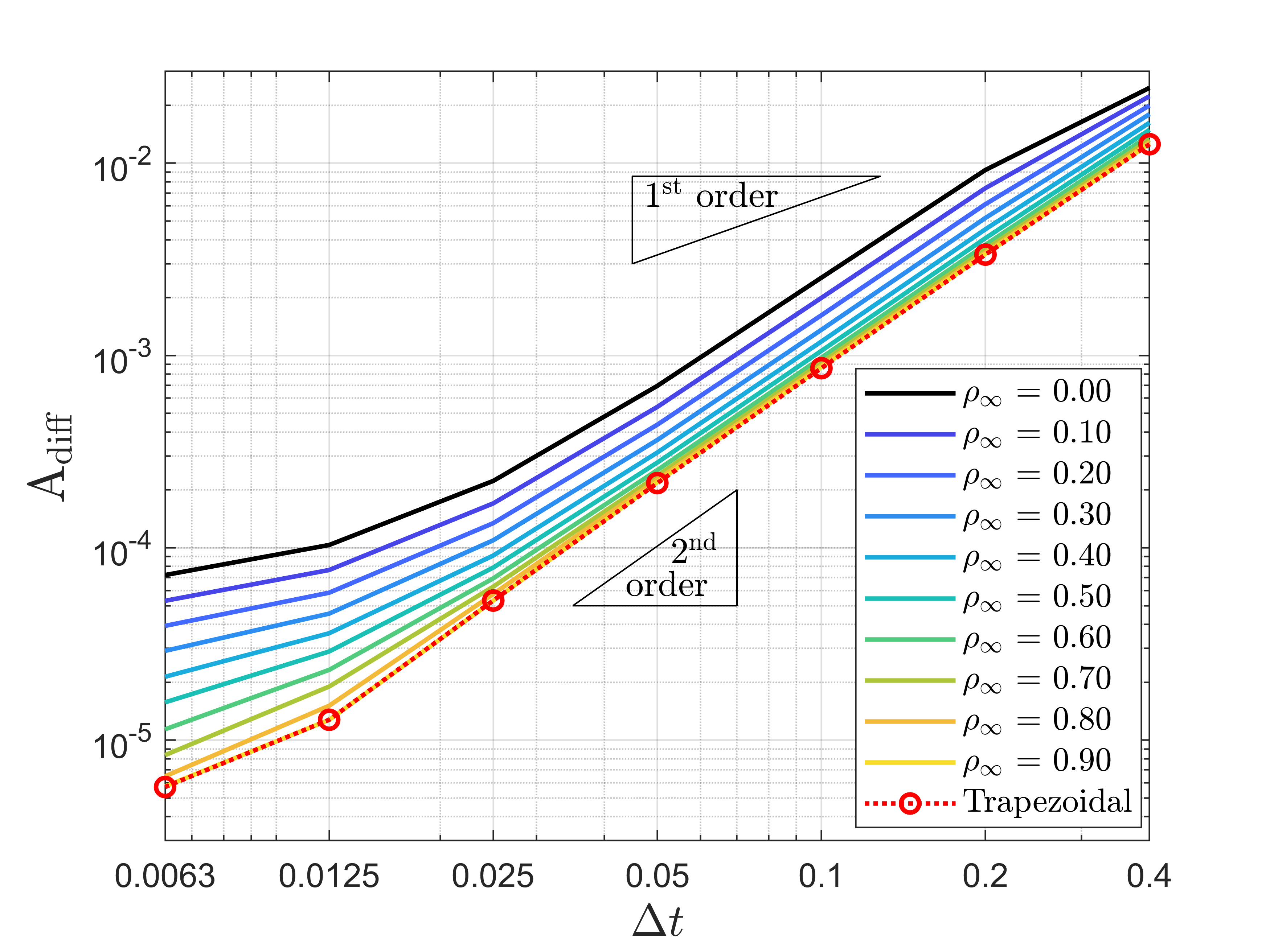}
		\caption{}
		\label{Fig:SS_Error_b}
	\end{subfigure}
	\\
	\vspace{-73mm}
	\hspace{-83mm} (\textbf{a}) \hspace{83mm} (\textbf{b})
	\vspace{60mm}
\caption{Results of the fluid sloshing. Error between reference elevation $h_\mathrm{ref}$ and simulated elevation $h$ at coordinate $x_b$. Simulations obtained using (a) Backward Euler, Newmark's Method (NM), Trapezoidal rule, and (b) Generalized-$\alpha$ ($\GAa$).}
\label{Fig:SS_Error}
\end{figure}

Now, the Backward Euler, Newmark, Generalized-$\alpha$ and Trapezoidal rule are considered in a velocity-based formulation. For Newmark, 6 problems are solved that differ in the value of $\gamma$, which are set to 0.9, 0.8, 0.7, 0.6, 0.55 and 0.52. The Generalized-$\alpha$ method considers 10 values of $\rho_\infty$ ranging from 0.0 to 0.9 with an increment of 0.1. Each time integration scheme is solved for 7 time steps $\Delta t$, ranging from 0.00625 to 0.4 s. That is, 126 problems are solved in total using different time integration schemes and time steps. Each one is compared against the reference curve to calculate the $A_\mathrm{diff}$ error. Results are shown in Figs.~\ref{Fig:SS_Error_a} and \ref{Fig:SS_Error_b}. Performance of time integration schemes is consistent with the literature. Among all the schemes, Backward Euler exhibits the lowest accuracy in the free surface height (black curve in Fig.~\ref{Fig:SS_Error_a}). Accuracy is improved by reducing time step $\Delta t$ but with an order of convergence approaching 1. The Trapezoidal rule is the most accurate and its error decreases with second order with respect to time step (red curve in Fig.~\ref{Fig:SS_Error_a}). Newmark's Method (NM) interpolates both Backward Euler and Trapezoidal schemes. However, $\gamma$ must be chosen very close to 0.5 to exhibit second-order convergence. In the case of Generalized-$\alpha$, all curves exhibit second order convergence for large time steps (Fig.~\ref{Fig:SS_Error_b}). \review{This expected result} is due to the parameterization of $\gamma$, $\beta$, $\alpha_m$ and $\alpha_f$ in terms of the spectral radius $\rho_\infty$, which ensures stability and second-order accuracy for a linear problem. 

%\review{
%It can be observed that error curves in Fig.~\ref{Fig:SS_Error} deteriorate at small time steps. The reason is that the amplitude curves (Fig.~\ref{Fig:SS_time}) of the reference solution, and those obtained with small time step, present a staircase pattern, either because small time steps are not able to catch significant displacements, because high frequency modes pollute results, or because of numerical inaccuracies. These sources of error prevent convergence of error curves towards 0 and make it meaningless to further reduce time step size in the error analysis.
%}

After validating our implementations of time integration schemes in PFEM with the results of Avancini and Sanches \citep{avancini2020total}, and obtaining an anticipated behavior of these for a problem with small displacements and deformations, the following examples compare the different implementation approaches of the Generalized-$\alpha$ method and incorporate the displacement-based formulation.

\subsection{Solitary wave propagation}

The geometry of the free-surface problem is illustrated in Fig.~\ref{Fig:SingleWavePropagation_TestCase_a}. This consists of simulating a wave of amplitud $A$ propagating in a rectangular container of constant depth $H$. Instead of displacing a fluid volume at the beginning of the simulation to generate the wave, another common approach \citep{radovitzky1998lagrangian,avancini2020total,ramaswamy1990numerical,staroszczyk2011simulation} is adopted here: to impose as initial condition the analytical approximation proposed Wehausen and Laitone \citep{wehausen1960surface}. Taking their system of equations with second order approximation (equation 31.37 of reference \citep{wehausen1960surface}), the free surface height, velocity and pressure are defined as follows: 
\begin{subequations}\label{EQ:SWP_hvp}
\begin{align}
h(x,t) 
&= \mathrm{H} + A \: \mathrm{sech}^2 (\xi) - \frac{3A^2}{4H}\mathrm{sech}^2 (\xi)
\left( 1 - \mathrm{sech}^2(\xi)\right) \label{EQ:SWP_hvp_a}
\\[2ex]
\frac{\vel_x(x,y,t)}{\sqrt{gH}} 
&= \frac{h - H}{H} - \frac{A^2}{2H^2}
\left( 1 + 6 \:  \frac{y-H}{H} + 3 \: \frac{(y-H)^2}{H^2} 
\right) 
\: 
\mathrm{sech}^2(\xi) + \frac{A^2}{2H^2} 
\left(1 + 9 \: \frac{y-H}{H} +
9\:\frac{(y-H)^2}{2H^2}
\right)  
\mathrm{sech}^4(\xi)
\\[2ex]
\frac{\vel_y(x,y,t)}{\sqrt{gH}} &= \frac{y \sqrt{3} A^{3/2}}{H^{5/2}} \mathrm{sech}^2(\xi) \mathrm{tanh}(\xi) 
\left[
1 + \frac{A}{H}
\left(
\frac{1}{8}
-
\frac{y}{H}
-
\frac{(y-H)^2}{2H^2}
\right)
-
\frac{A}{\mathrm{2H}}
\left( 1 - 6 \: \frac{y-H}{H} - 3 \: \frac{(y - H)^2}{H^2}
\right)
\mathrm{sech}^2(\xi)
\right]
\\[1ex]
\frac{\pre(x,y,t)}{\rho g H} &= 1 + \frac{h - H - y}{H} - \frac{3A^2}{4\mathrm{H^2}}
\left(
\frac{y^2}{H^2}
-
1
\right)
\left(
2\: \mathrm{sech}^2(\xi) - 3\: \mathrm{sech}^4(\xi) 
\right)
\end{align}
\end{subequations}
\vspace{1mm}

\noindent where the time-dependent parameter $\xi$ is defined as:
\begin{equation}\label{EQ:SWP_xi}
\xi(x,t) = \frac{x - c\:t}{H} \left(\frac{3A}{4H}\right)^{1/2} 
\left(
1 - \frac{5A}{8H} 
\right)
\end{equation}
\noindent and the wave celerity given by:
\begin{equation}\label{EQ:SWP_c}
\frac{c}{\sqrt{gH}} = 1 + \frac{A}{2H} - \frac{3A^2}{20H^2} 
\end{equation}
\vspace{1mm}

\begin{figure}[t] \captionsetup[sub]{font=normalsize}\captionsetup[subfigure]{labelformat=empty}
	\centering 
	\begin{subfigure}[b]{0.48\textwidth}
		\includegraphics[width=1.0\linewidth]{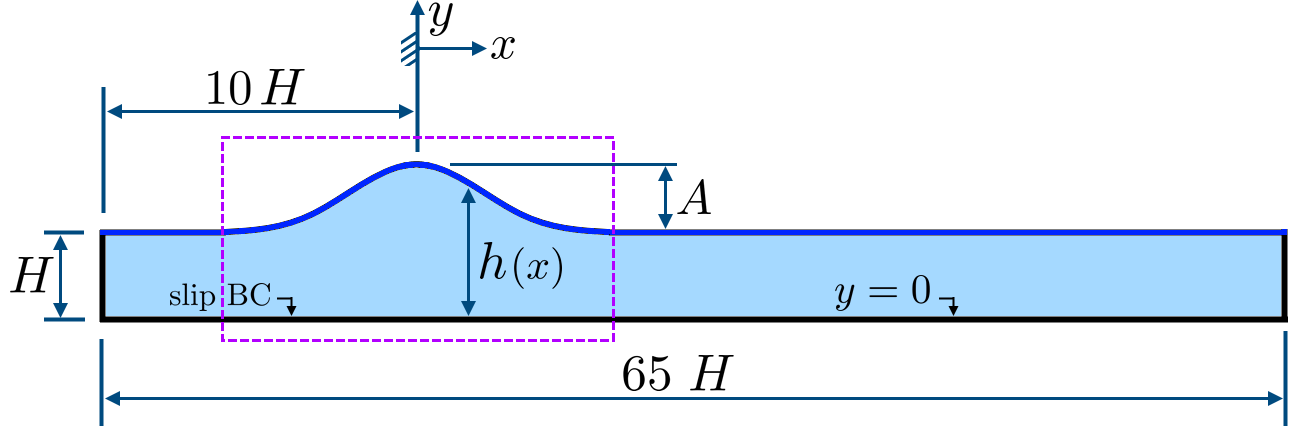}
		\caption{}
		\label{Fig:SingleWavePropagation_TestCase_a}
	\end{subfigure}
	~
	\begin{subfigure}[b]{0.44\textwidth}
		\includegraphics[width=1.0\linewidth]{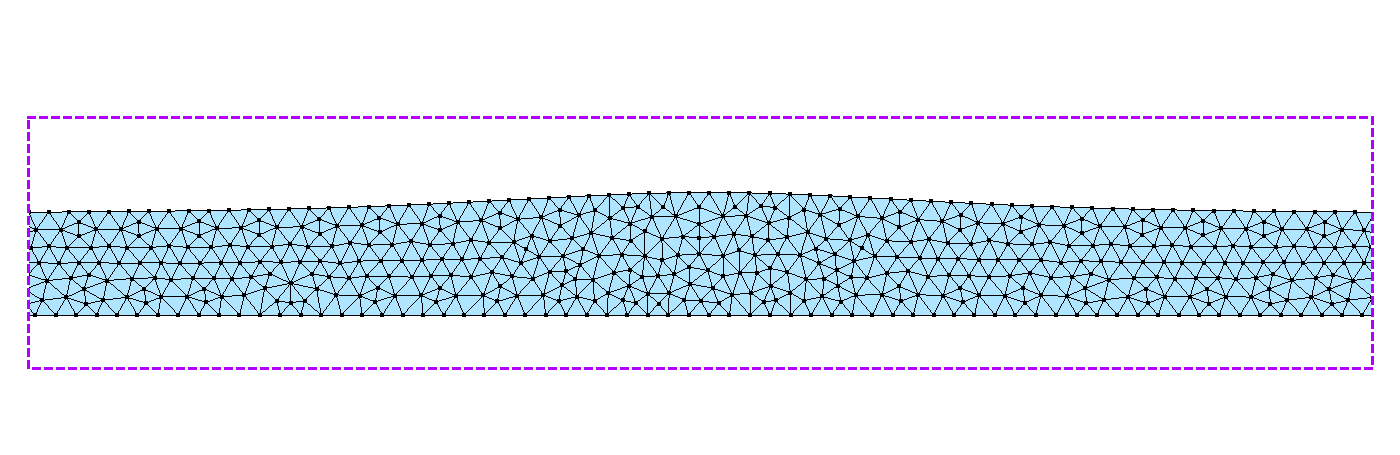}
		\caption{}
		\label{Fig:SingleWavePropagation_TestCase_b}
	\end{subfigure}
	\\
	\vspace{-35mm}
	\hspace{-83mm} (\textbf{a}) \hspace{83mm} (\textbf{b})
	\vspace{22mm}
\caption{Solitary wave propagation in constant water depth. (a) Geometry illustration and (b) initial Finite element discretization around the wave. Subfigure (b) represents a section of the model that is indicated by dashed lines in (a).}
\label{Fig:SingleWavePropagation_TestCase}
\end{figure}

Due to the position of the coordinate system (see Fig.~\ref{Fig:SingleWavePropagation_TestCase_a}), initial condition for velocity, pressure and free surface height is obtained by setting $t=0$ in Eq.~\eqref{EQ:SWP_xi}. This produces a wave displacement to the right, whose position in time is defined by Eq.~\eqref{EQ:SWP_hvp_a}. The chosen fluid has viscosity $\mu$ = 0.001 Pa~s and density $\rho$ = 1000 kg/m$^3$. The container depth is $H$ = 10 m, the amplitude $A$ is 2 m and the gravity acceleration is $g = 9.81 $m/s$^2$. The initial domain is discretized using 2400 particles and 4090 finite elements of 2 m average size, as shown in Fig.~\ref{Fig:SingleWavePropagation_TestCase_b}.

\begin{figure}[t] \captionsetup[sub]{font=normalsize}\captionsetup[subfigure]{labelformat=empty}
	\centering 
	\begin{subfigure}[b]{0.47\textwidth}
		\includegraphics[trim=3 0 32 17,clip,width=1.00\linewidth]{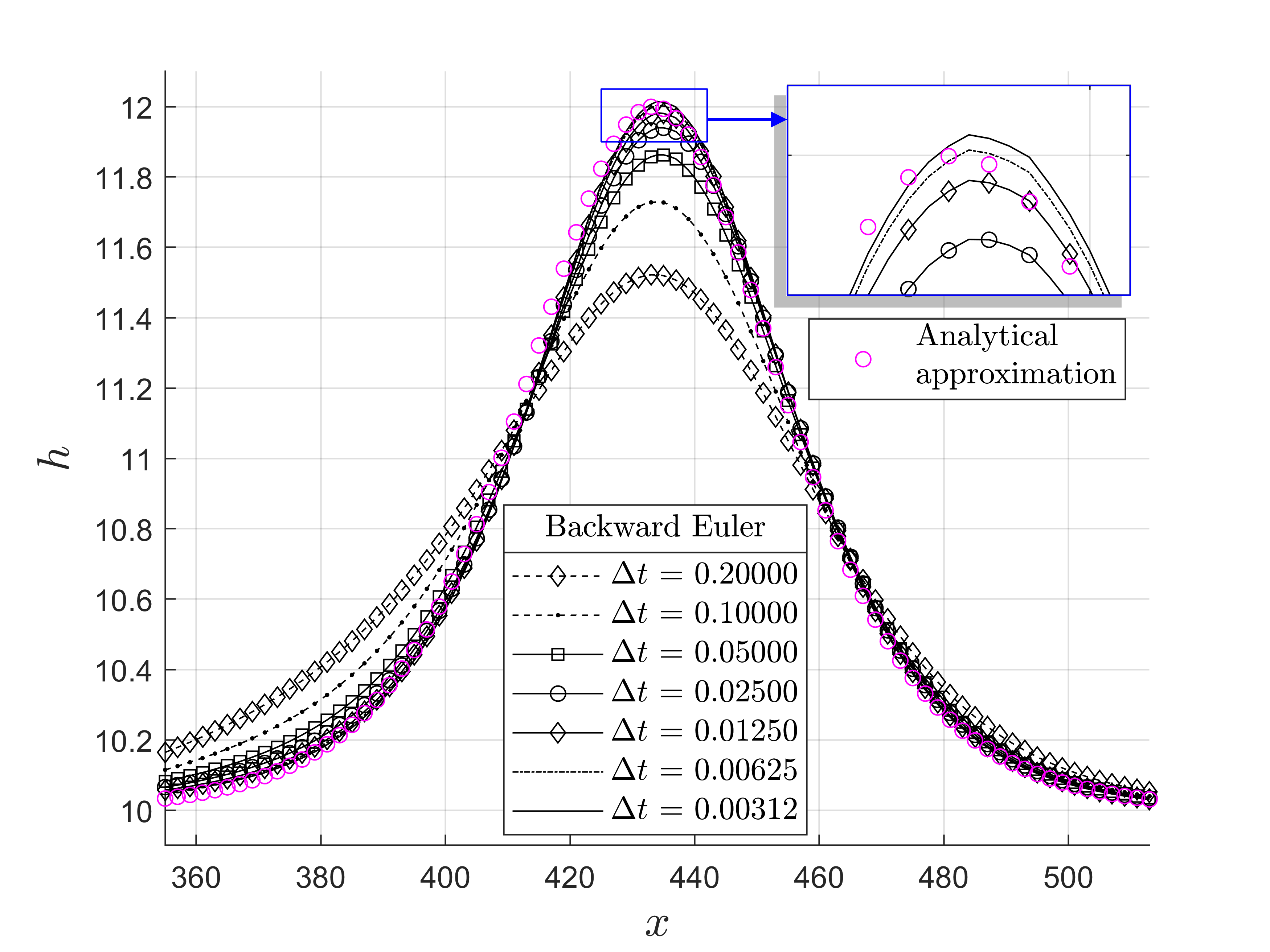}
		\caption{}
		\label{Fig:SWP_time_a}
	\end{subfigure}
	~\hspace{5mm}
	\begin{subfigure}[b]{0.47\textwidth}	
		\includegraphics[trim=3 0 32 17,clip,width=1.00\linewidth]{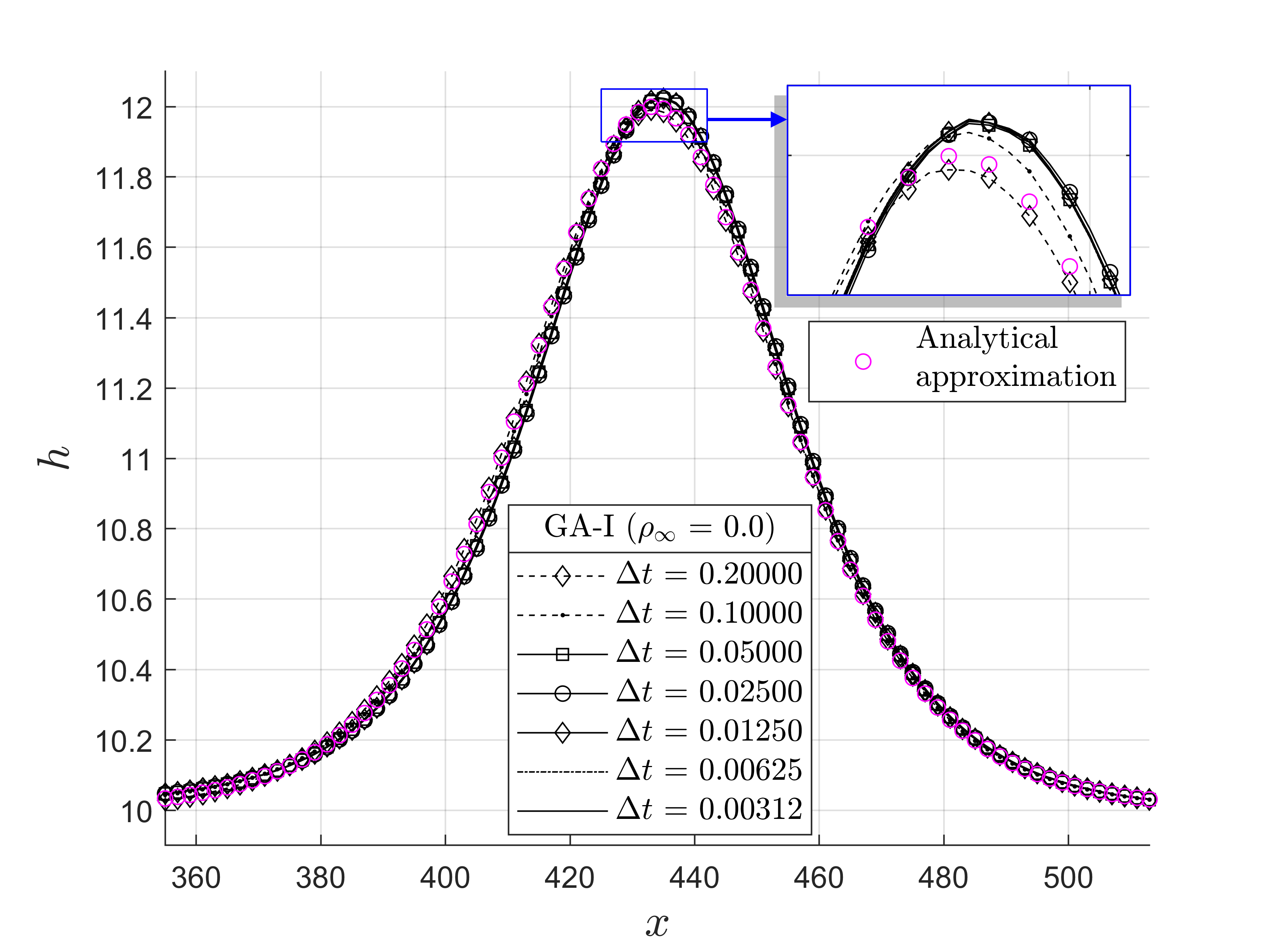}
		\caption{}
		\label{Fig:SWP_time_b}
	\end{subfigure}
	\\
	\vspace{-72mm}
	\hspace{-83mm} (\textbf{a}) \hspace{83mm} (\textbf{b})
	\vspace{59mm}
\caption{Wave profile after 40 s using (a) Backward Euler and (b) Generalized-$\alpha$ ($\GAa$) and different time steps.}
\label{Fig:SWP_time}
\end{figure}

\paragraph{\underline{Performance of time integration schemes}}\vspace{2mm}

Wave profiles after 40 s of simulation are shown in Figs.~\ref{Fig:SWP_time_a} and \ref{Fig:SWP_time_b} using Backward Euler and Generalized-$\alpha$, respectively. Both in a velocity-based formulation. Again, results show a strong dependence of Backward Euler on the time step $\Delta t$, whose excessive numerical damping for large time steps results in a decrease of wave amplitude. Figs.~\ref{Fig:SWP_ERROR_123_a} and \ref{Fig:SWP_time_b} also include the analytical approximation of Eqs.~\eqref{EQ:SWP_hvp}-\eqref{EQ:SWP_c}, which is in good agreement with the simulations. However, the maximum height does not completely match. This is because analytical equations are limited to linear conditions, such as waves with small amplitude $A$. As shown in \citep{staroszczyk2011simulation} and \citep{avancini2020total}, the analytical approximations are rather accurate if $A/H \leq $ 0.1, but less if $A/H \geq$ 0.2. Therefore, for error analysis of time integration schemes, the maximum wave height at 40 seconds of simulation is compared with that obtained from a simulation using a time step $\Delta t = 0.00078125$, the Generalized-$\alpha$ method $(\rho_\infty = 0.0)$ and a discretization with finite elements of 1 m size. The obtained reference height is $h_\mathrm{ref} = $ 12.0255. 

\begin{figure}[t] \captionsetup[sub]{font=normalsize}\captionsetup[subfigure]{labelformat=empty}
\centering
	\begin{subfigure}[b]{0.99\textwidth}
		\includegraphics[width=1.00\linewidth]{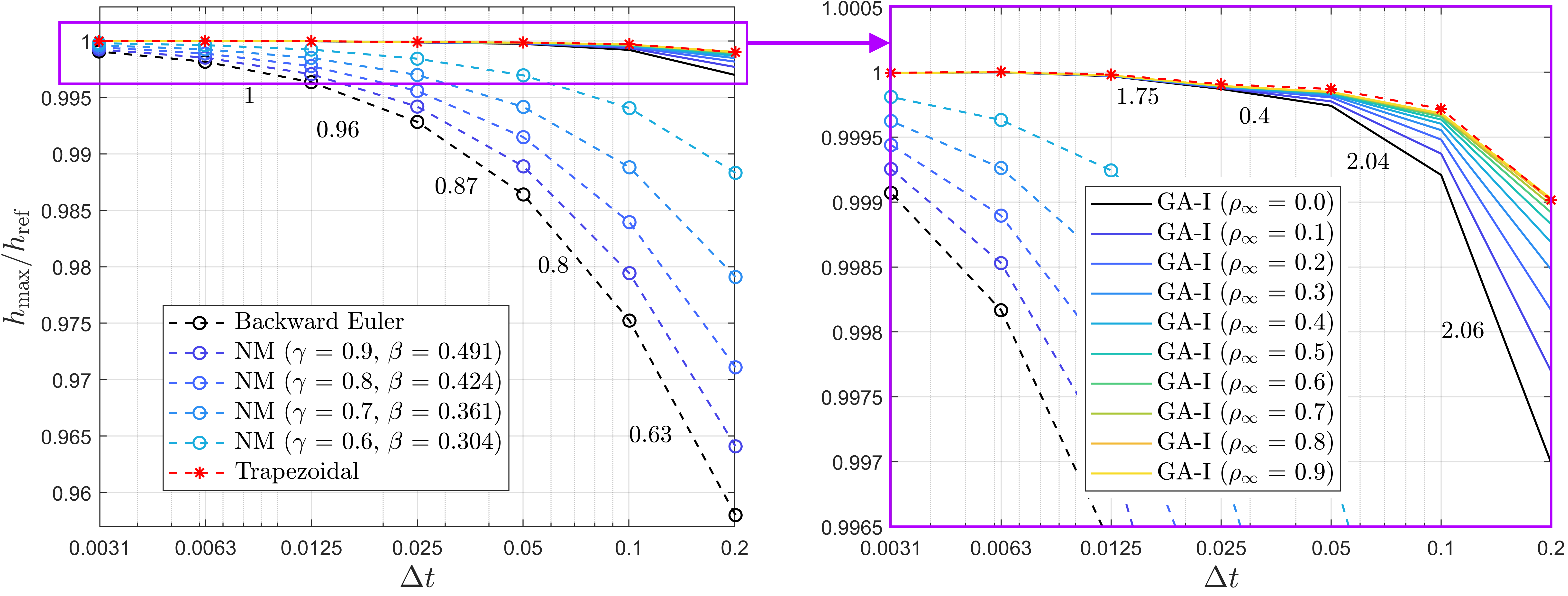}
		\caption{}
		\label{Fig:SWP_ERROR_a}
	\end{subfigure}
	~
	\begin{subfigure}[b]{0.001\textwidth}
		\caption{}
		\label{Fig:SWP_ERROR_b}
	\end{subfigure}
	\\
	\vspace{-83mm}
	\hspace{-83mm} (\textbf{a}) \hspace{81mm} (\textbf{b})
	\vspace{64mm}
\caption{Maximum height of the solitary wave normalized with respect to the reference height. (a) Curves obtained with Backward Euler, Trapezoidal, Newmark and Generalized-$\alpha$ ($\GAa$). Subfigure (b) is an enlarged view of (a). Numbers next to black curves (dashed and solid) indicate the order of convergence of these curves.}
\label{Fig:SWP_ERROR}
\end{figure}

The maximum wave height divided by the reference height is plotted next. The wave height is obtained from simulations using different time integration schemes and time steps. Specifically, seven time steps $\Delta t$ are considered, ranging from 0.003125 to 0.2. The Backward Euler, Trapezoidal, Newmark and Generalized-$\alpha$ schemes are considered. For Newmark, four combinations of parameters are chosen from defining $\gamma$ as 0.9, 0.8, 0.7 and 0.6. For Generalized-$\alpha$, implementation approach $\GAa$ is used and ten combinations of parameters are chosen from defining $\rho_\infty$ as 0.0 to 0.9 with an increment of 0.1.  Thus, 112 solitary wave propagation problems are solved in total. Results are summarized in Fig.~\ref{Fig:SWP_ERROR_a}, which show a pattern similar to the previous example. Backward Euler exhibits the lowest wave height accuracy, which improves as time step is reduced but with an order of convergence close to 1.0 (as shown by numbers next to the black dashed curve in Fig.~\ref{Fig:SWP_ERROR_a}). The Trapezoidal scheme is the most accurate while Newmark's Method (NM) interpolates Backward Euler and Trapezoidal accuracy. On the other hand, the Generalized-$\alpha$ method shows wave heights very close to the reference value, which gathers the GA-I curves in the upper zone of Fig.~\ref{Fig:SWP_ERROR_a}. For better visualization, such a zone is enlarged and is shown in Fig.~\ref{Fig:SWP_ERROR_b}. Numbers next to the black curve indicate the order of convergence for $\rho_\infty$ = 0.0. It can be seen that for large time steps ($\Delta t \geq 0.05$), the maximum wave height converges with 2nd order with respect to $\Delta t$. For smaller time steps, the order of convergence is reduced. This is attributed to a spatial discretization error resulting from a coarse discretization, which was chosen to extend the range of values of $\Delta t$ without compromising convergence of the nonlinear algorithm. Space discretization error becomes predominant as the maximum height of the simulation approaches the reference value, which is reflected in the Generalized-$\alpha$ and Trapezoidal curves when $\Delta t \leq 0.05$.

\paragraph{\underline{Implementation approaches of the Generalized-$\alpha$ method}}\vspace{2mm}

The next analysis incorporates results obtained with $\GAa$, $\GAb$ and $\GAc$, i.e., it compares the Generalized-$\alpha$ implementation approaches that write the pressure at $t_{n+\alpha_f}$. The same 7 time step sizes and 10 values of $\rho_\infty$ of previous analysis are considered. Results are summarized in Fig.~\ref{Fig:SWP_ERROR_123_a}. Color scale indicates the value of $\rho_\infty$ while marker type distinguishes implementation schemes. The relative difference between the 3 schemes is shown in Fig.~\ref{Fig:SWP_ERROR_123_b}. There, percentage difference is computed with respect to the scheme $\GAa$, taken as a reference.

It is observed that for $\rho_\infty$ = 0.0, all three schemes produce the same result when $\Delta t \leq 0.1$ (difference of $\approx$ 0$\%$ in Fig.~\ref{Fig:SWP_ERROR_123_b}, black color). In addition, the $\GAa$ and $\GAb$ schemes are coincident for all $\rho_\infty$ (all circle markers are at 0$\%$ in Fig.~\ref{Fig:SWP_ERROR_123_b}). Thus, at least in this problem, the algebraic treatment of the linear interpolation of state variables (Eq.~\ref{EQ:StateVariables_AlphaMethod_all}) is not relevant to the results. In case of $\GAc$, a difference with respect to GA-I is observed. The biggest difference is observed for $\Delta t$ = 0.2 and $\rho_\infty \geq 0.7$, which is $\approx$ 0.007 $\%$. The difference between GA-I and GA-II reduces rapidly when decreasing $\rho_\infty$ or the time step $\Delta t$, and becomes negligible for $\Delta t \leq 0.05$. This observation is in line with previous comments of section (Eq.~\ref{EQ:GAI_GAII_K}), where it is stipulated that GA-I and GA-II should converge to the same result in the presence of small deformations or small time steps if mass conservation holds.

\begin{figure}[t]\captionsetup[sub]{font=normalsize}\captionsetup[subfigure]{labelformat=empty}
\centering 
	\begin{subfigure}[b]{0.48\textwidth}
		\includegraphics[trim=0 0 36 10,clip,width=1.00\linewidth]{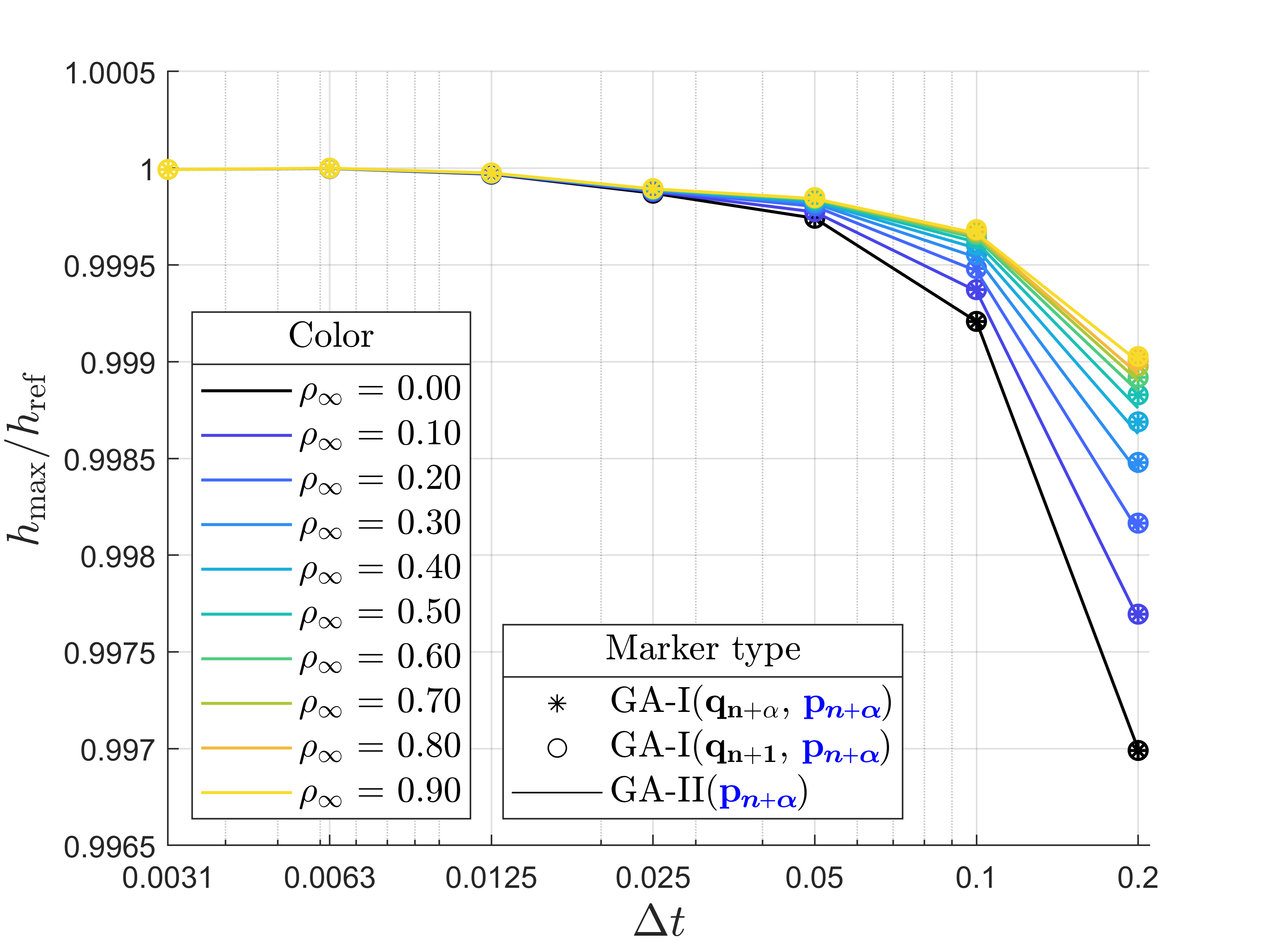}
		\caption{}
		\label{Fig:SWP_ERROR_123_a}
	\end{subfigure}
	~\hspace{5mm}
	\begin{subfigure}[b]{0.48\textwidth}	
		\includegraphics[trim=0 0 36 10,clip,width=1.00\linewidth]{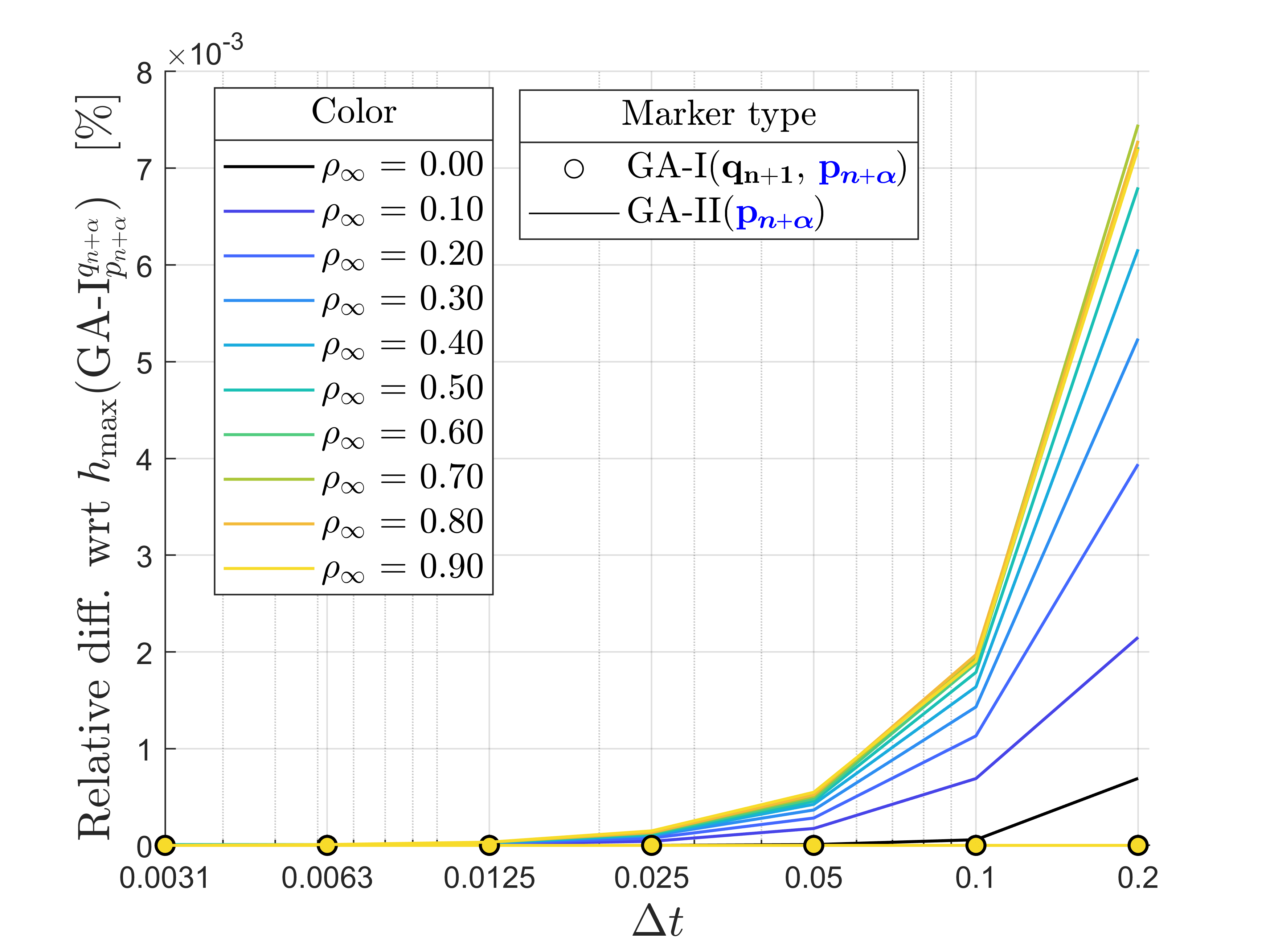}
		\caption{}
		\label{Fig:SWP_ERROR_123_b}
	\end{subfigure}
	\\
	\vspace{-71mm}
	\hspace{-83mm} (\textbf{a}) \hspace{83mm} (\textbf{b})
	\vspace{58mm}
\caption{(a) Normalized maximum height of the solitary wave using different implementation approaches of the Generalized-$\alpha$ method. (b) Relative difference between approaches, with $\GAa$ used as reference.}
\label{Fig:SWP_ERROR_123}
\end{figure}

Next, results from implementation approaches of the Generalized-$\alpha$ that write the pressure at $t_{n+1}$ instead of $t_{n+\alpha_f}$ are presented. The three implementation alternatives are considered, $\GAd$, $\GAe$ and $\GAf$. In addition, two scenarios are taken into account, one that integrates the gradient matrix at $t_{n+1}$ (as in Eq.~\ref{EQ:MomentumEquation_AlphaMethod_noPre}) and another that integrates it at $t_{n+\alpha}$ (as in Eq.~Eq.~\ref{EQ:MomentumEquation_AlphaMethod_noPre_2}). Results are shown in Figs.~\ref{Fig:SWP_ERROR_456789_a} and \ref{Fig:SWP_ERROR_456789_b}, respectively (plots appear on different scales). Figures also include scheme $\GAa$ that follows the rationale of the generalized alpha scheme and writes the pressure at $t_{n+\alpha_f}$. 

Results reveal that there is no significant difference between the GA-I and GA-II implementation approaches that write the pressure in $t_{n+1}$, since in both figures, $\GAd$, $\GAe$ and $\GAf$ are coincident. However, there is a big difference for when the gradient matrix is integrated. If $\mathbf{D}^\intercal_{n+1}$ is used (Fig.~\ref{Fig:SWP_ERROR_456789_a}), then performance of Generalized-$\alpha$ is reduced, especially for large values of $\rho_\infty$ since performance becomes similar to that of Backward Euler. In contrast, if $\mathbf{D}_{n+\alpha_f}^\intercal$ is used (Fig.~\ref{Fig:SWP_ERROR_456789_b}), then no detriment is seen on the time integration scheme, regardless of the value of $\rho_\infty$ and the implementation approach ($\GAd$, $\GAe$ or $\GAf$). Remarkably, all implementation schemes lead to the same result if $\rho_\infty$ = 0.

\begin{figure}[t]\captionsetup[sub]{font=normalsize}\captionsetup[subfigure]{labelformat=empty}
\centering 
	\begin{subfigure}[b]{0.48\textwidth}
		\includegraphics[trim=3 0 32 10,clip,width=1.00\linewidth]{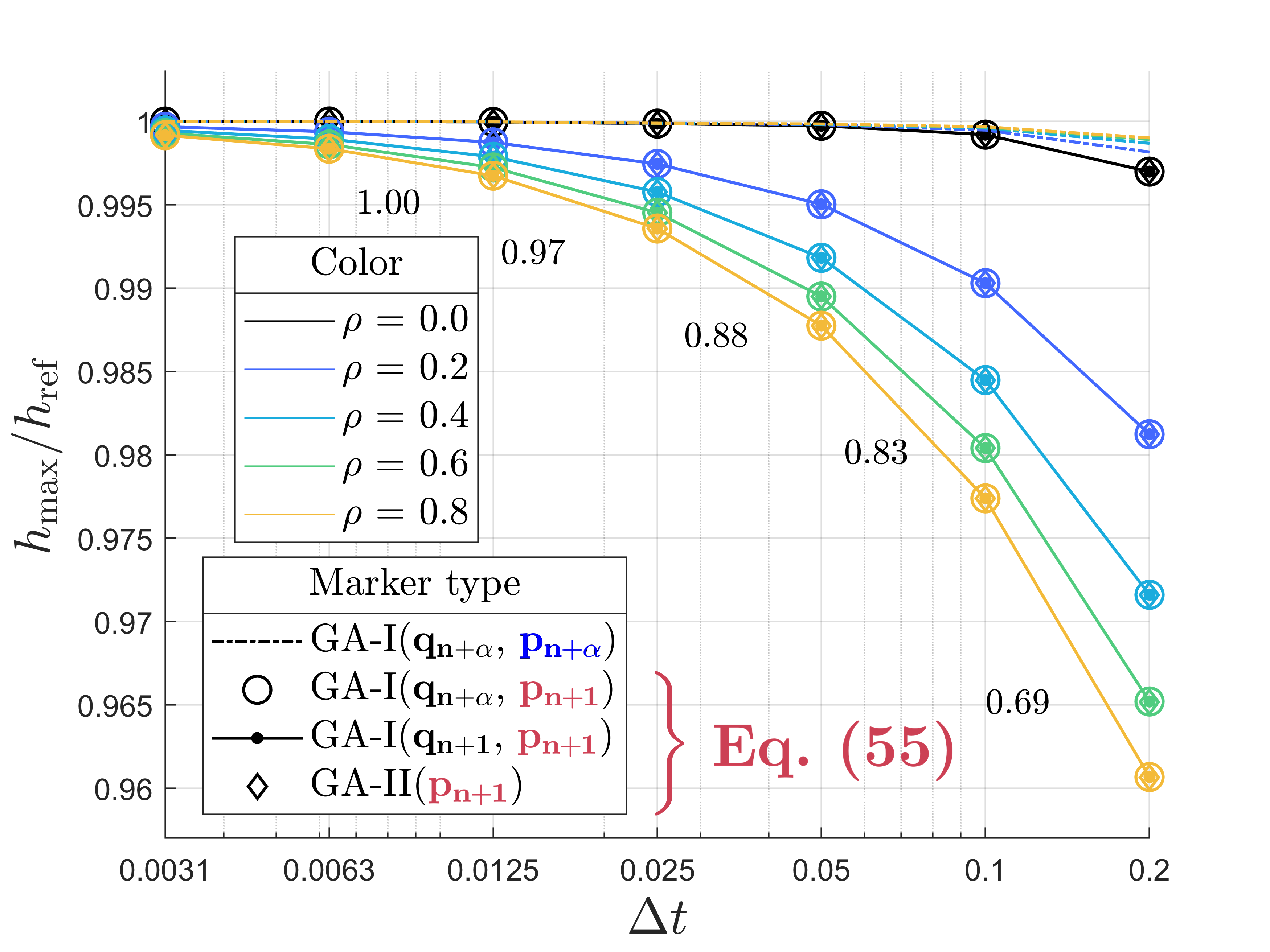}
		\caption{}
		\label{Fig:SWP_ERROR_456789_a}
	\end{subfigure}
	~\hspace{5mm}
	\begin{subfigure}[b]{0.48\textwidth}	
		\includegraphics[trim=3 0 32 30,clip,width=1.00\linewidth]{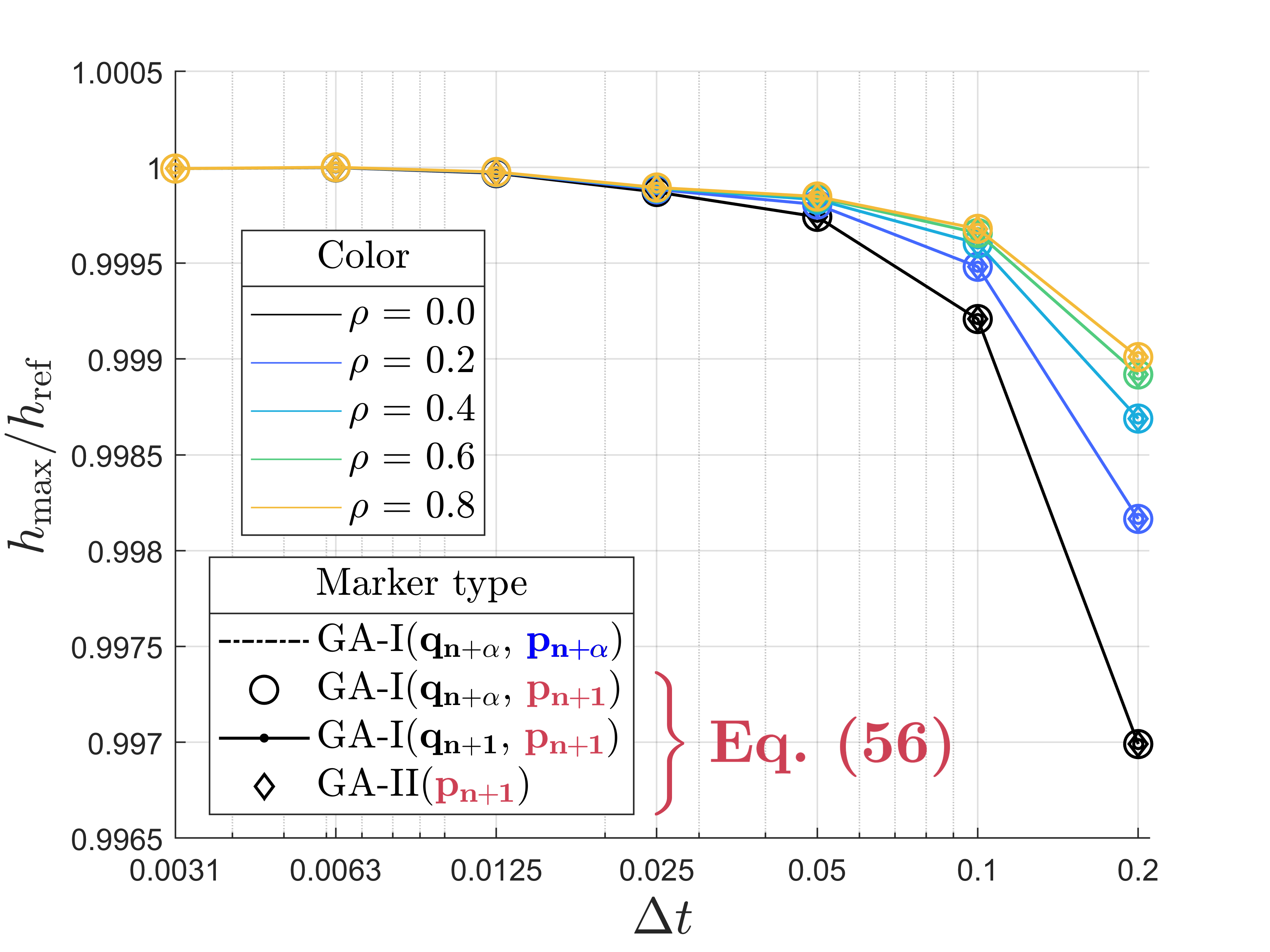}
		\caption{}
		\label{Fig:SWP_ERROR_456789_b}
	\end{subfigure}
	\\
	\vspace{-71mm}
	\hspace{-83mm} (\textbf{a}) \hspace{83mm} (\textbf{b})
	\vspace{58mm}
\caption{Normalized maximum height of the solitary wave. Subfigures compare implementation approaches of the Generalized-$\alpha$ method that write pressure at $t_{n+1}$. Schemes $\GAd$, $\GAe$ and $\GAf$ compute the gradient matrix $\mathbf{D}^\intercal$ in (a) $\Omega(t_{n+1})$ and (b) $\Omega(t_{n+\alpha_f})$.}
\label{Fig:SWP_ERROR_1456789}
\end{figure}

\paragraph{\underline{Velocity-based and Displacement-based formulations}}\vspace{2mm}

A final analysis using the solitary wave propagation problem compares results between velocity-based and displacement-based formulations. This is conducted for Backward Euler, Newmark and Generalized-$\alpha$ ($\GAa$). To avoid overloading graph results, only two sets of parameters are used for Newmark ($\gamma$ equal to 0.8 and 0.6), and five for Generalized-$\alpha$ ($\rho_\infty$ equal to 0.0, 0.2, 0.4, 0.6, and 0.8). Results are given in Fig.~\ref{Fig:SWP_ERROR_VP_DP}. It is observed that both formulations lead to practically the same results. That is, the change of unknown variable in the system of equations does not seem to have an impact on the accuracy of results. Furthermore, we did not observe significant changes in computation time between the two formulations (although computation time was not precisely tracked in the simulations) nor in the number of iterations used by the nonlinear algorithm. Therefore, besides adding additional terms on the right-hand side of the system of equations, it seems that there is no significant computational impact from switching to a displacement-pressure formulation.

Observations up to this point have been obtained from two problems that feature small deformations and exhibit quasi linear behavior. Furthermore, they are analyzed from a kinematic point of view while pressure has been excluded from the analysis. Therefore, the following example of the flow around a cylinder is chosen due to the presence of large velocity gradients and because comparison criterion is based on the force exerted on the cylinder, which is highly dependent on the pressure field.

\begin{figure}[t]
\centering 
\includegraphics[width=0.80\linewidth]{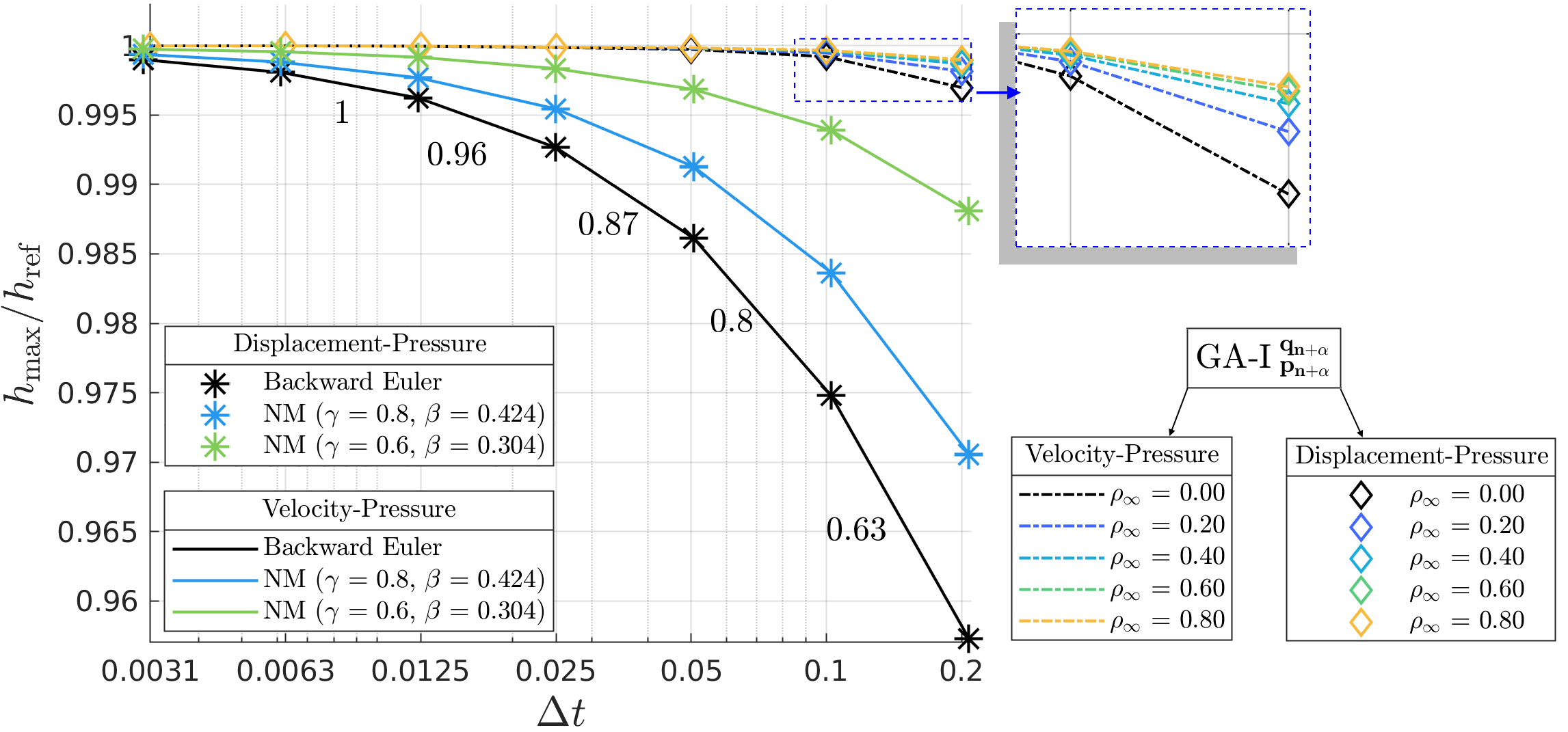}
\caption{Normalized maximum wave height using Backward Euler, Newmark and Generalized-$\alpha$ in velocity-based and displacement-based formulations.}
\label{Fig:SWP_ERROR_VP_DP}
\end{figure}

\subsection{Flow around a cylinder}\label{sec:FAC}

This example involves the flow past a circular cylinder. The computational domain is designed as in Idelsohn et.al\citep{idelsohn2012large, idelsohn2013fast}. Therefore, fluid particles cross a rectangular computational domain, as illustrated in Fig.~\ref{Fig:flowArroundCylinder_TestCase_a}. A horizontal velocity $\vel_0$ is imposed on particles located at the left, top and bottom boundaries. Particles that cross the right boundary are deleted from the model. A rigid cylinder of 1 m diameter is spaced 5 m from the left, top and bottom boundary, which leaves the cylinder at 15 m from the right boundary. 

\begin{figure}[t] \captionsetup[sub]{font=normalsize}\captionsetup[subfigure]{labelformat=empty}
\centering 
	\begin{subfigure}[b]{0.47\textwidth}
		\includegraphics[width=1.00\linewidth]{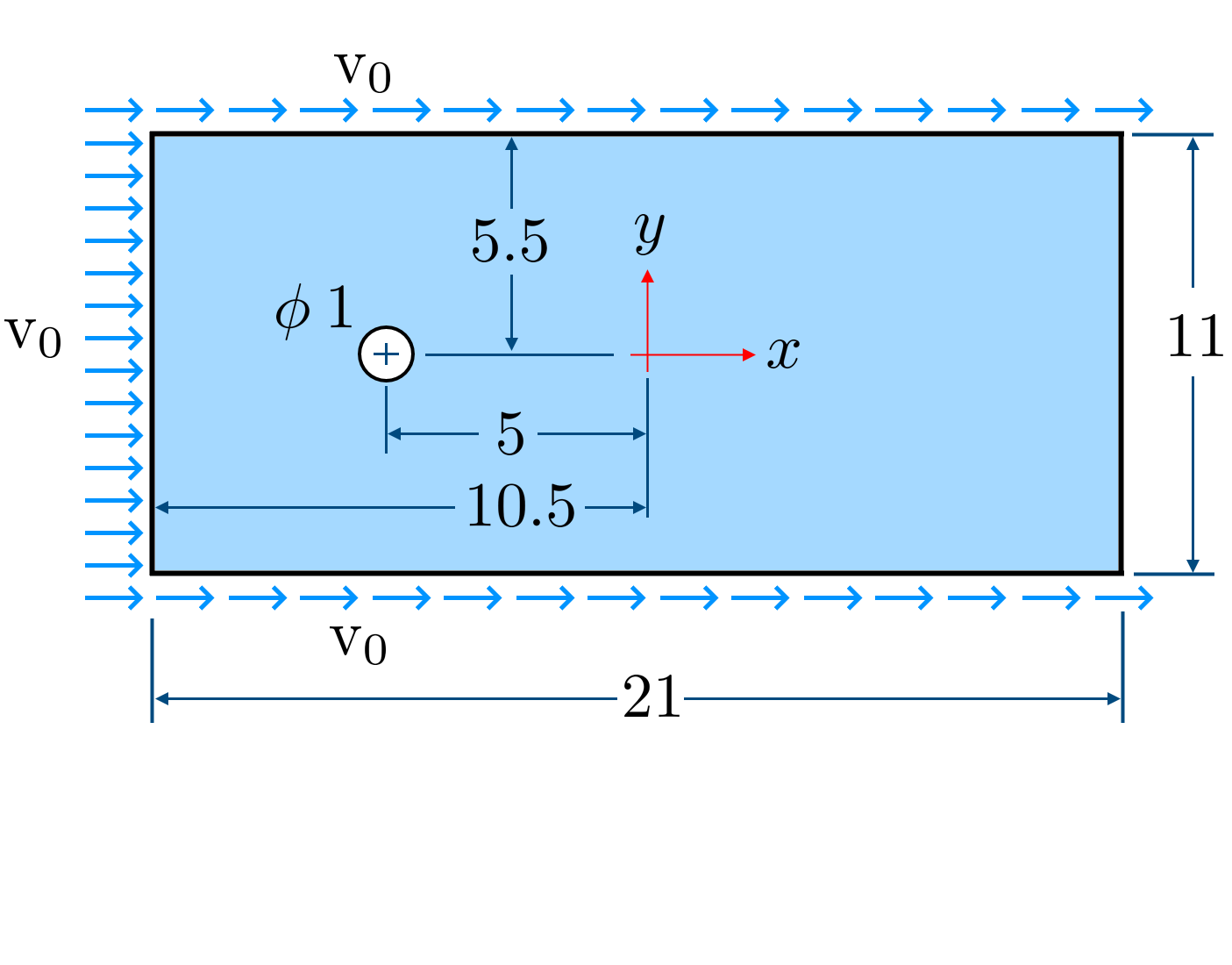}
		\caption{}
		\label{Fig:flowArroundCylinder_TestCase_a}
	\end{subfigure}
	~\hspace{5mm}
	\begin{subfigure}[b]{0.47\textwidth}	
		\includegraphics[width=0.95\linewidth]{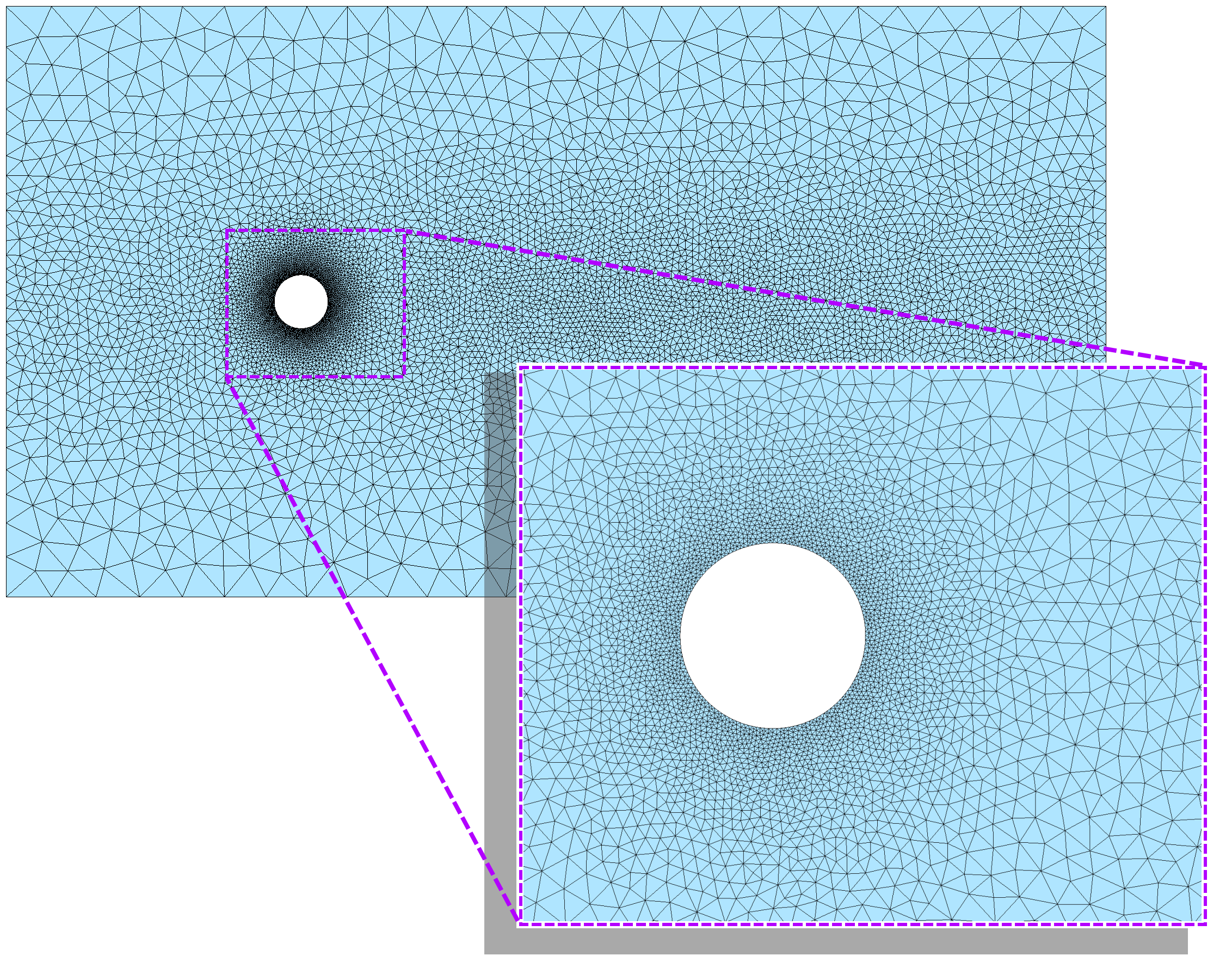}
		\caption{}
		\label{Fig:flowArroundCylinder_TestCase_b}
	\end{subfigure}	
	\\
	\vspace{-72mm}
	\hspace{-88mm} (\textbf{a}) \hspace{75mm} (\textbf{b})
	\vspace{58mm}
\caption{Flow around a cylinder. (a) Geometry and (b) initial finite element discretization.}
\label{Fig:flowArroundCylinder_TestCase}
\end{figure}

Results for two Reynolds (Re) numbers are reported in this example, 100 and 1000. The first is included in the study of Dettmer et.al.\citep{dettmer2003analysis}, who solve the flow around a cylinder problem using the Generalized-$\alpha$ scheme with various values of $\rho_\infty$ in an Eulerian framework. The second, Re = 1000, is included for comparing results with Idelsohn et al.\citep{idelsohn2012large, idelsohn2013fast} who are, to date, the only ones to present results for the flow around a cylinder using PFEM, to the best of our knowledge.  

Reynolds numbers 100 and 1000 are imposed by defining the viscosity $\mu$ as 0.01 and 0.001, respectively. Whereas, fluid density and inlet velocity are set to $\rho = 1$ kg/m$^3$ and $\vel_0 = 1$ m/s, respectively. Problems with Reynolds 100 and 1000 are simulated for 160 and 90 s of physical time, respectively. For both problems, a non-uniform mesh size is used, which is defined by the following function: 

\begin{equation}\label{EQ:hchar_FAC}
 h_\mathrm{elem} = \mathrm{min} 
 \left[ 
 \left( 0.05 \: (x + 5)^2 + 0.05 \: y^2 + 0.01 \right)
 \: , \:
 \left( 0.001\:(x - 1)^2 + 0.02*y^2 + 0.12 \right)
 \right]
\end{equation}

\noindent where $h_\mathrm{elem}$ is the element size at coordinates $x$ and $y$. The adopted mesh generator (Gmsh\citep{geuzaine2009gmsh}) produces 8018 particles with Eq.~\eqref{EQ:hchar_FAC}, places 140 on the cylinder surface and generates 15777 finite elements in the initial computational domain, as shown in Fig.~\ref{Fig:flowArroundCylinder_TestCase_b}. During PFEM simulation, each remeshing process ensures that spatial distribution of particles agrees with Eq.~\eqref{EQ:hchar_FAC}, otherwise particles are added or removed. 

For this example, the Strouhal number, the amplitude of the lift coefficient and the mean value of the drag coefficient are reported. These are respectively denoted as $S_t$, Amp($C_L$) and Mean($C_D$). The Strouhal number ($S_t$), lift ($C_L$) and drag ($C_D$) coefficients are defined as:
\begin{equation}
S_t = \frac{f_L \: D}{\vel_0} 
\;\;\;\; , \;\;\;\;\;
C_L(t) = \frac{F_\mathrm{lift}(t)}{\frac{1}{2} \rho \: \vel_0^2 \: D}
\;\;\;\; , \;\;\;\;\;
C_D(t) = \frac{F_\mathrm{drag}(t)}{\frac{1}{2} \rho \: \vel_0^2 \: D} 
\end{equation}
\vspace{1mm}

\noindent where $F_\mathrm{lift}$ and $F_\mathrm{drag}$ are the lift and drag forces, $D$ is the cylinder diameter (equal to 1 m) and $f_L$ is the frequency of the lift force. Typical lift and drag coefficients curves, for problems with Reynolds 100, are shown in Figs.~\ref{Fig:FAC_Coeffs_a} and \ref{Fig:FAC_Coeffs_b}, respectively. Curves for problems with Reynolds 1000 are shown in Figs.~\ref{Fig:FAC_Coeffs_c} and \ref{Fig:FAC_Coeffs_d}. The frequency of the lift force $f_L$, equivalent to the frequency of the lift coefficient, is obtained using a Fast Fourier Transform (FFT) on signal segments shown in Figs.~\ref{Fig:FAC_Coeffs_a} and \ref{Fig:FAC_Coeffs_c}.

\begin{figure}[t] \captionsetup[sub]{font=normalsize}\captionsetup[subfigure]{labelformat=empty}
\centering 
	\begin{subfigure}[b]{0.47\textwidth}
		\includegraphics[trim=15 55 30 70,clip,width=1.00\linewidth]{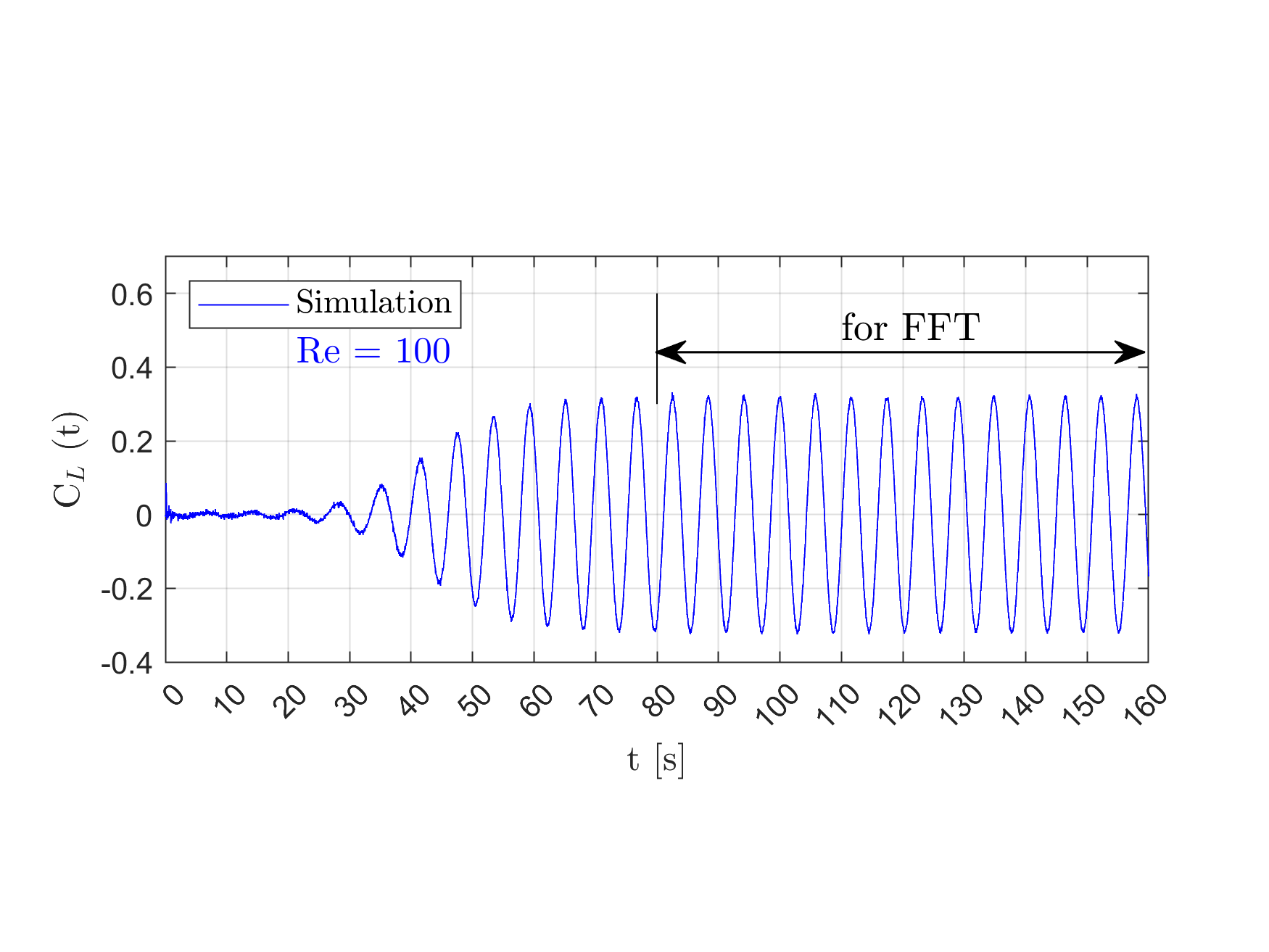}
		\caption{}
		\label{Fig:FAC_Coeffs_a}
	\end{subfigure}
	~
	\begin{subfigure}[b]{0.47\textwidth}	
		\includegraphics[trim=15 52 32 70,clip,width=0.96\linewidth]{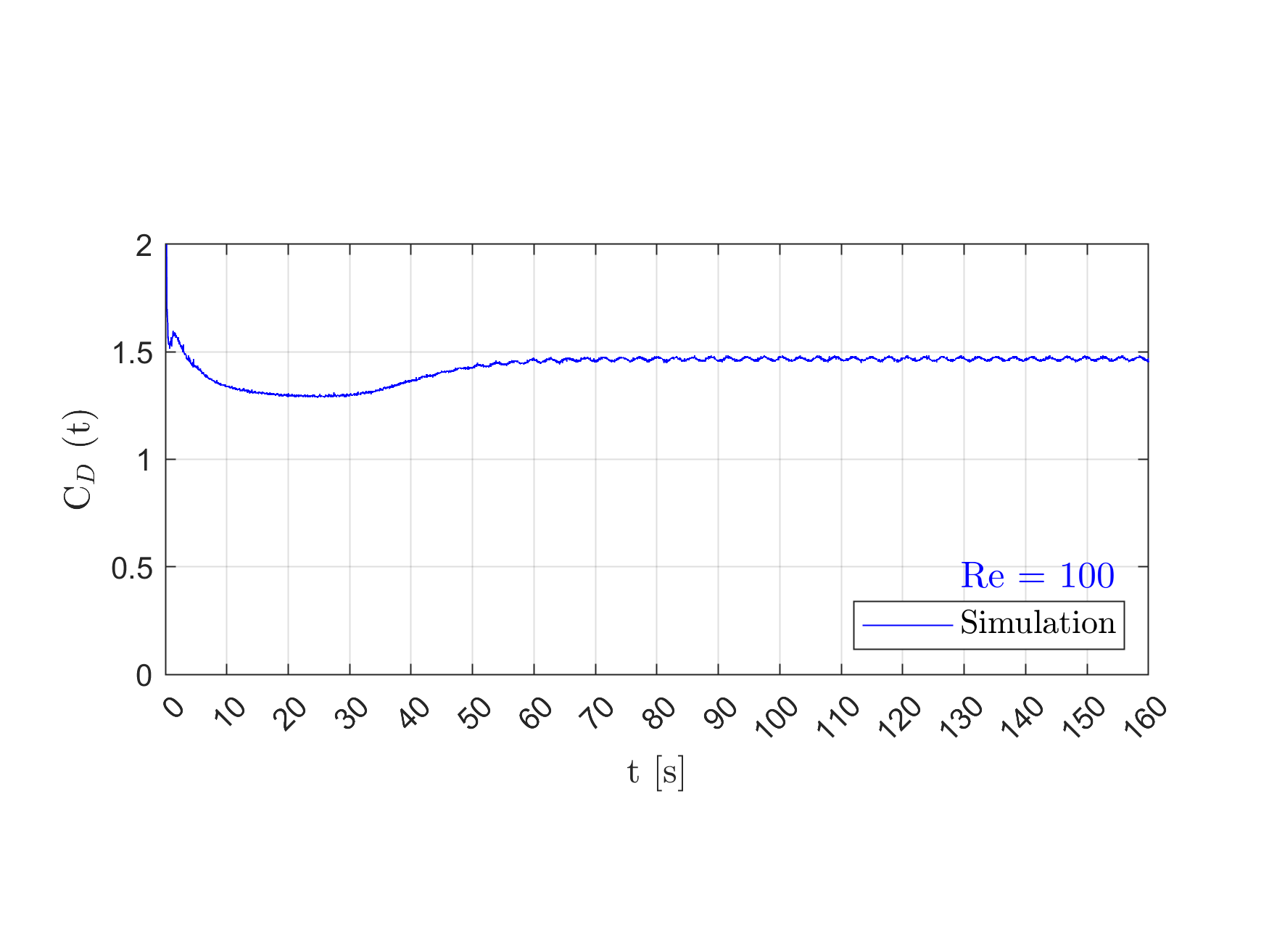}
		\caption{}
		\label{Fig:FAC_Coeffs_b}
	\end{subfigure}	
	\\[-3ex]
	\begin{subfigure}[b]{0.47\textwidth}
		\includegraphics[trim=15 55 30 70,clip,width=1.00\linewidth]{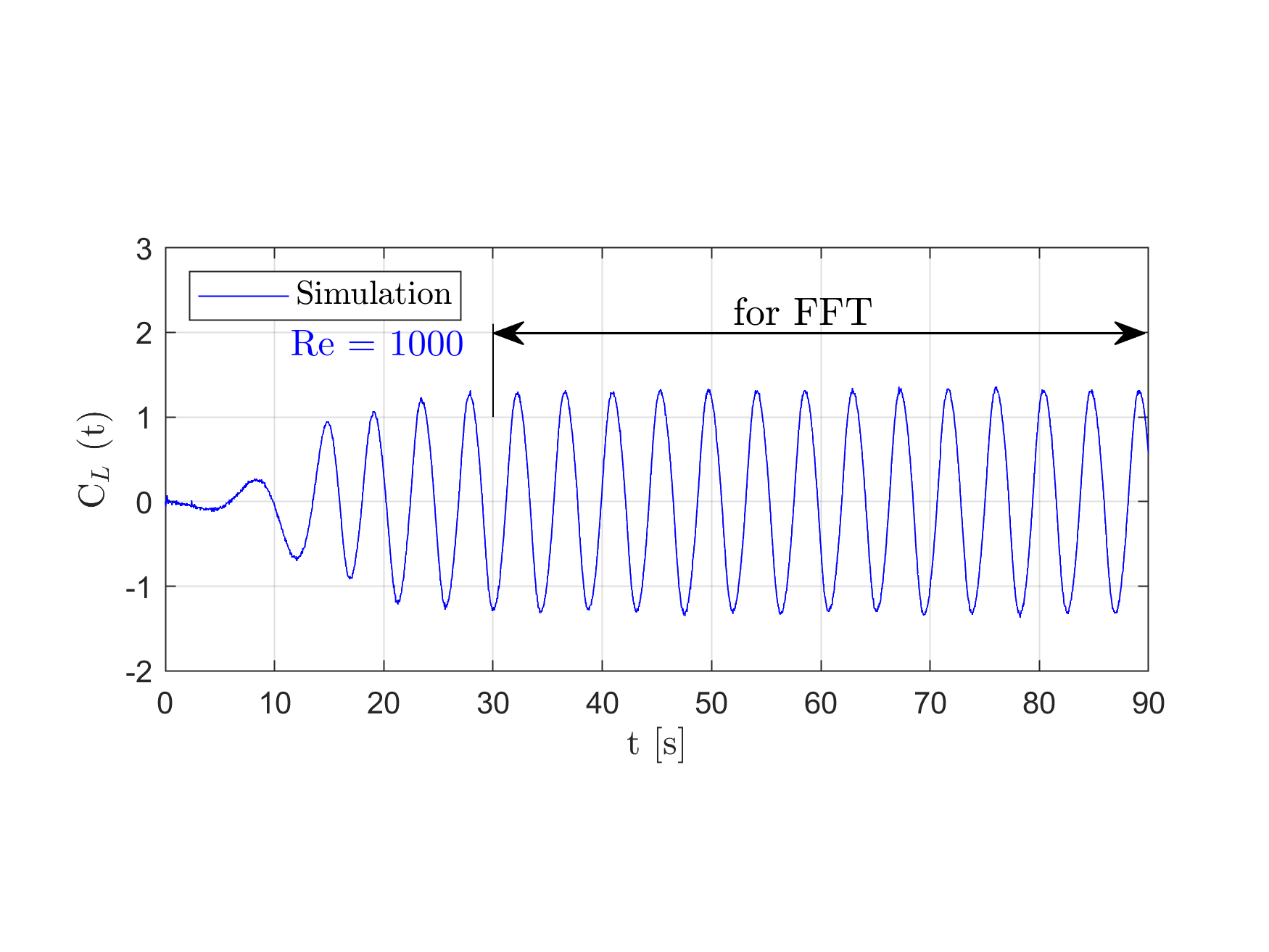}
		\caption{}
		\label{Fig:FAC_Coeffs_c}
	\end{subfigure}
	~
	\begin{subfigure}[b]{0.47\textwidth}	
		\includegraphics[trim=15 52 32 70,clip,width=0.98\linewidth]{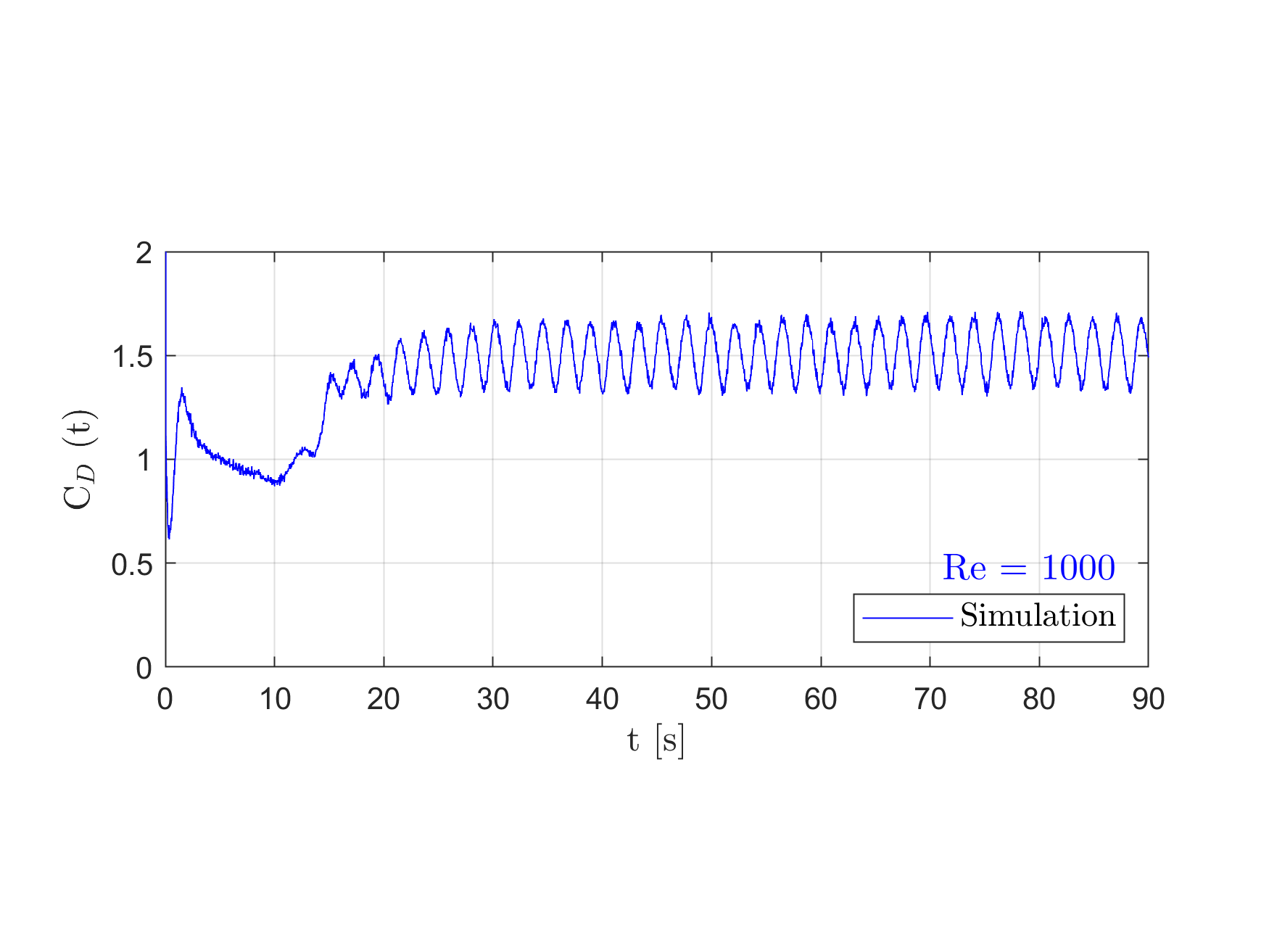}
		\caption{}
		\label{Fig:FAC_Coeffs_d}
	\end{subfigure}	
	\\
	\vspace{-94mm}
	\hspace{-80mm} (\textbf{a}) \hspace{81mm} (\textbf{b})
	\\
	\vspace{40mm}
	\hspace{-80mm} (\textbf{c}) \hspace{81mm} (\textbf{d})
	\\
	\vspace{35mm}
\caption{Simulation of the flow around a cylinder for Reynolds 100 (subfigures a,b) and 1000 (subfigures c,d). Evolution of the lift (subfigures a,c) and drag (subfigures b,d) coefficients. Results obtained with Backward Euler and $\Delta t = 0.016$.}
\label{Fig:FAC_Coeffs}
\end{figure}

\paragraph{\underline{Verification against the literature}}\vspace{2mm}

Several problems of the flow around a cylinder with variations in the time integration scheme and time step are solved. Specifically, for problems with Re = 100, six time steps are considered ranging from 0.002 to 0.064. Backward Euler, Trapezoidal, Newmark and Generalized-$\alpha$ are used as time integration schemes. For Newmark, four combinations of parameters are used by defining $\gamma$ as 0.9, 0.8, 0.7 and 0.6. For Generalized-$\alpha$, implementation $\GAa$ is used and six values of $\rho_\infty$, which are 0.0, 0.1, 0.2, 0.5, 0.7, and 0.9. Therefore, 72 problems of the flow around a cylinder are solved for Re = 100. The coefficients $S_t$, Amp($C_L$) and Mean($C_D$) are summarized in Fig.~\ref{Fig:FAC_Results_Re100}. On the left side (Figs.~\ref{Fig:FAC_Results_Re100_a}, \ref{Fig:FAC_Results_Re100_c} and \ref{Fig:FAC_Results_Re100_e}) are results with Backward Euler, Newmark's Method (NM), and Trapezoidal rule, while on the right side (Figs.~\ref{Fig:FAC_Results_Re100_b}, \ref{Fig:FAC_Results_Re100_d} and \ref{Fig:FAC_Results_Re100_f}) are those obtained with Generalized-$\alpha$. In addition, some references are included in the graphs, which appear as horizontal lines for ease of comparison, but it does not indicate that references are insensitive to time step $\Delta t$.

\begin{figure}[t!] \captionsetup[sub]{font=normalsize}\captionsetup[subfigure]{labelformat=empty}
\centering 
	\begin{subfigure}[b]{0.46\textwidth}
		\includegraphics[trim=0 0 32 25,clip,width=1.00\linewidth]{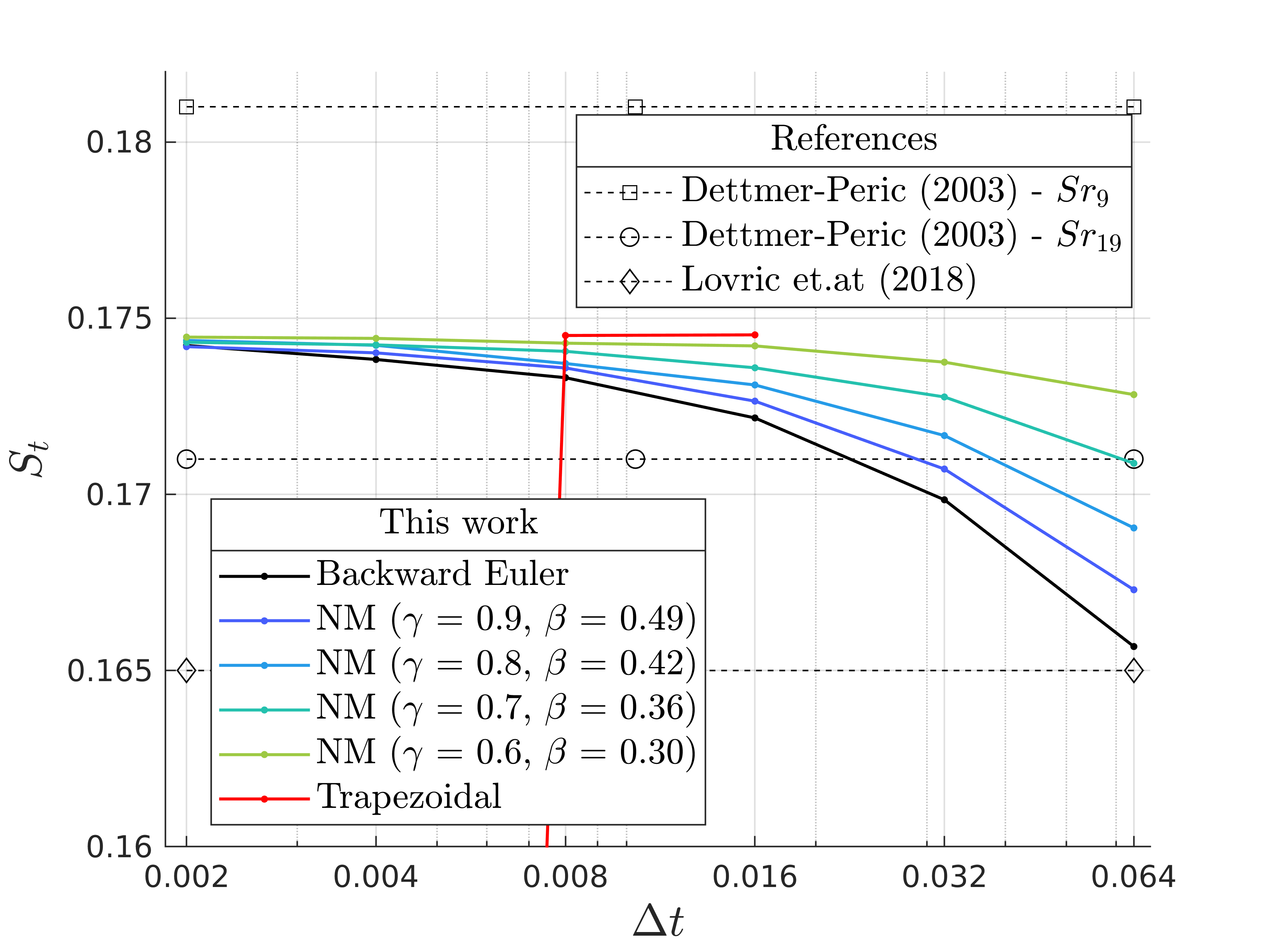}
		\caption{}
		\label{Fig:FAC_Results_Re100_a}
	\end{subfigure}
	~\hspace{2mm}
	\begin{subfigure}[b]{0.46\textwidth}	
		\includegraphics[trim=0 0 32 25,clip,width=1.00\linewidth]{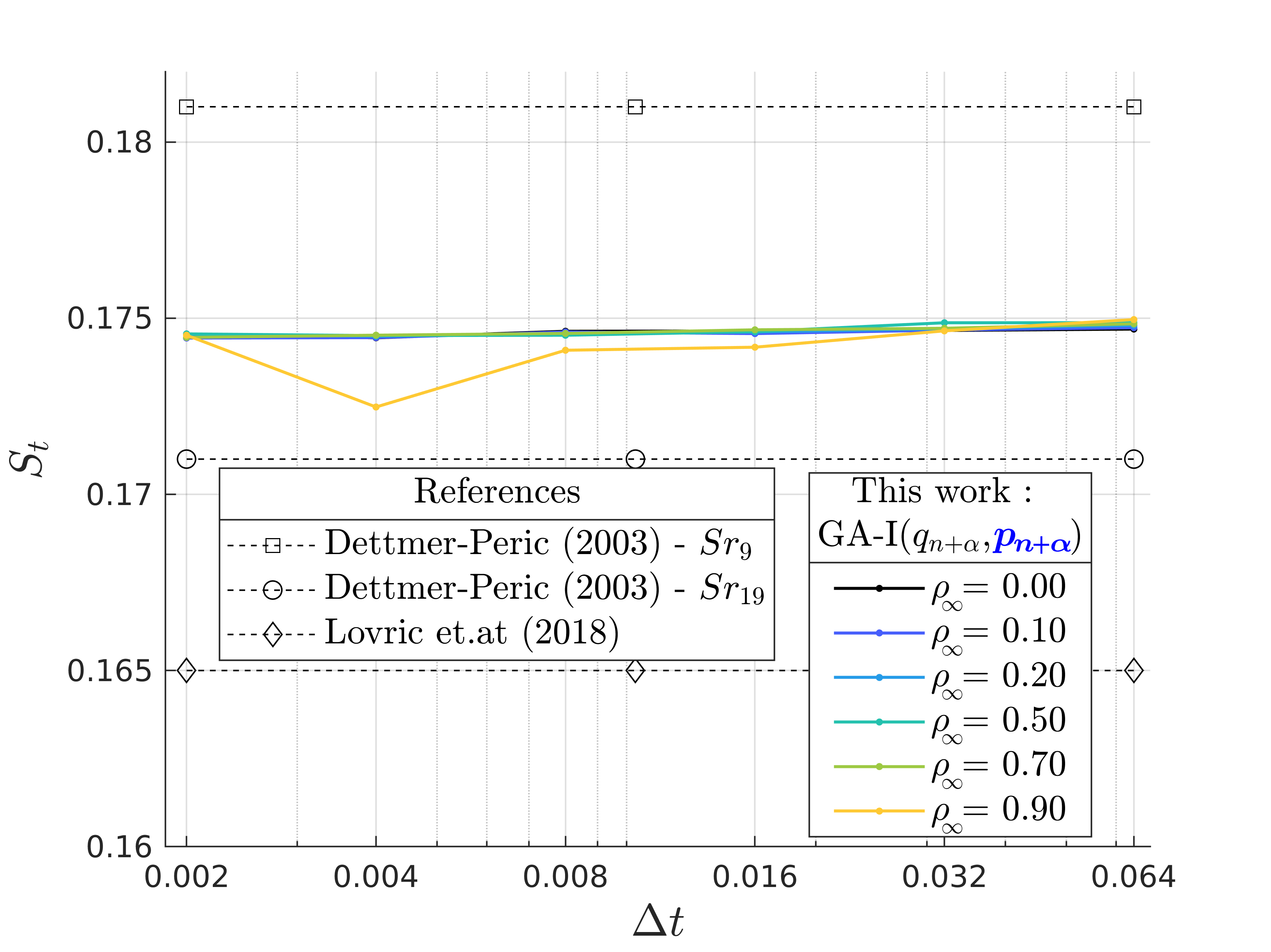}
		\caption{}
		\label{Fig:FAC_Results_Re100_b}
	\end{subfigure}	
	\\[-4ex]
	\begin{subfigure}[b]{0.46\textwidth}
		\includegraphics[trim=0 0 32 17,clip,width=1.00\linewidth]{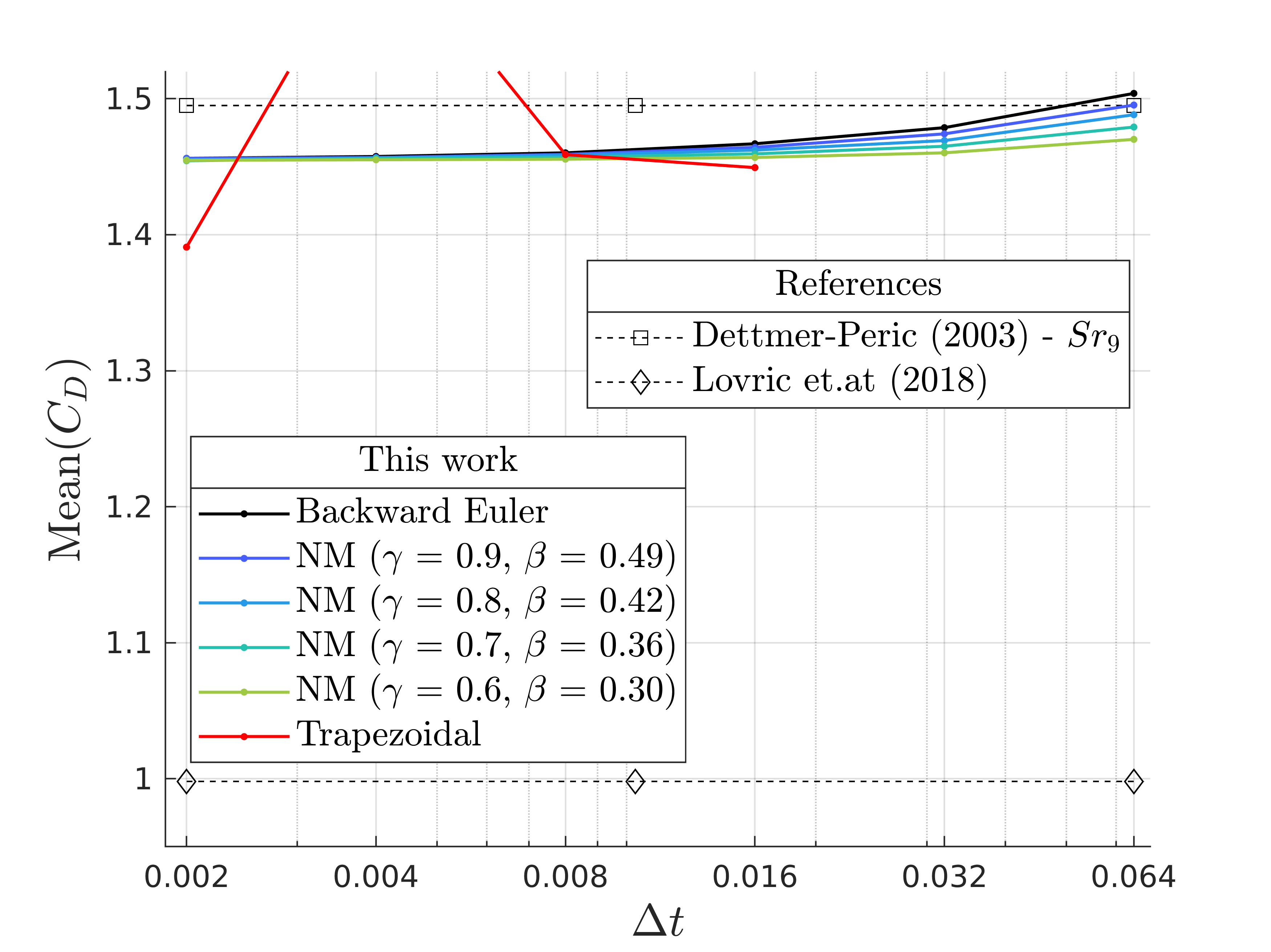}
		\caption{}
		\label{Fig:FAC_Results_Re100_c}
	\end{subfigure}
	~
	\begin{subfigure}[b]{0.46\textwidth}	
		\includegraphics[trim=0 0 32 17,clip,width=1.00\linewidth]{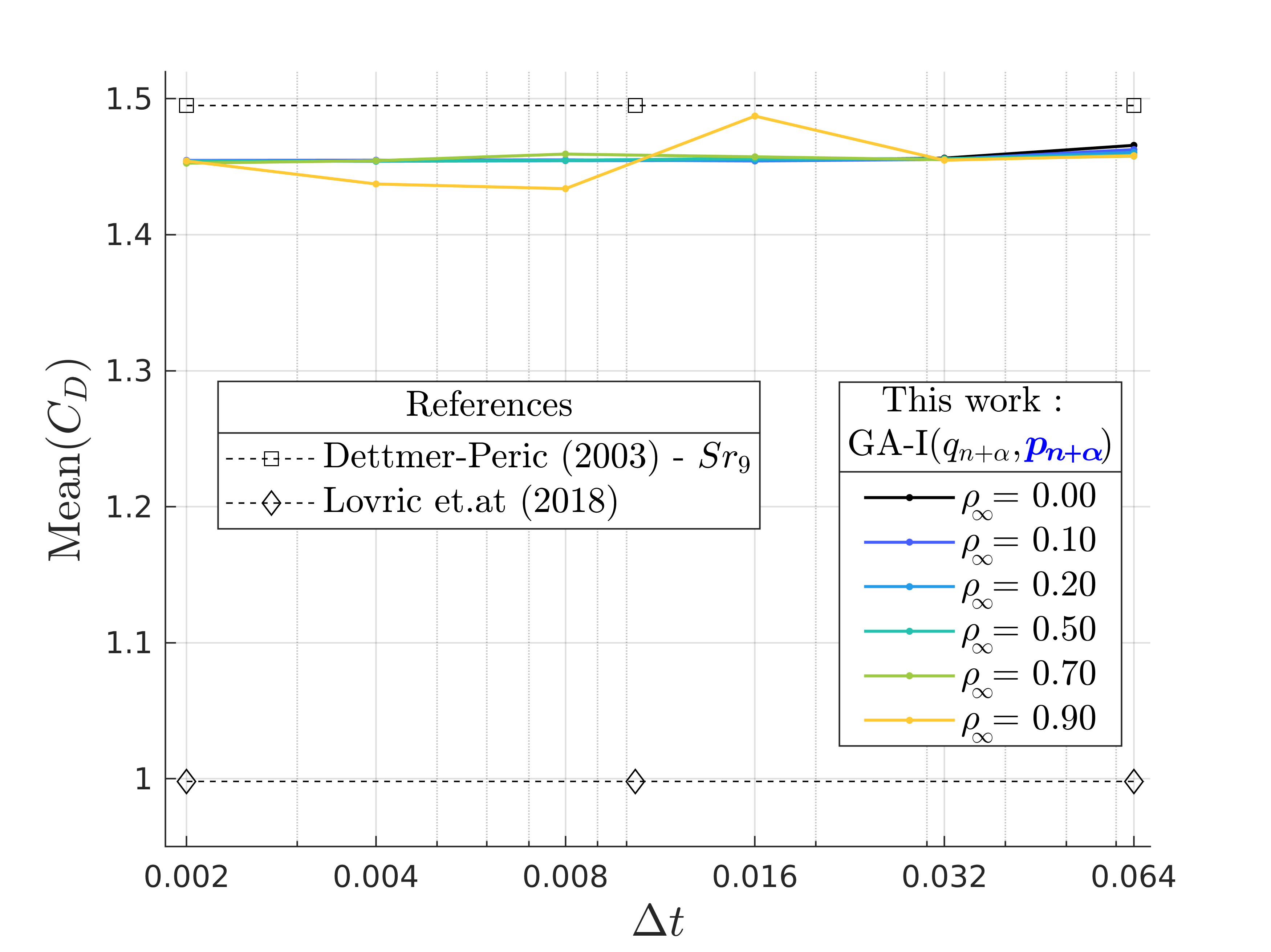}
		\caption{}
		\label{Fig:FAC_Results_Re100_d}
	\end{subfigure}	
	\\[-4ex]
	\begin{subfigure}[b]{0.46\textwidth}
		\includegraphics[trim=0 0 32 17,clip,width=1.00\linewidth]{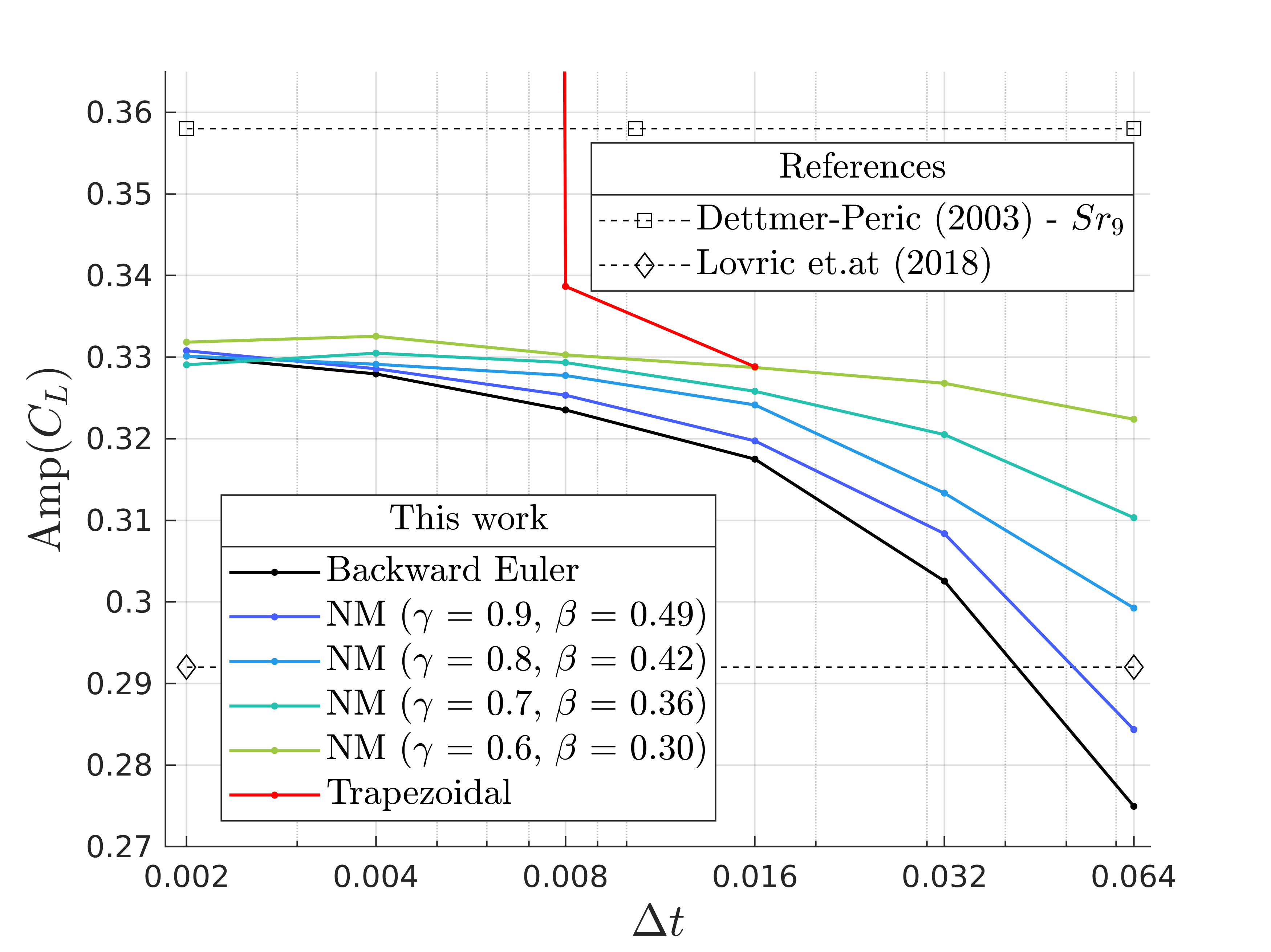}
		\caption{}
		\label{Fig:FAC_Results_Re100_e}
	\end{subfigure}
	~
	\begin{subfigure}[b]{0.46\textwidth}	
		\includegraphics[trim=0 0 32 17,clip,width=1.00\linewidth]{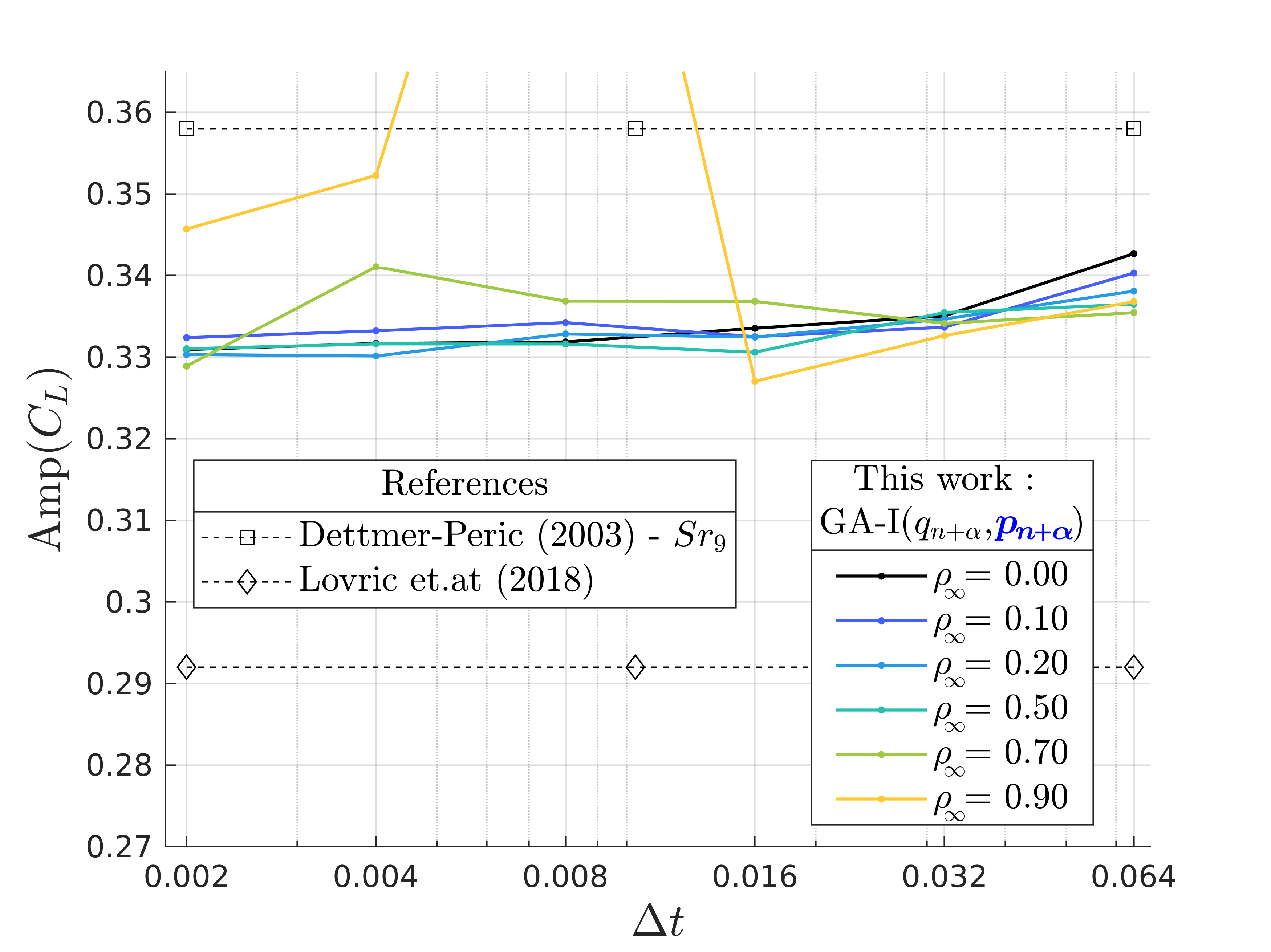}
		\caption{}
		\label{Fig:FAC_Results_Re100_f}
	\end{subfigure}	
	\\
	\vspace{-195mm}
	\hspace{-83mm} (\textbf{a}) \hspace{83mm} (\textbf{b})
	\\
	\vspace{60mm}
	\hspace{-83mm} (\textbf{c}) \hspace{83mm} (\textbf{d})
	\\
	\vspace{57mm}
	\hspace{-83mm} (\textbf{e}) \hspace{83mm} (\textbf{f})
	\vspace{56mm}
\caption{Coefficients of the flow around a cylinder at Re = 100. (a,b) Strouhal number (c,d) mean value of the drag coefficient, and (e,f) amplitude of the lift coefficient. Results obtained with (a,c,d) Backward Euler, Newmark's Method (NM) and Trapezoidal rule, and (b,d,f) Generalized-$\alpha$ ($\GAa$).}
\label{Fig:FAC_Results_Re100}
\end{figure}

\begin{figure}[t!] \captionsetup[sub]{font=normalsize}\captionsetup[subfigure]{labelformat=empty}
\centering 
	\begin{subfigure}[b]{0.46\textwidth}
		\includegraphics[trim=0 0 32 28,clip,width=1.00\linewidth]{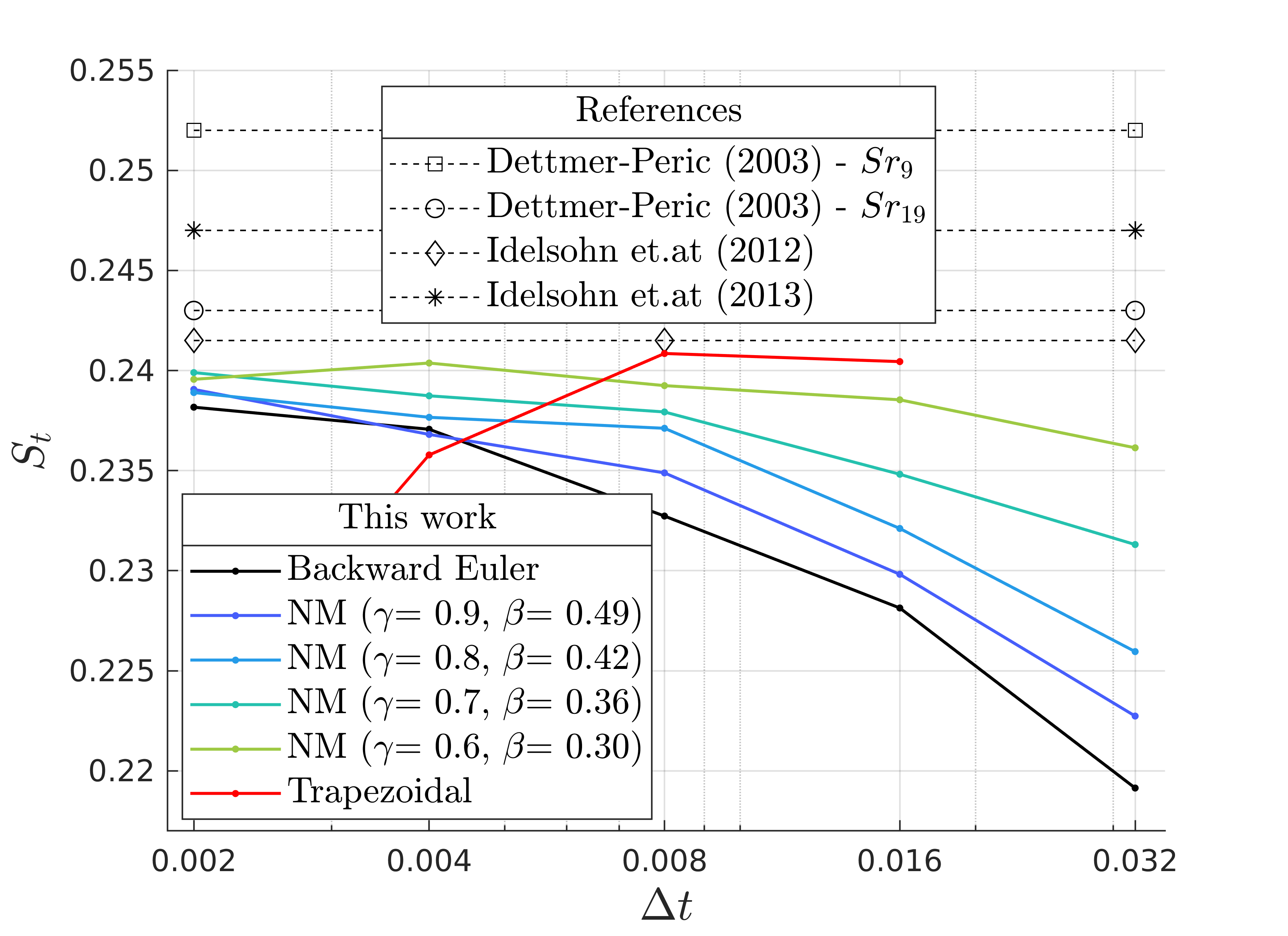}
		\caption{}
		\label{Fig:FAC_Results_Re1000_a}
	\end{subfigure}
	~
	\begin{subfigure}[b]{0.46\textwidth}	
		\includegraphics[trim=0 0 32 28,clip,width=1.00\linewidth]{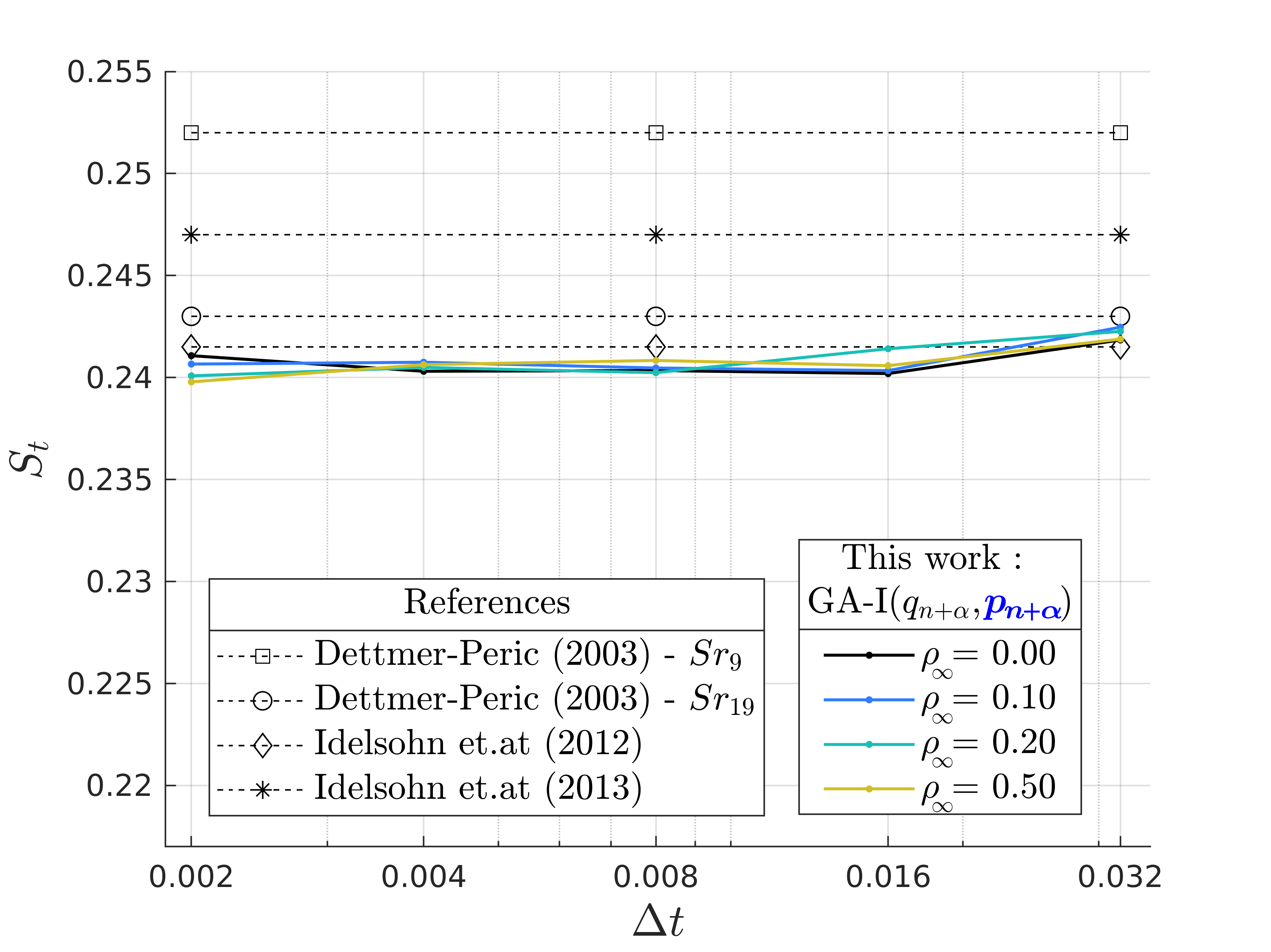}
		\caption{}
		\label{Fig:FAC_Results_Re1000_b}
	\end{subfigure}	
	\\[-3ex]
	\begin{subfigure}[b]{0.46\textwidth}
		\includegraphics[trim=0 0 32 17,clip,width=1.00\linewidth]{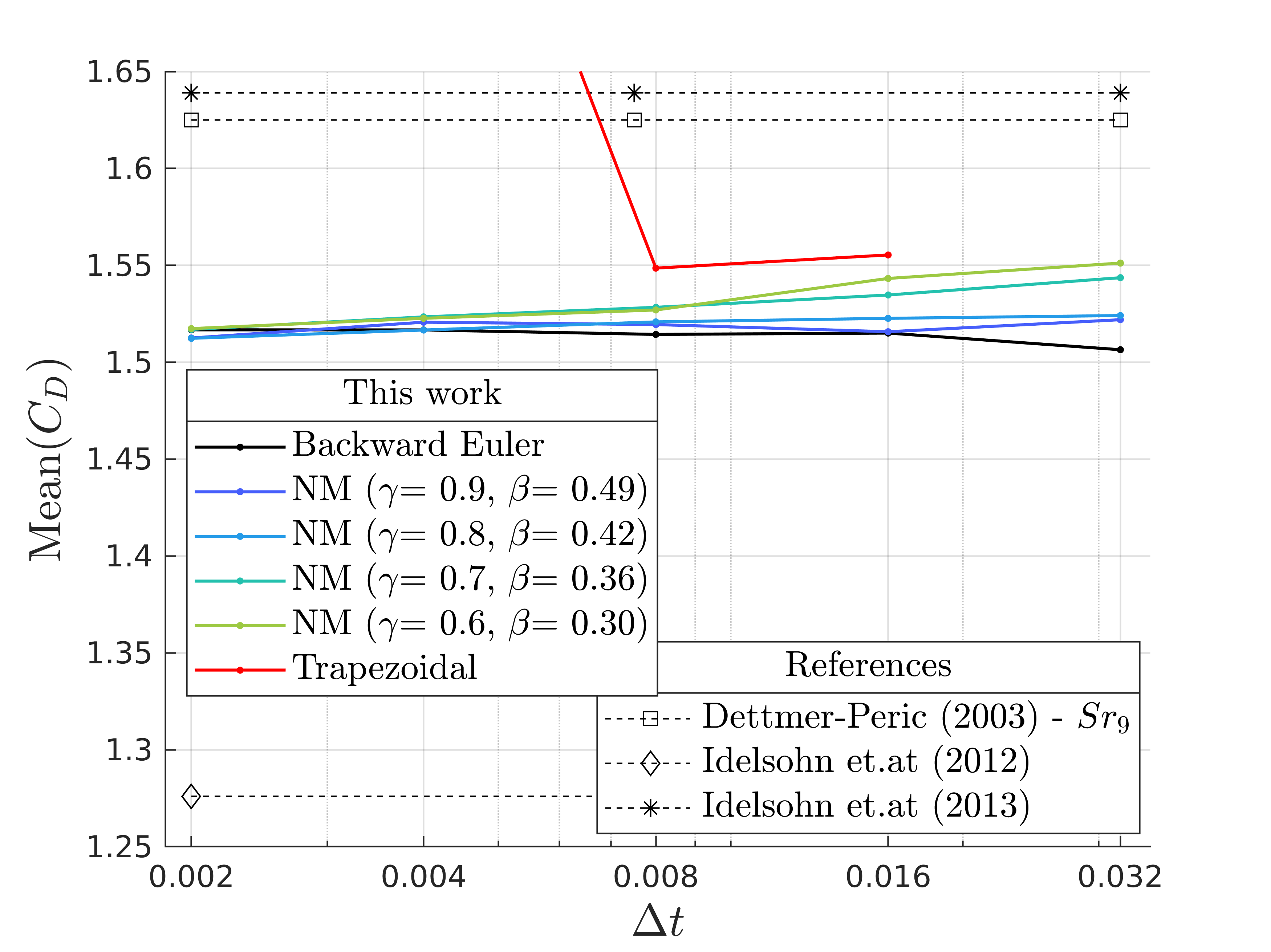}
		\caption{}
		\label{Fig:FAC_Results_Re1000_c}
	\end{subfigure}
	~
	\begin{subfigure}[b]{0.46\textwidth}	
		\includegraphics[trim=0 0 32 17,clip,width=1.00\linewidth]{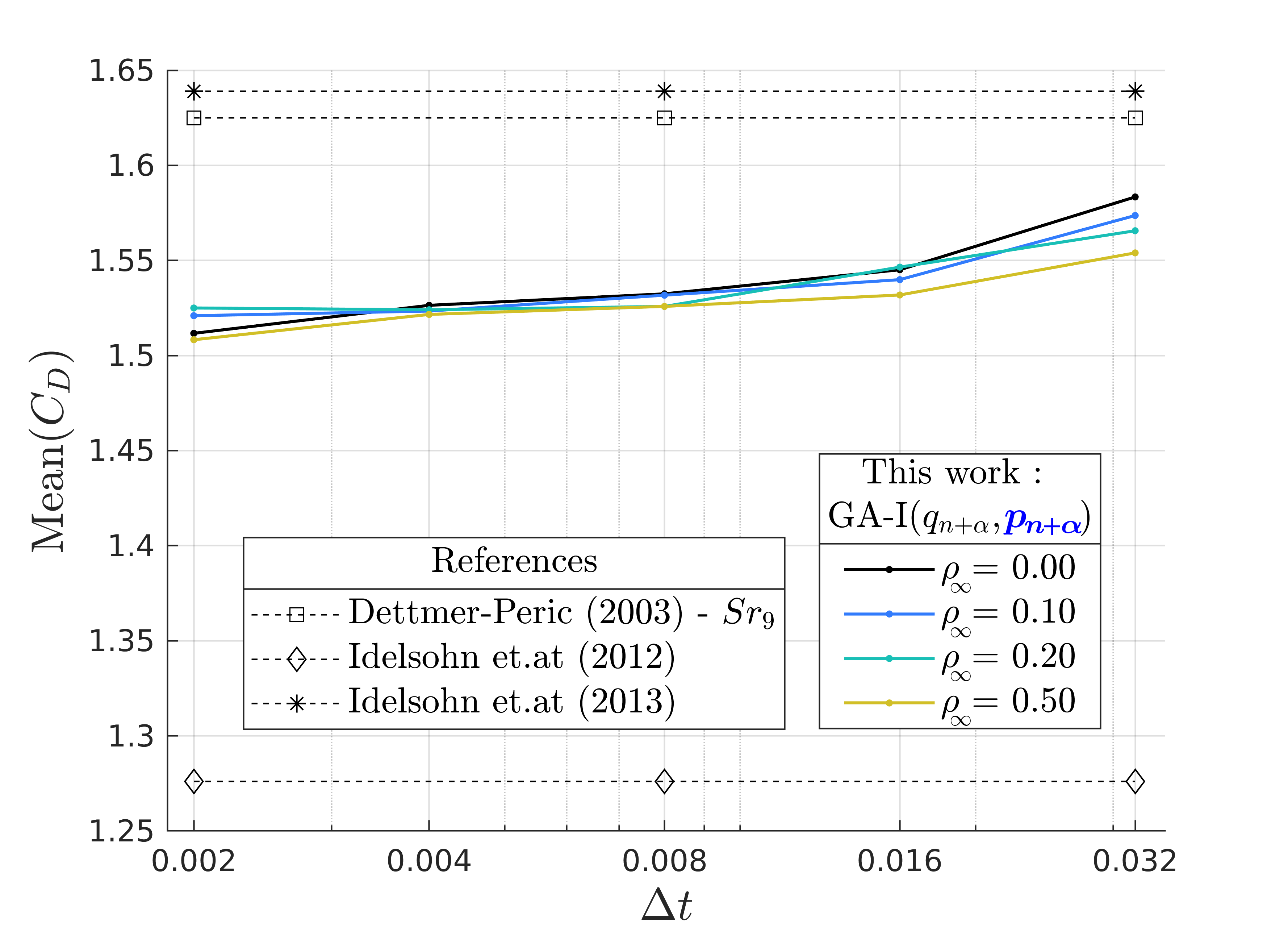}
		\caption{}
		\label{Fig:FAC_Results_Re1000_d}
	\end{subfigure}	
	\\[-3ex]
	\begin{subfigure}[b]{0.46\textwidth}
		\includegraphics[trim=0 0 32 17,clip,width=1.00\linewidth]{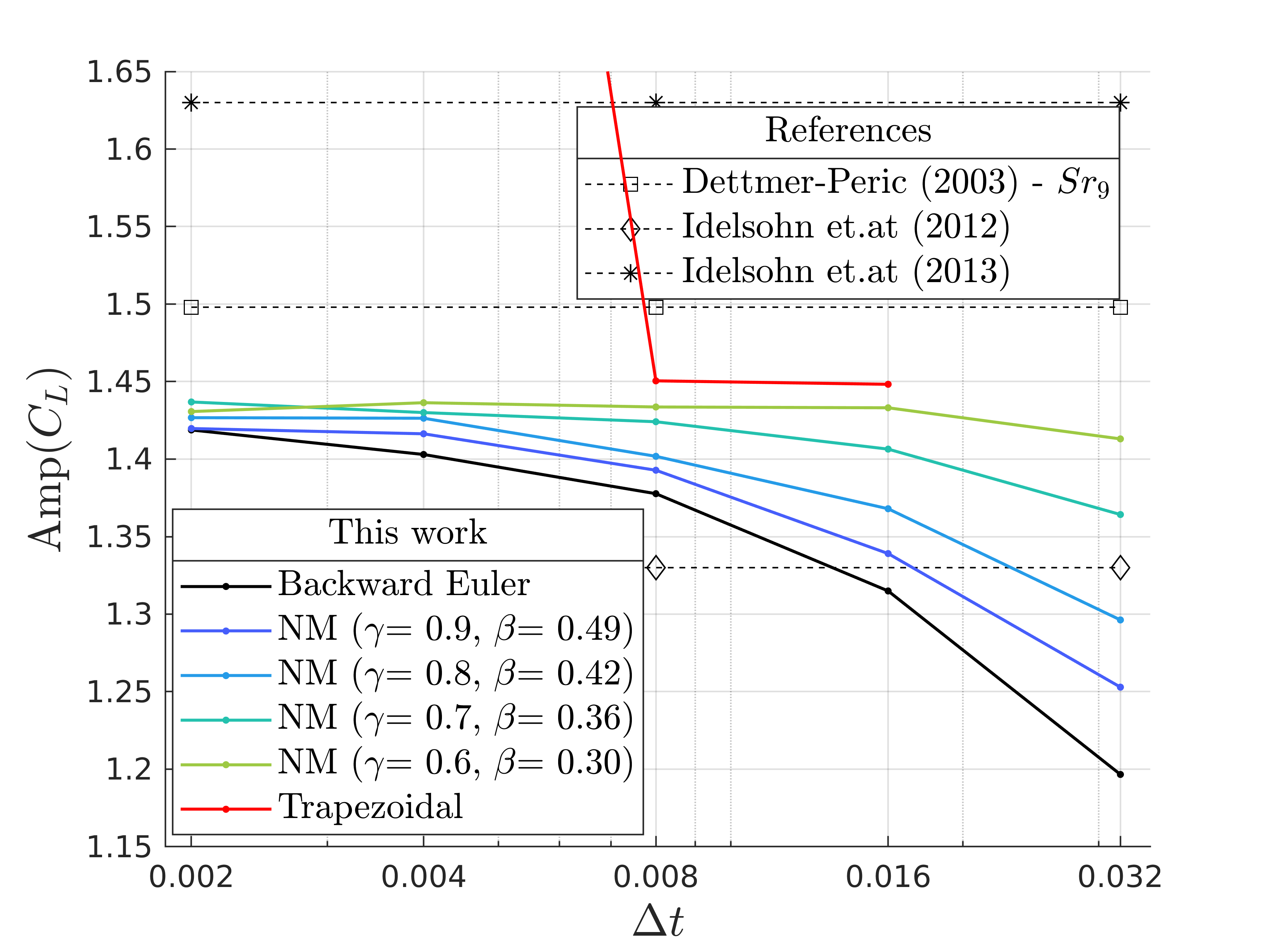}
		\caption{}
		\label{Fig:FAC_Results_Re1000_e}
	\end{subfigure}
	~
	\begin{subfigure}[b]{0.46\textwidth}	
		\includegraphics[trim=0 0 32 17,clip,width=1.00\linewidth]{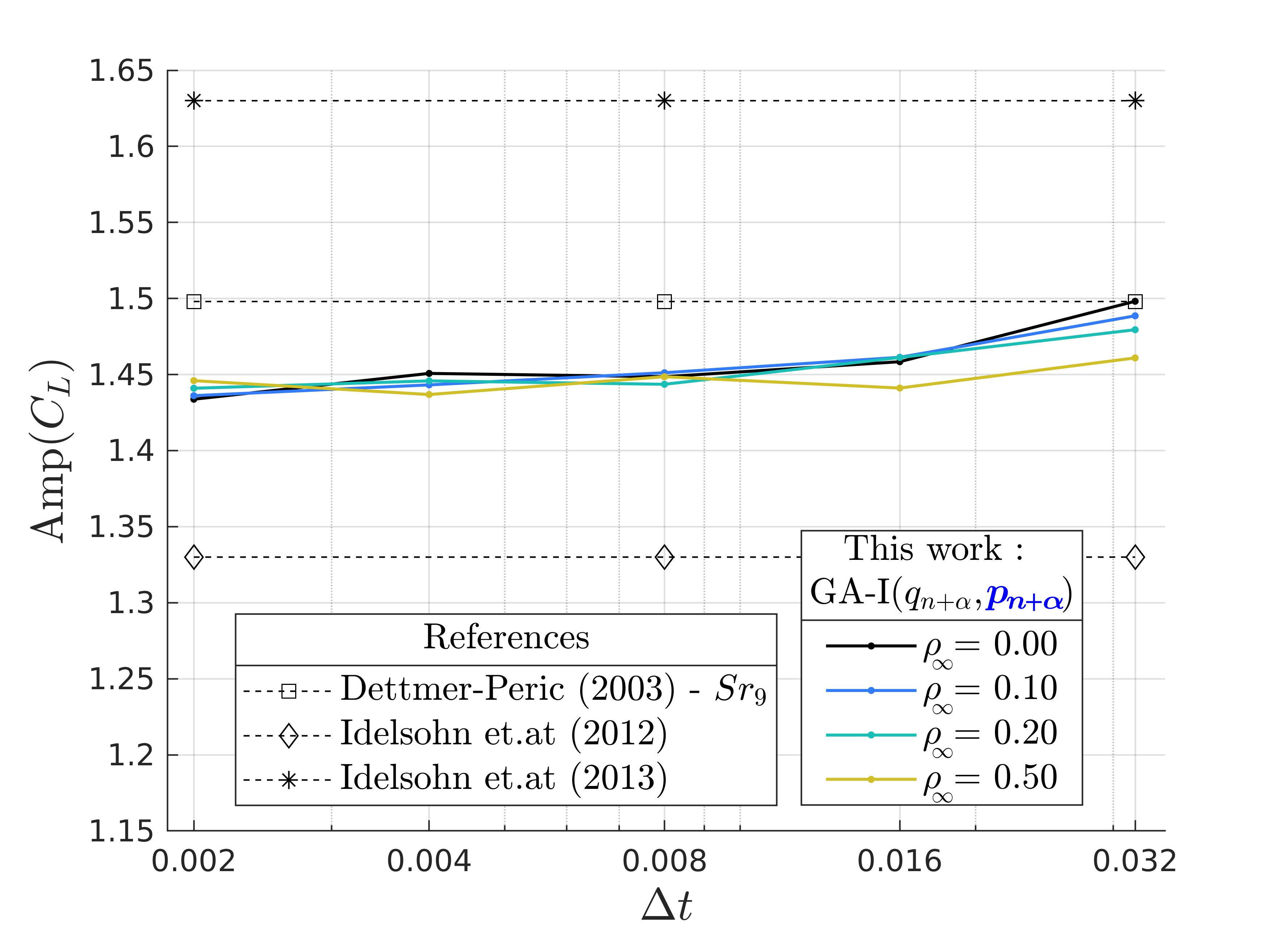}
		\caption{}
		\label{Fig:FAC_Results_Re1000_f}
	\end{subfigure}	
	\\
	\vspace{-200mm}
	\hspace{-83mm} (\textbf{a}) \hspace{83mm} (\textbf{b})
	\\
	\vspace{60mm}
	\hspace{-83mm} (\textbf{c}) \hspace{83mm} (\textbf{d})
	\\
	\vspace{61mm}
	\hspace{-83mm} (\textbf{e}) \hspace{83mm} (\textbf{f})
	\vspace{58mm}
\caption{Coefficients of the flow around a cylinder at Re = 1000. (a,b) Strouhal number (c,d) mean value of the drag coefficient, and (e,f) amplitude of the lift coefficient. Results obtained with (a,c,d) Backward Euler, Newmark's Method (NM) and Trapezoidal rule, and (b,d,f) Generalized-$\alpha$ ($\GAa$).}
\label{Fig:FAC_Results_Re1000}
\end{figure}

For problems with Re = 1000, five time steps are used ranging from 0.002 to 0.032. Similarly as for Re = 100, Backward Euler, Trapezoidal, Newmark and Generalized-$\alpha$ are used. For Newmark, $\gamma$ is set to 0.9, 0.8, 0.7 and 0.6. For Generalized-$\alpha$ ($\GAa$), $\rho_\infty$ is set to 0.0, 0.1, 0.2, 0.5. Thus, 50 problems are solved in total for Re = 1000.  Results are summarized in Fig.~\ref{Fig:FAC_Results_Re1000} using same layout as for Re = 100 (Fig.~\ref{Fig:FAC_Results_Re100}).

Results in Figs.~\ref{Fig:FAC_Results_Re100} and \ref{Fig:FAC_Results_Re1000} show that $S_t$, Amp($C_L$) and Mean($C_D$) obtained with stable time integration schemes converge to the same value as time step $\Delta t$ is reduced. Converged values are listed in Table \ref{Tab:FAC_References}. These are obtained from averaging coefficients obtained with $\Delta t$ = 0.002. Table \ref{Tab:FAC_References} also includes values from references that use Eulerian\citep{dettmer2003analysis,lovric2018new} or semi-Lagrangian approaches (PFEM-2) \citep{idelsohn2012large,idelsohn2013fast}. Notably, coefficients obtained in this work are in agreement with the literature. 

\paragraph{\underline{Performance of time integration schemes}}\vspace{2mm}

As in previous examples, Backward Euler is the least accurate at large time steps $\Delta t$. Its excessive numerical damping for high frequencies reduces both the magnitude and frequency of the lift force. In contrast, Newmark's Method (NM) offers better accuracy as $\gamma$ is reduced, as long as $\gamma$ and $\beta$ do not match the Trapezoidal rule, which leads to some problems in the present case. Noteworthy, the Trapezoidal curve (red) does not completely appear in Figs.~\ref{Fig:FAC_Results_Re100} and \ref{Fig:FAC_Results_Re1000} because either values are far outside the range of the plot (values for $\Delta t$ < 0.008) or because convergence was not achieved with the imposed time step ($\Delta t$ > 0.016). The instability of the trapezoidal rule in the absence of numerical damping at high frequencies is illustrated in Fig.~\ref{Fig:FAC_curves_vp_dt_04}. This figure depicts 4 cycles of the lift coefficient (for Re = 100) obtained with different time integration schemes and $\Delta t = 0.004$. The curves are aligned on the left side and abscissa is tabulated with respect to the converged frequency (equal to the converged $S_t$ number). Due to the lack of numerical damping, the lift coefficient shows huge values in the Trapezoidal scheme, thus it had to be scaled by 1/25 to make it fit into Fig.~\ref{Fig:FAC_curves_vp_dt_04}. 

Regarding the Generalized-$\alpha$ scheme, $\rho_\infty \leq 0.5$ exhibits high accuracy with respect to lift and drag coefficients, even for large time steps $\Delta t$. As numerical damping for high frequencies is reduced, specifically for $\rho_\infty > 0.5$, accuracy is lost because spurious oscillations become dominant in the system. This can be seen in Fig.~\ref{Fig:FAC_curves_vp_dt_04} and \ref{Fig:FAC_curves_vp_dt_64}.

\begin{figure}[t] \captionsetup[sub]{font=normalsize}
\centering 
	\begin{subfigure}[b]{0.48\textwidth}
		\includegraphics[trim=50 40 39 50,clip,width=1.00\linewidth]{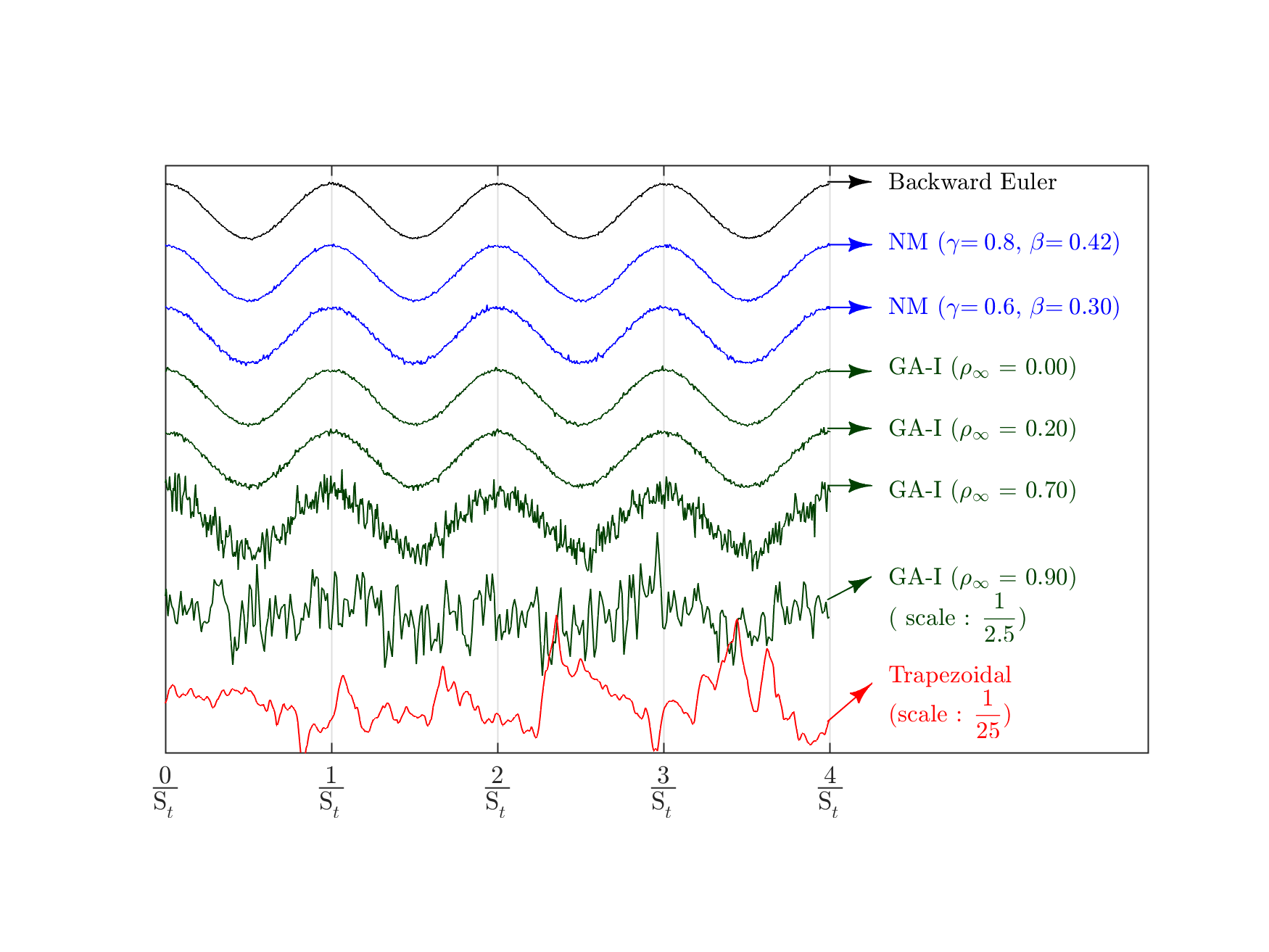}
		\caption{Obtained with $\Delta t = 0.004$.}
		\label{Fig:FAC_curves_vp_dt_04}
	\end{subfigure}
	~
	\begin{subfigure}[b]{0.48\textwidth}	
		\includegraphics[trim=50 40 39 50,clip,width=1.00\linewidth]{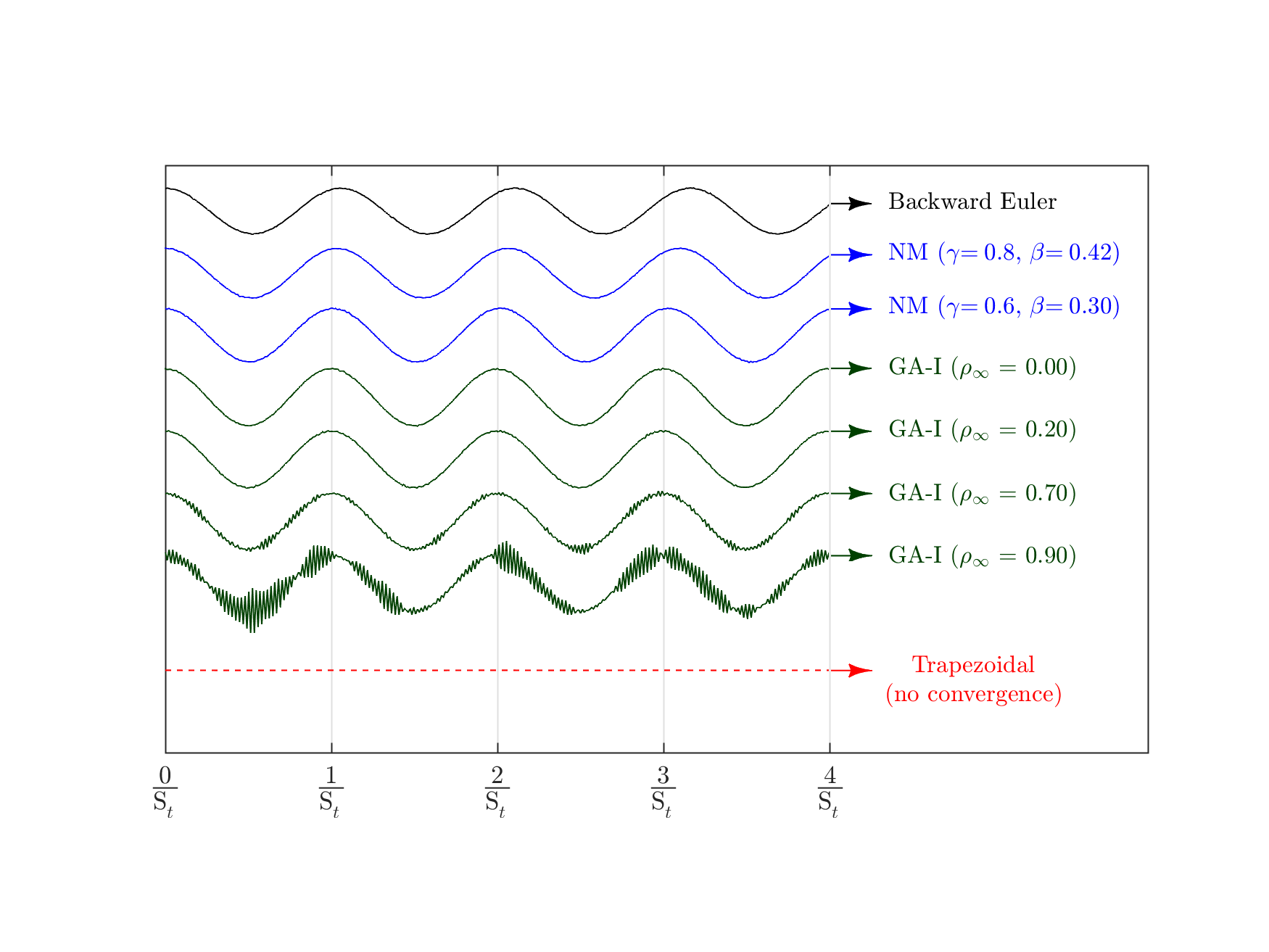}
		\caption{Obtained with $\Delta t = 0.064$.}
		\label{Fig:FAC_curves_vp_dt_64}
	\end{subfigure}	
\caption{Flow around a cylinder at Re = 100. Four cycles from the lift coefficient curve obtained with Backward Euler, Newmark's Method (NM), Generalized-$\alpha$ ($\GAa$) and Trapezoidal rule. Abscissa is scaled with respect to the reference Strouhal number $S_t = 0.174$. In (a), lift coefficient curves of GA-I ($\rho_\infty = 0.90$) and Trapezoidal rule are scaled by 1/2.5 and 1/25, respectively.}
\label{Fig:FAC_curves_Re100}
\end{figure}

A remarkable aspect of the Generalized-$\alpha$ method is that results are very similar for $\rho_\infty \leq 0.5$. This holds for both Reynolds (100 and 1000) and for both forces (lift and drag), as shown in Figs.~\ref{Fig:FAC_Results_Re100_b}, \ref{Fig:FAC_Results_Re100_d}, \ref{Fig:FAC_Results_Re100_f}, \ref{Fig:FAC_Results_Re1000_b}, \ref{Fig:FAC_Results_Re1000_d}, \ref{Fig:FAC_Results_Re1000_f}. On one hand, this is in line with observations of Jansen et.al.\citep{jansen2000generalized} who state "\textit{... the period and amplitude of both the lift and the drag are very weak functions of $\rho_\infty$ (which  might  be  expected  from  the  observation  that  30  points  per  wavelength is adequate for second-order accurate method)}". On the other hand, it is in contradiction with results by Dettmer and Peri\'{c} \citep{dettmer2003analysis}, who report a notorious influence of the lift force with respect to the spectral radius $\rho_\infty$. A possible explanation is that time step used in the present work and in that of Jansen et.al.\citep{jansen2000generalized} is significantly smaller than the one used by Dettmer and Peri\'{c} \citep{dettmer2003analysis}. At the largest time step ($\Delta t = 0.064$), the period of the lift force is discretized in $\approx$90 points in our work, in $\approx$60 points in the work of Jansen et.al.\citep{jansen2000generalized} (with $\Delta t = 0.1$), while in $\approx$18 points in that of Dettmer and Peri\'{c} \citep{dettmer2003analysis} (with $\Delta t = 0.3$). In the latter case, a coarse time discretization could be standing out a dependence of results on $\rho_\infty$.

Time steps used in this work are much smaller than those used by reference\citep{dettmer2003analysis} because a finer mesh around the cylinder is used here in order to reduce instabilities of the pressure gradient caused by the spatial discretization and the remeshing. If a coarser mesh were used, the lift and drag force would contain bigger spurious oscillations (see figure 10.11 of reference \citep{idelsohn2012large}) and it would be difficult to identify numerical damping associated to the time integration schemes.

\begin{center}
\begin{table}[t]%
\centering
\caption{Results from numerical simulations of the flow around a cylinder for Reynolds 100 and 1000. An em dash symbol indicates that value is not found in the reference.  \label{Tab:FAC_References}}%
\begin{tabular*}{500pt}{@{\extracolsep\fill}lcccccc@{\extracolsep\fill}}
\toprule
\multirow{2}{*}{\textbf{Source}} & \multicolumn{3}{c}{Re = 1000}   &  \multicolumn{3}{c}{Re = 100} \\
\cmidrule{2-7}
& $S_t$ & Amp($C_L$) & Mean($C_D$) & $S_t$ & Amp($C_L$) & Mean($C_D$)
\\
\midrule
This work 
& \textbf{0.241} & \textbf{1.427}$^{\color{white}*}$  &  \textbf{1.512}$^{\color{white}*}$ 
& \textbf{0.174} & \textbf{0.331}$^{\color{white}*}$ & \textbf{1.455}$^{\color{white}*}$
\\
Dettmer-Peri\'{c} ($Sr_9$) \citep{dettmer2003analysis} 
& 0.252  & 1.498$^*$ &  1.625$^*$
& 0.181  & 0.358$^*$ &  1.495$^*$
\\
Dettmer-Peri\'{c} ($Sr_{19}$) \citep{dettmer2003analysis} 
& 0.243  & \textemdash   &  \textemdash  
& 0.171  & \textemdash   &  \textemdash 
\\
Lovri\'{c} et.al.\citep{lovric2018new} 
& \textemdash  & \textemdash  &  \textemdash 
& 0.165        & 0.292$^{\color{white}*}$        & 0.998$^*$
\\
Idelsohn et.al.\citep{idelsohn2012large} 
& 0.2415  & 1.330  &  1.276 & \textemdash & \textemdash & \textemdash
\\
Idelsohn et.al.\citep{idelsohn2013fast} 
& 0.2475  & 1.630  &  1.639 & \textemdash & \textemdash & \textemdash
\\
\bottomrule
\end{tabular*}
\begin{tablenotes}
\item[$^*$] Obtained from figure
\end{tablenotes}
\end{table}
\end{center}

\paragraph{\underline{Velocity-based and Displacement-based formulations}}\vspace{2mm}

So far all results for the flow around a cylinder have been obtained using the velocity-pressure formulation. Now these are compared with results using the displacement-pressure formulation at Re = 100. Results for both formulations are shown in Fig.~\ref{Fig:FAC_vp_dp}, using the Backward Euler, Newmark (Fig.~\ref{Fig:FAC_vp_dp_a}) and Generalized-$\alpha$ (Fig.~\ref{Fig:FAC_vp_dp_b}) time integration schemes.

Subtle differences can be seen in Figs.~\ref{Fig:FAC_vp_dp_a} and \ref{Fig:FAC_vp_dp_b}. The largest difference between formulations in the Newmark method is $0.09\%$, obtained with $\Delta t = 0.016$ and $\gamma = 0.6$. In the Generalized-$\alpha$, the largest difference is likewise $0.09\%$ and is found for $\Delta t = 0.032$ and $\rho_\infty$ = 0.5. Typically, the larger the numerical damping for high frequencies, the smaller the difference between velocity- and displacement-based formulations. We believe, that this subtle difference between the two formulations is mainly caused by high-frequency perturbations, which stand out in problems with long simulation times and several remeshing, such as the flow around a cylinder.

\begin{figure}[t] \captionsetup[sub]{font=normalsize}\captionsetup[subfigure]{labelformat=empty}
	\centering 
	\begin{subfigure}[b]{0.48\textwidth}
		\includegraphics[trim=2 0 33 17,clip,width=1.00\linewidth]{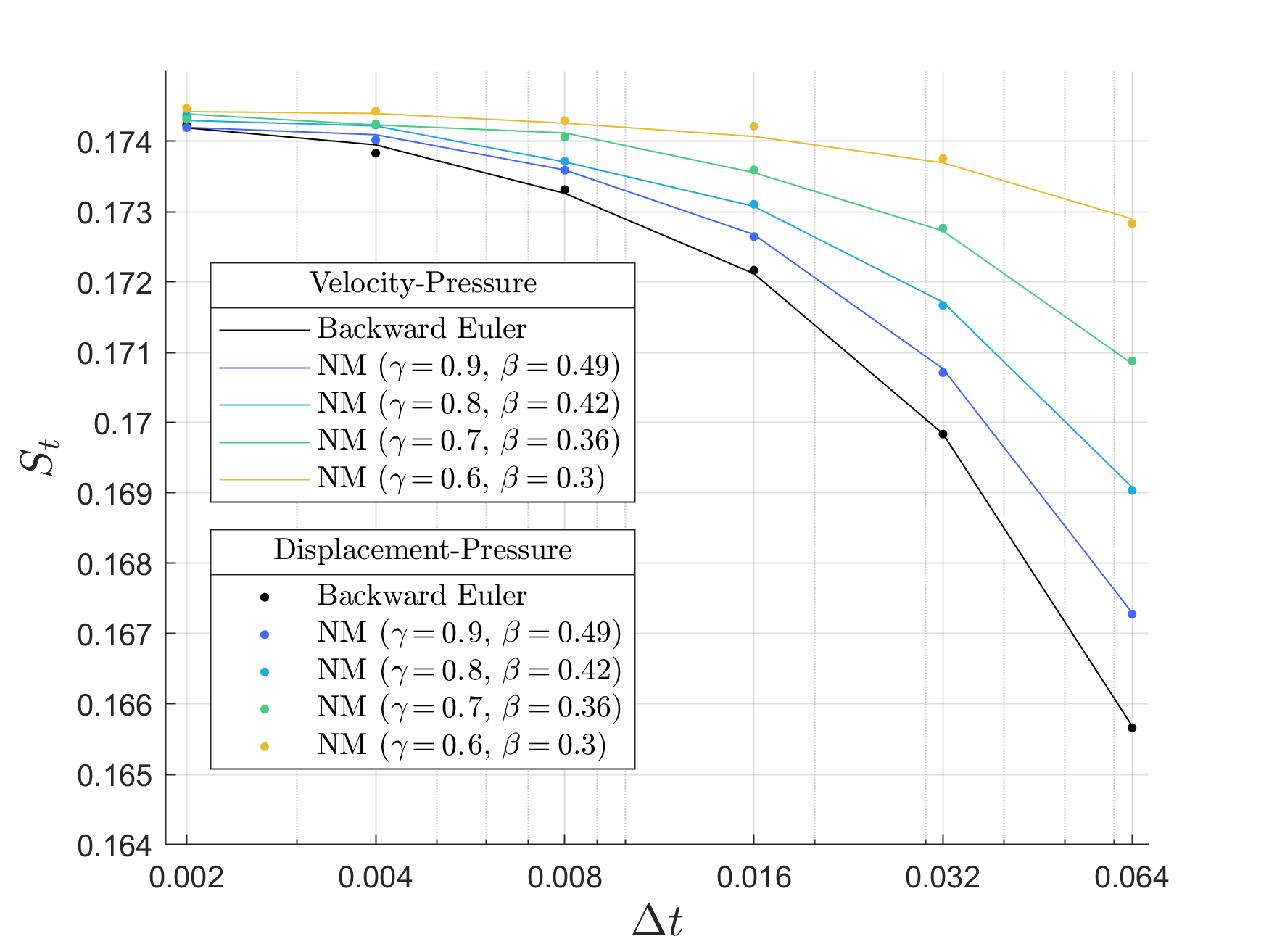}
		\caption{}
		\label{Fig:FAC_vp_dp_a}
	\end{subfigure}
	~\hspace{1mm}
	\begin{subfigure}[b]{0.48\textwidth}	
		\includegraphics[trim=2 0 33 17,clip,width=1.00\linewidth]{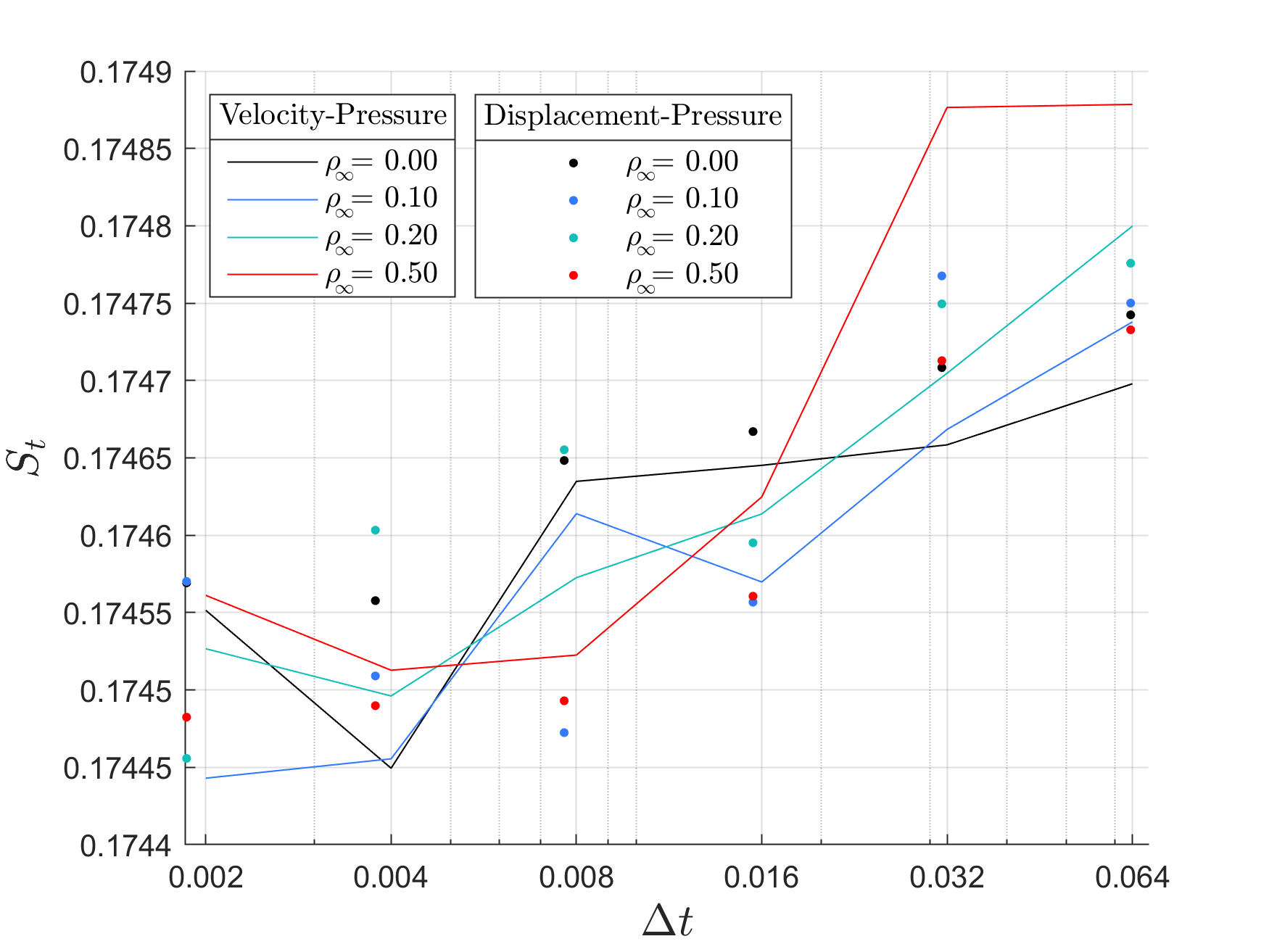}
		\caption{}
		\label{Fig:FAC_vp_dp_b}
	\end{subfigure}
	\\
	\vspace{-72mm}
	\hspace{-83mm} (\textbf{a}) \hspace{83mm} (\textbf{b})
	\vspace{60mm}
\caption{Strouhal number of the flow around a cylinder at Re = 100. Results for velocity-based and displacement-based formulations using (a) Backward Euler and Newmark's Method (NM) and (b) Generalized-$\alpha$ method.}
\label{Fig:FAC_vp_dp}
\end{figure}

\paragraph{\underline{Implementation approaches of the Generalized-$\alpha$ method}}\vspace{2mm}

The last analysis using the flow around a cylinder considers the different implementation approaches of the Generalized-$\alpha$ scheme. The comparison is performed for Re = 100 and using five time steps $\Delta t$ ranging from 0.004 to 0.064. First, schemes that write the pressure at $t_{n+\alpha_f}$ are compared, i.e., $\GAa$, $\GAb$, and $\GAc$. Two values for the spectral radius are considered, 0.0 and 0.5. Results for the Strouhal number are summarized in Fig.~\ref{Fig:FAC_vp_123456_a}. This shows a small difference between implementation approaches, which is about $0.1\%$ as indicated in Fig.~\ref{Fig:FAC_vp_123456_a}. This is also seen in the amplitude of the lift coefficient and the mean of the drag coefficient, although no plots are given here to avoid overextending the manuscript. Thereby, as expected, the hypotheses of scaling state variables (GA-I) or forces of momentum balance (GA-II) lead to practically the same result, especially when small time steps are used. 

\begin{figure}[t] \captionsetup[sub]{font=normalsize}\captionsetup[subfigure]{labelformat=empty}
	\centering 
	\begin{subfigure}[b]{0.47\textwidth}
		\includegraphics[trim=0 0 32 17,clip,width=1.00\linewidth]{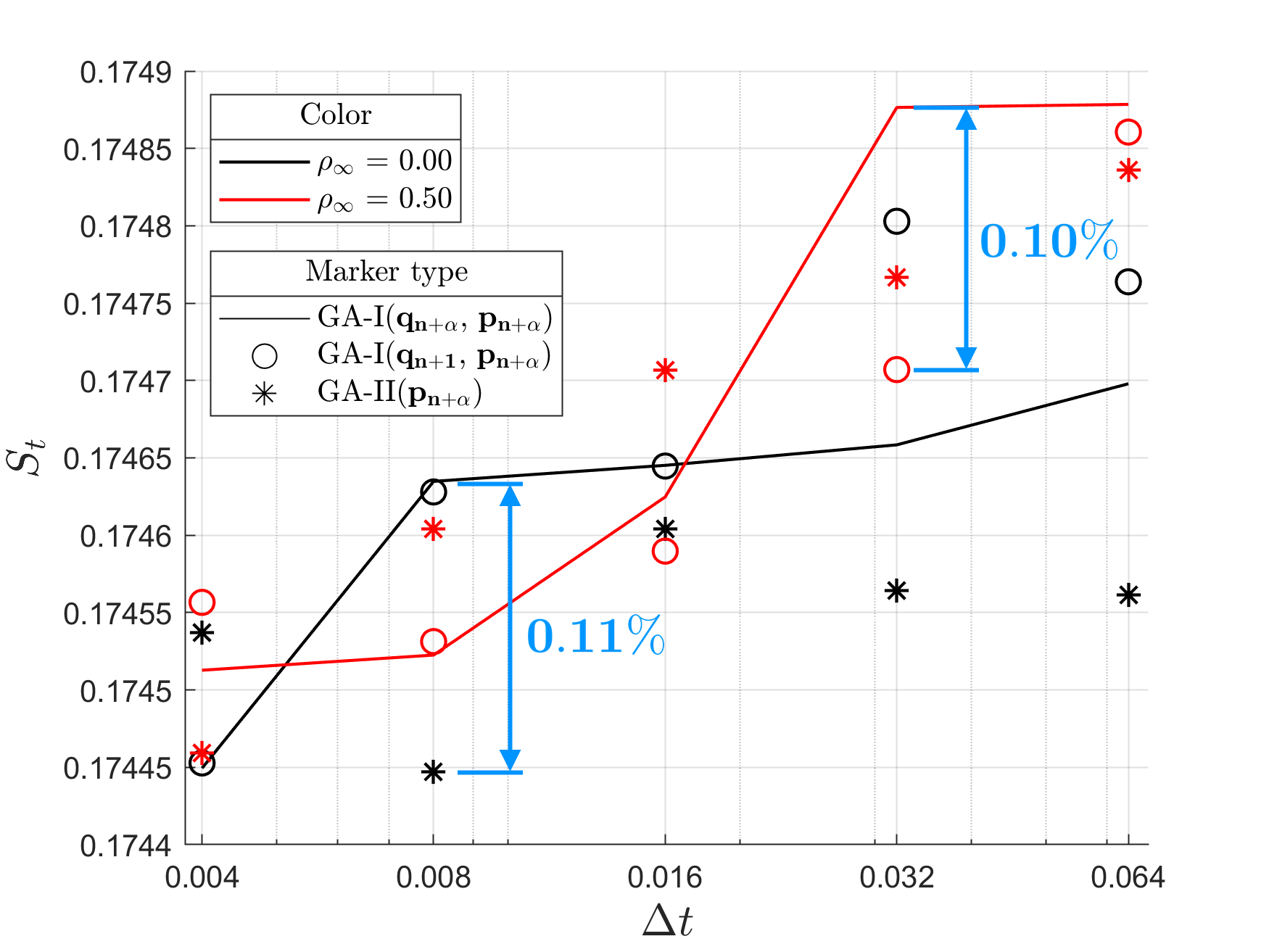}
		\caption{}
		\label{Fig:FAC_vp_123456_a}
	\end{subfigure}
	~\hspace{2mm}
	\begin{subfigure}[b]{0.47\textwidth}	
		\includegraphics[trim=0 0 32 17,clip,width=1.00\linewidth]{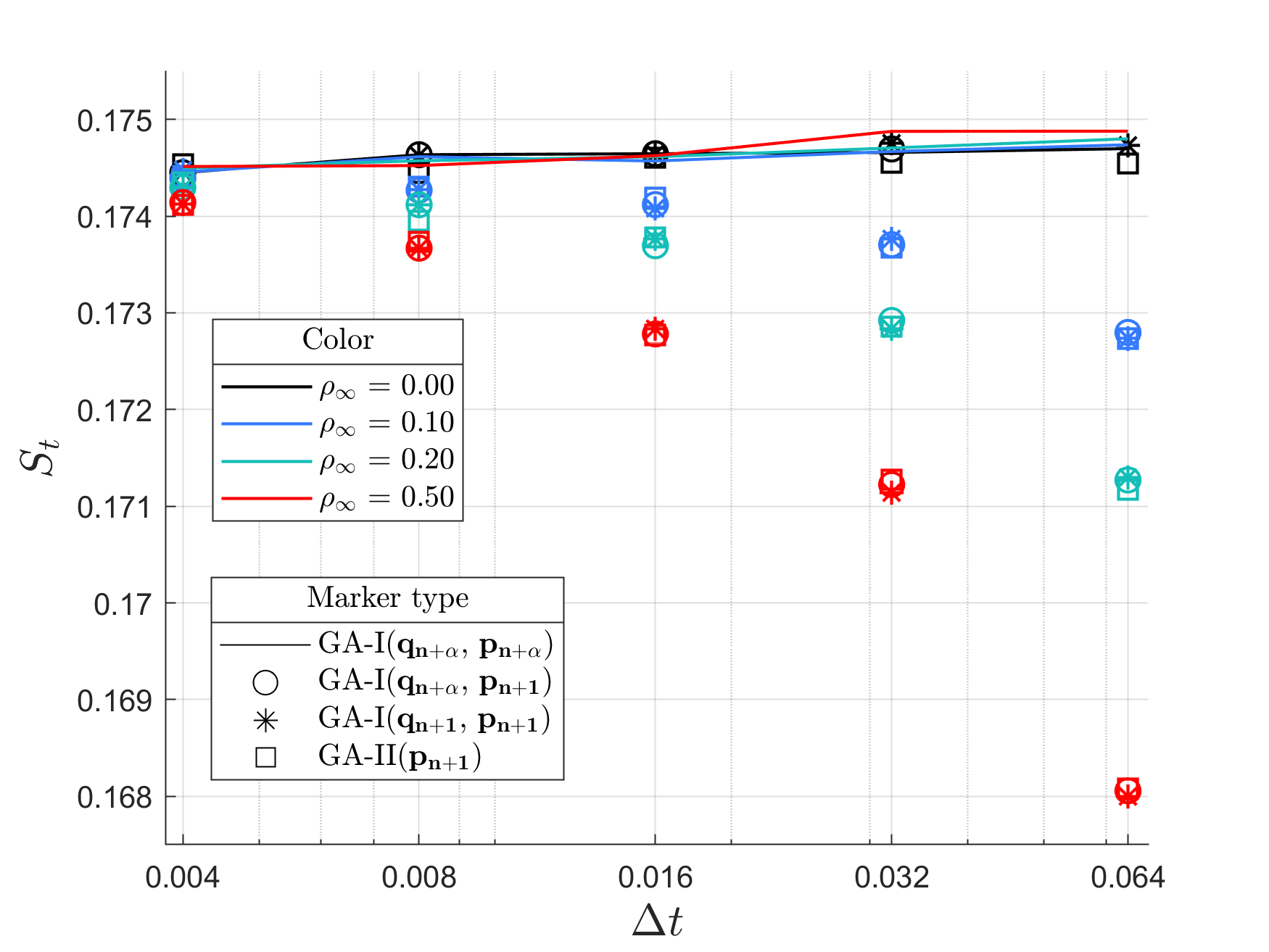}
		\caption{}
		\label{Fig:FAC_GA_1456_vp_rho_St_Re100}
	\end{subfigure}
	\\
	\vspace{-72mm}
	\hspace{-83mm} (\textbf{a}) \hspace{83mm} (\textbf{b})
	\vspace{60mm}
\caption{Strouhal number of the flow around a cylinder at Re = 100. Results using different implementation approaches of the Generalized-$\alpha$ scheme. In (a) are implementation approaches that follow the principle of the Generalized-$\alpha$ method and impose the pressure at $t_{n+\alpha_f}$. In (b) are implementation approaches that write the pressure at $t_{n+1}$ and integrate the discrete gradient operator ($\mathbf{D}^\intercal$) in $\Omega (t_{n+1})$.}
\label{Fig:FAC_vp_123456}
\end{figure}

A comparison is now made between schemes that write the pressure at $t_{n+1}$ in the momentum equation, i.e., $\GAd$, $\GAe$ and $\GAf$. It is worth recalling that two scenarios are presented, one that integrates the gradient matrix in $\Omega(t_{n+1})$ (Eq.~\ref{EQ:MomentumEquation_AlphaMethod_noPre}) and another that does so in $\Omega(t_{n+\alpha_f})$ (Eq.~\ref{EQ:MomentumEquation_AlphaMethod_noPre_2}). Results for the first case are presented in Fig.~\ref{Fig:FAC_GA_1456_vp_rho_St_Re100}. The figure also includes results with $\GAa$ (solid line) that are used as reference. It is apparent from Fig.~\ref{Fig:FAC_GA_1456_vp_rho_St_Re100} that results are similar for all implementation approaches when $\rho_\infty$ = 0.0. However, difference between the reference $\GAa$ and schemes that write the pressure at $t_{n+1}$ increases with the value of $\rho_\infty$, in detriment of the schemes that do not follow the principle of the Generalized-$\alpha$ method. Similar observations are obtained when analyzing the amplitude of the lift coefficient or the mean value of the drag coefficient (results not reported here). This behavior was observed for the kinematics of the solitary wave propagation problem and it is now found in the pressure field of the flow around a cylinder.

\begin{figure}[t] \captionsetup[sub]{font=normalsize}\captionsetup[subfigure]{labelformat=empty}
	\centering
	\includegraphics[width=0.31\linewidth]{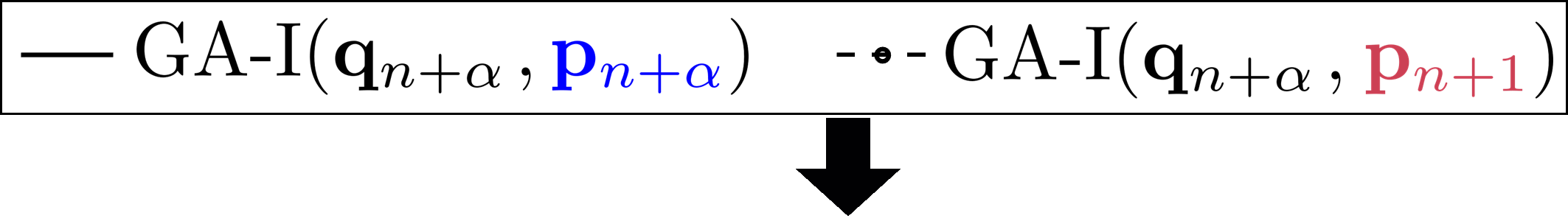} 
	~\hspace{2mm}
	\includegraphics[width=0.31\linewidth]{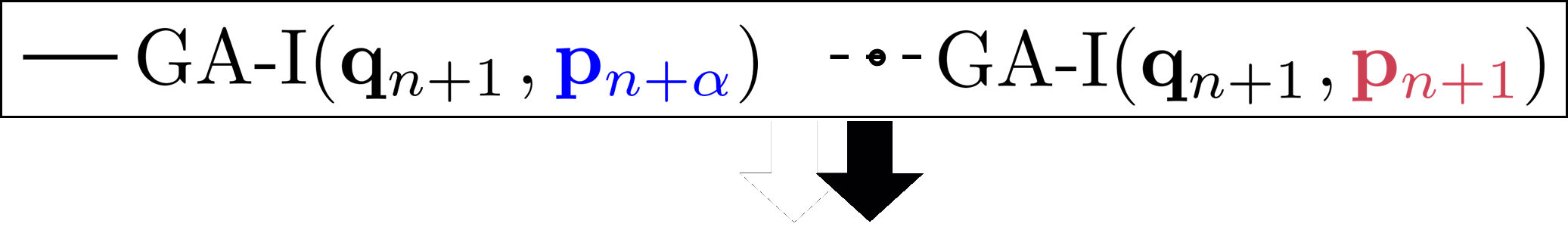} 
	~\hspace{2mm}
	\includegraphics[width=0.31\linewidth]{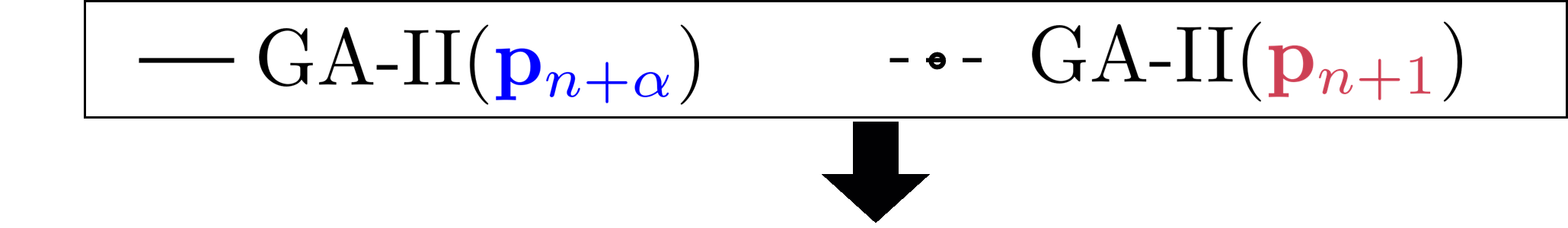}  
	\\[1ex]
	\begin{subfigure}[b]{0.32\textwidth}
		\includegraphics[trim=2 60 25 90,clip,width=1.00\linewidth]{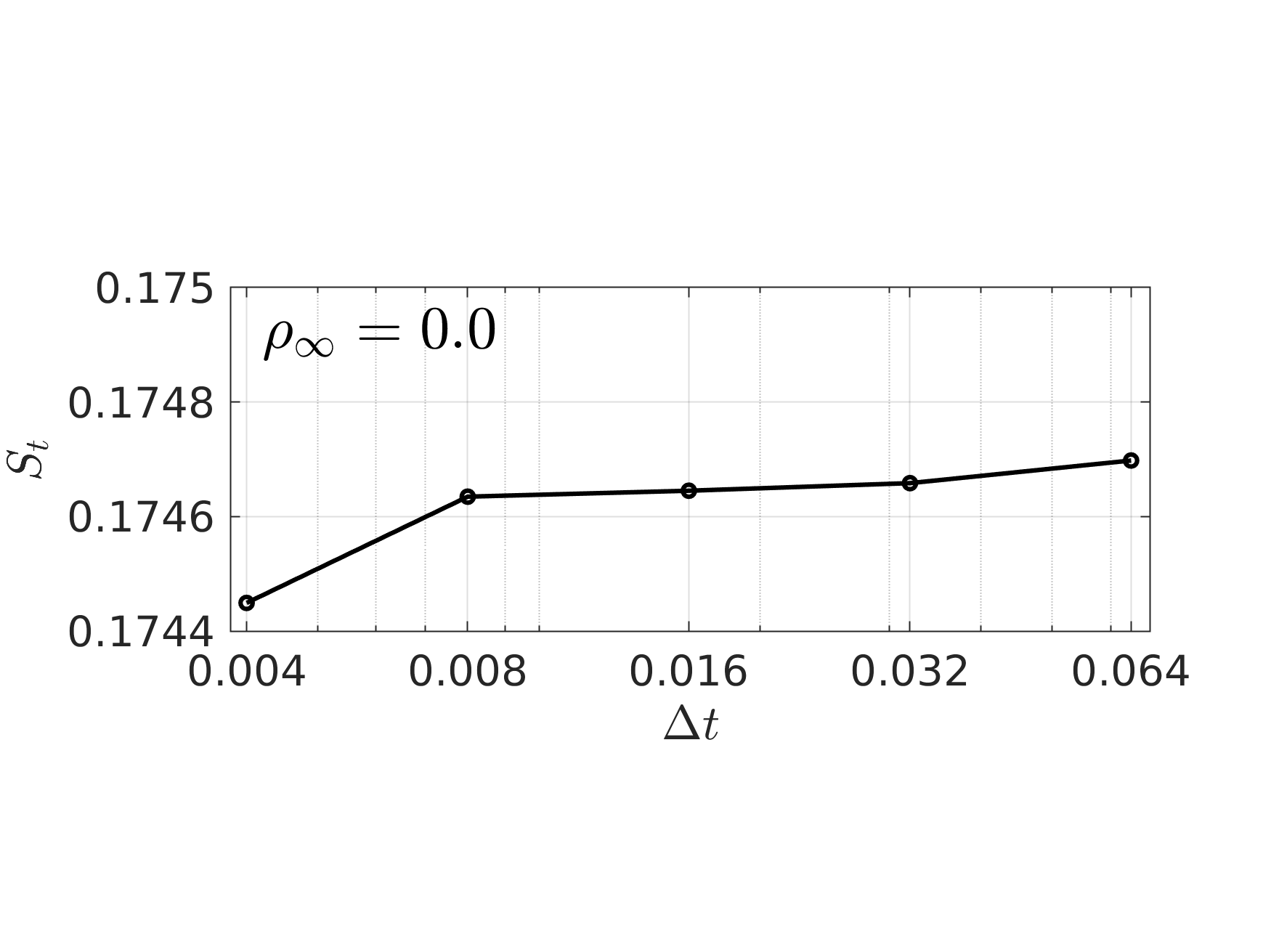}
		\caption{}
		\label{Fig:FAC_GAg_123456_a}
	\end{subfigure}
	~
	\begin{subfigure}[b]{0.32\textwidth}	
		\includegraphics[trim=2 60 25 90,clip,width=1.00\linewidth]{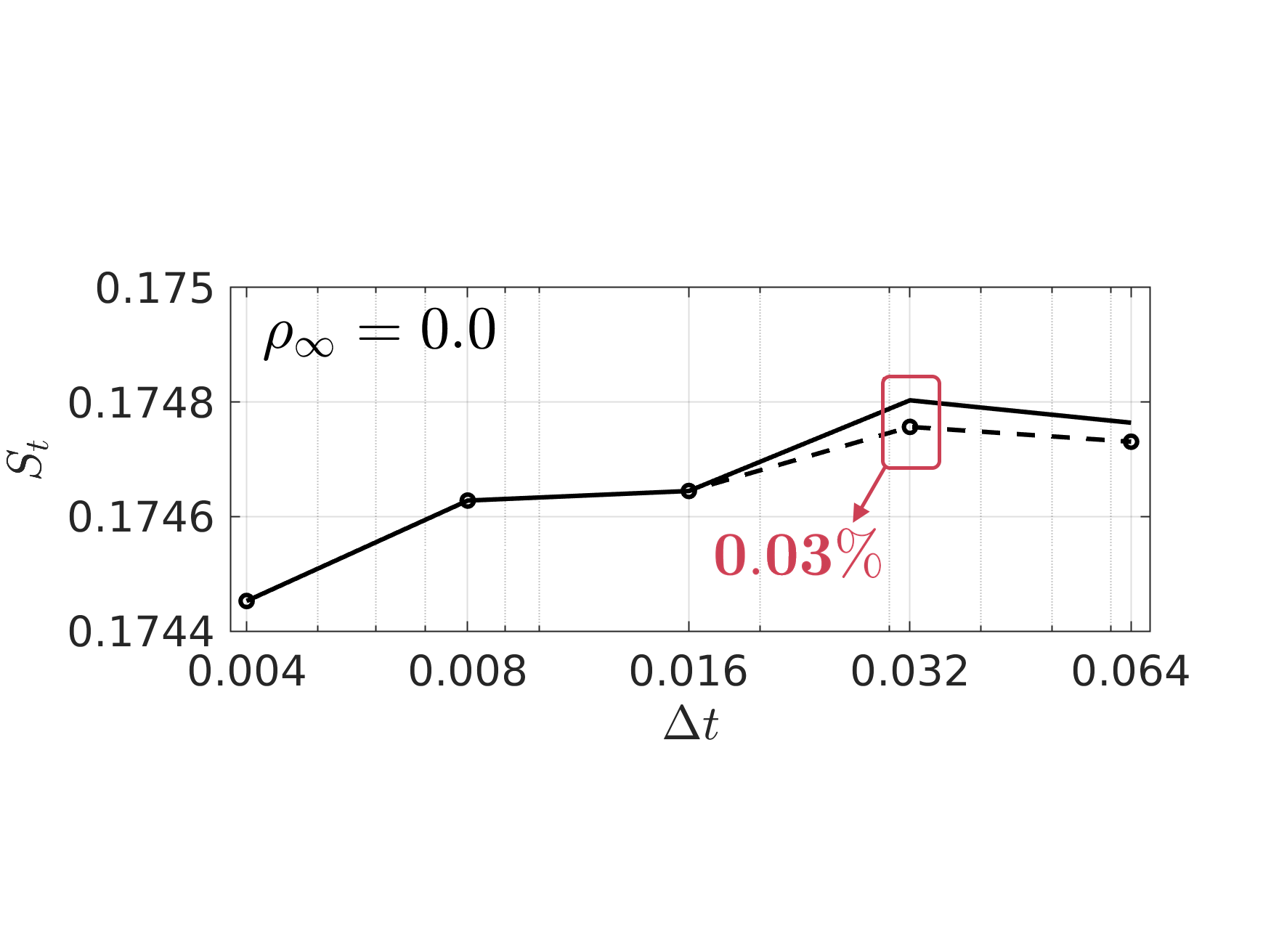}
		\caption{}
		\label{Fig:FAC_GAg_123456_b}
	\end{subfigure}
	~
	\begin{subfigure}[b]{0.32\textwidth}	
		\includegraphics[trim=2 60 25 90,clip,width=1.00\linewidth]{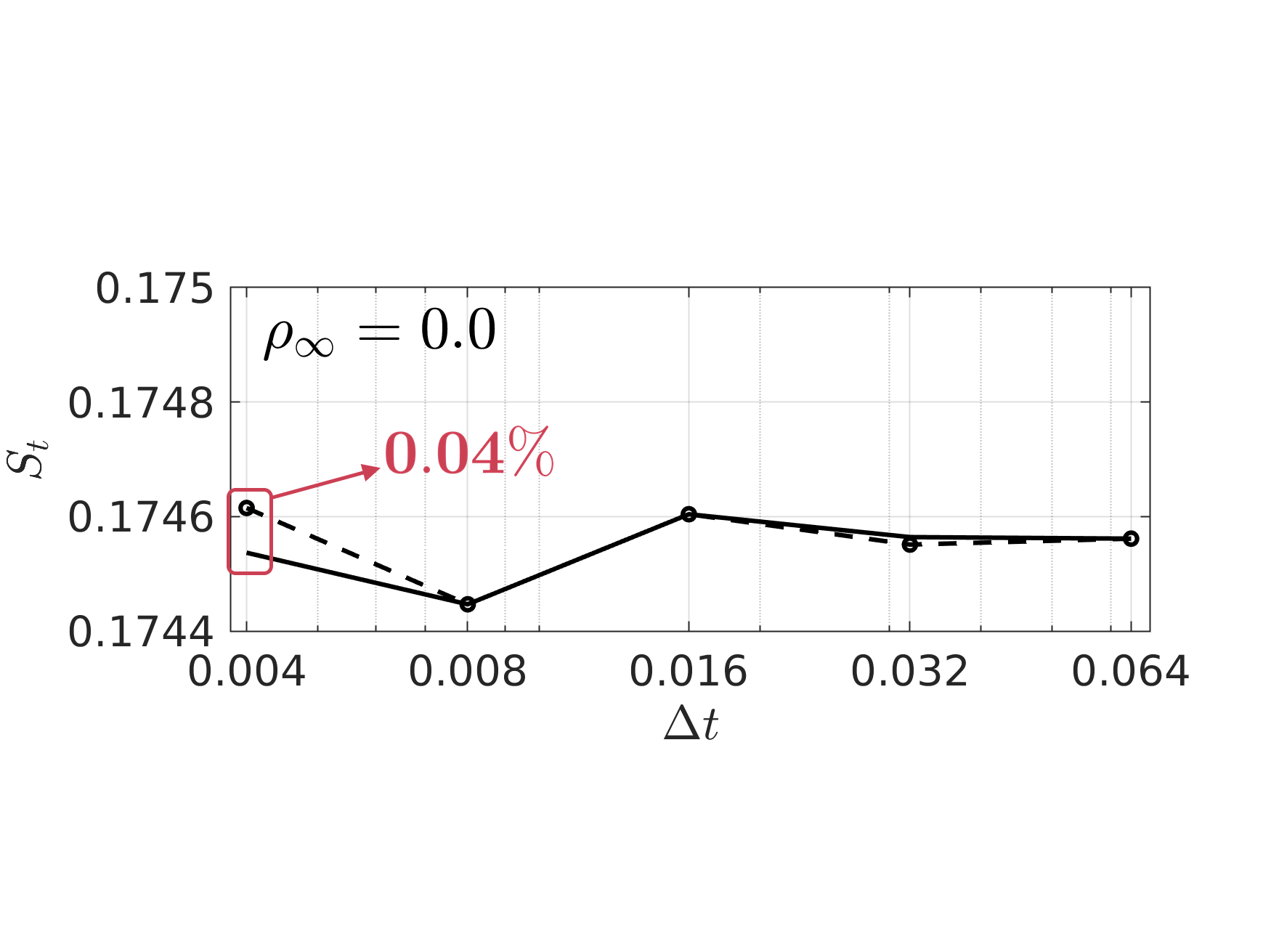}
		\caption{}
		\label{Fig:FAC_GAg_123456_c}
	\end{subfigure}
	\\[-4ex]
	\begin{subfigure}[b]{0.32\textwidth}
		\includegraphics[trim=2 60 25 90,clip,width=1.00\linewidth]{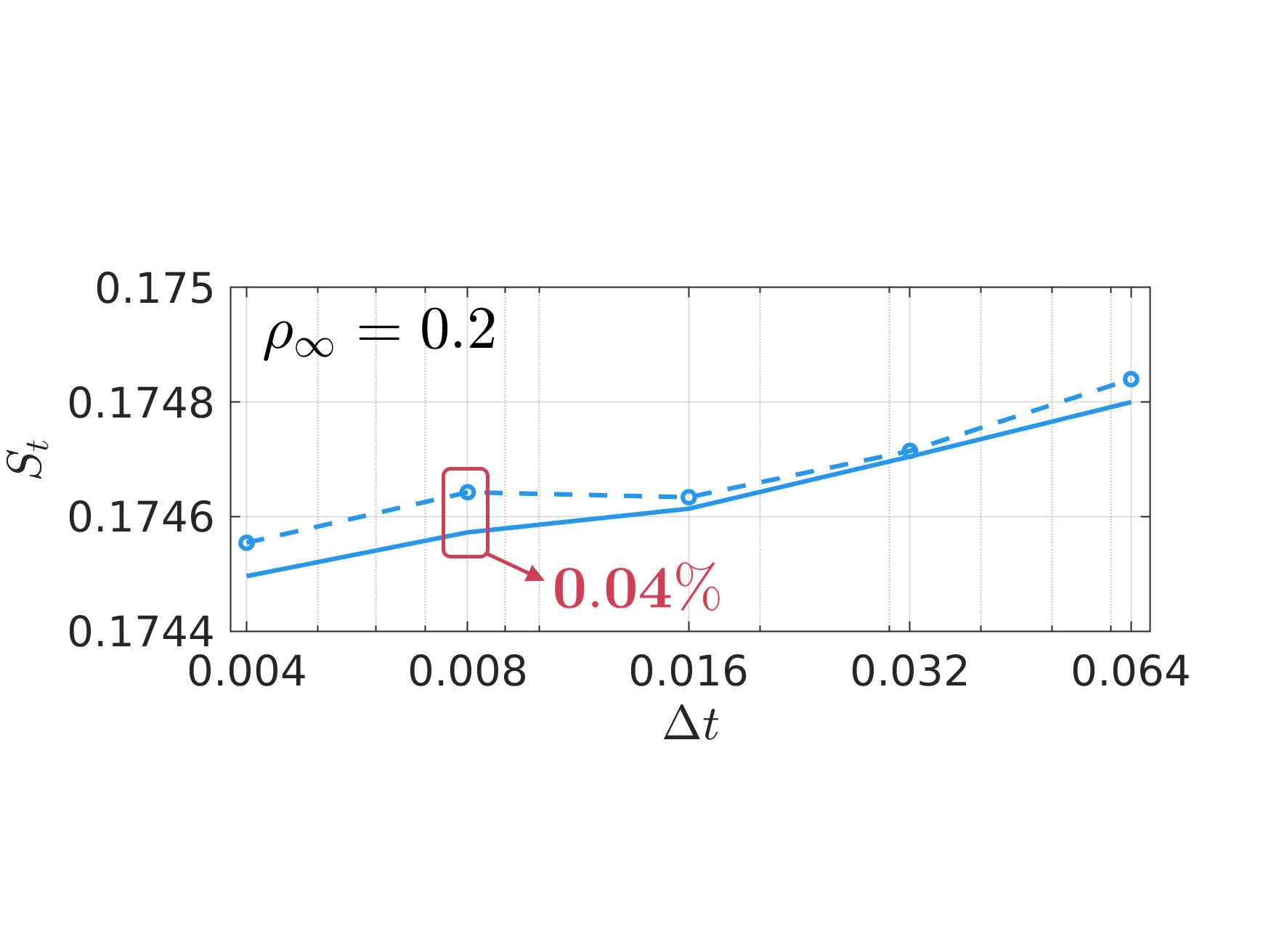}
		\caption{}
		\label{Fig:FAC_GAg_123456_d}
	\end{subfigure}
	~
	\begin{subfigure}[b]{0.32\textwidth}	
		\includegraphics[trim=2 60 25 90,clip,width=1.00\linewidth]{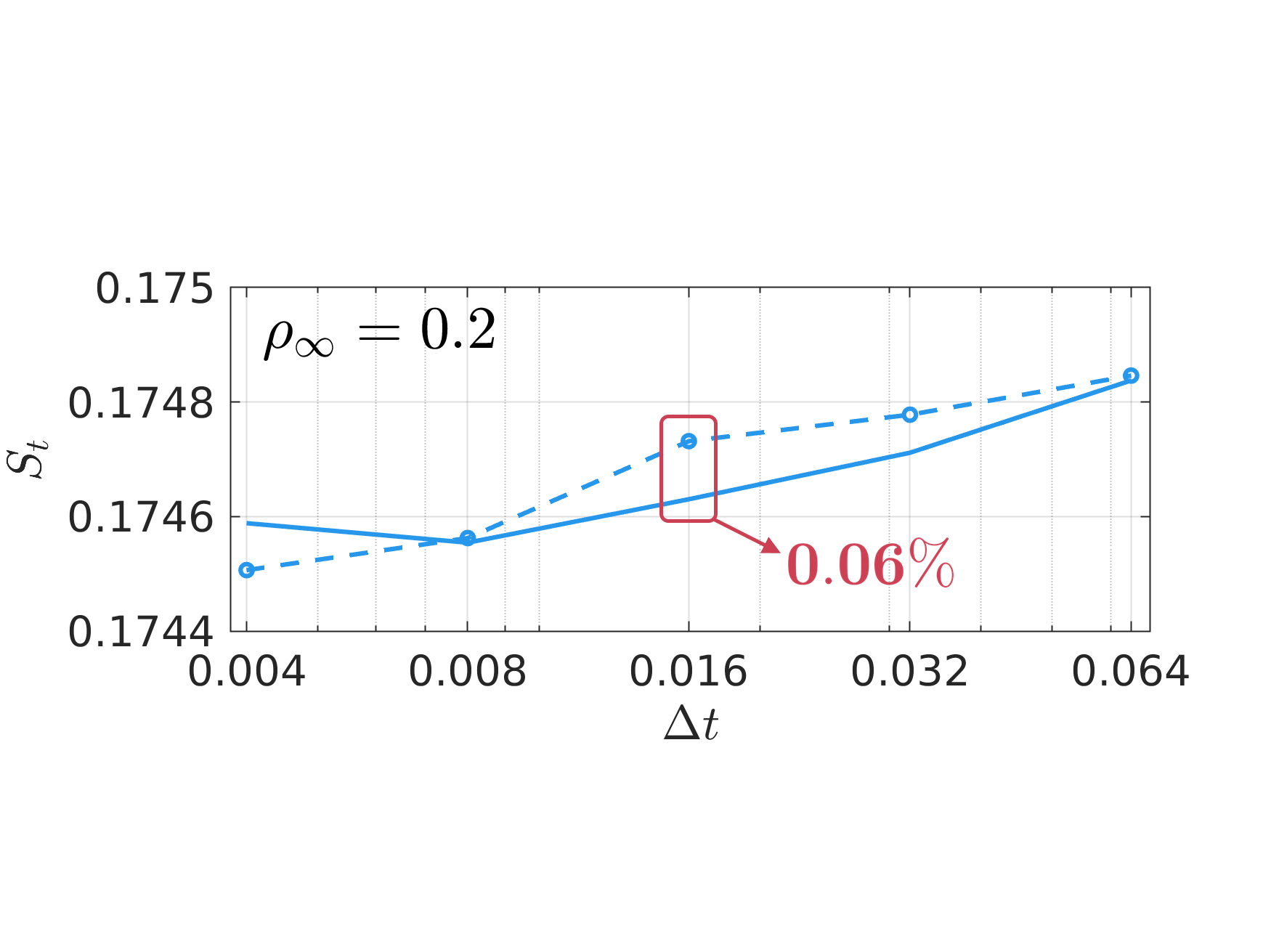}
		\caption{}
		\label{Fig:FAC_GAg_123456_e}
	\end{subfigure}
	~
	\begin{subfigure}[b]{0.32\textwidth}	
		\includegraphics[trim=2 60 25 90,clip,width=1.00\linewidth]{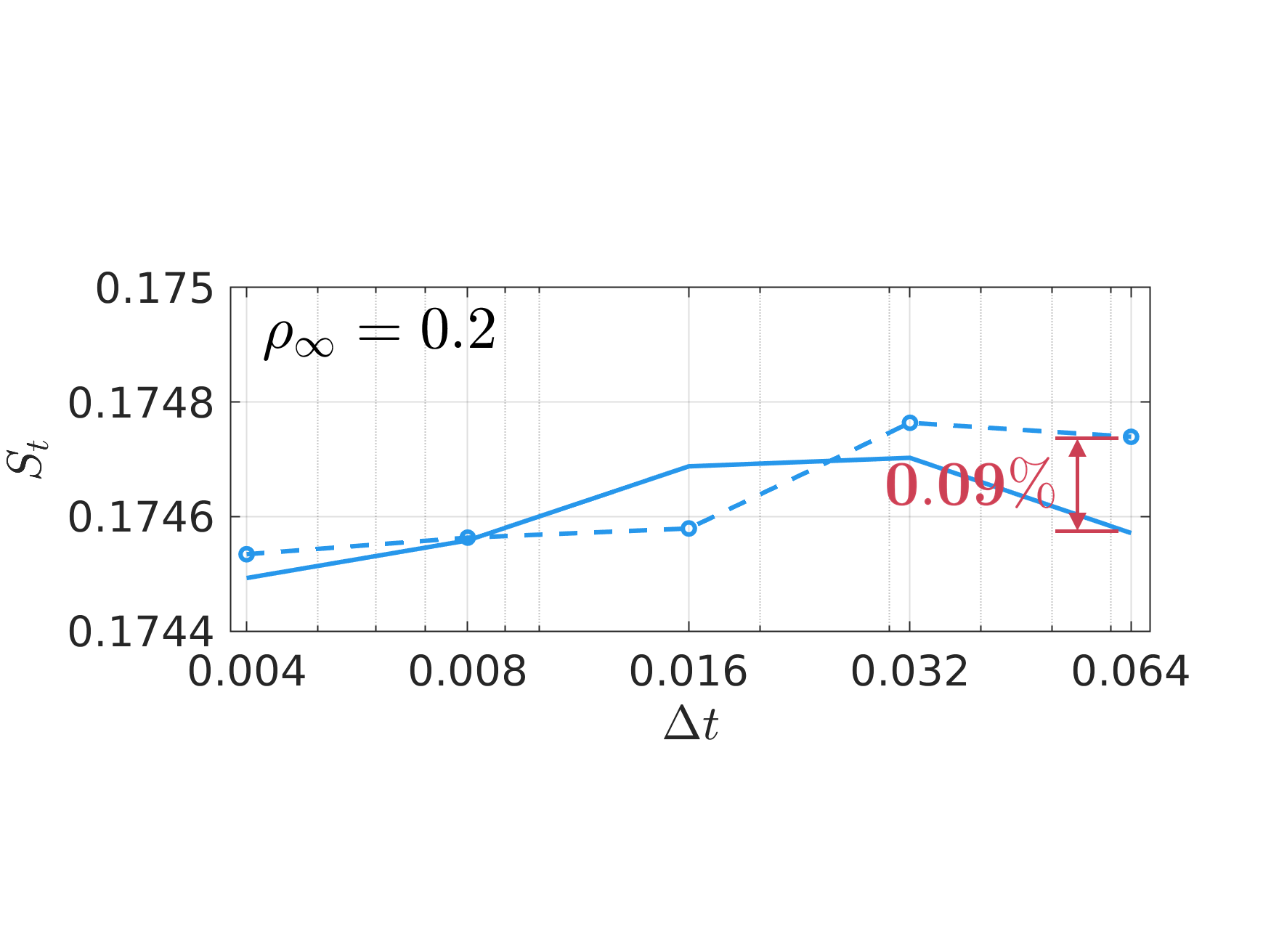}
		\caption{}
		\label{Fig:FAC_GAg_123456_f}
	\end{subfigure}
	\\[-4ex]
	\begin{subfigure}[b]{0.32\textwidth}
		\includegraphics[trim=2 60 25 90,clip,width=1.00\linewidth]{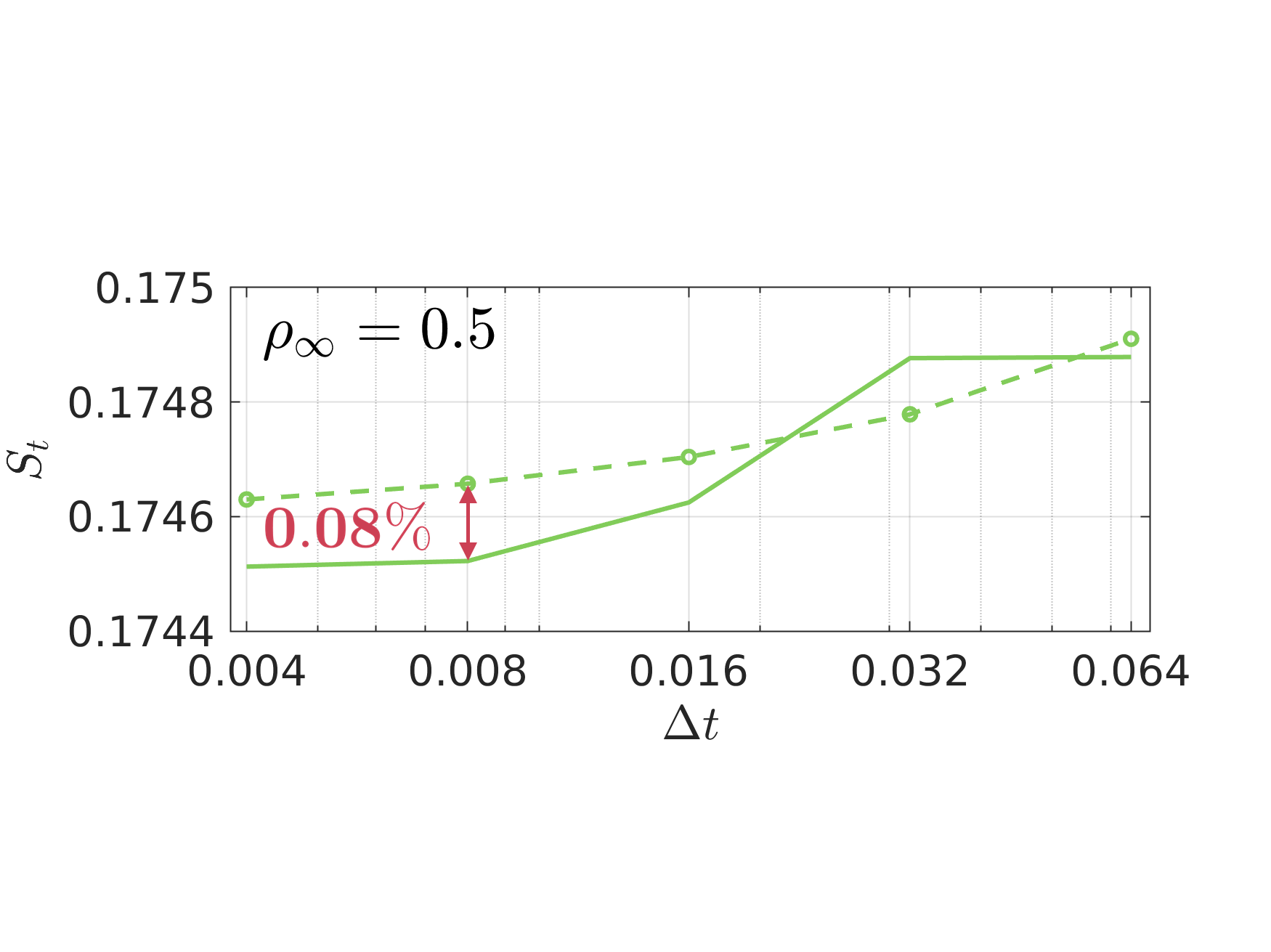}
		\caption{}
		\label{Fig:FAC_GAg_123456_g}
	\end{subfigure}
	~
	\begin{subfigure}[b]{0.32\textwidth}	
		\includegraphics[trim=2 60 25 90,clip,width=1.00\linewidth]{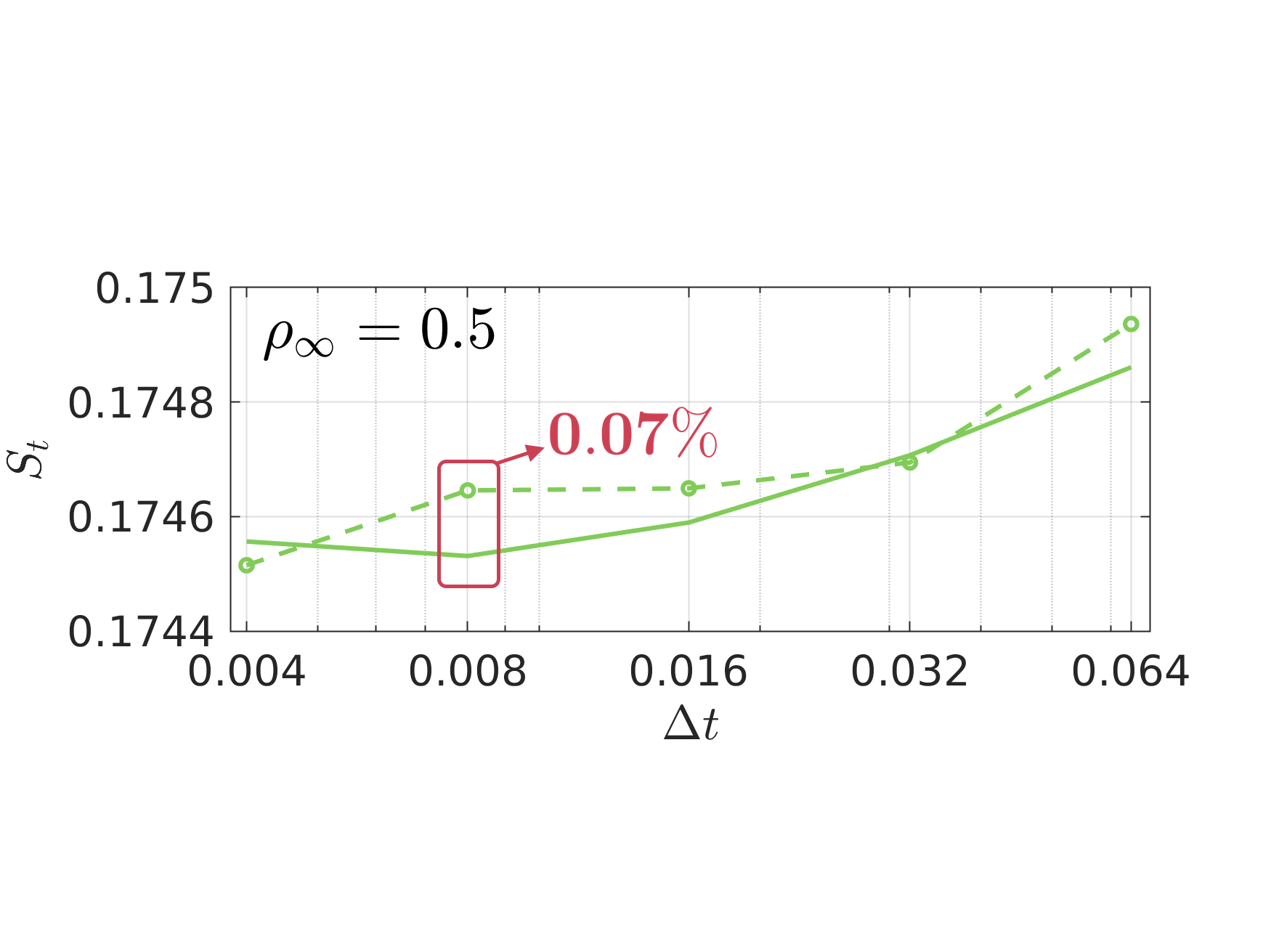}
		\caption{}
		\label{Fig:FAC_GAg_123456_h}
	\end{subfigure}
	~
	\begin{subfigure}[b]{0.32\textwidth}	
		\includegraphics[trim=2 60 25 90,clip,width=1.00\linewidth]{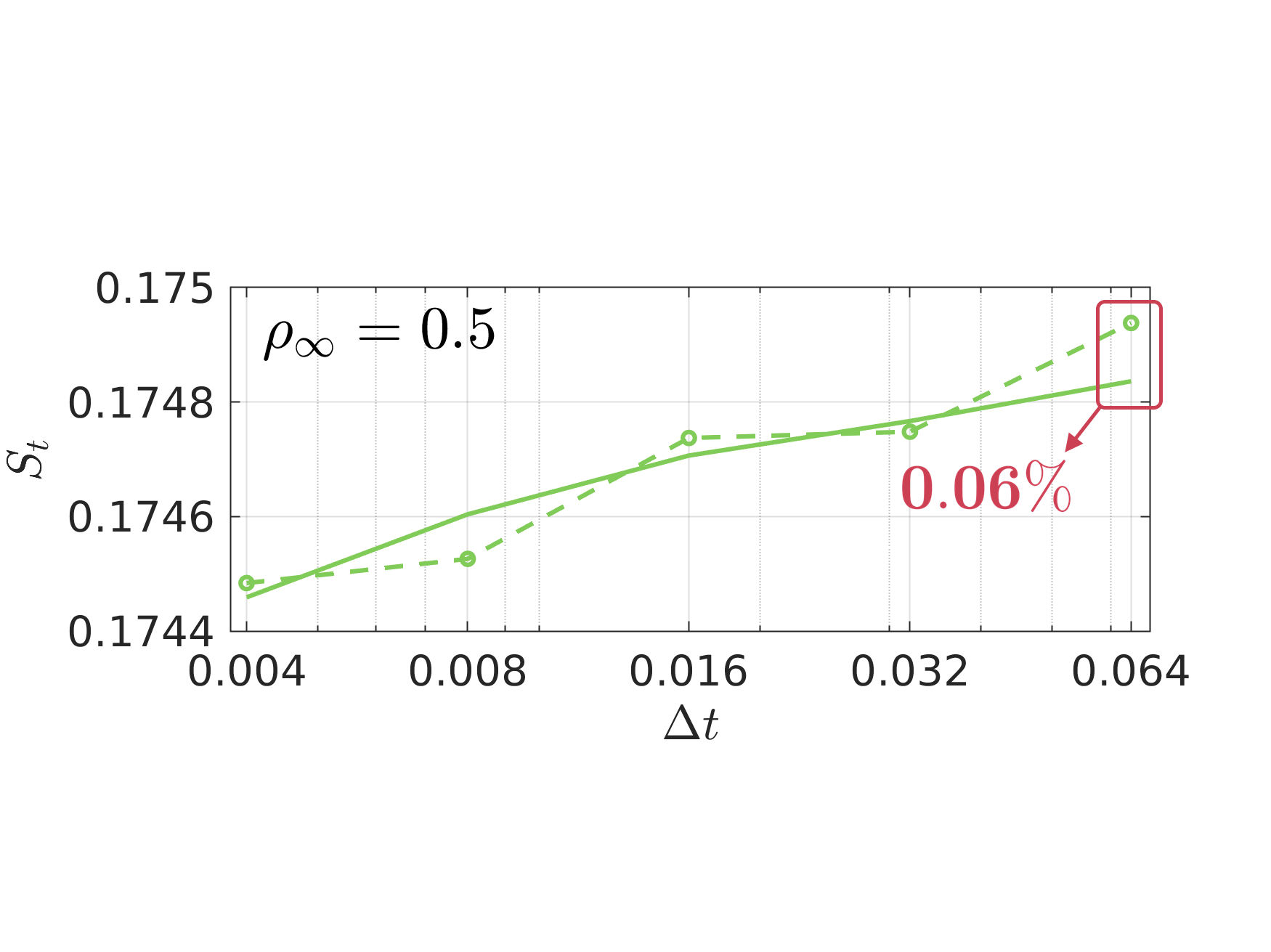}
		\caption{}
		\label{Fig:FAC_GAg_123456_i}
	\end{subfigure}
	\\
	\vspace{-79mm}
	\hspace{6mm} (\textbf{a}) \hspace{54mm} (\textbf{b}) \hspace{54mm} (\textbf{c})
	\\
	\vspace{20mm}
	\hspace{6mm} (\textbf{d}) \hspace{54mm} (\textbf{e}) \hspace{54mm} (\textbf{f})
	\\
	\vspace{19mm}
	\hspace{6mm} (\textbf{g}) \hspace{54mm} (\textbf{h}) \hspace{54mm} (\textbf{i}) 
	\\
	\vspace{18mm}
\caption{Strouhal number ($S_t$) from the flow around a cylinder at Re = 100. Results obtained with different implementation approaches of the Generalized-$\alpha$ method. Continuous and dashed lines represent approaches that write pressure at $t_{n+\alpha_f}$ and $t_{n+1}$, respectively. Different spectral radii are considered: (a-c) $\rho_\infty = 0.0$, (d-f) $\rho_\infty = 0.2$, and (g-i) $\rho_\infty = 0.5$. }
\label{Fig:FAC_GAg_123456}
\end{figure}
\begin{figure}[t] \captionsetup[sub]{font=normalsize}\captionsetup[subfigure]{labelformat=empty}
	\centering
	\includegraphics[width=0.31\linewidth]{Figures/results/legend_f_18_19_1.png} 
	~\hspace{2mm}
	\includegraphics[width=0.31\linewidth]{Figures/results/legend_f_18_19_2.png} 
	~\hspace{2mm}
	\includegraphics[width=0.31\linewidth]{Figures/results/legend_f_18_19_3.png} 
	\\[1ex]
	\begin{subfigure}[b]{0.32\textwidth}
		\includegraphics[trim=2 60 25 82,clip,width=1.00\linewidth]{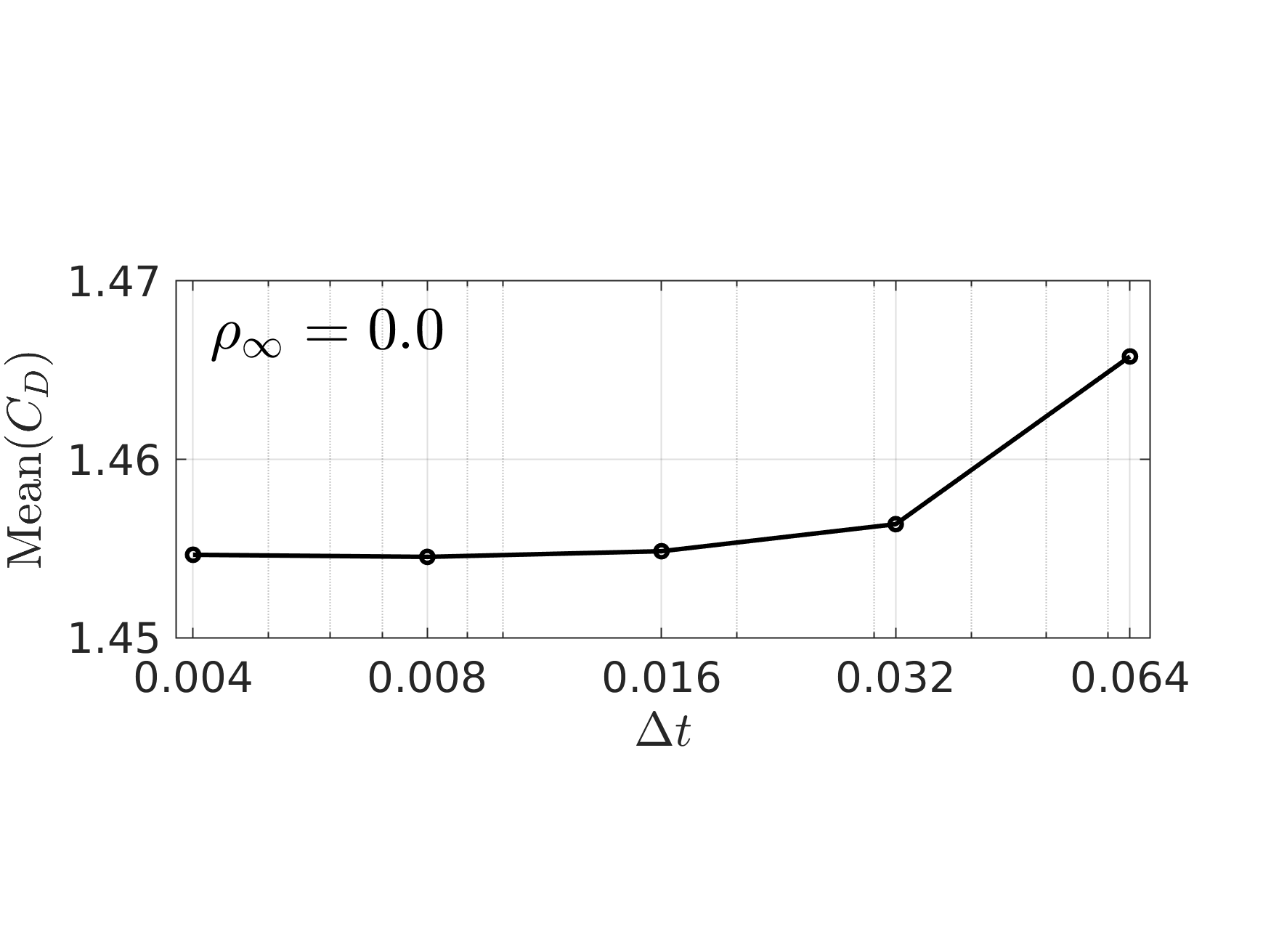}
		\caption{}
		\label{Fig:FAC_GAg_123456_CD_a}
	\end{subfigure}
	~
	\begin{subfigure}[b]{0.32\textwidth}	
		\includegraphics[trim=2 60 25 82,clip,width=1.00\linewidth]{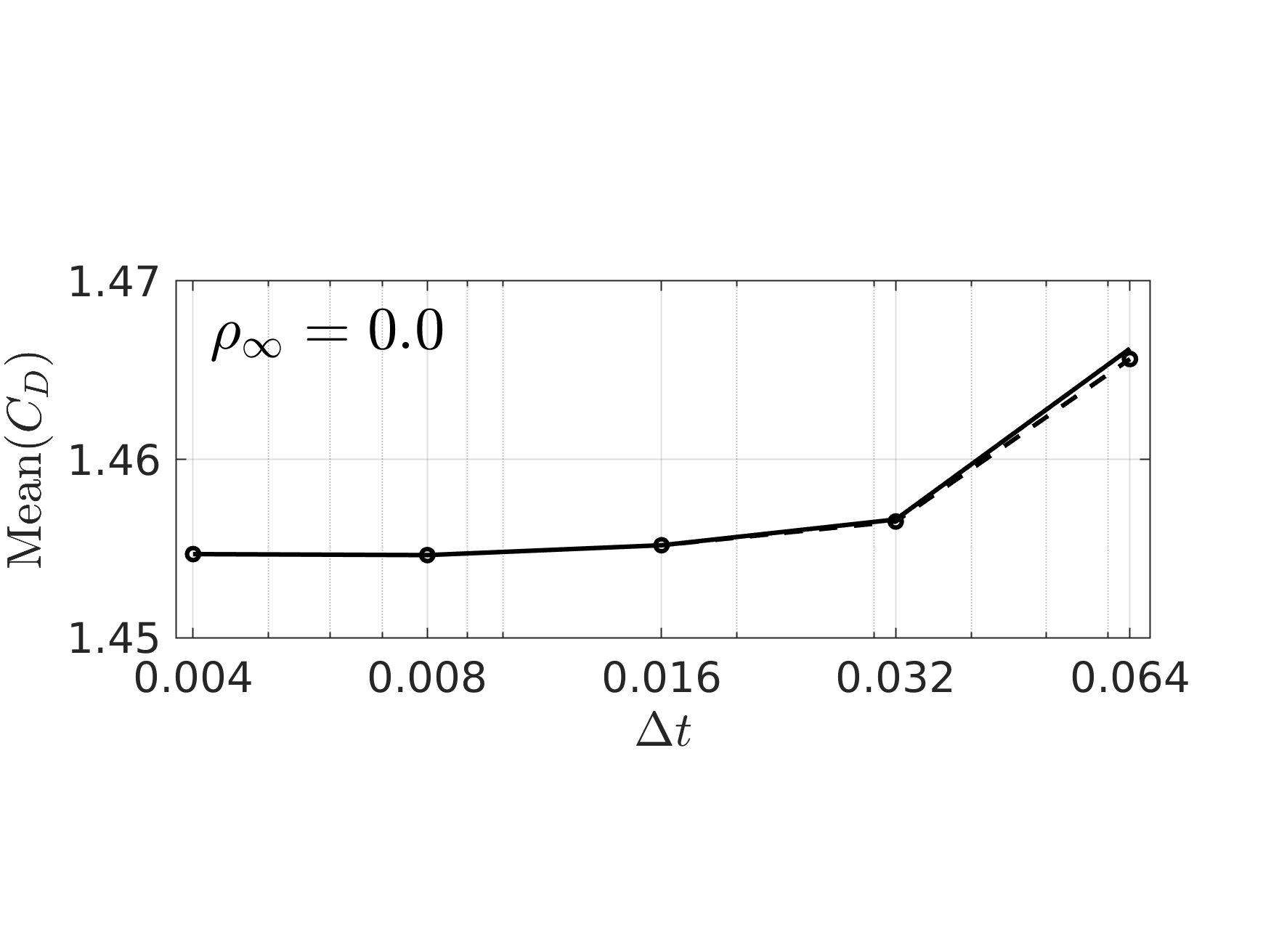}
		\caption{}
		\label{Fig:FAC_GAg_123456_CD_b}
	\end{subfigure}
	~
	\begin{subfigure}[b]{0.32\textwidth}	
		\includegraphics[trim=2 60 25 82,clip,width=1.00\linewidth]{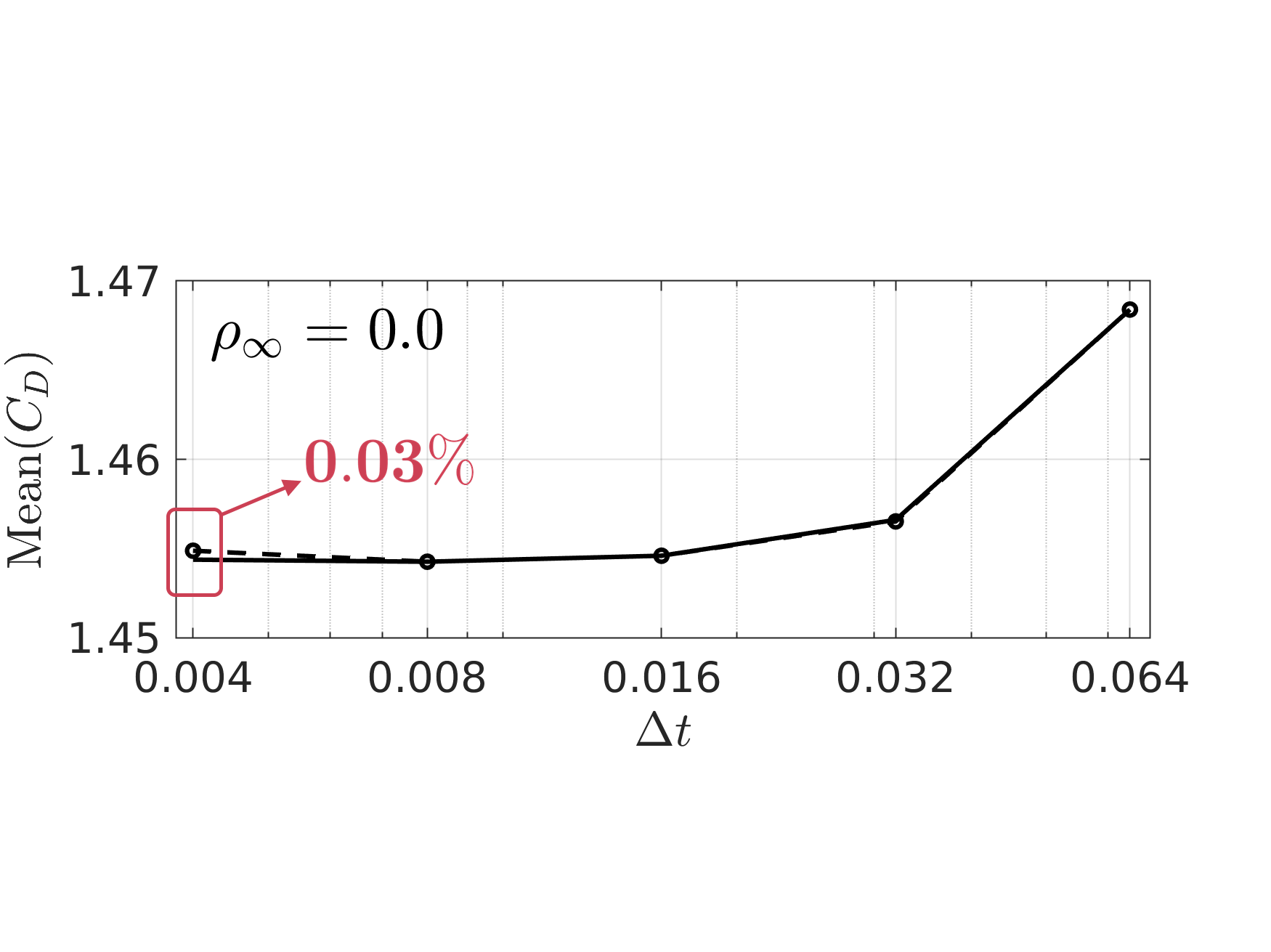}
		\caption{}
		\label{Fig:FAC_GAg_123456_CD_c}
	\end{subfigure}
	\\[-4ex]
	\begin{subfigure}[b]{0.32\textwidth}
		\includegraphics[trim=2 60 25 82,clip,width=1.00\linewidth]{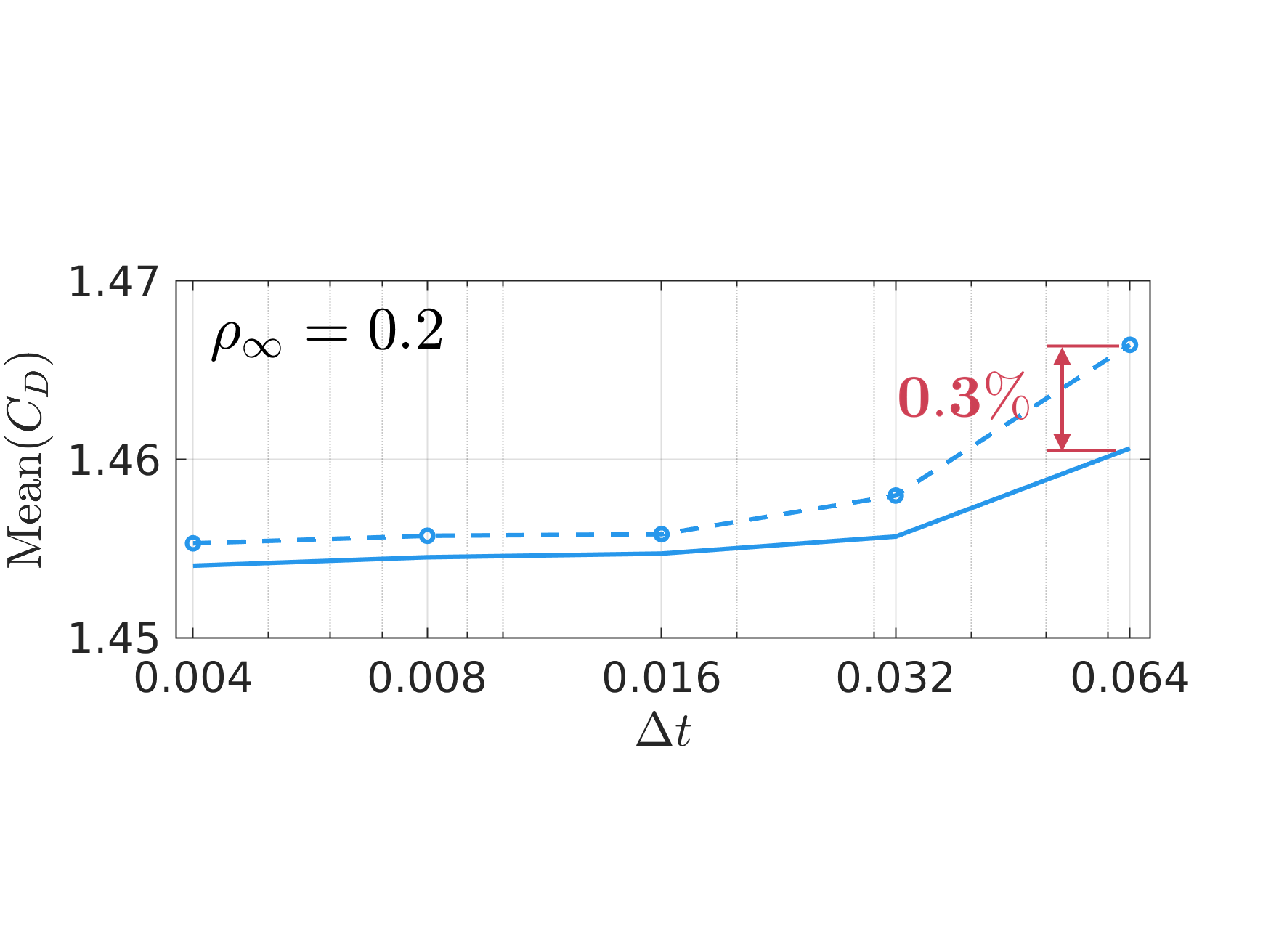}
		\caption{}
		\label{Fig:FAC_GAg_123456_CD_d}
	\end{subfigure}
	~
	\begin{subfigure}[b]{0.32\textwidth}	
		\includegraphics[trim=2 60 25 82,clip,width=1.00\linewidth]{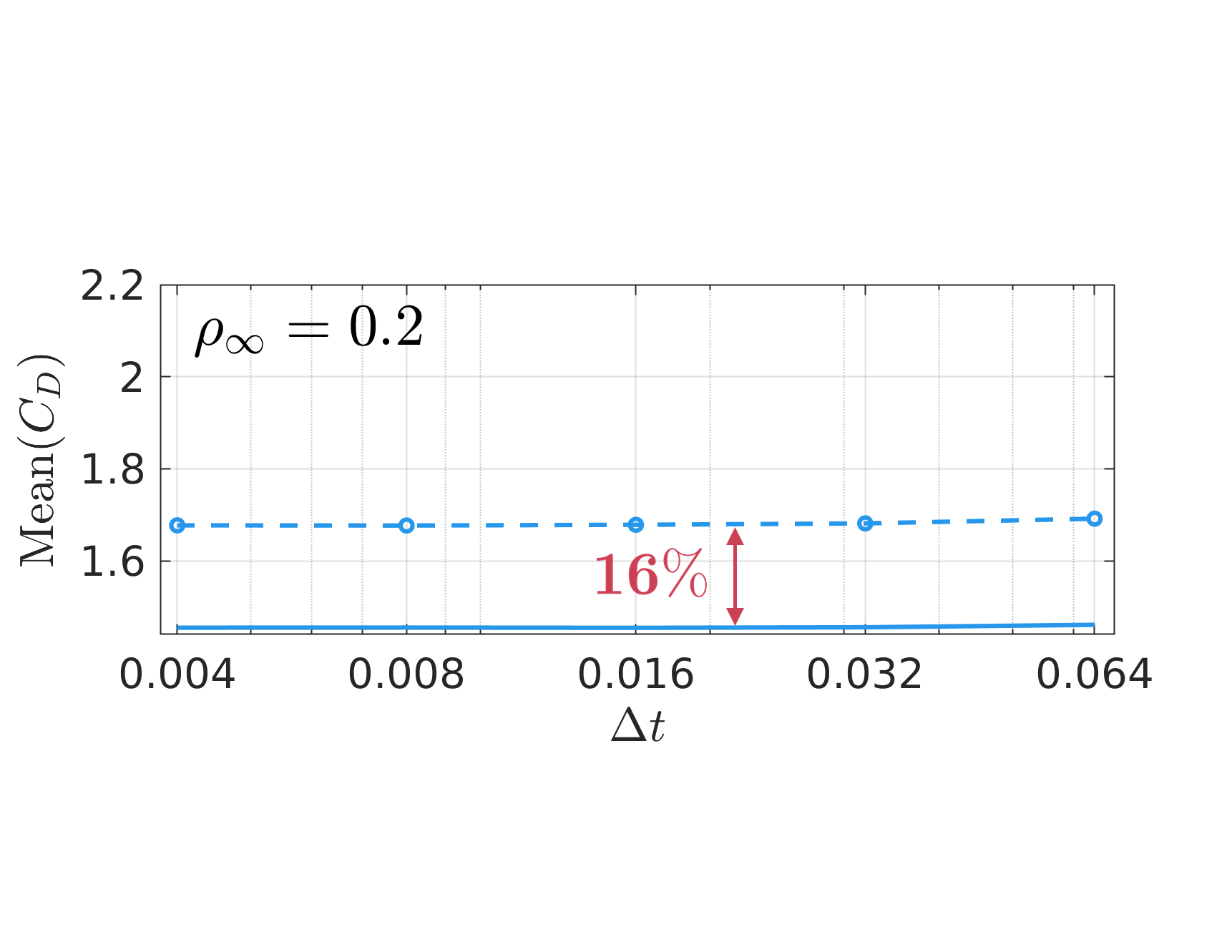}
		\caption{}
		\label{Fig:FAC_GAg_123456_CD_e}
	\end{subfigure}
	~
	\begin{subfigure}[b]{0.32\textwidth}	
		\includegraphics[trim=2 60 25 82,clip,width=1.00\linewidth]{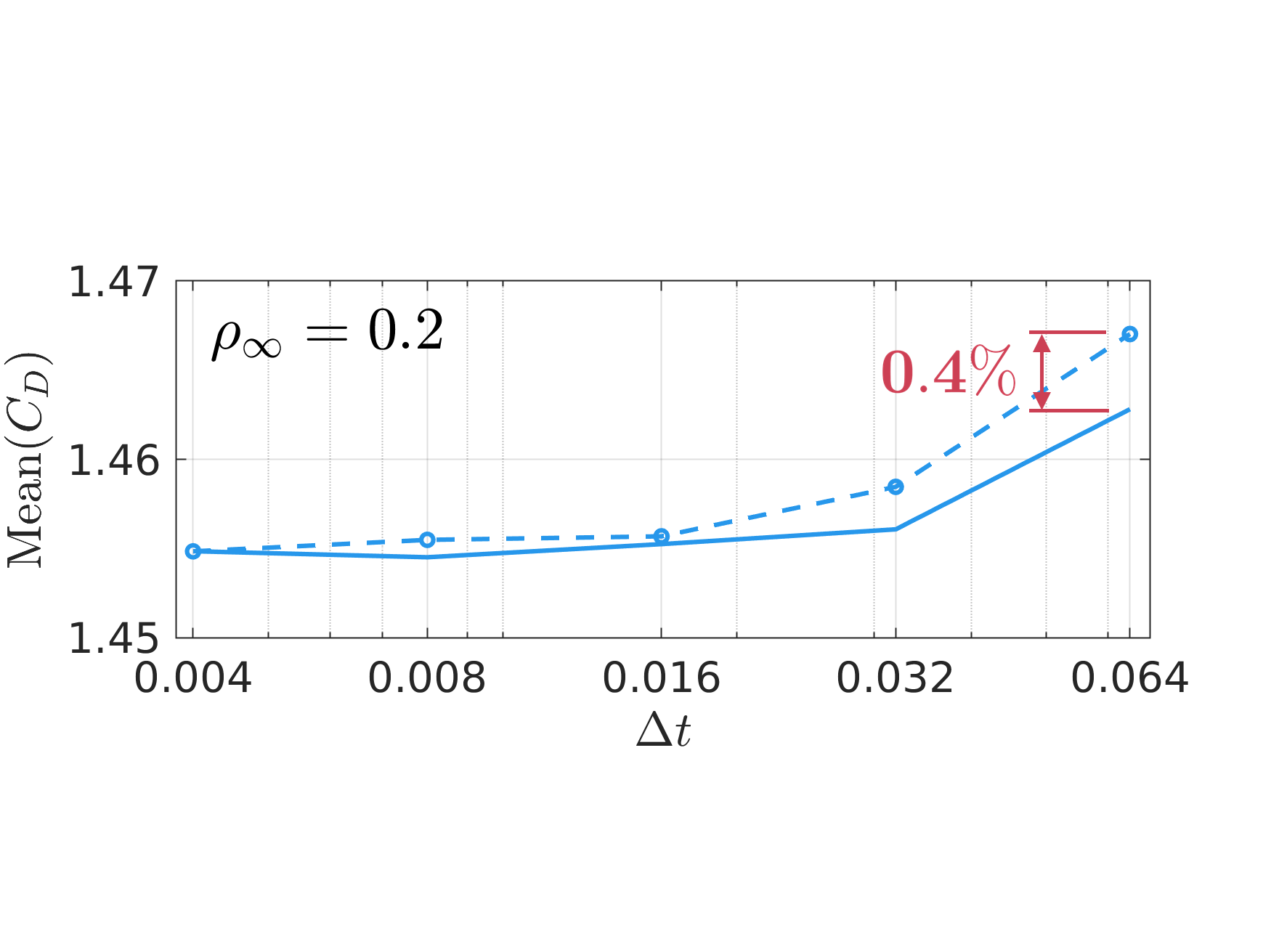}
		\caption{}
		\label{Fig:FAC_GAg_123456_CD_f}
	\end{subfigure}
	\\[-4ex]
	\begin{subfigure}[b]{0.32\textwidth}
		\includegraphics[trim=2 60 25 82,clip,width=1.00\linewidth]{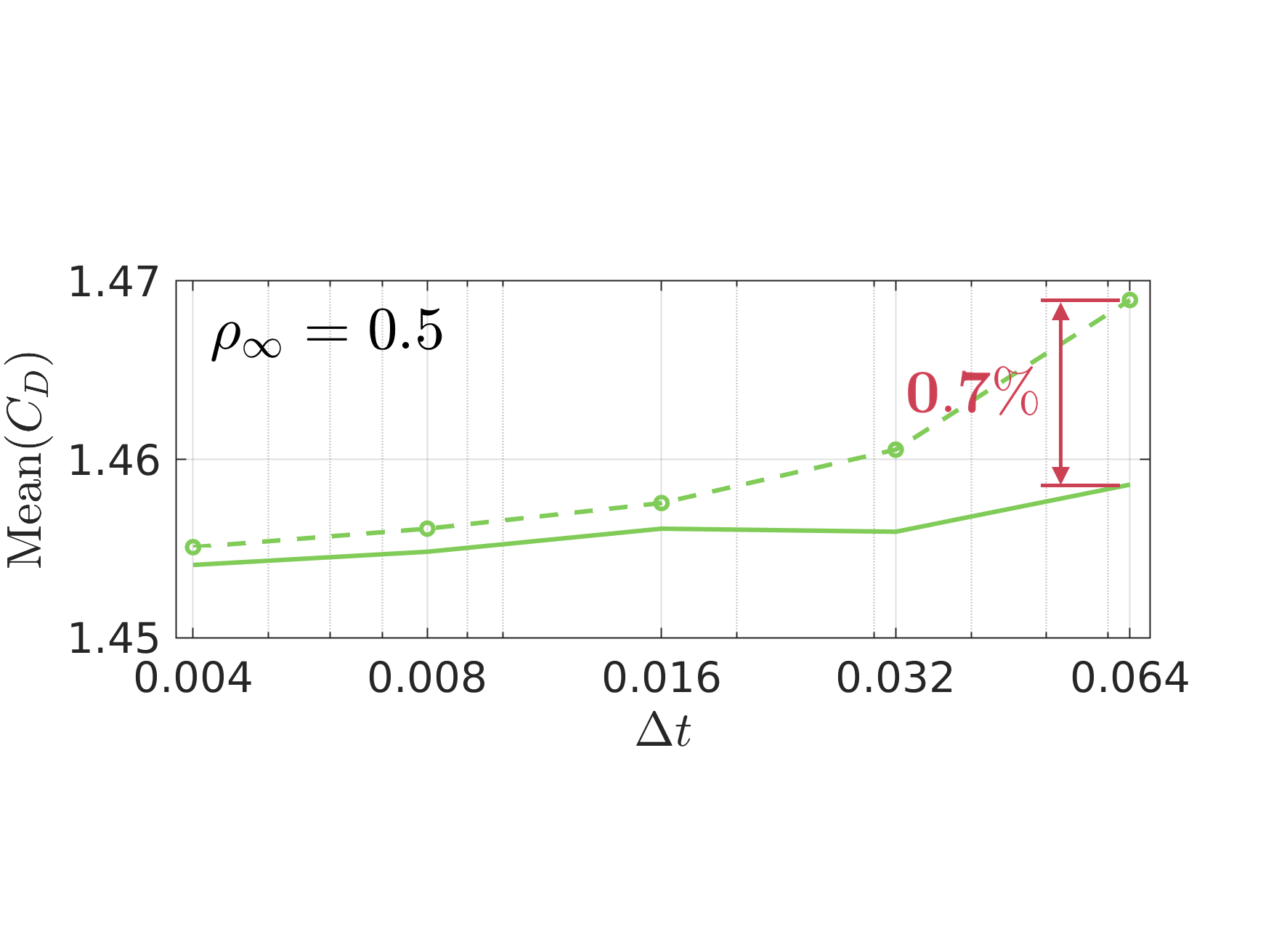}
		\caption{}
		\label{Fig:FAC_GAg_123456_CD_g}
	\end{subfigure}
	~
	\begin{subfigure}[b]{0.32\textwidth}	
		\includegraphics[trim=2 60 25 82,clip,width=1.00\linewidth]{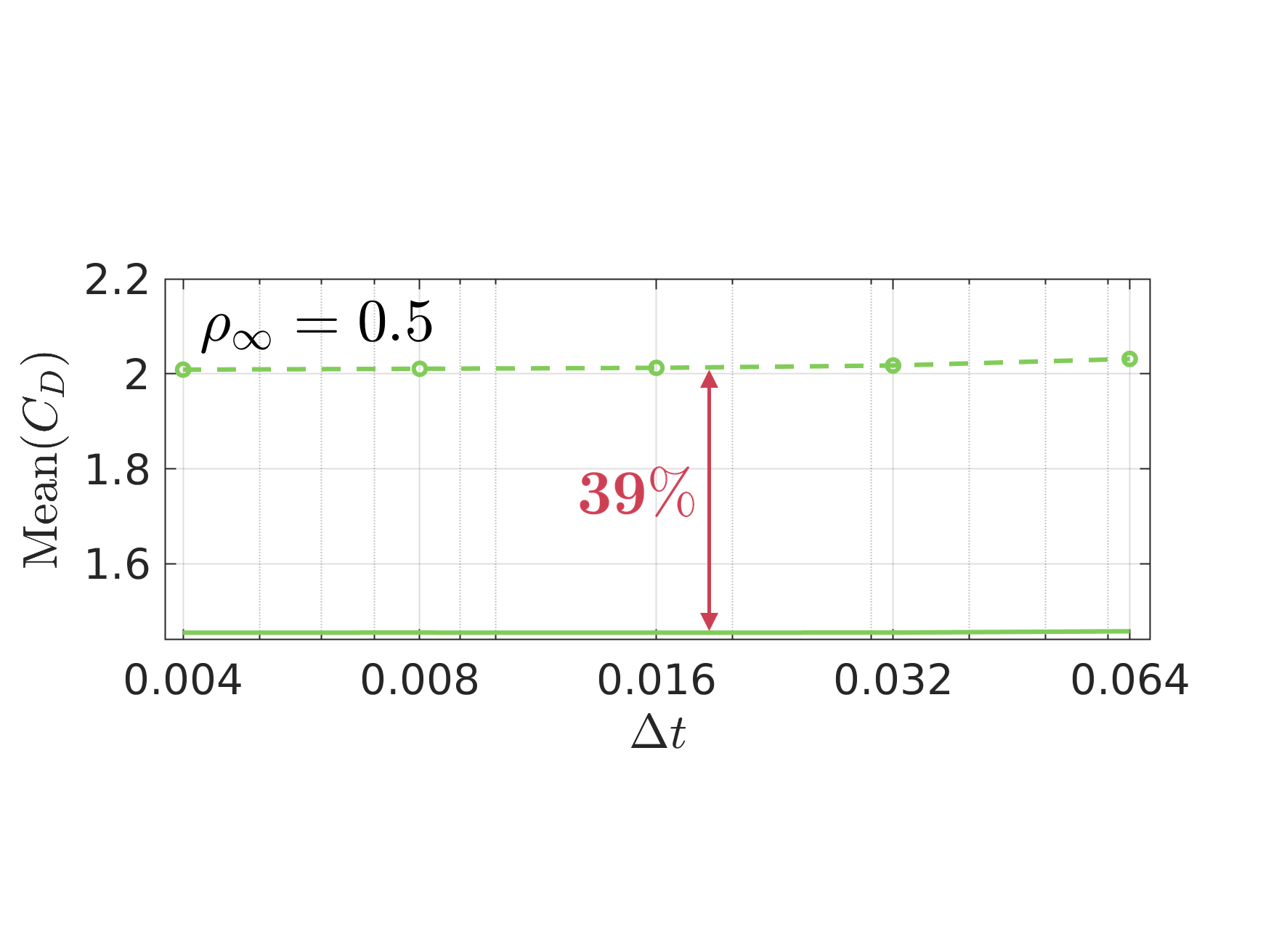}
		\caption{}
		\label{Fig:FAC_GAg_123456_CD_h}
	\end{subfigure}
	~
	\begin{subfigure}[b]{0.32\textwidth}	
		\includegraphics[trim=2 60 25 82,clip,width=1.00\linewidth]{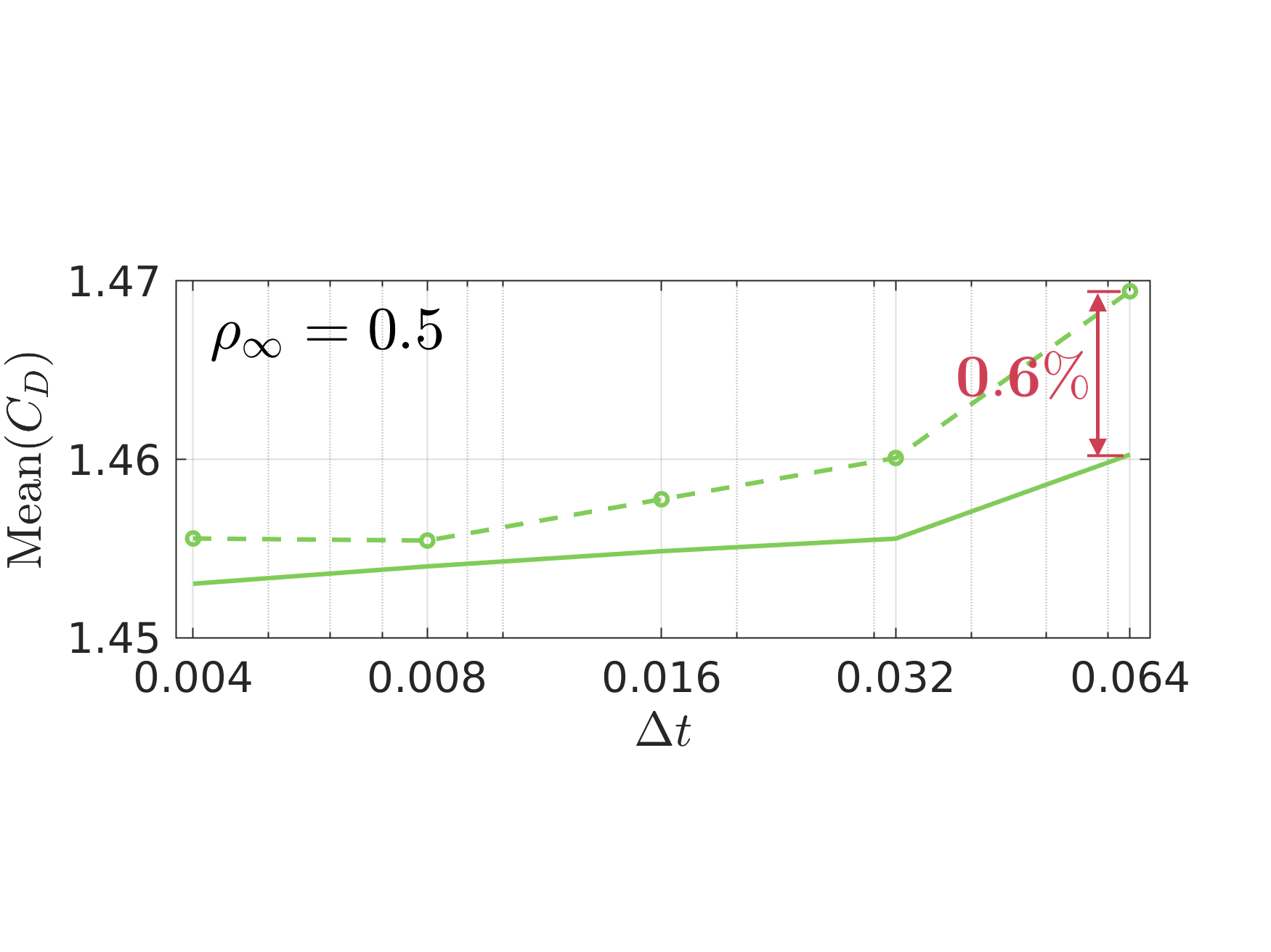}
		\caption{}
		\label{Fig:FAC_GAg_123456_i}
	\end{subfigure}
	\\
	\vspace{-83.5mm}
	\hspace{6mm} (\textbf{a}) \hspace{54mm} (\textbf{b}) \hspace{54mm} (\textbf{c})
	\\
	\vspace{22mm}
	\hspace{6mm} (\textbf{d}) \hspace{54mm} (\textbf{e}) \hspace{54mm} (\textbf{f})
	\\
	\vspace{21mm}
	\hspace{6mm} (\textbf{g}) \hspace{54mm} (\textbf{h}) \hspace{54mm} (\textbf{i}) 
	\\
	\vspace{18mm}
\caption{Mean value of drag coefficient ($C_D$) from the flow around a cylinder at Re = 100. Results obtained with different implementation approaches of the Generalized-$\alpha$ method. Continuous and dashed lines represent approaches that write pressure at $t_{n+\alpha_f}$ and $t_{n+1}$, respectively. Different spectral radii are considered: (a-c) $\rho_\infty = 0.0$, (d-f) $\rho_\infty = 0.2$, and (g-i) $\rho_\infty = 0.5$. }
\label{Fig:FAC_GAg_123456_CD}
\end{figure}

The second scenario for $\GAd$, $\GAe$ and $\GAf$ integrates the gradient matrix in $\Omega(t_{n+\alpha_f})$, as done in Eulerian-based formulations that adopt the modified approach for the pressure. Three values of $\rho_\infty$ are considered, 0.0, 0.2 and 0.5. Results for the Strouhal number are shown in Fig.~\ref{Fig:FAC_GAg_123456}. This summarizes all the implementation approaches and plots on each graph the schemes that write the pressure at $t_{n+\alpha}$ (solid line) and $t_{n+1}$ ( dashed line). Unlike the previous scenario (using $\mathbf{D}_{n+1}$) here the Strouhal number is not significantly influenced. The difference with respect to $\GAa$ is about 0.1 $\%$ for $\rho_\infty$ = 0.5. Something similar is observed for the amplitude of the lift coefficient (not reported here). However, a different behavior is observed in the drag force. Fig.~\ref{Fig:FAC_GAg_123456_CD} shows the mean of the drag coefficient. As expected, for $\rho_\infty$ = 0.0 all 6 approaches ($\GAa$, $\GAb$, $\GAc$, $\GAd$, $\GAe$ and $\GAf$) yield very similar results. However, as $\rho_\infty$ increases, schemes start to differ. The trend is that the mean drag coefficient grows in the schemes that write the pressure at $t_{n+1}$. In addition, one observes a notorious difference between implementation approaches $\GAd$, $\GAe$, and $\GAf$, which has not been exhibited in the previous example. Interestingly, $\GAe$ features a high dependence on the value of $\rho_\infty$, even more, the converged value ($\Delta t \to 0$) seems to be dependent on $\rho_\infty$. Although this behaviour is observed only in the drag force, it is an inconsistency of the Generalized-$\alpha$ method that arises from writing the pressure at $t_{n+1}$ instead of $t_{n+\alpha_f}$, and integrating the gradient matrix in $\Omega(t_{n+\alpha_f})$.

\subsection{Collapse of a cylindrical water column}\label{sec:CCWC}

The final problem consists of simulating the collapse of a water column whose initial condition follows the experimental setting of Martin and Moyce \citep{martin1952part}. The purpose is to validate performance of the analyzed time integration schemes on a problem with a different simulation setting. For this reason, the problem is modelled in 3D, as shown in Fig.~\ref{Fig:cylindricalWatterColumn_TestCase_a}. Slip boundary condition is used in the base (plane $x$-$y$) and in the symmetry plane ($y$-$z$). The initial water column radius is $r_0 = 0.05715$ m and its height is $h_0 = 2\:r_0$. The acceleration of gravity is $g = 9.81$ m/s$^2$, the density $\rho = 1000$ kg/m$^3$ and the viscosity $\mu = 0.001$ Pa s. A non-uniform discretization is chosen for refining elements on the free surface, as shown in Fig.~\ref{Fig:cylindricalWatterColumn_TestCase_b}. The characteristic element size is 5 mm at the free surface and 10 mm at a distance $r_0$ from the free surface, while a linear interpolation defines the size of elements in between. \review{For some problems,} the time step size ($\Delta t$) is set as a function of a maximum CFL number set at the beginning of the simulation. The CFL number of an element is computed as:

\begin{equation}\label{EQ:CFL}
\text{cfl} = \frac{\mathrm{max}(\:\lVert \velbf \rVert\:) \: \Delta t}{\mathrm{d}_\mathrm{eq}}
\end{equation}
\vspace{1mm}

\noindent where $\mathrm{max}(\:\lVert \velbf \rVert\:)$ is the maximum velocity norm of the element nodes and $\mathrm{d}_\mathrm{eq}$ is the equivalent spherical diameter of the tetrahedral element. \review{The maximum CFL number is denoted as CFL$^*$ and is computed as CFL$^*$ = max(\textbf{cfl})}, max(\textbf{cfl}) being the vector that gathers the cfl of all elements.

\begin{figure}[t] \captionsetup[sub]{font=normalsize}\captionsetup[subfigure]{labelformat=empty}
\centering 
	\begin{subfigure}[b]{0.30\textwidth}
		\includegraphics[width=0.75\linewidth]{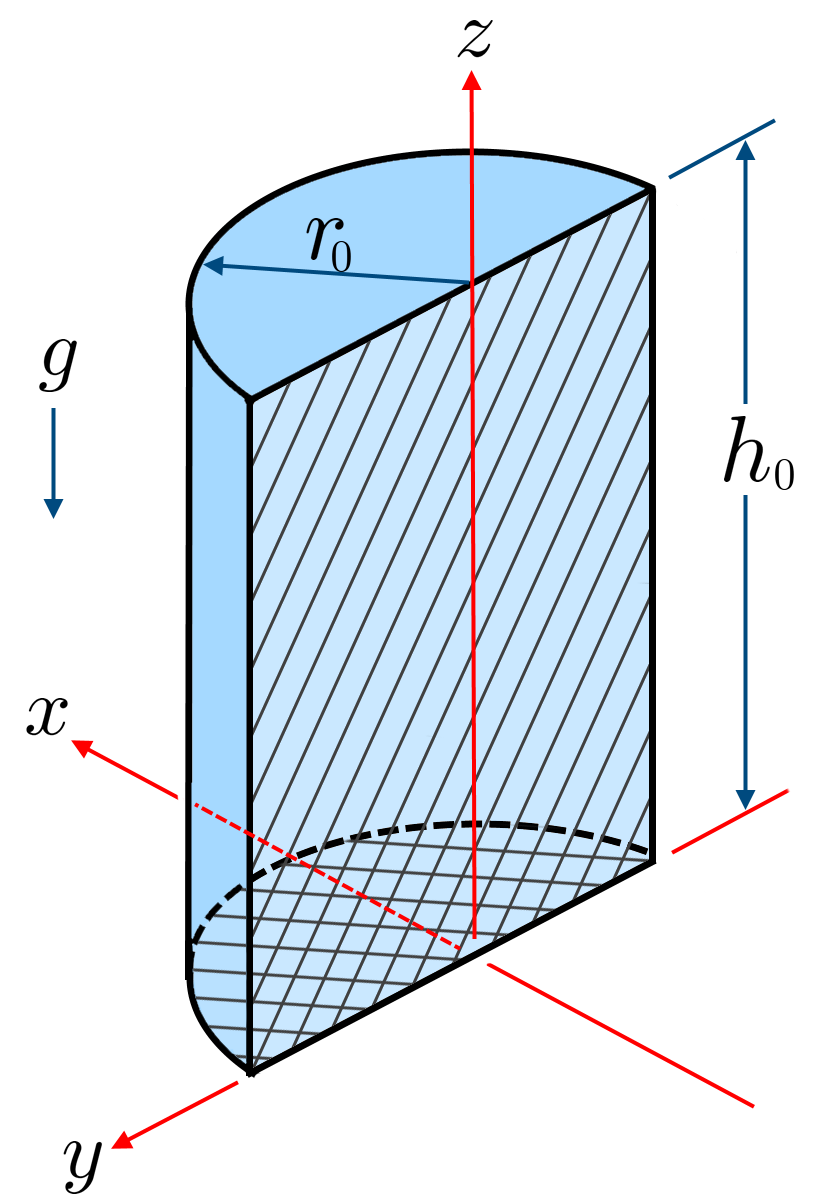}
		\caption{}
		\label{Fig:cylindricalWatterColumn_TestCase_a}
	\end{subfigure}
	~
	\hspace{10mm}
	\begin{subfigure}[b]{0.30\textwidth}		
		\includegraphics[width=0.67\linewidth]{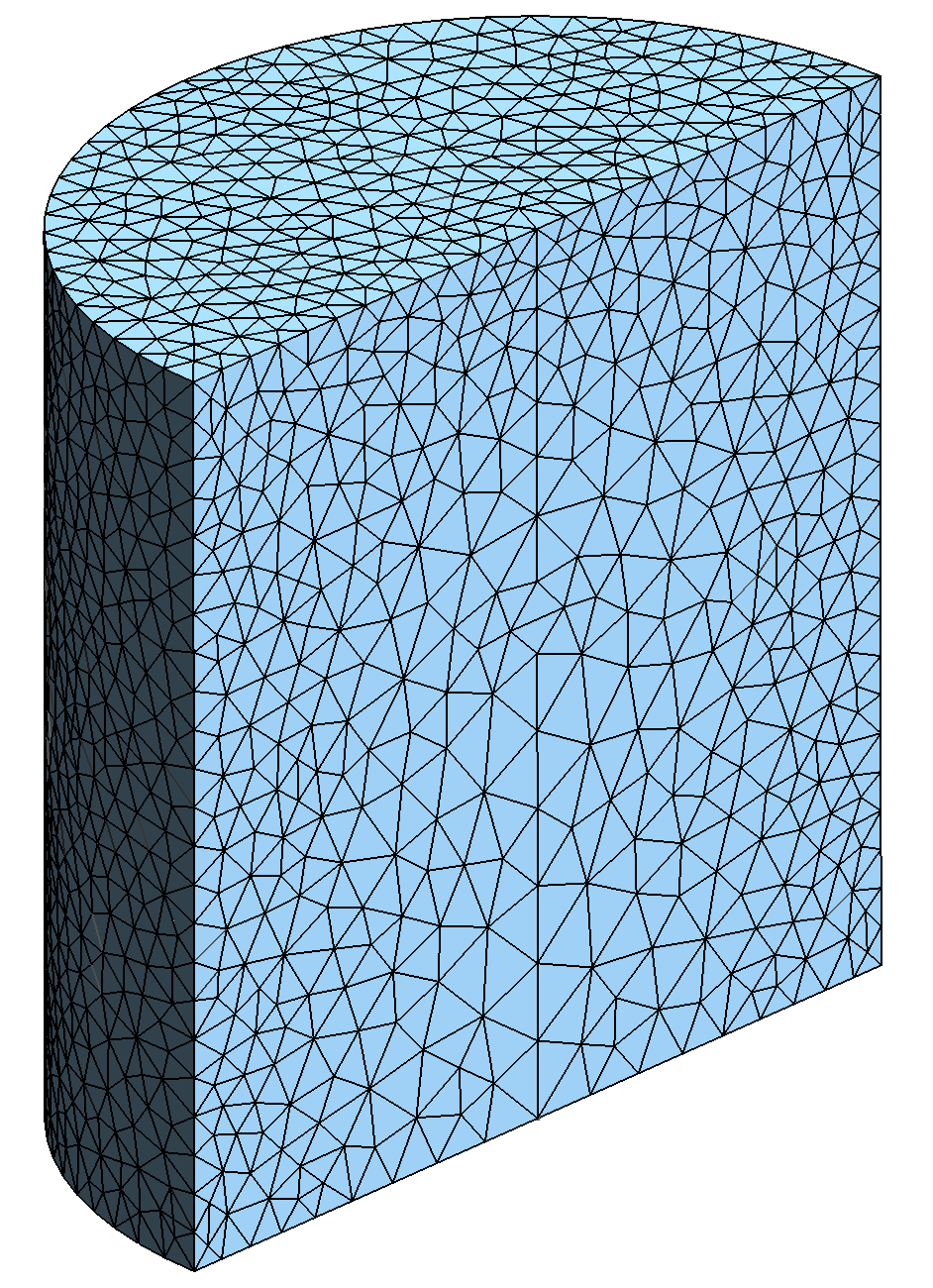}
		\caption{}
		\label{Fig:cylindricalWatterColumn_TestCase_b}
	\end{subfigure}	
	\\
	\vspace{-65mm}
	\hspace{-63mm} (\textbf{a}) \hspace{63mm} (\textbf{b})
	\vspace{53mm}	
\caption{Collapse of a cylindrical water column problem. (a) Geometry and (b) initial finite element discretization.}
\label{Fig:cylindricalWatterColumn_TestCase}
\end{figure}

Six collapse of a water column problems are solved. \review{One using a fixed time step $\Delta t = 0.00005$ and Backward Euler. The other five consider a time step set by a maximum CFL$^*$ = 0.5}. These five problems use Backward Euler, Trapezoidal, and Generalized-$\alpha$ ($\GAa$) with $\rho_\infty$ = 0.0, 0.25 and 0.50. From each problem, height ($h$) and base radius ($r_b$) of the water column versus dimensionless time ($t^*$) are reported. The latter is defined as $t^* = t\:(2g/r_0)^{0.5}$. The base radius is computed using particles located in the basal semicircle and taking the average distance from these particles to the origin [0 0 0]. The height is measured at the center of the water column (along the $z$-axis). Results are summarized in Fig.~\ref{Fig:CWC_R} along with references using 3D Eulerian-based \citep{akin2007computation,cruchaga2010surface,battaglia2010simulation}, 3D Lagrangian FEM \citep{tang2008viscous}, and 2D Lagrangian PFEM \citep{cerquaglia2019development} formulations. Legend of Fig.~\ref{Fig:CWC_R_a} is shown in Fig.~\ref{Fig:CWC_R_b}.

\begin{figure}[t] \captionsetup[sub]{font=normalsize}\captionsetup[subfigure]{labelformat=empty}
	\centering 
	\begin{subfigure}[b]{0.53\textwidth}
		\includegraphics[trim=0 0 0 0,clip,width=1.00\linewidth]{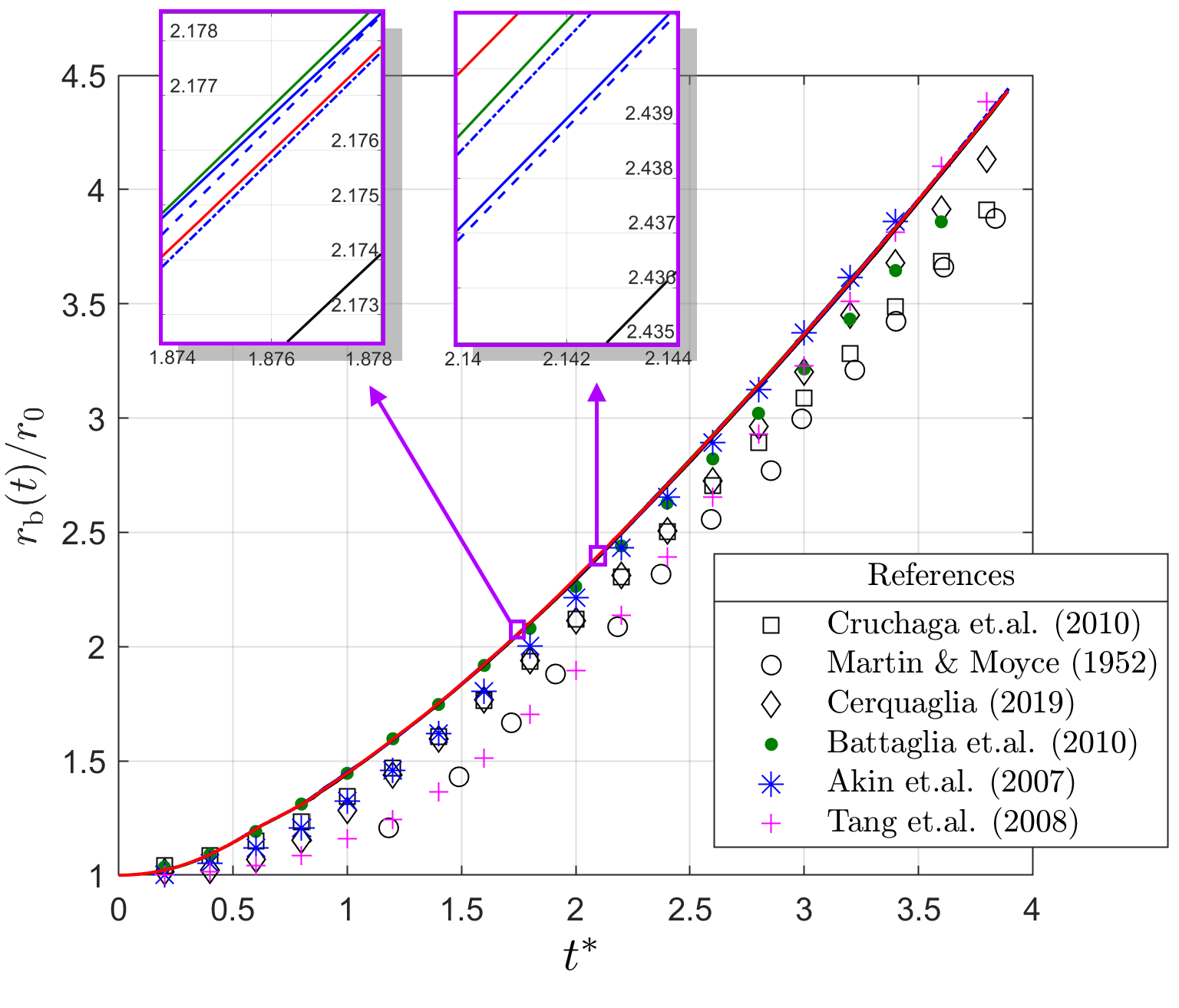}
		\caption{}
		\label{Fig:CWC_R_a}
	\end{subfigure}
	~\hspace{0mm}
	\begin{subfigure}[b]{0.43\textwidth}	
		\includegraphics[trim=0 0 0 0,clip,width=1.00\linewidth]{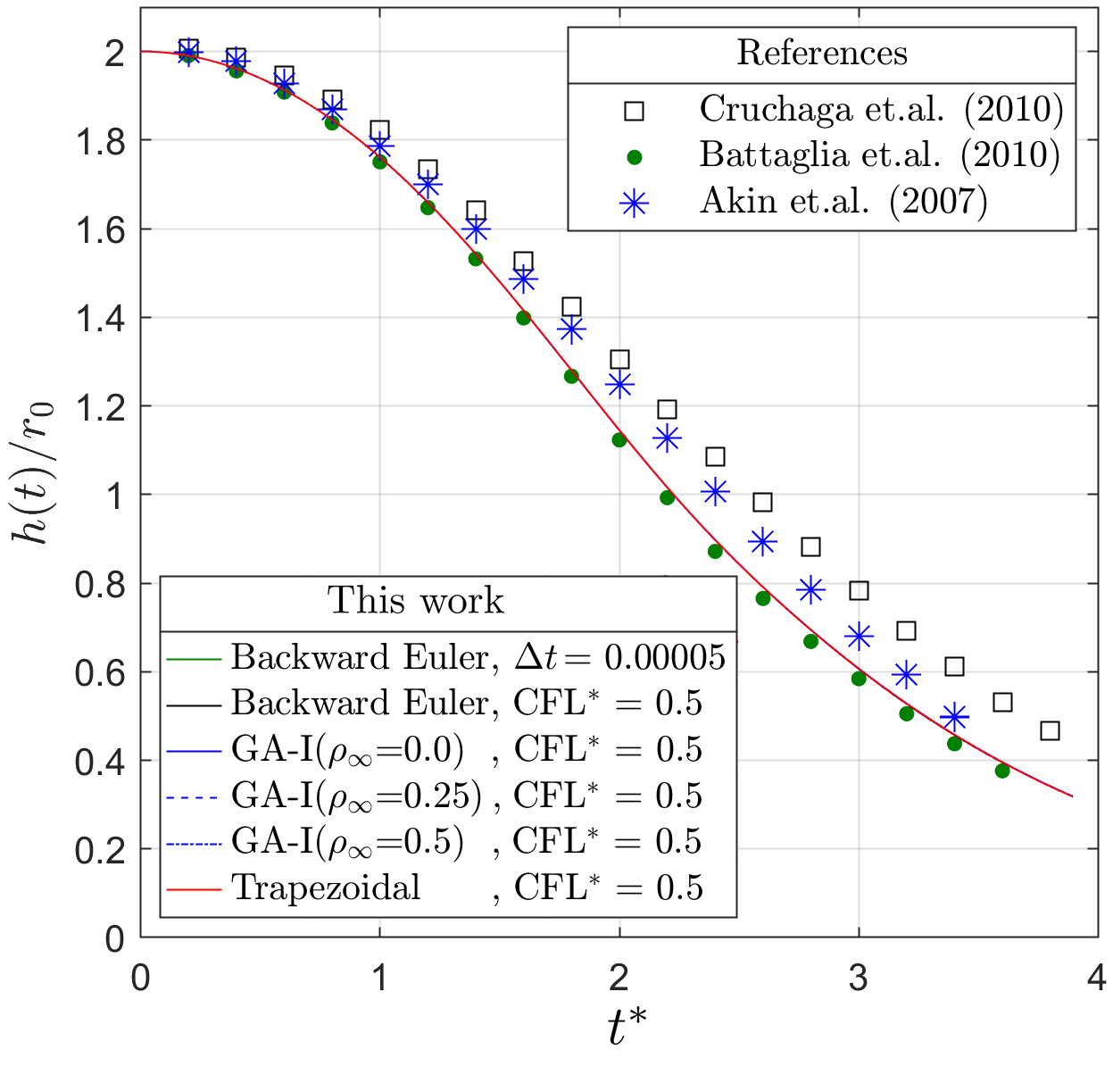}
		\caption{}
		\label{Fig:CWC_R_b}
	\end{subfigure}
	\\
	\vspace{-83mm}
	\hspace{-83mm} (\textbf{a}) \hspace{83mm} (\textbf{b})
	\vspace{68mm}
\caption{Collapse of a water cylindrical column problem. Evolution of (a) base radius $r_b$ and (b) height $h$ of the water column. The legend of curves in (a) is found in subfigure (b).}
\label{Fig:CWC_R}
\end{figure}

\review{
Fig.~\ref{Fig:CWC_R_a} shows that the experimental reference (Martin and Moyce \citep{martin1952part}) exhibits a similar pattern to that of the present work, but with the experimental points shifted to the right. The fluid front delay of Martin and Moyce has also been noted in experimental studies (e.g., see Fig.~12 in reference \citep{lobovsky2014experimental}), which suggest that the discrepancy with Martin and Moyce may be due to their experimental technique used for dam gate removal.} In particular, this work is closer to the numerical results of Battaglia et.al.\citep{battaglia2010simulation} and Akin et.al.\citep{akin2007computation} for both fluid front position ($r_b$) and height column ($h$). \review{ The small differences with respect to those references may be due to the fact that they use a formulation based on a fixed regular mesh and larger time steps than the present work (approximately one order of magnitude). Due to the chosen space discretization, we cannot use larger time steps since tetrahedral elements could be skewed after updating nodal position. Enlarging element size in favor of larger time steps would impair the geometric resolution of the fluid front, since the Alpha Shape algorithm eliminates those highly distorted elements in the fluid front. In other words, the element and time step sizes were chosen by balancing the trade-off between computation time and resolution.
}

As for the time integration schemes, no significant differences are observed. \review{At least from a kinematic point of view, results are insensitive to the time integration scheme, presumably because the problem is solved with time steps small enough to avoid numerical damping. Looking in more detail,} left and right close-up views in Fig.~\ref{Fig:CWC_R_a} show, respectively, that Backward Euler with \review{fixed and} small time step (green curve), and Trapezoidal (red curve) lead the front position ($r_b$). This alternation between curves leading the front position is seen everywhere in the graph, although Backward Euler with large time step (CFL$^*$ = 0.5, black curve) always exhibits the major delay, \review{which is a sign of numerical damping.}

As it was not possible to find references presenting pressure results for this example, it is decided to report the pressure at the origin point [x y z] = [0 0 0]. Results for each time integration scheme are shown in Figs.~\ref{Fig:CWC_p_a} and \ref{Fig:CWC_p_b}. These represent the same plot but from different perspectives. Results show that all time integration schemes deliver the same order of magnitude for the pressure, but different quantities of spurious oscillations. A simple look at Fig.~\ref{Fig:CWC_p_a} allows to classify results by the amount of spurious oscillations. The trapezoidal scheme (CFL$^*$ = 0.5) has the largest number of oscillations, which grow significantly from $t^*$ = 2. In second place is the generalized alpha (CFL$^*$ = 0.5) with $\rho_\infty$ = 0.5. Third is Backward Euler with small time step $\Delta t = 0.00005$. Note that this $\Delta t$ is much smaller than the one imposed by CFL$^*$ = 0.5, as seen in Figs.~\ref{Fig:CWC_p_c} and \ref{Fig:CWC_p_d}. In fourth place is generalized alpha (CFL$^*$ = 0.5) with $\rho_\infty$ = 0.25. Finally, schemes with the smallest amount of spurious oscillation are Backward Euler (CFL$^*$ = 0.5) and Generalized-$\alpha$ (CFL$^*$ = 0.5) with $\rho_\infty$ = 0.0. 

\begin{figure}[t] \captionsetup[sub]{font=normalsize}\captionsetup[subfigure]{labelformat=empty}
	\centering 
	\begin{subfigure}[b]{0.47\textwidth}
		\includegraphics[trim=0 0 0 0,clip,width=1.00\linewidth]{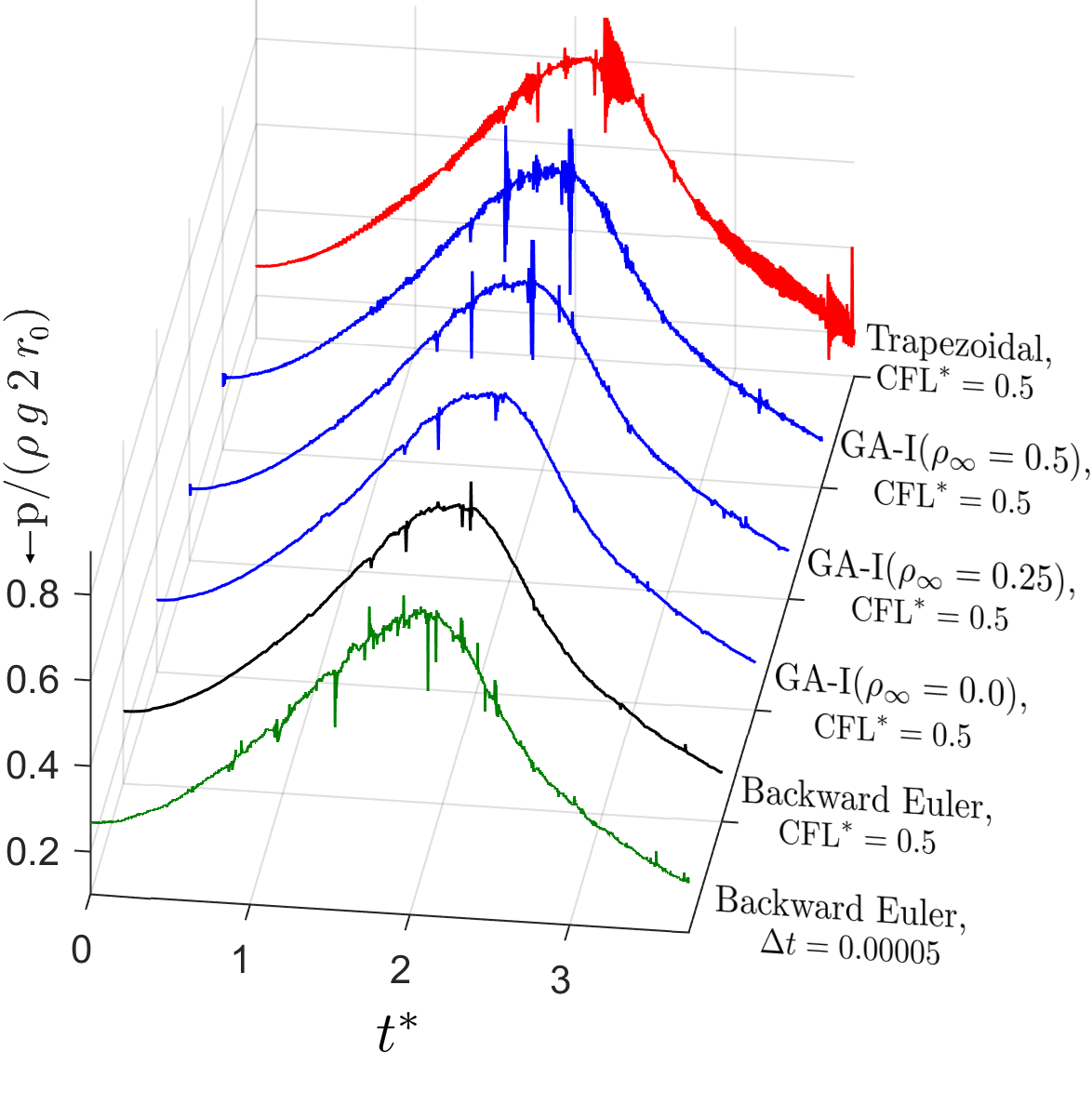}
		\caption{}
		\label{Fig:CWC_p_a}
	\end{subfigure}
	~\hspace{2mm}
	\begin{subfigure}[b]{0.47\textwidth}	
		\includegraphics[trim=0 0 32 17,clip,width=0.98\linewidth]{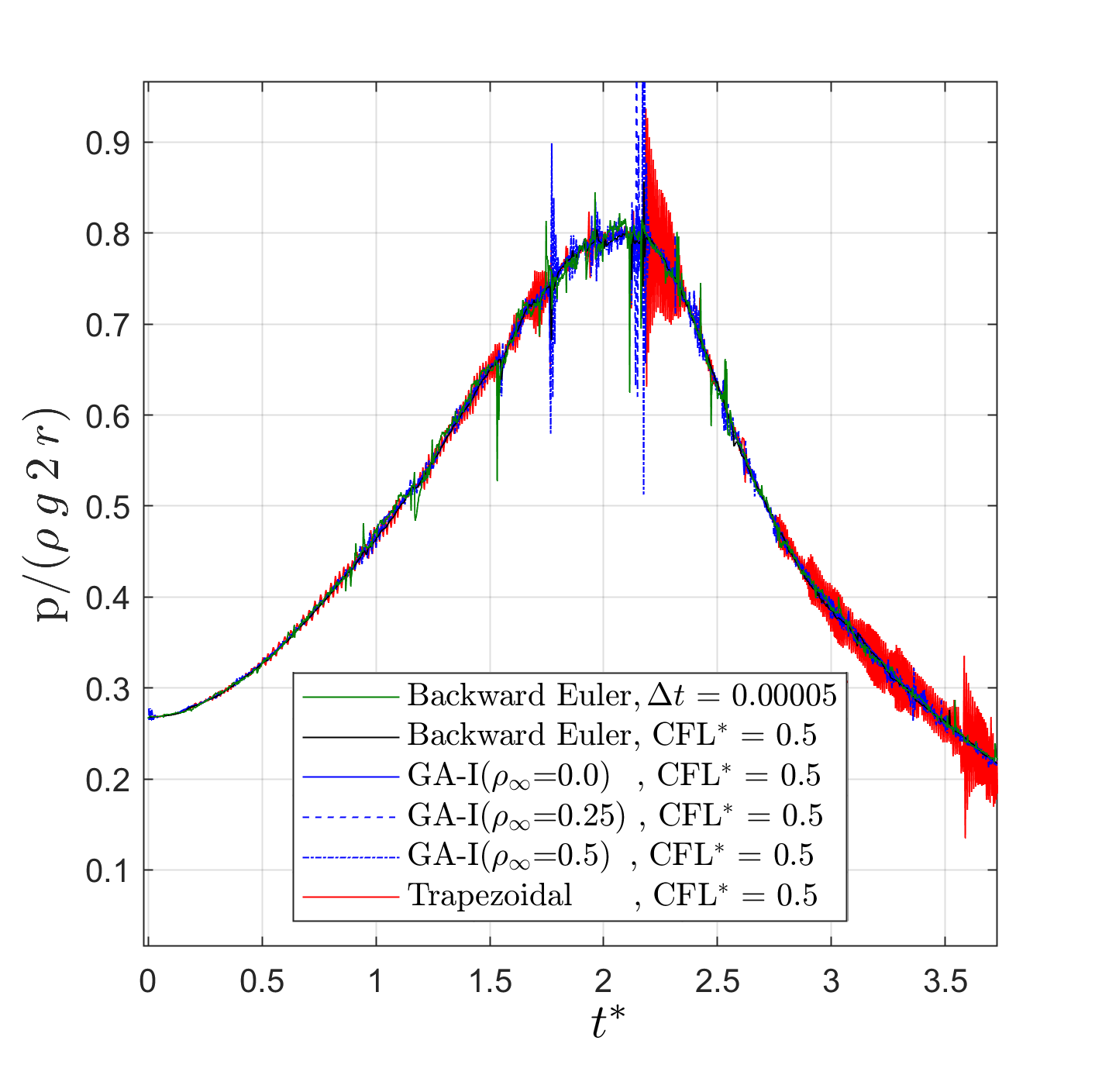}
		\caption{}
		\label{Fig:CWC_p_b}
	\end{subfigure}
	\\[-5ex]
	\begin{subfigure}[b]{0.47\textwidth}
		\includegraphics[trim=0 0 35 17,clip,width=1.00\linewidth]{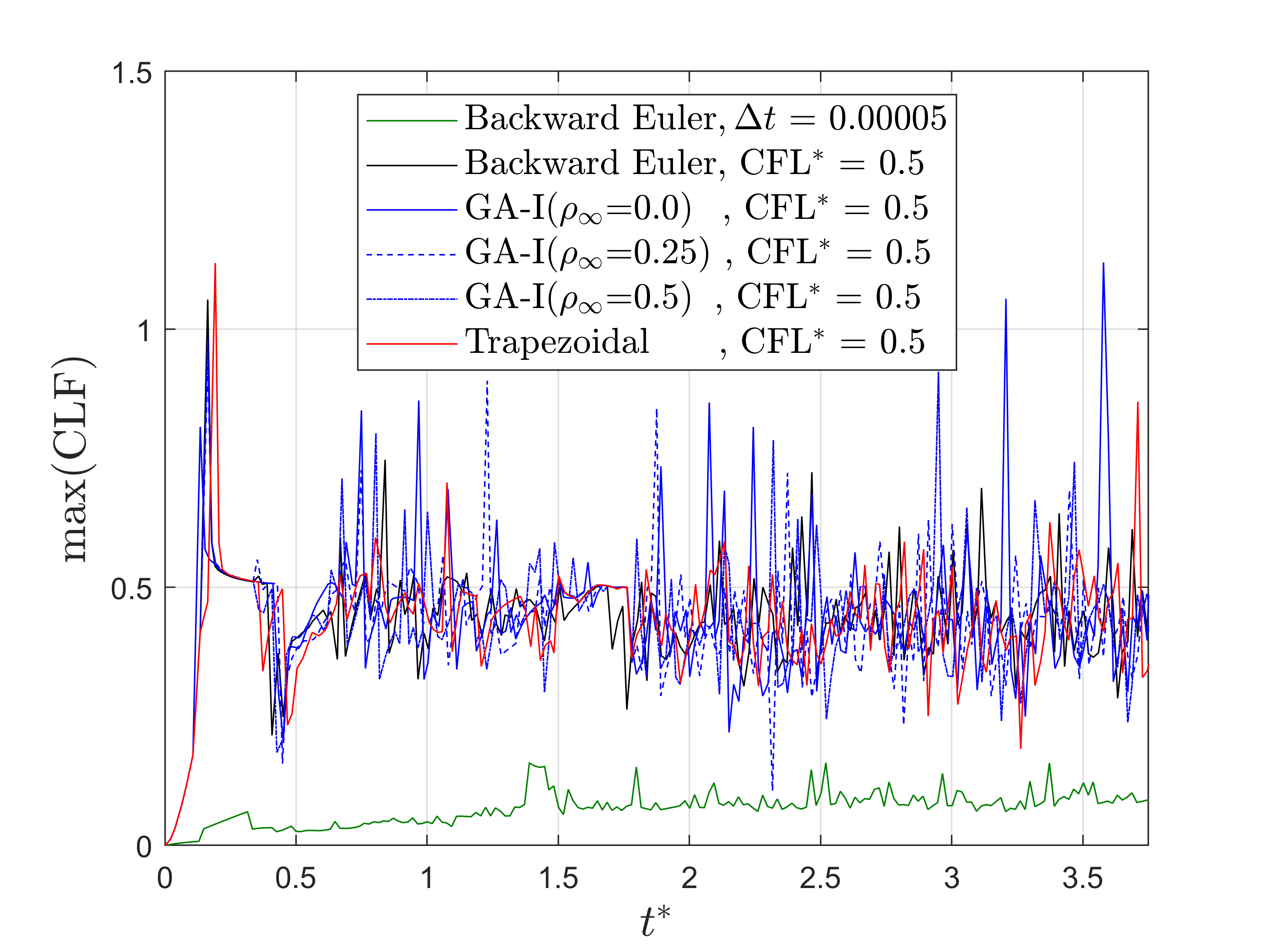}
		\caption{}
		\label{Fig:CWC_p_c}
	\end{subfigure}
	~\hspace{2mm}
	\begin{subfigure}[b]{0.47\textwidth}	
		\includegraphics[trim=0 0 35 12,clip,width=1.00\linewidth]{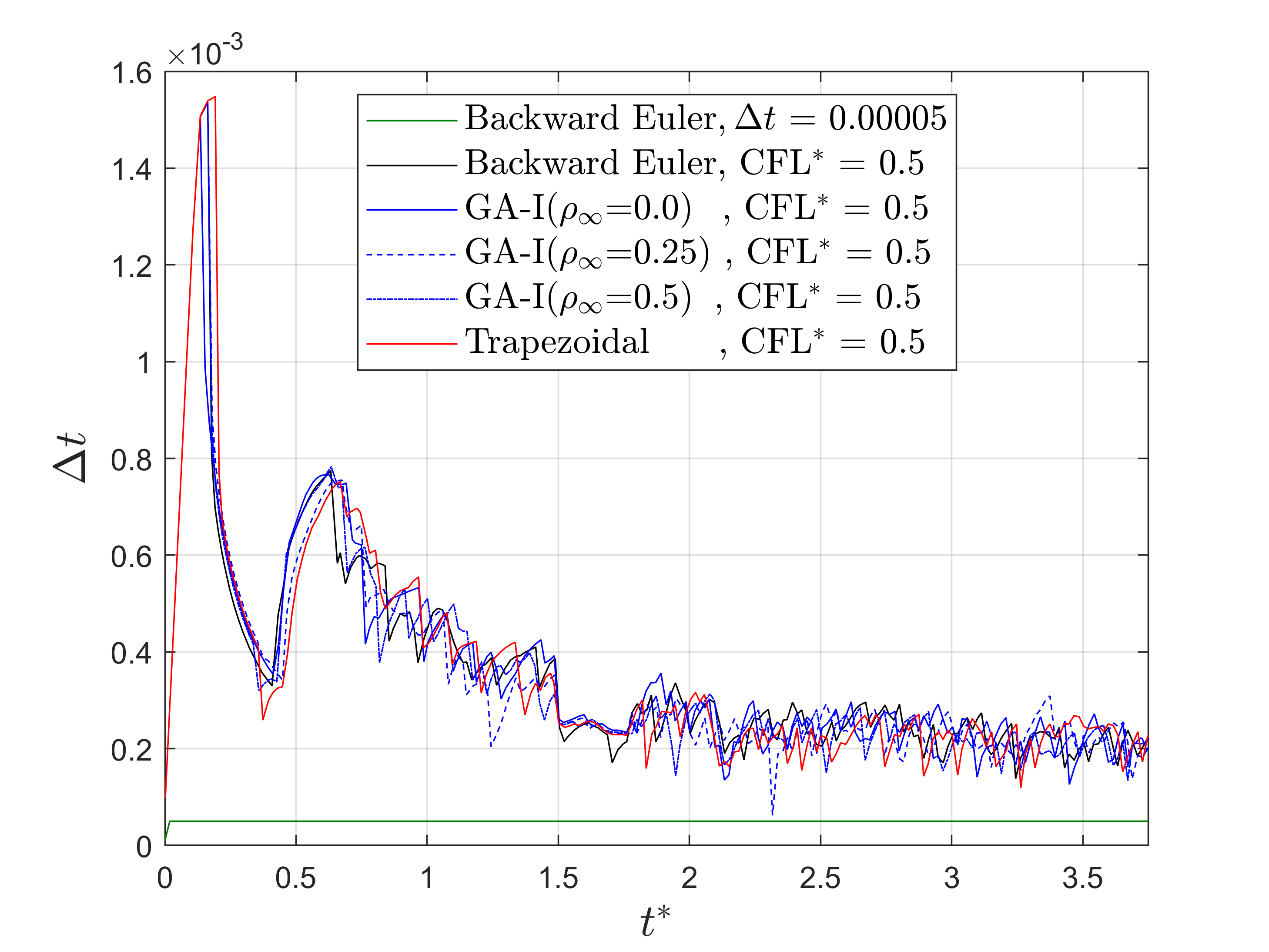}
		\caption{}
		\label{Fig:CWC_p_d}
	\end{subfigure}
	\\
	\vspace{-158mm}
	\hspace{-83mm} (\textbf{a}) \hspace{83mm} (\textbf{b})
	\vspace{81mm}
	\\
	\hspace{-83mm} (\textbf{c}) \hspace{83mm} (\textbf{d})
	\vspace{60mm}
\caption{Collapse of a cylindrical water column problem. (a) Pressure at [x y z] = [0 0 0], (b) side view of subfigure (a), (c) maximum CFL number and (d) time step $\Delta t$ during simulation.}
\label{Fig:CWC_p}
\end{figure}

Importantly, the choice of the best time integration scheme depends on a number of criteria that must be weighed by the user. For example, if high accuracy is required for fluid kinematics (position, velocity, acceleration), then the trapezoidal or Generalized-$\alpha$ scheme with $\rho_\infty \approx 1$ is suggested, as long as convergence of the nonlinear algorithm is achieved. On the other hand, if precision and stability in the pressure field is desired, for example for fluid-structure simulation, then schemes with more numerical damping are recommended such as Backward Euler and Generalized-$\alpha$ with $\rho_\infty \approx 0$. If time step size is relevant to the computational time, then Generalized-$\alpha$ with $\rho_\infty \approx 0$ is a good choice, since it does not present excessive numerical damping at large time steps in comparison to Backward Euler. In particular for the authors of this work, the recommendation is Generalized-$\alpha$ ($\GAa$) with $\rho_\infty \leq 0.5$.

\review{
The last remark concerns the computation time. Notably, the Generalized-$\alpha$ method allows using larger time steps than Backward Euler or Newmark and still achieve the same order of error. The ratio of time step sizes between time integration schemes depends on each problem. For instance, the reader can verify from the flow around a cylinder problem that the Generalized-$\alpha$ reaches the same accuracy than Backward Euler, even with a $\Delta t$ that is 32 times larger than that used with Backward Euler. Therefore, a reduction in computation time is expected with Generalized-$\alpha$ due to the use of larger time steps. 

To provide the reader with an approximate computation cost of our Generalized-$\alpha$ implementation, the computation time of three collapse of cylindrical water column problems are reported below. The problems consider the Backward Euler (BE), $\GAa$, and $\GAc$ as time integration schemes. All three problems use the same time step $\Delta t = 0.00005 s$.

The average computation times are compared in Fig.~\ref{Fig:CWC_time_a}. These are computed as the total computation time divided by the total amount of iterations of the nonlinear algorithm (time / $n$). This average computation time is normalized by that obtained with Backward Euler (time$_{BE}$ / $n_{BE}$). Thus, Fig.~\ref{Fig:CWC_time_a} shows that the three problems take the same computation time per iteration of the nonlinear solver. However, the nonlinear solver requires more iterations in the GA schemes (about 7$\%$ more iterations), which is reflected in the total computation time shown in Fig.~\ref{Fig:CWC_time_b}. The increased computation time in problems using GA-I and GA-II is due to the Finite Element Analysis (FEA), as depicted in Fig.~\ref{Fig:CWC_time_c}. On the contrary, the time spent in remeshing and other tasks (e.g.~data input/output) is the same as that required by Backward Euler. This is reasonable since all three problems use the same time step size, so the amount of remeshing (and other tasks per time step) are the same. 

Fig.~\ref{Fig:CWC_time_d} shows the time spent in different tasks of the FEA code. There, the time to build the system of equations (matrix $\mathbf{A}$ and vector $\mathbf{b}$) is slightly longer in problems using Generalized-$\alpha$. This is because GA-I and GA-II present more terms in the governing equations, and because there are about 7$\%$ more system builds in problems using Generalized-$\alpha$ schemes. However, most of the computation time is spent in solving the system $\mathbf{A} \mathbf{q} = \mathbf{b}$, which is a well known drawback of monolithic formulations.

In summary, for the same set of simulation parameters, there is no significant difference in computation time and memory usage between the Backward Euler and Generalized-$\alpha$, because the bottleneck is the routine that solves the system $\mathbf{A} \mathbf{q} = \mathbf{b}$. This observation also holds for the 2D problems reported in this work. 
}

\begin{figure}[t] \captionsetup[sub]{font=normalsize}\captionsetup[subfigure]{labelformat=empty}
	\centering 
	\begin{subfigure}[b]{0.229\textwidth}
		\includegraphics[width=1.00\linewidth]{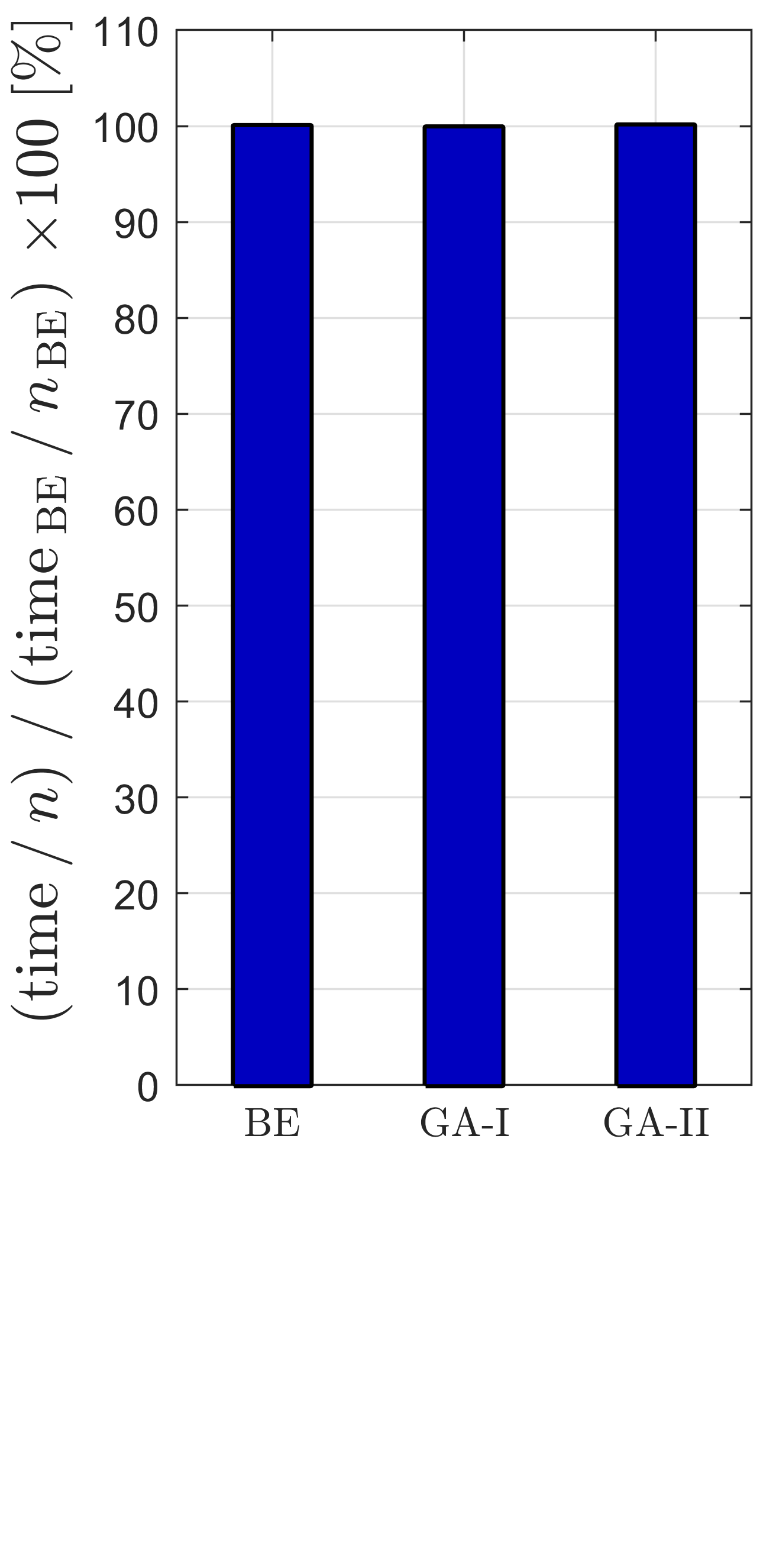}
		\caption{}
		\label{Fig:CWC_time_a}
	\end{subfigure}
	~
	\begin{subfigure}[b]{0.67\textwidth}	
		\includegraphics[width=1.00\linewidth]{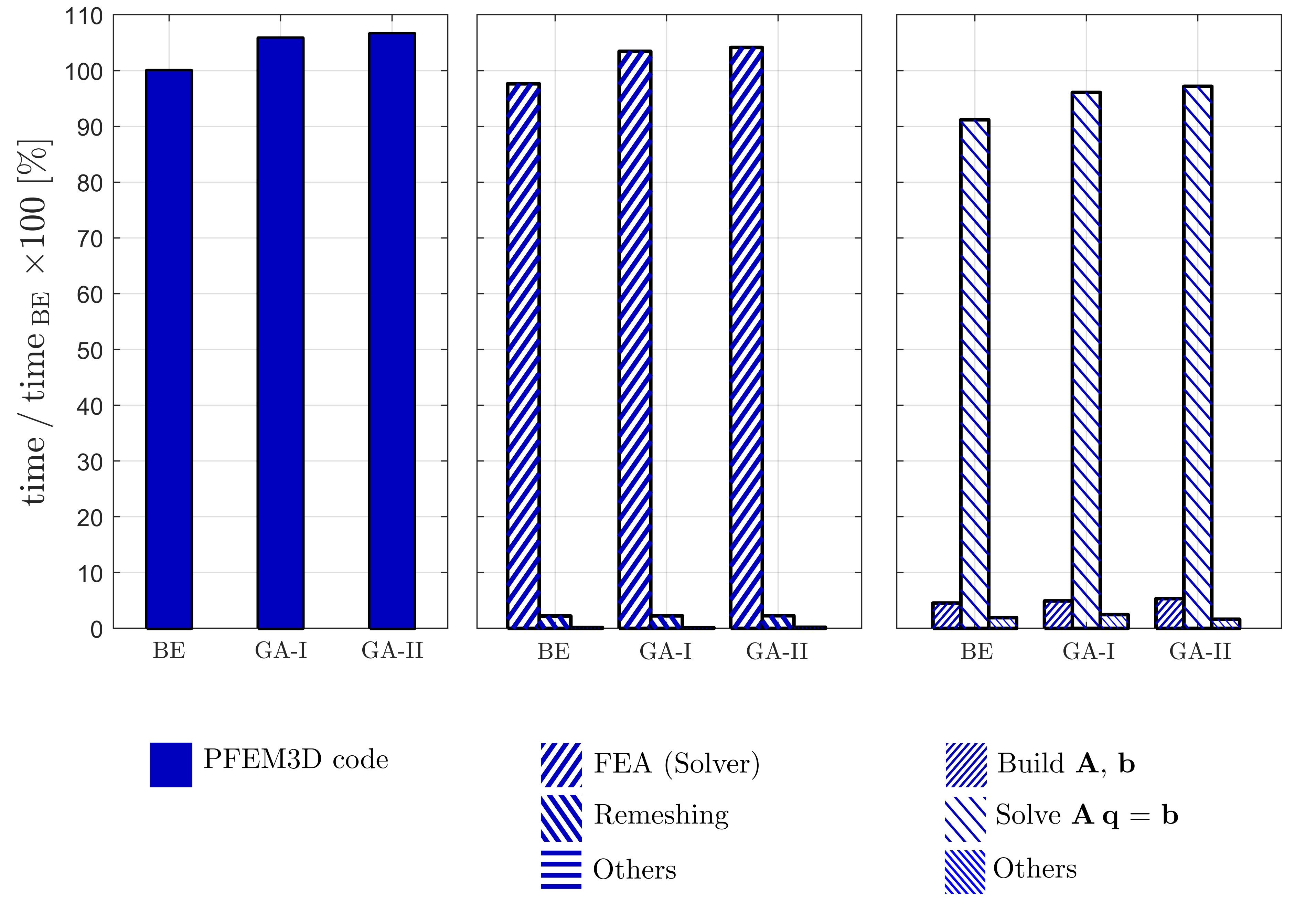}
		\caption{}
		\label{Fig:CWC_time_b}
	\end{subfigure}
	~
	\begin{subfigure}[b]{0.0\textwidth}	
		\caption{}
		\label{Fig:CWC_time_c}
	\end{subfigure}
	~
	\begin{subfigure}[b]{0.0\textwidth}	
		\caption{}
		\label{Fig:CWC_time_d}
	\end{subfigure}
	\\
	\vspace{-30mm}
	\hspace{-2mm} (\textbf{a}) Time per iteration \hspace{16mm} (\textbf{b}) Total time \hspace{13mm} (\textbf{c}) Time in PFEM code \hspace{2mm} (\textbf{d}) Time in FEA code 
	\vspace{14mm}
\caption{\review{Breakdown of the computation time in the collapse of a water column problem. Three time integration schemes are considered: Backward Euler (BE), $\GAa$ and $\GAc$. Graphs are scaled with respect to the total time of Backward Euler, denoted as $\mathrm{time}_\mathrm{BE}$. In (a), $n$ denotes the total amount of iterations carried out by the nonlinear solver. In (d), the task \textit{Build} represents the computation of triplets ($i$, $j$, A$_{i,j}$), doublets ($j$, b$_j$), and matrix/vector assembly; \textit{Solve} involves a pattern analysis of matrix $\mathbf{A}$, an LU factorization of $\mathbf{A}$, and solving a linear system; and \textit{Others} consists of updating nodal state variables according to the time integration scheme, and verification of convergence criteria.}}
\label{Fig:CWC_time}
\end{figure}

\section{Conclusions}\label{sec:5}

This work incorporates the Generalized-$\alpha$ time integration scheme into the Particle Finite Element Method (PFEM) to solve the incompressible Navier-Stokes equations in a Lagrangian framework. The discretized system of equations is derived for both velocity-pressure and displacement-pressure formulations. The time integration scheme is compared with the Backward Euler and Newmark, which are well established in the PFEM literature. Numerical comparisons of several benchmark problems and a literature review of the Generalized-$\alpha$ method for computational fluid mechanics lead to the following conclusions:

\begin{itemize}
\item The Generalized-$\alpha$ method in PFEM outperforms the classical Backward Euler and Newmark schemes, since it exhibits less numerical damping at large time steps and less spurious oscillations than the Trapezoidal rule. These observations are in line with studies carried out in Eulerian-based frameworks \citep{dettmer2003analysis}.

\item The literature presents different implementation approaches of the Generalized-$\alpha$ method. A first classification can be made based on the assumption for stating the momentum equation at the intermediate time $t_{n+\alpha}$. One approach (GA-I) assumes that state variables scale linearly between $t_n$ and $t_{n+1}$, while the other (GA-II) assumes linear scaling for the forces in the momentum equation. These two approaches produce similar results in PFEM, especially if \review{linear problems are considered}, or small time steps are used. However, the approach that scales forces (GA-II) demands more matrix operations in our PFEM implementation. Therefore, the first implementation (GA-I) approach is preferred.

\item A second classification of implementation approaches considers the way in which the linear scaling of state variables is used in the momentum equation. One approach solves the system of equations for state variables at $t_{n+\alpha_f}$ and then uses the linear interpolation assumption to recover state variables at $t_{n+1}$. The other approach substitutes the linear interpolation equations into the momentum equation to write unknown states at $t_{n+1}$. Both approaches lead to nearly identical results, but the first scheme is preferred in this work since the algebraic substitution introduces additional terms in the momentum equation, and hence demands more matrix operations.

\item A third classification refers to the pressure. Some authors write the momentum equation with pressure at $t_{n+1}$  instead of $t_{n+\alpha_f}$. This modified scheme results in a degradation of the Generalized-$\alpha$ method, as found by Liu et.al.\citep{liu2021note}. Although two cases are distinguished here. The first case integrates the discrete gradient matrix in the intermediate configuration $\Omega$($t_{n+\alpha_f}$). In this case, an inconsistency of the time integration scheme is observed for the pressure. In the flow around a cylinder, implementation approaches GA-I and GA-II differ, and the drag coefficient becomes highly dependent on the spectral radius. The second case integrates the gradient matrix in the current configuration $\Omega$($t_{n+1}$), which results in a high degradation of the time integration scheme in terms of pressure and velocity.

\item All the implementation approaches mentioned above are equivalent if the spectral radius is set to 0.0. Such a case yields to a highly damped system for high frequencies but still results in a time integration scheme significantly superior to Backward Euler. Moreover, it is much simpler to implement than a generalized code script containing $\alpha_m$ and $\alpha_f$, since $\rho_\infty = 0.0$ yields $\alpha_f = 1.0$, thus it avoids code routines to compute equations at the intermediate time $t_{n+\alpha_f}$. 

\item Numerical examples of this work do not reveal significant differences in accuracy between the velocity-based and displacement-based formulations. The latter does present more terms in the system of equations, however, no perceptible increase in computational time is observed in the academic examples of this manuscript. Also, the displacement-based formulation draws the same conclusions mentioned for the velocity-based formulation.

\end{itemize}

\review{
This work focused on four benchmark problems widely reported in the literature, which allowed validation and analysis of the presented methods. A perspective of this work is to consider more complex problems, for example, those that present wave breaking, splashing and fluid-structure interaction, among others. In addition, it would be interesting to analyse the effect of remeshing parameters on fluid kinematics. For example, the effect of the Alpha-Shape algorithm on the temporal integration and positioning of the free surface. These and other relevant issues will be the subject of future studies. 
}

\subsection*{Financial disclosure}
This work was supported by the ALFEWELD project --- \textit{Amélioration et modélisation du FMB (Friction Melt Bonding) pour le soudage par recouvrement de l’aluminium et de l’acier} --- (convention 1710162) funded by the WALInnov program of the Walloon Region of Belgium.

\subsection*{Conflict of interest}
The authors declare no potential conflict of interests.

\vspace{-5mm}
\appendix

\clearpage
\section{: Momentum equations for Velocity-Pressure formulation}\label{TAB:Velocity_Pressure}

\checkoddpage
\ifoddpage
\begin{minipage}{\textwidth}
  \includepdf[pages=2,clip,trim=3.5cm 0cm 2cm 1.6cm,scale=0.91,offset=-41 -35]{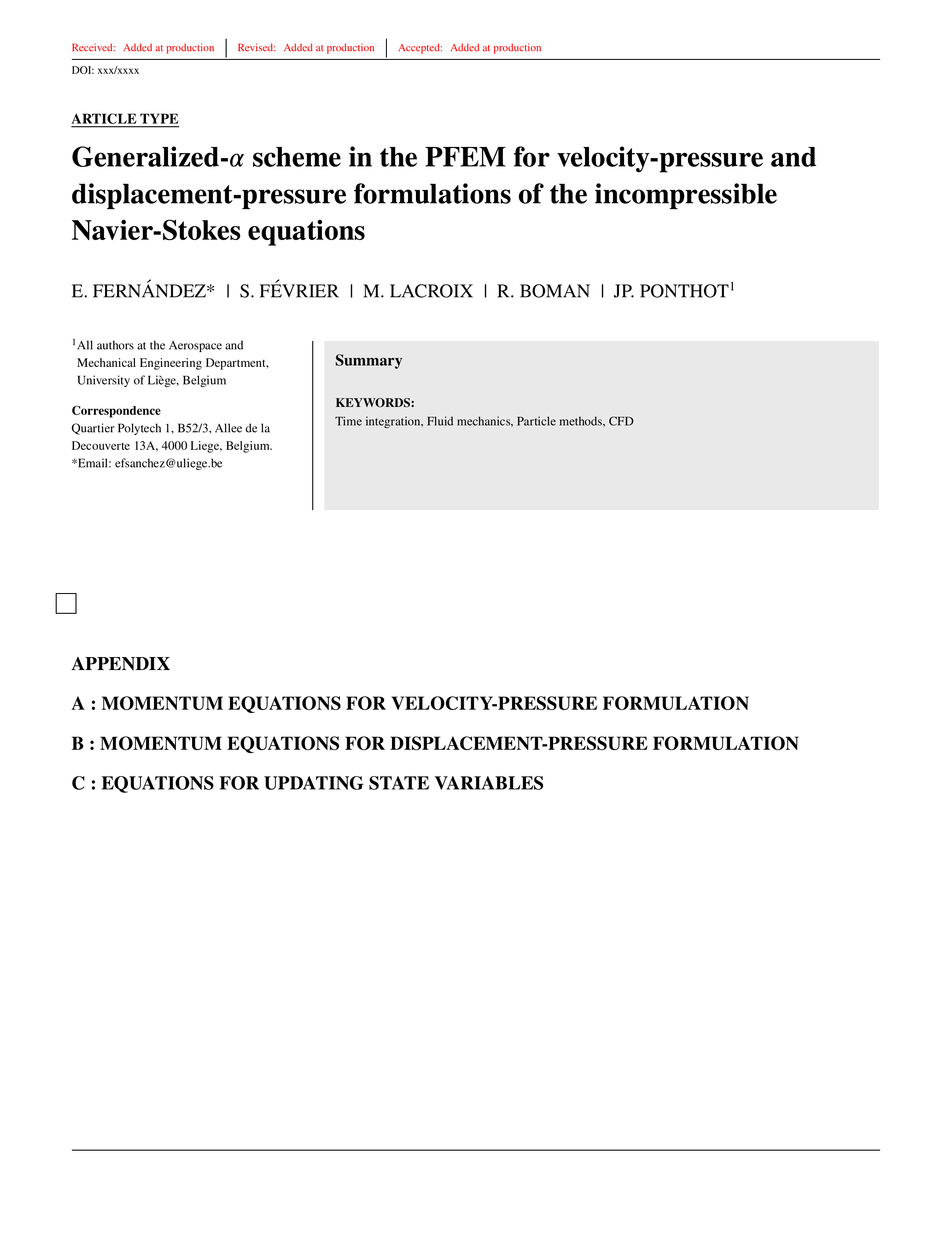}
\end{minipage}
\else
\begin{minipage}{\textwidth}
  \includepdf[pages=2,clip,trim=3.5cm 0cm 2cm 1.6cm,scale=0.91,offset=43 -35]{Tables/Tables_VP_DP_Update.pdf}
\end{minipage}
\fi

\clearpage
\section{: Momentum equations for Displacement-Pressure formulation}\label{TAB:Displacement_Pressure}

\checkoddpage
\ifoddpage
\begin{minipage}{\textwidth}
  \includepdf[pages=3,clip,trim=4.0cm 0cm 2cm 1.6cm,scale=0.9,offset=-47 -32]{Tables/Tables_VP_DP_Update.pdf}
\end{minipage}
\else
\begin{minipage}{\textwidth}
  \includepdf[pages=3,clip,trim=4.0cm 0cm 2cm 1.6cm,scale=0.9,offset=47 -32]{Tables/Tables_VP_DP_Update.pdf}
\end{minipage}
\fi

\clearpage
\section{: Equations for updating state variables}\label{TAB:Update_States}

\checkoddpage
\ifoddpage
\begin{minipage}{\textwidth}  
	\includepdf[pages=4,clip,trim=0cm 0cm 0cm 1.6cm,scale=0.95,offset=-16 -28]{Tables/Tables_VP_DP_Update.pdf}
\end{minipage}
\else
\begin{minipage}{\textwidth}  
	\includepdf[pages=4,clip,trim=0cm 0cm 0cm 1.6cm,scale=0.95,offset=16 -28]{Tables/Tables_VP_DP_Update.pdf}
\end{minipage}
\fi

\clearpage
\bibliography{wileyNJD-AMA}%

\end{document}